\documentclass[11pt,a4paper,preprintnumbers]{article}
\pdfoutput=1 
\usepackage{jheppub, hyperref, cleveref}
\usepackage{amsfonts, amsmath, amssymb, latexsym}
\usepackage{graphicx, caption, subcaption}
\usepackage{enumitem}
\usepackage{colortbl,bm}

\usepackage[normalem]{ulem}
\usepackage[dvipsnames]{xcolor}
\definecolor{alizarin}{rgb}{0.82, 0.1, 0.26}
\newcommand{\GWB}{SGWB}
\newcommand{\GWBs}{SGWBs}
\newcommand{\Om}[1]{{\Omega_{\rm GW}} }
\newcommand{\seCL}{68 \% CL }
\newcommand{\nfCL}{95 \% CL }
\newcommand{\snCL}{68 and 95 \% CL }

\preprint{CERN-TH-2024-072}

\title{Gravitational waves from inflation in LISA: reconstruction pipeline and physics interpretation}

\author[1, a]{Matteo Braglia\note{Corresponding author:  \href{mailto:mb9289@nyu.edu}{mb9289@nyu.edu}}}
\affiliation[a]{Center for Cosmology and Particle Physics, New York University, 726 Broadway, New York, NY 10003, USA}

\author[b]{\!\!, Gianluca Calcagni}
\affiliation[b]{Instituto de Estructura de la Materia, CSIC, Serrano 121, 28006 Madrid, Spain}

\author[c]{\!\!, Gabriele Franciolini}
\affiliation[c]{Theoretical Physics Department, CERN, 1211 Geneva 23, Switzerland}

\author[2, d]{\!\!, Jacopo Fumagalli\note{Corresponding author:  \href{jfumagalli@fqa.ub.edu}{jfumagalli@fqa.ub.edu}}}
\affiliation[d]{Departament de F\'isica Quàntica i Astrofísica i 
Institut de Ciències del Cosmos,
Universitat de Barcelona, Martí i Franquès 1, 08028 Barcelona, Spain}

\author[3,e]{\!\!, Germano Nardini\note{Project coordinator: \href{mailto:germano.nardini@uis.no}{germano.nardini@uis.no}}}
\affiliation[h]{Department of Mathematics and Physics, University of Stavanger, NO-4036 Stavanger, Norway}

\author[4,f,g]{\!\!, Marco Peloso\note{Project coordinator: \href{mailto:marco.peloso@pd.infn.it}{marco.peloso@pd.infn.it}}}
\affiliation[f]{Dipartimento di Fisica e Astronomia “Galileo Galilei”, Universit\`a di Padova, 35131 Padova, Italy}
\affiliation[g]{INFN, Sezione di Padova, 35131 Padova, Italy}

\author[5, c]{\!\!, Mauro Pieroni\note{Corresponding author:  \href{mauro.pieroni@cern.ch}{mauro.pieroni@cern.ch}}}

\author[h]{\!\!, S\'ebastien Renaux-Petel}
\affiliation[h]{Institut d’Astrophysique de Paris, UMR 7095 du CNRS et de Sorbonne Universit\'e, 98 bis bd Arago, 75014 Paris, France}

\author[i,j,f]{\!\!, Angelo Ricciardone}
\affiliation[i]{Dipartimento di Fisica ``Enrico Fermi'', Universit\`a di Pisa, Largo Bruno Pontecorvo 3, Pisa I-56127, Italy}
\affiliation[j]{INFN, Sezione di Pisa, Largo Bruno Pontecorvo 3, Pisa I-56127, Italy}

\author[l,m]{\!\!, Gianmassimo Tasinato}
\affiliation[l]{Physics Department, Swansea University, SA28PP, United Kingdom}
\affiliation[m]{Dipartimento di Fisica e Astronomia, Universit\`a di Bologna, and \\
INFN, Sezione di Bologna, I.S.~FLAG, viale B.~Pichat 6/2, 40127 Bologna, Italy}

\author[f,g,n]{\!\!, Ville~Vaskonen}
\affiliation[n]{Keemilise ja bioloogilise f\"u\"usika instituut, R\"avala pst. 10, 10143 Tallinn, Estonia}

\author[]{\\ \centering \texttt{(For the LISA Cosmology Working Group)}}

\abstract{Various scenarios of cosmic inflation enhance the amplitude of the stochastic gravitational wave background (SGWB)  at frequencies detectable by the LISA detector.  We develop tools for a template-based analysis of the SGWB and introduce a template databank to describe well-motivated signals from inflation, prototype their template-based searches, and forecast their reconstruction with LISA. Specifically, we classify seven templates based on their signal frequency shape, and we identify 
representative fundamental physics models leading to them. By running a template-based analysis, we forecast the accuracy with which LISA can reconstruct the template parameters of representative benchmark signals, with and without galactic and extragalactic foregrounds. We identify the parameter regions that can be probed by LISA within each template. Finally, we investigate how our signal reconstructions shed light on fundamental physics models of inflation: we  discuss their impact for measurements of  \emph{e.g.,} ~the couplings of inflationary axions  to gauge fields;  the graviton mass during inflation;  the fluctuation seeds of primordial 
black holes;  the consequences of excited states  during  inflation, and the presence of small-scale spectral features.  
}

\begin{document}
\begin{figure}
\begin{flushright}
\href{https://lisa.pages.in2p3.fr/consortium-userguide/wg_cosmo.html}{\includegraphics[width = 0.2 \textwidth]{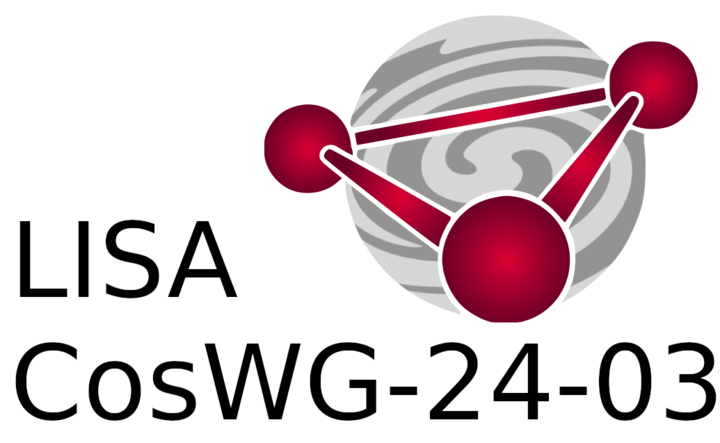}}\\[5mm]
\end{flushright}
\end{figure}

\maketitle

\section{Introduction}
\label{sec:intro}
The Laser Interferometer Space Antenna (LISA)~\cite{LISA:2017pwj} is a planned space-based gravitational wave (GW) detector designed for the mHz frequency band. This mission, led by the European Space Agency (ESA), involves ESA member countries and substantial contributions from NASA and other international space agencies. The mission was adopted in 2024. 
During this phase, established Figures of Merit 
 bind the LISA sensitivity and ensure its main mission objectives~\cite{Colpi:2024xhw}.  With fewer variables affecting mission performance, the scientific community can design and adjust the data analysis pipelines of signal searches and science interpretations toward the established experimental configuration. The core of the adjustments should be finalized over the next six years or so. This in fact provides ample time for ESA to validate and properly code the analyses, before the planned launch in the mid 2030s.

The search for the cosmological Stochastic Gravitational Wave Background (\GWB) poses a major challenge for LISA due to the similarities between its statistical properties and those of the instrumental noise and of unresolved astrophysical events. Therefore, it will be challenging to extract and characterise the contribution of these different components from the data stream. This problem is further complicated by the fact that there is no unique expectation for the cosmological \GWB, neither concerning its amplitude  nor its spectral (frequency) shape, as there are several, non mutually exclusive, potential phenomena in the early universe that could generate it. However, there is a vast amount of literature on possible cosmological GW signals, and one can then envisage classifying and compiling them into a list of signal predictions to search for.

If the primordial \GWB\ signal is not sufficiently strong to dominate over most of the transient events (see \emph{e.g.,} ~ref.~\cite{Braglia:2024siw} for a study of the strong background case), its reconstruction at LISA is planned to be achieved via a ``global fit'', where all sources are simultaneously reconstructed in an iterative manner~\cite{Cornish:2005qw, Vallisneri:2008ye, MockLISADataChallengeTaskForce:2009wir, Littenberg:2023xpl}. Based on current estimates, the global fit is too computationally expensive to be repeated for the whole list of primordial \GWB\ templates~\cite{Littenberg:2023xpl}. A more feasible approach is to consider a global fit where the primordial \GWB\ is firstly isolated with an approximate template-free approach (see \emph{e.g.,} ~refs.~\cite{Caprini:2019pxz, Flauger:2020qyi, Karnesis:2019mph}), and then use this reconstruction to shortlist the theoretically-motivated templates (\emph{i.e.,}~parameterisation of the expected \GWB\ power spectra) that best suit the reconstructed signal. When several  theoretically well-motivated templates exhibit minor differences from each other, it is more efficient to group them into approximate templates, and to introduce an intermediate step between the template-free analysis and the refined theory-based template analysis.  At this point, pursuing the global fit with a few branches of the cosmological \GWB\ templates becomes affordable. In light of the above reasoning, a comprehensive list of cosmological \GWB\ templates is required. 

The present work is part of a series of three papers that have the purpose to start the collection of this `template \GWB\ databank'. While in the works \cite{Caprini:2024hue, Blanco-Pillado:2024aca}, templates motivated by, respectively, phase transitions and topological defects are considered, here we propose and study templates that are motivated by broad classes of \GWB\ production mechanisms associated with primordial inflation. Inflation \cite{Guth:1980zm,Starobinsky:1980te,Linde:1981mu,Albrecht:1982wi} solves several shortcomings in cosmology, and can naturally produce primordial fluctuations with properties in agreement with observations \cite{Mukhanov:1981xt,Hawking:1982cz,Starobinsky:1982ee,Guth:1982ec,Bardeen:1983qw}. However the precise inflationary model, and even the energy scale at which inflation took place, are still unknown. The most minimal models 
 of inflation predict a slightly red-tilted \GWB\ spectrum, and that, once the cosmic microwave background (CMB) bounds are taken into account, the amplitude of this signal is too low to be observed at LISA (as well as all current and next generation GW detectors). However, several well motivated inflationary models have been proposed that can lead to an observable signal; see \emph{e.g.,} ~refs.~\cite{Bartolo:2016ami, LISACosmologyWorkingGroup:2022jok} for an extensive list of references. In this work we present a set of templates that can mimic a large number of these proposed signals, motivated by theoretical considerations on existing inflationary scenarios. For each of these templates, we build a dedicated LISA data analysis pipeline, reconstructing the cosmological \GWB\ signal as well as the expected astrophysical foregrounds and instrumental noise. With this pipeline, we forecast the accuracy at which LISA will reconstruct the parameters characterising each template,  if the signal is drawn from it. For some illustrative cases, we show how the reconstruction can shed light on the fundamental parameters of such inflationary setups.

Due to the nature of the global fit, definitive conclusions on the LISA capabilities for the detection and reconstruction of the cosmological \GWB\ require  data analysis pipelines for all possible LISA sources. In particular, quantifying such capabilities requires a global-fit pipeline exploiting all possible means to isolate the cosmological \GWB\ from the other sources, \emph{e.g.,} ~tools based on its statistical properties (as for instance Gaussianity and stationarity), anisotropic features \cite{LISACosmologyWorkingGroup:2022kbp}, or polarisation \cite{Domcke:2019zls}. There is no doubt that the pipeline presented in this paper is just one of the first steps towards the final result. It is however sufficient for our main purposes which are:
\begin{enumerate}
    \item To initiate a LISA template bank of well-motivated \GWB\ models justified by inflationary mechanisms.
    \item To quantify the ballpark of inflation-model parameter space that LISA can probe with excellent scientific insight,  \textit{i)} if ESA delivers a very accurate LISA noise model, \textit{ii)} if the astrophysical community~\cite{LISA:2022yao} manages to precisely model the astrophysical foregrounds, 
    \textit{iii)} and if the data analysis and waveform communities~\cite{LISAConsortiumWaveformWorkingGroup:2023arg} achieve binary waveform reconstruction with residuals that do not mimic too strongly a \GWB\ signal. 
    \item  To estimate the accuracy at which LISA can reconstruct the \GWB\ templates, provided the three ``ifs'' in 2) are met, as a function of their parameters. 
    \end{enumerate}
The latter is particularly important for the theory community, as it gives some accuracy maps that the \GWB\ theoretical predictions should aim at. Indeed, 
 theoretical uncertainties above the estimated accuracies would risk introducing biases, 
  and then jeopardising the reconstruction of both the cosmological \GWB\ and all the other GW sources.

The outline of the paper is as follows. In \cref{sec:models} we propose a list of templates that approximate wide classes of \GWB\ originated by inflationary mechanisms. For each template, we single out some specific inflationary setups that illustrate concrete examples of the physics and their fundamental-parameter versus template-parameter maps. \Cref{sec:SGWBinner} describes how we extend the \texttt{SGWBinner} code~\cite{Caprini:2019pxz, Flauger:2020qyi} in order to perform searches and parameter estimations for the \GWB\ templates. In \cref{sec:results} we use the extended \texttt{SGWBinner} to identify the parameter space  that LISA can probe for the considered templates in the presence of galactic and extragalactic foregrounds. We also forecast the striking accuracy at which the inflationary \GWB\ can be reconstructed,  provided sufficient knowledge on three LISA areas is reached (specifically, the modelling of the LISA noise,  the foregrounds and the primordial \GWB). In \cref{sec:interpr}  we map the forecast reconstructions of the template parameters to some of the parameter space of  inflationary setups. 
We devote \cref{sec:conclusions} to our conclusions. In \cref{app:variation_of_template_parameters} we show how changes in the parameters of each template influence the spectral shape of the \GWB. In \cref{app:from_P_to_Omega} we include further technical details on SGWB sourced at second order by scalar fluctuations.

\section{\GWB\ template databank for inflationary mechanisms}
\label{sec:models}

The inflationary mechanism can be realised in a variety of different models. In the following subsections, we introduce a selection  of templates that describe well-motivated  inflationary scenarios able to amplify the SGWB spectrum in the LISA frequency band. 
The resulting list of templates  does not exhaust all possibilities, but aims at initiating the first LISA template databank for \GWBs\ from inflation. 

It is worth noting that the templates we present do not have a unique parameterisation. However, for efficient numerical analysis, it is convenient to choose parameterisations that minimize degeneracies and correlations. One way to achieve this is by focusing on parameters that better characterise the 
 frequency shape and the  `geometrical' features (such as \emph{e.g.,} ~the high or low frequency tilt, the position of a peak, etc) of the SGWB signal:
 we  follow this guiding principle in our selection of templates below. In \cref{app:variation_of_template_parameters} we show how the frequency profile of the different templates is affected by the change in their parameters.
  This is aimed at offering a visual understanding of how the template depends on its parameters.

 In this paper, we will express the \GWB\ power spectrum in terms of $\Omega_{\rm GW}$, defined as the GW fractional energy density per unit frequency logarithmic interval,
\begin{equation}
\Omega_{\rm GW} \left( f \right) \equiv \frac{1}{\rho_{0,\rm crit}} \, \frac{d \rho_{\rm GW}}{d \ln f} \,, 
\label{eq:energy density}
\end{equation}
where $\rho_{0,\rm crit}\equiv3 H_0^2/8\pi G$ is the present time critical energy density of a flat universe, and $H_0 \equiv h \times 100 \, {\rm km} \, {\rm s}^{-1} \, {\rm Mpc}^{-1}$ is the present time value of the Hubble parameter. As customary, we multiply $\Omega_{\rm GW} \left( f \right)$ by the dimensionless  Hubble parameter squared $h^2$ so as not to propagate uncertainties related to its estimate in our analysis.

\subsection{Power law}\label{sec:pl}

The first template  we consider is for a power-law (PL) signal

\begin{equation}
   h^2 \Omega^{\rm PL}_{\textrm{GW}}(f, \vec{\theta}_{\rm cosmo}) = h^2\Omega_* \left( \frac{f}{f_*} \right)^{n_t}\;, 
   \label{eq:PL_template}
\end{equation}
parameterised in terms of the parameters $\vec{\theta}_{\rm cosmo}=\{\Omega_*,  
f_*, n_t\}$. Here, $n_t$ is the tilt of the power law, $f_*$ is a pivot frequency, 
$\Omega_*$  parameterises the amplitude of
the spectrum at the frequency $f_*$. The two parameters $f_*$ and 
$\Omega_*$ are degenerate so one of them should be fixed in the analysis. We fix $f_*=1\,$mHz. This choice minimizes the correlation between $\Omega_*$ and $n_t$ in the analysis around $f_*$, \emph{i.e.,} close to the frequency at which LISA reaches its best sensitivity ($\simeq 3\,$mHz). 

The PL template has a simple parameter dependence and can be analyzed relatively quickly. For this reason, we see no need for setting a stringent prior to optimize the search. We thus take a 
log-uniform prior on $\Omega_*\in[10^{-30},\,10^{-5}]$
and a flat prior on $n_t \in[-10,\,10]$, keeping however in mind that signals with $h^2 \Omega_{\textrm{GW}}  \gtrsim \mathcal{O} \left( 10^{-6} \right)$ would be in tension with the big-bang nucleosynthesis (BBN) bound\footnote{ In order for our priors to be as uninformative as possible, hereafter we will always provide a broad prior range on the amplitude of the SGWB, including values that are excluded by the BBN+CMB bound $h^2\,\int_{f_{\rm min}}^{f_{\rm max}}\,{\rm d}\ln f\, \Omega_{\rm GW}(f)\,<\,1.2\times10^{-6}$ at 95\%  CL, see \cite{Pagano:2015hma}. In practice, however, we will never adopt benchmarks that are ruled out by this bound and therefore we will not discuss this issue anymore. }~\cite{LISACosmologyWorkingGroup:2022jok}. The prior on the tensor spectral index is motivated by the fact that $|n_t|\lesssim {\cal O}(1)$ in all models we consider. Although, strictly speaking, for our purposes the parameterisation in \cref{eq:PL_template} needs to be valid only in the LISA observational window (so that $n_t$ might be negative within this window) we stress that an overall growing trend of the spectrum with increasing frequency 
is needed to have an inflationary signal visible in LISA, while being compatible with CMB bounds \cite{LISACosmologyWorkingGroup:2022jok}.

In general, we expect that any \GWB\   resembles a PL at frequencies far away from those corresponding to the typical scales involved in the processes sourcing the GWs, or those diluting their energy density throughout cosmic history. In the following, we link explicitly the PL template and some illustrative, well-motivated inflationary mechanisms.

\begin{description}
\item[Axion inflation.] In this setup, the inflaton $\phi$ is an axion coupled to a gauge field through the interaction $\phi F {\tilde F}/(4 f_\phi)$, with $f_\phi$ being the axion decay constant. The rolling axion strongly amplifies the gauge field, which in turn generates a  \GWB\ with large amplitude~\cite{Barnaby:2010vf,Sorbo:2011rz}. The production is exponentially sensitive to the parameter $\xi \equiv {|\dot{\phi}|}/({2 f_\phi H})$ (from now on, a dot denotes differentiation with respect to cosmic time) whose growth during inflation strongly tilts the GW spectrum to blue. The overall frequency shape of the \GWB\ from CMB to interferometer scales is not a single power-law. But the PL template is adequate for the part of the signal falling in the LISA frequency window, with amplitude and spectral tilt given by \cite{Barnaby:2010vf,Sorbo:2011rz,Bartolo:2016ami} 
\begin{eqnarray} 
	h^2\Omega_*   &\simeq& 1.5 \times 10^{-13} \, \frac{H_*^4}{M_{\rm Pl}^4} \, \frac{{\rm e}^{4 \pi \xi_*}}{\xi_*^6} \;, \nonumber\\ 
	n_T &\simeq& - 4 \epsilon_* + \left( 4 \pi \xi_* - 6 \right) \left( \epsilon_* - \eta_* \right) \;.
	\label{axion-inf}
\end{eqnarray} 
Here $\epsilon\equiv-\dot{H}/H^2$ and $\eta\equiv-\ddot{\phi}/(H \dot{\phi})$ are the first and second slow-roll parameter respectively, and the star indicates that they are evaluated when the mode of pivot frequency $f_*$ leaves  the horizon during inflation. Namely, at $k_* \equiv {2 \pi f_*} = a_* H_*$, with $a_*$ being the scale factor of the universe at that time. The expression for the spectral tilt in eq.~(\ref{axion-inf}) assumes a steady-state evolution of the parameter $\xi$, which holds in the regime of weak backreaction of the produced gauge fields on the inflaton dynamics, but also at the transition between the weak and the strong backreaction regime, during which the \GWB\ spectrum might rise to observable levels.\footnote{The result for modes that leave the horizon during the strong backreaction regime is still under investigation~\cite{Garcia-Bellido:2023ser,Caravano:2022epk,Figueroa:2023oxc} due to the nontrivial evolution of $\xi$ in this regime~\cite{Peloso:2022ovc}. Depending on the parameters of the model, it is possible that this uncertain regime manifests itself at high frequencies outside the LISA window.}   Note that, in the limit in which the two slow-roll parameters $\epsilon_*$ and $\eta_*$ are hierarchical, eq.~\eqref{axion-inf} provides a one-to-one correspondence between $(\Omega_*,\,n_T)$ and the relevant input parameters:
\begin{eqnarray} 
&& \xi_* \simeq -\frac{3}{2\pi}\, W\left(-\frac{0.015}{\sqrt[6]{\frac{h^2 \Omega _* M_{\text{pl}}}{H_*}}}\right) 
\;,\;\; \left\{ 
\begin{array}{l}
\eta_* \simeq -\frac{n_T}{2 \left(2 \pi  \xi _*-3\right)} \;\;,\;\; \eta_* \gg \left\vert \epsilon_* \right\vert \\ 
\epsilon_* \simeq \frac{n_T}{2 \left( 2 \pi \xi_* -5\right)} \;\;,\;\; \left\vert \epsilon_* \right\vert \gg \eta_* 
\end{array} \right. \;,
\label{axion-inf-inverse}
\end{eqnarray} 
where $W(x)$ is the Lambert function.

\item[Massive graviton during inflation.] The breaking of space diffeomorphisms can give rise to a 
massive graviton  during the inflationary epoch; see \emph{e.g.,} ~refs.~\cite{Endlich:2012pz,Ricciardone:2016lym,Bartolo:2015qvr,Cannone:2014uqa,Cai:2014uka,Lin:2015cqa,Cannone:2015rra,Graef:2015ova,Domenech:2017kno,Ricciardone:2017kre,Celoria:2017idi,Dimastrogiovanni:2018uqy,Fujita:2018ehq}. 
Such a graviton mass $m_h$ has the effect of tilting the \GWB\ spectrum towards the blue, making the signal potentially detectable with  LISA, while below the current bound at CMB frequencies. Such a tilt is related to the graviton mass by
\begin{eqnarray}
\label{eqsgl}
n_T\,=3-2\sqrt{\frac{9}{4}-\frac{m_h^2}{H^2}}\,\,,
\end{eqnarray} 

where we have focused on the vanishing $\epsilon$ limit (see \emph{e.g.,}  ref.~\cite{Ricciardone:2016lym}).
A connection between the amplitude of the SGWB and the model parameter is, however, more complicated to draw. Indeed, while the tilt is robustly given by eq.~\eqref{eqsgl}, the overall normalization of the SGWB is model dependent and controlled by several free parameters~\cite{Endlich:2012pz,Ricciardone:2016lym,Bartolo:2015qvr,Cannone:2014uqa,Cai:2014uka,Lin:2015cqa,Cannone:2015rra,Graef:2015ova,Domenech:2017kno,Ricciardone:2017kre,Celoria:2017idi,Dimastrogiovanni:2018uqy,Fujita:2018ehq}.

  Finally, let us also emphasize that, in order for the square root in eq.~\eqref{eqsgl}, not to be imaginary the graviton mass only needs to satisfy $m_h/H<3/2$. We also point out that values much smaller than $3/2$ are  apparently  in tension with
 the Higuchi bound $m_h/H>\sqrt{2}$ \cite{Higuchi:1986py}.
 Nevertheless, the Higuchi bound  
relies on exact de  Sitter isometries: the latter can be spontaneously broken in the aforementioned 
 scenarios where scalars acquire spatial vacuum expectation values, 
  hence allowing
 for the Higuchi bound to be violated~\cite{Ricciardone:2017kre}.

\item[Time-dependent tensor sound speed.]  
The sound speed of gravitational tensor modes (or other helicity-2 degrees of freedom coupled to it) can exhibit a time dependence in inflationary setups with  non-minimally coupled scalars \cite{Kobayashi:2011nu,Creminelli:2014wna} or with extra helicity-2 component of spin-2 fields~\cite{Iacconi:2020yxn,Dimastrogiovanni:2021mfs}. These phenomena have implications for CMB polarisation experiments \cite{Raveri:2014eea}, but they can also have interesting consequences for GW detections, leading \emph{e.g.,} ~to a growth of the GW spectrum at small scales (see \emph{e.g.,} ~ref.~\cite{Bartolo:2016ami} and references therein).
 
In these scenarios, it is assumed that the tensor speed becomes equal to the speed of light at the end of inflation. For the case of single-field slow-roll inflation plus heavy helicity-2 components, the produced \GWB\ in the LISA band approximately follows eq.~\eqref{eq:PL_template} with $\Omega_*$ being a non-trivial function of the sound speed~\cite{Iacconi:2020yxn, Dimastrogiovanni:2021mfs} and $n_t$ reading
\begin{equation}
n_t\simeq |s_2| (2\nu -1)  \;,
\end{equation}
where $s_2 = \dot{c}_2/(H c_2)$, a parameter
which we assume constant during inflation~\cite{Iacconi:2020yxn},
 and $\nu=\sqrt{9/4 - m_h^2/H^2}$.
 We recall that here $m_h$ is the graviton mass,  while $c_2$ is the 
time-dependent sound speed of the additional helicity-2 field.

\end{description}

\subsection{Log-normal bump}
\label{sec_LNB}

For a spectral shape with a log-normal bump (LN), we adopt the template 
\begin{equation}
	h^2\Omega^{\rm LN}_{\textrm{GW}}(f, \vec{\theta}_{\rm cosmo}) =  h^2 \Omega_* \, {\rm exp} \left[ - \frac{1}{2 \rho^2 } \, 
	\log_{10}^2 \left( \frac{f}{f_*} \right) \right] \;, 
	\label{eq:master_modelIII_bum}   
\end{equation}
where the parameters $\vec{\theta}_{\rm cosmo}=\{\Omega_*, f_*, \rho \}$ control, respectively, the height, the position, and the width of the bump.

Due to the simplicity of the template, we assume the log-uniform priors $h^2 \Omega_*\in[10^{-30},\,10^{-5}]$, 
$\rho \in[0.01,\,10]$ (ranging from a narrow to a broad bump, including values that have been considered in the literature; see below for one specific example), and   $f_* \in[10^{-5},\,10^{-1}]$\,Hz, so that the bump is detectable in the LISA window.  

The LN template typically covers inflationary phenomena where the GW production is maximal at the scale $f_*$, which then persists for some time during inflation. The bump is narrow (respectively, broad) for small (respectively, large) values of $\rho$, corresponding to a shorter (respectively, longer) stage of GW production. We now discuss a concrete model motivating  the LN template. 

\begin{description}

\item[Axion spectator.] While the inflaton slowly rolls throughout inflation, there might also exist a `spectator' axion field $\chi$, which undergoes rolling for a certain number of e-folds $\Delta N$ of inflation~\cite{Namba:2015gja}. During this phase, the axion can excite gauge fields through a $\chi F {\tilde F}$ interaction: in turn, the excited gauge fields can source a sizeable \GWB. This production mechanism is similar to the axion inflation mechanism described above, with the distinction that now the \GWB\ exhibits a peak at the frequency $f_*$, corresponding to the typical wavenumber scale $k_*$ that left the horizon during the period of fastest rolling of the field $\chi$. The \GWB\ spectrum can be approximated by the shape (\ref{eq:master_modelIII_bum}), where the height of the peak is exponentially sensitive to $\dot{\chi}$, while the width is an increasing function of $\Delta N$~\cite{Namba:2015gja,Campeti:2022acx}.\footnote{References~\cite{Dimastrogiovanni:2016fuu,Thorne:2017jft,Putti:2024uyr} studied variations of the same class of models with a GW signal that can be well described by this template.}

In particular, based on the calculations and conventions of ref.~\cite{Namba:2015gja}, if we fix the maximum parameter $\xi_*=5$ and the inflaton slow roll parameter $\epsilon_\phi = 10^{-3}$, we obtain $\left\{ \log_{10}\Omega_* ,\, \log_{10}\rho \right\} \simeq \left\{ -9.9, -0.13 \right\}$ for the broader bump studied in ref.~\cite{Namba:2015gja}, while $\left\{\log_{10}\Omega_* ,\, \log_{10}\rho \right\} \simeq \left\{ -11.8, -0.37 \right\}$ for the narrower one. The moment at which the axion experiences the maximum speed, sets $f_*$, and the case $f_*=10^{-3}$\,Hz can be easily achieved. 

\end{description}

\subsection{Broken power law}
\label{subsec:BPL}
The template
\begin{eqnarray}
	h^2 \Omega_{\rm GW}^{\rm BPL}(f, \vec{\theta}_{\rm cosmo})&=& h^2 \Omega_* 
 \frac{\left(\frac{f}{f_*}\right)^{n_{{t},1}}}{\left\{\frac{1}{2}\left[1+\left(\frac{f}{f_*}\right)^{1/\delta}\right]\right\}^{(n_{t,1}-n_{t,2})\delta}}\;,
 \label{eq:master_modelIV_stBBN}
\end{eqnarray}
is adequate for \GWBs\ with frequency shapes looking like a smooth broken power law (BPL). Its parameters are $\vec{\theta}_{\rm cosmo}=\{\Omega_*, f_*, n_{t,1}, n_{t,2}, \delta \}$, with $\Omega_*$ parameterising the amplitude of the peak located at frequency $f_*$, while $n_{t,1}$ and $n_{t,2}$ are, respectively, the tilt of the signal tail at $f\ll f_*$ and  $f\gg f_*$, with the final parameter $\delta$ controlling the sharpness of the transition between the two regimes.

For sufficiently large and positive  $n_{t,1}$, and a negative and sufficiently large (in absolute value) $n_{t,2}$, the 
amplitude of the spectrum can be localized within the LISA frequency band, and be negligible at the frequencies probed by other GW detectors. We take the log-uniform prior $h^2 \Omega_*\in[10^{-30},\,10^{-5}]$. For the other parameters, we take the flat priors $n_{t,1}\in[-10,\,10]$ and  $n_{t,2} \in[-10,\, 10]$, and the log-uniform prior  $f_* \in[10^{-5},\,10^{-1}]\,$Hz. We finally take flat priors 
$\delta\in[0.1,10]$ to describe both the case in which the slope changes softly, or abruptly.

The template is motivated by, \emph{e.g.,}  the concrete examples we discuss next.

\begin{description}

\item[Second slow-roll phase.] 
When GWs are induced at second-order by curvature perturbations sourced by a broad and flat primordial curvature power spectrum, they lead to an SGWB featuring a BPL frequency shape with $n_{t,2} \simeq  0$ (see~\cref{app:from_P_to_Omega} for details). 
Such a scenario can be achieved, for example, through an ultra-slow-roll (USR) phase followed by a second slow-roll regime
generating the required plateau~\cite{Wands:1998yp,Leach:2000yw, Leach:2001zf, Biagetti:2018pjj,Franciolini:2022pav}.
This model, originally suggested in ref.~\cite{DeLuca:2020agl}, could potentially connect
the pulsar timing array frequency range (where recent data provide evidence for the existence of a SGWB~\cite{NANOGrav:2023gor,EPTA:2023fyk,Reardon:2023gzh,Xu:2023wog} of yet-to-be determined nature) to signatures in the LISA frequencies associated with the formation of primordial black holes \cite{Sasaki:2018dmp} with masses in the range $(10^{-15} \,-\, 10^{-11})M_\odot$, where they may comprise the totality of the dark matter in the universe~\cite{Vaskonen:2020lbd,Franciolini:2022pav}.

	\item[Hybrid inflation with mild waterfall stage.] 
The template above can also describe a broad bump, which is asymmetric around $f_*$ if the primordial curvature power spectrum takes the shape of a broad bump. 
Popular models within this category
are multifield setups such as hybrid inflation~\cite{Linde:1993cn,Garcia-Bellido:1996mdl,Clesse:2015wea} or other types of models featuring a slow-turn in the field space together with a tachyonic instability of isocurvature perturbations~\cite{Fumagalli:2020adf,Braglia:2020eai,Braglia:2020taf}. We focus on hybrid inflation for illustration. In this setup, there are two dynamical scalar fields, the inflaton and a so-called hybrid field. The potential of these fields is designed to induce a second stage of inflation during which isocurvature perturbations grow tachyonically on superhorizon scales, and are converted into adiabatic perturbations, leading to a peak in the primordial curvature power spectrum~\cite{Garcia-Bellido:1996mdl,Clesse:2015wea}. The induced \GWB\,  typically takes the form of a  BPL with $n_{t,1}>0$ and $n_{t,2}<0$~\cite{Clesse:2018ogk,Spanos:2021hpk,Braglia:2022phb}. The relationship between the BPL parameters and the parameters of the two-fields potential cannot generally be expressed in a closed form, but some features can be generically related to the underlying dynamics during inflation. For example, let us consider the potential of the
	$\alpha$-attractor case \cite{Braglia:2022phb}. We have the following features: the linear term of the hybrid field controls $\Omega_*$; the value of the hybrid field at the minimum of the potential sets $f_{*}$; 
	the first derivative along the hybrid direction modulates $n_{t, 2}$; and the time profile of the squared mass of the hybrid field controls how much isocurvature modes with $f<f_{\rm peak}$ are amplified,  and hence the size of $n_{t, 1}$.  For instance, for the parameter values in Eq.~(5.3) of ref.~\cite{Braglia:2022phb}, one numerically obtains a \GWB\ resembling the BPL with $\log_{10} (h^2\Omega_*)= -9.3$, $f_* = 1\, {\rm mHz}$, $n_{t,1}= 2.65$, $n_{t,2}=-2.1,\,\delta=5.3$.

\end{description}

\subsection{Double peak}\label{sec:double_peak}

The template
\begin{equation} 
\begin{aligned}
h^2\Omega^{\rm DP}_{\rm GW}(f, \vec{\theta}_{\rm cosmo}) =& 
\,\,h^2 \Omega_*
\Bigg [\beta \, \left( \frac{f}{\kappa_1 f_*} \right)^{\!n_{p}}  \left[\frac{c_1 - f/ f_*}{c_1 - \kappa_1}\right]^{\!\!\frac{n_{p}}{\kappa_1} \left(c_1-\kappa_1\right)} \Theta\!\left(c_1 - \frac{f}{f_*} \right) \\
&+ \exp\!\left[-\frac{1}{2\rho^2} \log_{10}^2\!\left(\frac{f}{\kappa_2 f_*}\right)\right] \left\{1 + {\rm erf} \left[ -\gamma \log_{10}\!\left(\frac{f}{\kappa_2 f_*}\right) \right] \right\}
\Bigg ] 
\end{aligned}
\label{eq:templateDP} \;,
\end{equation}
is aimed at reconstructing a double-peak (DP) \GWB\ signal.  The free parameters controlling the shape are $\vec{\theta}_{\rm cosmo}=\{h^2\Omega_*, f_*, \beta, \kappa_1, \kappa_2, \rho, \gamma \}$,
whereas $c_1=\sqrt{2/3}$ and $n_p\simeq 2.5$ are kept constant, thanks to universal infrared  properties of the GW background, as discussed  in
  \cref{app:lognormal}. The parameter $f_*$ sets the frequency of the transition from the low-frequency to the high-frequency peak via the Heaviside step function $\Theta$. The high-frequency peak is a skewed log-normal with its maximal amplitude $h^2 \Omega_*$ occurring at $f\simeq\kappa_2 f_*$, with $\kappa_2>1$.
Its skewness and width are set by $\gamma$ and $\rho$, respectively. In contrast, the low-frequency peak reaches its maximum amplitude $\beta h^2 \Omega_*$, with $0<\beta<1$, at frequency $f\simeq \kappa_1 f_*$ with $\kappa_1 < 1$. At $f\ll \kappa_1 f_*$, the template is a power law  $\propto f^{n_p}$. A minimum between the two peaks arises at $f\simeq
c_1 f_*$ when $\rho$ is sufficiently small.\footnote{In principle, a DP frequency shape can be described by functional forms different from the one in eq.~\eqref{eq:templateDP}. As alternatives, we  considered a ratio of polynomials and a sum of power laws times exponentials of polynomials. We discard these alternatives in favor of \cref{eq:templateDP}. By performing a Bayesian analysis, both with a uniform search (equally weighted simulated data points) and a weighted search (data weighted with the LISA sensitivity curve), we indeed found that our choice (\ref{eq:templateDP}) better describes the DP \GWB\ signal predicted in the inflationary models that motivate the template.}

The DP \GWB\ profile is expected from second-order emission triggered by enhanced scalar perturbations (see  \cref{app:from_P_to_Omega} for details). A signal  described by the DP template represents the characteristic \GWB\ profile induced by large (compared to CMB scales) inflationary scalar fluctuations re-entering the horizon during a radiation-dominated phase~\cite{Tomita:1975kj,Matarrese:1993zf,Matarrese:1997ay,Acquaviva:2002ud,Mollerach:2003nq,Nakamura:2004rm,Ananda:2006af,Baumann:2007zm,Espinosa:2018eve, Kohri:2018awv,Inomata:2019yww,Domenech:2019quo}.  This enhancement of scalar fluctuations can be generated in single- and multi-field inflationary models featuring a short period of non-attractor evolution, which, in turn, amplifies the \GWB\ in the corresponding frequency range. 

In addition, if the support of the scalar spectrum is narrow
in momentum space, a resonance between the transfer function of the scalar modes and the Green function in the tensor evolution, occurring when the frequency of the scalar source coincides with the frequency of the GWs, introduces a second pronounced peak on  the \GWB\ profile~\cite{Ananda:2006af,Saito:2008jc}.

This broad class of scenarios 
can be generated by one of the following approximated primordial curvature power spectra:
\begin{eqnarray}
\mathcal{P}_\zeta^{\rm ln}(k) &=& 
{\cal A}_s \exp\!\left[-\frac{1}{2\Delta^2}\ln^2\!\left(\frac{k}{k_*}\right)\right] \qquad \textrm{(log-normal)}\,, \label{eq:Pk_LN1} \\
\label{eq:Pk_bpl1}
\mathcal{P}_\zeta^{\rm bpl}(k) &=&
\frac{{\cal A}_s (p_1+p_2)}{\left[p_2 \left( \frac{k}{k_*}\right)^{-p_1} + p_1\left(\frac{k}{k_*}\right)^{p_2} \right]}  \qquad \textrm{(broken power-law)}\;,
\end{eqnarray} 
where $\Delta,\, p_1,\, p_2 > 0$. We prove in \cref{app:lognormal} and \cref{app:UVcutoff} that both spectra lead to a \GWB\  frequency shape that is well fitted by the DP template, and we use them to set the priors on the DP template. Specifically, we determine the overall ranges of DP template parameters that allow us to fit the SGWBs derived from $\mathcal{P}_\zeta^{\rm ln}(k)$ when $\Delta \in [0.1,1]$ and $\mathcal{P}_\zeta^{\rm bpl}(k)$ when $p_2 \in [0.5,4]$ and $p_1 = 4$ (see motivations below)
 within the ranges 
$h^2 \Omega_* \in [10^{-30},10^{-5}]$, $f_*\in [10^{-5},10^{-1}]\,\textrm{Hz}$, $\beta\in [0.05,0.6]$, $\kappa_1\in [0.05,0.9]$,
$\kappa_2\in [1.1,3]$, $\rho\in [0.04,0.5]$, and $\gamma \in [0.02,20]$. We then assume log-uniform priors on $h^2 \Omega_*$ and $f_*$, and flat priors on the other DP parameters.

Let us now motivate the two spectral shapes in eqs.~\eqref{eq:Pk_LN1} and \eqref{eq:Pk_bpl1} in terms of concrete scenarios.

\begin{description}
\item[Log-normal $\mathcal{P}_{\zeta}^{\rm ln}$.]

One example of peaks in the primordial curvature power spectrum that can be parameterised by a narrow log-normal shape is provided by single-field models where the enhancement of scalar fluctuations is given by a resonant mechanism. 
This is for instance realised in scenarios with 
an oscillating speed of sound, with amplitude and frequency parameterised by $(\xi,k_*)$, that modifies the standard evolution equation for the perturbations~\cite{Cai:2018tuh,Cai:2019jah,Chen:2019zza,Chen:2020uhe,Addazi:2022ukh}. The Mukhanov--Sasaki equation can thus be approximated as a Mathieu equation which, in turn, presents a parametric instability for certain ranges of modes located around the oscillatory frequency $k_*$ with width $\sim \xi k_*$.
In principle, this feature provides a direct link between the position and width of the peak in the primordial power spectrum, and the frequency and amplitude of the speed-of-sound oscillations.

\item[Broken power-law $\mathcal{P}_{\zeta}^{\rm bpl}$.]
A period of USR in single-field models of inflation produces a primordial curvature power spectrum with a shape that can be approximated as in eq.~\eqref{eq:Pk_bpl1}. The powers are determined by the second slow-roll parameter $\eta \equiv - \ddot H/(2 H \dot H)$, where $H$ is the Hubble parameters and dot indicates derivatives with respect to cosmic time. 
Denoting it $\eta_1$ during the slow-roll period preceding USR, 
and $\eta_3$ during the constant-roll period typically following the USR, one finds~\cite{Karam:2022nym}: $p_1 = 5-|1-2\eta_1|$ and $p_2 = 2\eta_3$. Typically, the conditions $|\eta_1| \ll 1$ and $\eta_1 < 0$ arise, so that the growth of the spectrum is $\mathcal{P}_\zeta \propto k^4$~\cite{Byrnes:2018txb}, though a steeper growth can also be realised in some scenarios~\cite{Carrilho:2019oqg,Fumagalli:2020adf,Tasinato:2020vdk,Braglia:2020taf,Davies:2021loj}. The peak amplitude of the curvature spectrum is determined by the second slow-roll parameter $\eta_2$ during USR and the duration $\Delta N_{\rm USR}$ of the USR period as $\mathcal{A}_s \sim e^{2\eta_2 \Delta N_{\rm USR}} \mathcal{P}_{\rm CMB}$, where $\mathcal{P}_{\rm CMB}$ denotes the amplitude at the CMB scales. Assuming Wands duality~\cite{Wands:1998yp} between the USR and the final constant-roll, which commonly holds for smooth potentials, the second slow-roll parameter during the USR period is $\eta_2 = 3 - \eta_3$. 
Assuming the enhanced spectrum is produced by an USR phase of inflation, one can perform a reverse engineering procedure to connect the spectral shape to the inflationary dynamics and the inflaton potential \cite{Franciolini:2022pav}.
The location of the peak $k_*$ is determined by the number of e-folds between the epoch of Hubble crossing of CMB modes and the onset of USR dynamics. 

The broken power-law spectrum in eq.~\eqref{eq:Pk_bpl1} is also realised in thermal inflation models where the fluctuations become large around the time when the inflaton potential turns from convex to concave. In this case, the growth is $\mathcal{P}_\zeta \propto k^3$ and the decay and the peak amplitude are determined by the tachyonic mass of the thermal inflaton in the false vacuum and by the potential energy of the false vacuum~\cite{Dimopoulos:2019wew}.

\end{description}

\subsection{Excited states}
\label{sec:excited}
The template 
\begin{equation}
\label{templateexcited}
h^2\Omega_{\mathrm{GW}}^{\rm ES}(f, \vec{\theta}_{\rm cosmo}) = 
\frac{h^2 \Omega_*}{0.052} \frac{1}{x^3}\left[1-\frac{x^2}{4\gamma_{\rm ES}^2}\right]^2\left[\sin(x) -4\frac{\sin^2(x/2)}{x}\right]^2  \Theta\!\left(x_{\rm cut}  - x \right)  ,
\end{equation}
where $x \equiv ( f \, \omega_{\rm ES} ) /2 $, parameterises a \GWB\ generated in the presence of a scalar excited state (ES) during inflation. Its parameters are $\vec{\theta}_{\rm cosmo} = \{\Omega_*, \gamma_{\rm ES}, \omega_{\rm ES}\}$, where $\Omega_*$ sets the amplitude of the primary peak at frequency $f_{\rm max}\simeq 6/\omega_{\rm ES}$,
$\omega_{\rm ES}$ sets the periodicity of the subsequent peaks (at $f>f_{\rm max}$), and $\gamma_{\rm ES}$ determines the largest frequency  $f_{\rm cut}= 2/3 \gamma_{\rm ES} f_{\rm max}$, corresponding to $x_{\rm cut} = 2 \gamma_{\rm ES}$, where the template is applicable.
As explained below, theoretical constraints leads us to expect 
$\gamma_{\rm ES} \gg 1$, but not arbitrarily large. 
We thus set a flat prior $\log_{10}(\gamma_{\rm ES}) \in [0.699, 2]$. We take the flat prior $\log_{10}(h^2 \Omega_*) \in [-30, -5]$, while we assume  $\log_{10}( \omega_{\rm ES} \, {\rm Hz} ) \in [2, 5]  $ because this range of values provides a signal peaked in the LISA frequency band (for signals with $\omega_{\rm ES}$ outside this range, simpler templates than the ES would be more suitable).

A \GWB\ following the ES template is expected in inflationary setups where a large number of scalar particles are produced dynamically with momenta peaked around a narrow range of scales \cite{Fumagalli:2021mpc}. Such an excited state dynamically arises in the presence of a transient non-adiabatic evolution during inflation. The energy-momentum tensor of the produced particles sources the tensor modes, giving rise to an ES SGWB. The typical time of particle production, or, equivalently, the associated frequency $f_{\rm out}$, sets the frequency of the periodicity, with $\omega_{\rm ES} \simeq 2/f_{\rm out}$, while
the deeper inside the horizon particle production took place, the larger $\gamma_{\rm ES}$, \emph{i.e.,}~the larger frequencies the signal extends to. Moreover, the higher the number of scalars in an excited state, the larger the amplitude of the signal.

A dynamical mechanism to generate an excited state can be found, for instance, in models of multifield inflation in which the background trajectory deviates strongly and for a brief period from a geodesic of the field-space manifold \cite{Palma:2020ejf,Fumagalli:2020adf}. Other realisations can be found in single-field models with a feature in the potential \cite{Inomata:2021uqj,Inomata:2021tpx}, parametric resonances due to a periodic structure in the potential \cite{Peng:2021zon,Inomata:2022yte}, or with a spectator field coupled to the inflaton via a periodic function \cite{Cai:2021wzd}. Quite generically, tensor modes with spectral shape as in \cref{templateexcited} can be generated whenever scalar fluctuations are amplified on sub-horizon scales by some mechanism during inflation \cite{Fumagalli:2021mpc} (see ref.~\cite{Inomata:2021zel} for upper bounds).
Let us delve into this topic further by exploring a specific scenario.

\begin{description}

\item[Strong turn in two-field inflation.]
A transient non-adiabatic evolution is provided by
a brief period during which the turn of the trajectory in field space is large. This  leads to the dynamical appearance of an excited state, and consequently to a signal following the ES template\footnote{This setup also predicts the formation of primordial black holes \cite{Palma:2020ejf,Fumagalli:2020adf}.}.
Starting from the generic multifield non-linear sigma models Lagrangian, one can derive the effective action for the fluctuations at second order around a given
homogeneous background. In the latter, adiabatic and entropic degrees of freedom are coupled by a term proportional to
$\eta_{\perp} \equiv -V_N/(M_{\mathrm{Pl}} H^2 \sqrt{2\epsilon})$, which is the dimensionless parameter measuring the turning rate of the trajectory, while $V_N$ is the projection of the first derivative of the potential over the entropic direction.  A top-hat profile for the time dependence of $\eta_\perp$, with strength $\eta_{\perp}\gg 1$ and duration less than one e-folds, provides a useful idealization of a strong and sharp turn. Note that, in order to satisfy backreaction and perturbativity bounds, $\eta_\perp$ cannot be arbitrarily large~\cite{Fumagalli:2020nvq}, motivating our previous priors choices. In this example, the ES template parameters are related to the fundamental ones as
\begin{equation}
\gamma_{\rm ES} \simeq \eta_{\perp},\quad \omega_{\rm ES} \simeq \frac{2}{f_{\rm out}} \,,
\end{equation}
where $f_{\rm{out}}$ is the frequency corresponding to the mode leaving the horizon at the end of a turn of strength  $\eta_{\perp}$. 
Furthermore, the overall amplitude can be expressed as 
\begin{equation}
h^2\Omega_* = 0.03\, r_i \,\mathcal{N}^4 \,\gamma_{\rm ES}^5\,|\beta|^4 \,H_*^4/(\pi M_{\rm Pl})^4,
\end{equation}
with $r_i$ a redshift factor, $H_*$ the value of the Hubble parameter at the turn, $\mathcal{N}$ the number of scalars affected by the excited state, and $|\beta|^2 \gg 1$ the large occupation number of particles~\cite{Fumagalli:2021mpc}.
\end{description}

\subsection{Linear oscillations}\label{sec:linear}

The linear oscillations (LO) template
\begin{equation}
    \label{eq:sharp-template} 
    h^2\Omega_{\textrm{GW}}^{\textrm{LO}}(f, \vec{\theta}_{\rm cosmo}) = \Big[1+ \mathcal{A}_\textrm{lin} \cos \big(\omega_\textrm{lin} f + \theta_\textrm{lin}  \big) \Big] h^2 \Omega_{\textrm{GW}}^{\,\rm env}(f, \vec{\theta}_{\rm env}) \;, \\
\end{equation}
describes oscillations periodic in $f$ that modulate a smooth envelope. Its parameters are $\vec{\theta}_{\rm cosmo}=\{\vec{\theta}_{\rm env},\mathcal{A}_{\mathrm{lin}},\omega_{\mathrm{lin}},\theta_{\mathrm{lin}}\}$, with $\vec{\theta}_{\rm env}$ describing the overall envelope $\Omega_{\textrm{GW}}^{\,\rm env}(f)$, and the other three controlling the relative amplitude, the period and the phase of the oscillations.
Since the oscillations and the envelope signal are theoretically associated with a common sourcing mechanism (see below),  we consider a profile for
$\Omega_{\textrm{GW}}^{\,\rm env}(f)$ 
 exhibiting a dominant, narrow peak. In particular, linear oscillations arise in the context of GWs induced at second order by scalar fluctuations, when the enhancement of the latter is narrowly peaked around a given scale. Thus, the natural envelope to use would be the double peak template of \cref{sec:double_peak} augmented with the fact that linear oscillations would appear only on the highest peak \cite{Fumagalli:2020nvq}. The latter being given by a log-normal bump (modulo the skewness parameter $\kappa_2$ that we neglect here for simplicity) we restrict ourselves, in the analysis below, to a log-normal envelope, \emph{i.e.,}~$\Omega_{\textrm{GW}}^{\,\rm env}(f) = \Omega_{\textrm{GW}}^{\rm LN}(f)$, with $\rho \equiv 10^{\Delta}$, although other envelopes
can be considered. As discussed in \cref{sec:double_peak}, $\rho$ has values within an order of magnitude, so there is no need for a log prior for the width of the envelope.

For the envelope, we make use of the log-normal bump template of \cref{sec_LNB} and we choose the flat priors $\Omega_* \in [-30,-5]$, $\log_{10}(f_*/{\rm Hz})\in [-5,-1]$, and $\rho \in [0.05,1]$.
For the oscillatory part we set the flat priors $\mathcal{A}_{\rm lin} \in [0.05,1]$ and
$\omega_{\rm lin}\, \in [0.1,10^3]\, \mathrm{mHz}^{-1}$.

The template \eqref{eq:sharp-template} well describes the \GWB\ generated at horizon re-entry of primordial density fluctuations with power spectrum \cite{Fumagalli:2020nvq} 
\begin{align}
\mathcal{P}_\zeta(f) = \overline{\mathcal{P}}_\zeta(f) \bigg[ 1 + A_{\textrm{lin}} \cos \Big( \omega f + \phi_\textrm{lin} \Big) \bigg] \, , 
\label{Power-spectrum-sharp-feature}
\end{align}
where we traded  wavenumbers $k$ for frequencies $f$. Despite this phenomenon being inherently nonlinear, the sinusoidal modulations in the power spectrum get processed into corresponding sinusoidal oscillations in $\Omega_\textrm{GW}$ with $\omega_\textrm{lin} = c_s^{-1} \omega$ \cite{Fumagalli:2020nvq,Witkowski:2021raz}, where $c_s$ is the propagation speed of density fluctuations, equal to $1/\sqrt{3}$ in the conventional scenario of radiation domination at horizon reentry. The oscillations do get averaged out, though, so that the relative amplitude $\mathcal{A}_\textrm{lin}$ in eq.~\eqref{eq:sharp-template} is typically $\mathcal{O}(20 \%)$, even for $A_{\textrm{lin}} \simeq 1$ (with a nonstandard expansion history, the amplitude can be larger up to $\mathcal{O}(40 \%)$  \cite{Witkowski:2021raz}). Finally, the envelope $\overline{\Omega}_{\textrm{GW}}(f)$, is determined by the envelope $\overline{\mathcal{P}}_\zeta(f)$, whose details differ between different inflationary models.

\begin{description}
\item[Sharp features.] The power spectrum \eqref{Power-spectrum-sharp-feature} is characteristic of inflationary models with a sharp feature, \emph{i.e.,}~a transition during inflation that occurs on a timescale approximately smaller than one e-fold \cite{Slosar:2019gvt}. We can unequivocally assign a scale $k_\textrm{f}$ to these sharp events, the momentum of modes crossing the Hubble radius at that time, or equivalently   the corresponding frequency $f_\textrm{f}$.
Generically,
 sharp features  generate particle production.  When the latter is significant, it leads to an enhancement of the primordial power spectrum at the corresponding scales, modulated by $\mathcal{O}(1)$  oscillations ($A_{\textrm{lin}} \simeq 1$) with frequency $\omega=2 / f_\textrm{f}$ \cite{Fumagalli:2020nvq}.
As the particle production is effective for scales inside the Hubble radius at the time of the features ($f \gg f_\textrm{f}$), the range of scales for which the envelope $\overline{\mathcal{P}}_\zeta(f)$ is significant spans several periods of the oscillations.

Sharp features are not associated with a single model of inflation, but can rather be realised through a variety of mechanisms \cite{Slosar:2019gvt}. Large features leading to significant particle production and to the power spectrum \eqref{Power-spectrum-sharp-feature}, and hence the template \eqref{eq:sharp-template}, can occur in single-field inflation, \emph{e.g.,}  caused by a step in the inflaton potential \cite{Inomata:2021uqj,Dalianis:2021iig}, or in multifield settings, \emph{e.g.,}  when the inflationary trajectory exhibits a sharp turn \cite{Palma:2020ejf,Fumagalli:2020adf,Aragam:2023adu}. It is difficult, in general, to give analytic expressions of the power spectrum in terms of model parameters and one has to resort to numerical computations. Hence, the parameters appearing in eq.~\eqref{Power-spectrum-sharp-feature} can be thought as the `fundamental' ones that one may be interested in reconstructing, and chief amongst them is the frequency $\omega$, indicative of the time at which
the feature arises.

The example in \cref{sec:excited} of a strong sharp turn in two-field inflation provides a useful illustration. With the same notations, one finds \cite{Palma:2020ejf,Fumagalli:2020nvq}
\begin{align}
\label{sharp-turn-result}
\overline{\mathcal{P}}_\zeta(f)&={\cal P}_0 \frac{e^{2 \sqrt{(2-\kappa)\kappa} \, \eta_\perp \delta} }{4 (2-\kappa) \kappa},\qquad A_{\textrm{lin}}=1, \quad \\ \nonumber
\omega&=\frac{2e^{-\delta/2}}{f_\textrm{f}},\qquad \phi_{\textrm{lin}}= 2 \arctan\left[\frac{\kappa}{\sqrt{(2-\kappa)\kappa}}\right] + \pi \;, 
\end{align}
which is valid for $\kappa\equiv f/(f_\textrm{f}\, \eta_\perp)< 2$, and where ${\cal P}_0$ denotes the amplitude of the power spectrum without turn (a large enhancement limit has already been taken in these formul\ae). In this example, the peak of the $\Omega_{\rm GW}$ envelope is given by $f_* = 2 c_s \eta_\perp f_\textrm{f} =4 c_s \eta_\perp e^{-\delta/2} / \omega$. Let us recall that the frequency $\omega_{\rm lin}$ of the oscillations in $\Omega_{\rm GW}$ depends generically on the one in the primordial power spectrum $\omega$ through $\omega_{\rm lin} = c_s^ {-1}\omega $. We thus have the following relation between the peak frequency of the envelope and $\omega_{\rm lin}$: 
\begin{equation}\label{strong turn relation}
\omega_{\rm lin} =4 \eta_\perp e^{-\delta/2}/ f_*.
\end{equation}
In general, the phenomenon of sharp feature leads to $\omega_{\rm lin} = \mathcal{O}(10-100) /f_*$.

\end{description}

\subsection{Logarithmic resonant oscillations}\label{sec:resonant}

\begin{figure}[t]
\centering
\includegraphics[width=0.8\textwidth]{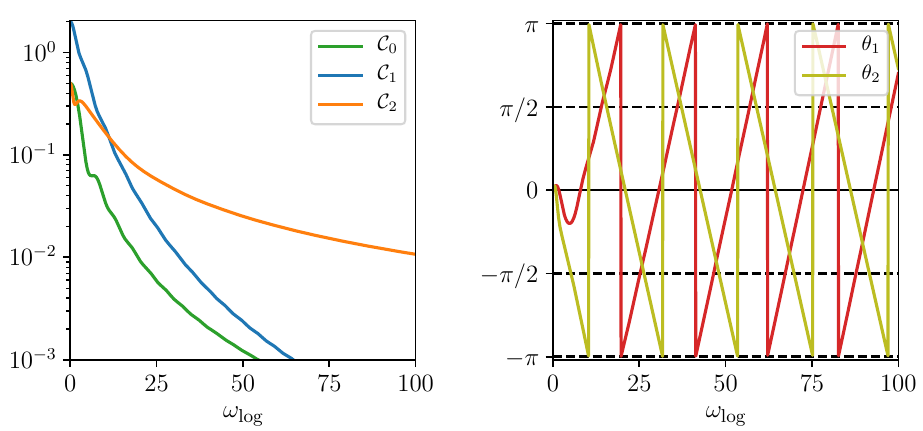}
\caption{\small Functions $\mathcal{C}_{0,1,2}(\omega_\textrm{log})$ and $\theta_{\textrm{log},1,2}(\omega_\textrm{log})$ appearing in the template \eqref{eq:resonant-template-const} and as computed in ref.~\cite{Fumagalli:2021cel} for GWs induced during a period of radiation domination. 
}
\label{fig:C012theta12}
\end{figure}

The logarithmic resonant oscillations (RO) template is given by 
\begin{eqnarray}
    h^2 \Omega_{\textrm{GW}}^{\textrm{RO}}(f,\vec{\theta}_{\rm cosmo}) &=&  \Big\{1+ \mathcal{A}_1(A_{\log},\omega_{\rm log}) \cos \big[\omega_\textrm{log} \ln (f/\textrm{Hz}) + \theta_{\textrm{log},1} \big]
    \label{eq:resonant-template}\\
    &&\hspace{.53cm}+ \mathcal{A}_2(A_{\log},\omega_{\rm log})  \cos \big[2 \omega_\textrm{log} \ln (f/\textrm{Hz}) + \theta_{\textrm{log},2} \big] \Big\} 
    h^2\Omega^{\rm env}_{\textrm{GW}}(f,\vec{\theta}_{\rm env})\,,\nonumber
    \label{eq:resonant-template2}
\end{eqnarray}
with \cite{Fumagalli:2021cel} %
\begin{align}
    \label{eq:resonant-template-const} 
    \mathcal{A}_{1} &= \frac{A_{\textrm{log}} \mathcal{C}_1(\omega_\textrm{log})}{1 + A_{\textrm{log}}^2 \mathcal{C}_0(\omega_\textrm{log})},\qquad
     \theta_{\textrm{log},1}= \phi_\textrm{log}+ \theta_{\textrm{log},1}(\omega_\textrm{log}),  \\
    \nonumber \mathcal{A}_{2} &= \frac{A_{\textrm{log}}^2 \mathcal{C}_2(\omega_\textrm{log})}{1 + A_{\textrm{log}}^2 \mathcal{C}_0(\omega_\textrm{log})},\qquad
    \theta_{\textrm{log},2}= 2 \phi_\textrm{log} + \theta_{\textrm{log},2}(\omega_\textrm{log}) \, ,
\end{align}
where $\mathcal{C}_{0,1,2}(\omega_\textrm{log})$ and $\theta_{\textrm{log},1,2}(\omega_\textrm{log})$ are numerical functions that depend 
on the cosmic expansion at the time the SGWB was produced \cite{Witkowski:2021raz}; these functions are shown in \cref{fig:C012theta12} under the assumption of a conventional radiation-domination era. For reasons explained below, we choose a flat envelope for $h^2 \Omega^{\rm env}_{\textrm{GW}}$.
The model parameters of the RO template are thus $
\vec{p}_{\textrm{RO}} =\{ \Omega_*, A_\textrm{log}, \omega_\textrm{log}, \phi_\textrm{log}\}$, with the latter three parameters controlling the relative amplitude, the period, and the phase of the oscillations.

The RO template describes log-periodic oscillations, of frequencies $\omega_\textrm{log}$ and $2 \omega_\textrm{log}$, that modulate a smooth envelope (scenarios
with more general logarithmic oscillations exist~\cite{Calcagni:2023vxg}). A noteworthy aspect of the template is the qualitative change in the oscillatory structure of \cref{eq:resonant-template} as a function of the frequency $\omega_\textrm{log}$. For values  smaller than the critical frequency $\omega_\textrm{log,c} \simeq 4.77$, the oscillation with frequency $\omega_\textrm{log}$ dominates over the one with double frequency, while the situation is reversed for large frequency values, with the precise cross-over value ${\cal O}(1)\omega_\textrm{log,c}$ depending on $A_{\textrm{log}}$ \cite{Fumagalli:2021cel}.
We set flat priors as follows: 
$\log_{10}(h^2\Omega_*) \in [-30,-5]$,   $\log_{10}(A_{\rm log}) \in [-3,0]$, $\log_{10}(\omega_{\rm log}) \in [0,2]$ and $\phi_{\rm log}\in [-\pi,\pi]$.\footnote{Note that $\omega_\textrm{log}$ is well below the upper bound $\omega_\textrm{log}^{\rm max}=1/\Delta\ln f\simeq f_*/\Delta f\approx 10^4$ due to the frequency resolution $\Delta f$ of LISA around $f_*\approx 2\times 10^{-3}\,{\rm Hz}$ \cite{Calcagni:2023vxg}. For the linear oscillations of  \cref{sec:linear}, one has $\omega_\textrm{lin}^{\rm max}=1/\Delta f\approx 5\times 10^6\,{\rm Hz}^{-1}$, again much larger than the theoretical prior reported above \cref{Power-spectrum-sharp-feature}.}

The RO template  captures well the properties of the \GWB\ induced by primordial density fluctuations with power spectrum 
\begin{align}
    \label{eq:P-resonant}
    \mathcal{P}_\zeta(f)=\overline{\mathcal{P}}_\zeta(f)  \Big[1 + A_{\textrm{log}} \cos \big(\omega_{\textrm{log}} \ln(f/f_*) + \phi_{\textrm{log}} \big) \Big] \, ,
\end{align}
especially for the scales where $\Omega_{\textrm{GW}}$ is maximal \cite{Fumagalli:2020nvq,Fumagalli:2021cel}. For simplicity, here we concentrate on a flat envelope $\overline{\mathcal{P}}_\zeta(f)$ (see ref.~\cite{Fumagalli:2021dtd} for motivations), in which case the mapping from eq.~\eqref{eq:P-resonant} to eq.~\eqref{eq:resonant-template} is exact.

Power spectra of the form \eqref{eq:P-resonant} are characteristic of inflationary models in which some components of the background oscillate with a frequency $\tilde\omega$ larger than the Hubble scale, inducing a resonance with the oscillations of the quantum modes of the density perturbations, and resulting in the spectrum \eqref{eq:P-resonant} with  $\omega_{\textrm{log}} = \tilde\omega/H > 1$ \cite{Chen:2008wn}. These so-called resonant features have been extensively studied on cosmological scales \cite{Slosar:2019gvt}, where observations mandate a small amplitude of oscillations $A_{\textrm{log}}$, a typical example being axion monodromy inflation \cite{Silverstein:2008sg,Flauger:2009ab,Parameswaran:2016qqq,Ozsoy:2018flq}. Scenarios with larger amplitude of  the oscillations, and of the power spectrum, have started to be studied only recently,
see \emph{e.g.,} ~refs.~\cite{Fumagalli:2021cel, Bhattacharya:2022fze, Mavromatos:2022yql}.

\begin{description}
\item[Axion-type inflation.] Explicit models with large resonant features have been studied in refs.~\cite{Bhattacharya:2022fze,Mavromatos:2022yql}. Both are multifield axion models that share the same qualitative features, namely, periodic modulations of the inflationary potential due to subleading nonperturbative corrections, which
lead to periodic violations of slow-roll conditions and  strong sharp turns of the inflationary trajectory,  resulting in the resonant amplification of fluctuations (see also ref.~\cite{Boutivas:2022qtl}). In both cases, an analytical understanding of the link between the microphysical parameters of the inflationary model and the power spectrum is lacking. However, just like the linear-oscillations template, the parameters appearing in eq.~\eqref{eq:P-resonant} can be thought of as the main ones to be reconstructed, since this expression encompasses several realisations of the resonance mechanism. In particular, the frequency $\omega_{\textrm{log}}$ is of prime interest, as it indicates in a model-independent manner the existence of periodic modulations of the inflationary Lagrangian with frequency $\omega_\textrm{log}$ in e-fold.
\end{description}

\subsection{Deformations of the above templates, due to additional physics}\label{sec:deformations}

The embedding of a given inflationary setup may include independent mechanisms that act during or after inflation and deform the \GWB\ frequency shape that is predicted in their absence. By testing such deformations, LISA can probe deviations from the standard model of cosmology and its particle content. 
Including such deformations in the LISA SGWB template bank would make it much larger, with benefits and costs that should be carefully assessed. For this reason, we here list some plausible deformations to remind us of their existence and implications for the template bank, but we leave their detailed analysis to the future.

\subsubsection{Pre-BBN cosmology with matter-domination or kination eras}

The PL template \eqref{eq:PL_template} and the BPL template \eqref{eq:master_modelIV_stBBN} can be used to describe the effect of the Hubble expansion rate after inflation. When the expansion evolution of the universe differs from the radiation-dominated phase, the primordial tilt $n_{t,\rm{prim}}$ is modified by the equation of state of the universe $w$ as~\cite{Seto:2003kc,Boyle:2005se,Boyle:2007zx,Nakayama:2008ip,Kuroyanagi:2011fy,Figueroa:2019paj, ValbusaDallArmi:2020ifo}
\begin{equation}
n_t = n_{t,\rm{prim}}+\frac{2 \left(3w-1 \right)}{3w+1} \;.
\label{nt-w}
\end{equation}
We consider the equation of state $w$
as an additional parameter, within a reasonable range  $w \in[-1/3,\,1]$. In fact $w<-1/3$ would correspond to an accelerated expansion during which the behavior of GWs is different from the one we consider, while $w>1$ is physically not motivated as it implies a sound speed faster than the speed of light. An interesting case is when the transition scale
between two different $w$'s falls within the frequency band in which case LISA may be able to observe the bend of the spectrum. In this case, the spectral tilts $n_{t,1}$ and $n_{t,2}$ are described by the equation of state after and before the transition $w_1$ and $w_2$ 
as in \cref{nt-w}. 
The transition frequency $f_*$ corresponds to the temperature of the universe $T_*$ when the transition takes place, 
\begin{equation}
f_* = 2.6\times 10^{-7} \, \frac{T_*}{\rm GeV} \, \left( \frac{g_*}{106.75} \right)^{1/2} \, \left( \frac{g_{*s}}{106.75} \right)^{-1/3} \, {\rm Hz} \;,
\label{fs-Ts}
\end{equation}
where $g_*$ and $g_{*s}$ are the effective number of relativistic degrees of freedom contributing to the radiation density and to the entropy density, respectively. We have normalized these quantities to the value they assume when all the Standard Model degrees of freedom are relativistic.
The relation (\ref{fs-Ts}) indicates that LISA can probe physics near the $100$ TeV scale.

Any change in the Hubble expansion rate generically affects GWs generated outside the horizon during inflation. Thus, in general, the \GWB\ can be a probe of a non-standard Hubble expansion history after inflation. We note that GWs generated at the horizon entry, \emph{e.g.,}  induced GWs, show different features~\cite{Domenech:2020kqm} because the source term is also affected by a change in the Hubble expansion rate, as mentioned above for the LO and RO templates \cite{Fumagalli:2021dtd}. The following are two examples of this mechanism; in both cases, we take $n_{t,1} \neq n_{t,2}$, and  both positive.

\begin{description}

\item[Early matter phase due to inflaton oscillation.]
An early matter-dominated phase can occur soon after inflation through the oscillation of the inflaton field at the bottom of the quadratic potential, where the field decays into particles and reheats the universe. Such matter phase can also be realised by any types of scalar field~\cite{Kuroyanagi:2013ns,DEramo:2019tit,Kuroyanagi:2020sfw} including the curvaton field~\cite{Nakayama:2009ce} and moduli particles in string theory~\cite{Durrer:2011bi}.
The GW modes which reenter the horizon during the matter-dominated phase have an  $f^{n_{t,\rm{prim}}-2 }$ dependence. If we assume a perturbative decay of the scalar field into fermionic particles, the bend of the spectrum can be described by the following fitting function \cite{Kuroyanagi:2014nba}:
\begin{equation}
\frac{\Omega_{\rm GW}(f)}{\Omega_{\rm GW,0}} =  (1 - 0.22 x_R^{1.5} + 0.65 x_R^2)^{-1} \,,
\label{fitting_MD}
\end{equation}
where $x_R \equiv f/f_R$ and $\Omega_{\rm GW,0}$ is the GW spectrum obtained assuming a radiation-dominated era. 
The transition frequency $f_R$ corresponds to the energy scale at the end of reheating, more precisely expressed akin to eq.~\eqref{fs-Ts} through substitution of $T_*$ with the temperature $T_R$ of the thermal bath at the end of reheating.

\item[Early kination phase.]
In some scenarios of the early universe, the radiation-dominated epoch is followed by a kination epoch, in which the kinetic energy of a scalar field dominates the energy density of the universe. Examples of this type of model are quintessential inflation \cite{Giovannini:1998bp,Peebles:1998qn, Giovannini:1999bh,Giovannini:1999qj,Tashiro:2003qp,Giovannini:2008tm,Ahmad:2019jbm} and particle-physics motivated scenarios \cite{Gouttenoire:2021jhk}. During the kination epoch, the energy density of the scalar field scales as $\rho_\phi \propto a^{-6}$ and gives a Hubble expansion rate $H \propto a^{-3}$, which results in the frequency dependence $\propto f^{n_{t,\rm{prim}}+1}$. In this case, the \GWB\ spectrum is given by
\begin{equation}
\frac{\Omega_{\rm GW}(f)}{\Omega_{\rm GW,0}} =  (1 - 0.5 x_{\rm kin}^{2/3} + 1.27 \, x_{\rm kin}) \,,
\label{fitting_KD}
\end{equation}
where $x_{\rm kin} \equiv f/f_{\rm kin}$ and we have assumed an instantaneous transition~\cite{Figueroa:2019paj, Duval:2024jsg}. The transition frequency $f_{\rm kin}$ is also expressed analogously to eq.~\eqref{fs-Ts} by replacing $T_*$ with $T_{\rm  kin}$, which is the temperature of the universe at the transition from the kination to the radiation-dominated epoch. Note that the detailed shape of the transition depends on the mechanism to end the kination phase. The second term of Eq.~\eqref{fitting_KD} controls the details of the transition and should be modified depending on the transition model. 
\end{description}

\subsubsection{Varying degrees of freedom}

When we consider a contribution of one particle with mass $m$ and degrees of freedom $g$, the feature can be fitted by a hyperbolic tangent function \cite{Caldwell:2018giq}:
\begin{equation}
  \Omega_{\rm GW}(f)=\Omega_{\rm GW,0}(f)F(f,g,m) \,,
\end{equation}
where 
\begin{equation}
  F(f,g,m) = \frac{1-\epsilon(g)\tanh[\ln f/f_0(m)]}{1+\epsilon(g)}\,,
\end{equation}
where $\epsilon(g)=(1-\Delta)/(1+\Delta)$ and $\Delta \simeq (1+g/g_{\rm SM})^{-1/3}$. The characteristic frequency is given by the particle mass as $2\pi f_0(m)=H_* a_*/a_0|_{T\simeq m/b}$ with $b=2.2/\Delta$, which has been determined empirically. Here $H_*$ and $a_*$ correspond to the Hubble rate and the scale factor when the particle became non-relativistic and can be expressed as $H(T_*)=\sqrt{\pi^2 g_*/90} ~ T_*^2/M_{\rm Pl}$ and $a(T_*)=(11 g_*/43)^{-1/3}T_0/T_*$ with $g_*=g_{\rm SM}+g$, where $M_{\rm Pl}$ is the reduced Planck mass.\footnote{Recently, the detectability of the impact of a hypothetical smooth crossover in the early universe beyond the Standard Model of particle physics on the scalar-induced gravitational wave was reported in ref.~\cite{Escriva:2024ivo}.} \\

\section{Template-based reconstruction with the SGWBinner pipeline}
\label{sec:SGWBinner}

This section summarizes the data analysis methods we employ in this work. Most of the analyses rely on the \texttt{SGWBinner} code~\cite{Caprini:2019pxz, Flauger:2020qyi}, which, for this work, has been modified to perform template-based analyses\footnote{We note that the name of our code was named "\texttt{SGWBinner}" after the algorithm employed in Refs.~\cite{Caprini:2019pxz, Flauger:2020qyi}, which was based on splitting the whole frequency range tested by LISA in smaller bins and fitting the signal with a free amplitude and tilt in each bin. The name of the code should not mislead the reader, as we do not use such algorithm in this paper, relying instead on the template-based analysis described in this Section.}. After briefly recalling some features of the LISA \GWB\ data analysis, we discuss the key ingredients of the algorithm and the updates on the astrophysical foregrounds, compared to ref.~\cite{Flauger:2020qyi}. \\

LISA will provide three time-domain data streams $d_i$, with $i$ running over the LISA channels, which we divide into segments of individual duration $\tau$. Then the frequency domain data $\tilde{d}_i(f)$ are defined via Fourier transform as
\begin{equation}
	\tilde{d}_i(f) = \int_{-\tau/2}^{\tau/2} d_i(t) \; \textrm{e}^{- 2 \pi i f t }  \; \textrm{d} t\; .
\end{equation}
where for simplicity the central time of that segment has been set to zero. In the following, we assume that appropriate methods, \emph{e.g.,}  some procedures to be integrated within the LISA global fit scheme~\cite{Cornish:2005qw, Vallisneri:2008ye, MockLISADataChallengeTaskForce:2009wir, Littenberg:2023xpl}, remove all transients, including loud deterministic signals and glitches, from the data stream, leaving `clean data' consisting only of stochastic components. Under this assumption, we express the data as a superposition of some noises $\tilde{n}^{\nu}_i$ and signals $\tilde{s}_i^\sigma$ as
\begin{equation}
 \tilde{d}_i(f) = \sum_\nu \tilde{n}^\nu_i(f) + \sum_\sigma \tilde{s}^\sigma_i(f) \; ,
\end{equation}
with $\nu$, $\sigma$ running respectively over the different noise and signal components.
We make the hypothesis that each noise and signal component is independent of the others and characterised solely by specific statistical properties. Additionally, our analysis is limited to signals that are Gaussian, isotropic, stationary, and non-chiral.\footnote{In principle, the real signal might violate all these assumptions. For the impact of anisotropies and non-Gaussianities see, \emph{e.g.,}  refs.~\cite{Bartolo:2019oiq, Bartolo:2019yeu, Contaldi:2020rht, LISACosmologyWorkingGroup:2022kbp,Dimastrogiovanni:2022eir,Malhotra:2022ply,Schulze:2023ich,Mentasti:2023icu,Mentasti:2023uyi,Perna:2024ehx}. While most early Universe mechanisms predict stationary signals, the presence of anisotropies, projected in the data by a time-varying and sky-dependent response function, will induce time modulations in the measurements. This effect is well-known to be present, \emph{e.g.,}  for the astrophysical foreground (see discussion around~\cref{eq:signal_spectrum}) due to the incoherent superposition of signals from compact binaries in our galaxy~\cite{Adams:2013qma}. While including this effect in the analysis might ease the separation of the different components, we ignore it in the present work. Finally, the problem of detecting chirality with a planar interferometer, like LISA, is highly non-trivial. By construction, planar interferometers are insensitive to chirality~\cite{Seto:2007tn, Seto:2008sr, Smith:2016jqs}. However, cross-correlating the measurements of different, and non-coplanar, detectors~\cite{Seto:2007tn, Seto:2008sr, Crowder:2012ik, Orlando:2020oko, ValbusaDallArmi:2023ydl}, or using the dipole induced by the detector motion with respect to the SGWB frame~\cite{Seto:2006hf, Seto:2006dz, Domcke:2019zls}, might help to overcome this limitation.} Stationarity and Gaussianity are assumed to hold for noise, too.\footnote{As for the signal, these hypotheses might be violated by real data. Transients (\emph{e.g.,} glitches) and other effects, such as modulations of the noise due to instrumental component degradation, might induce non-stationarities and non-Gaussianities in the noise. While transients will be systematically modeled and removed from the data stream (see, \emph{e.g.,} , refs.~\cite{Robson:2018jly, Baghi:2021tfd}), the real analysis will consistently keep track of long-term modulations in the noise model.} Thus, assuming both components to have zero mean, we obtain
\begin{equation}
    \label{eq:data_statistics}
	\langle  \tilde{d}_i(f) \rangle  = 0  \; ,  \qquad \langle  \tilde{d}_i(f) \tilde{d}^{*}_j(f^{\prime})  \rangle  = \frac{\delta(f -f')}{2} \left[  \sum_\nu P^\nu_{N,ij}(f)  + \sum_\sigma P^\sigma_{S,ij}(f) \right]   \; , 
\end{equation}
where $P^\nu_{N,ij}$ and $P^\sigma_{S,ij}$ denote the noise and signal power spectra respectively and the brackets denote an ensemble average.\footnote{The Dirac delta in frequency arises from stationarity, and, in reality, it would only be an exact Dirac delta in the limit of infinite observation time $T_{\rm obs}$. For finite observation time, a sinc$[T_{\rm obs} (f-f')]$ function appears in~\cref{eq:noise_spectrum}. For simplicity, we restrict ourselves to the limit where $T_{\rm obs} f \ll 1 $, where the sinc can be replaced by a Dirac delta.} Notice that, in reality, each noise and signal component is a combination of some `physical' spectrum and a response (or transfer) function, which projects the spectrum onto the data stream.  In the following, we will respectively denote with $T^{\nu}_{ij,lk}$ and $\mathcal{R}_{ij}$ the transfer functions for the noise and signal components (see refs.~\cite{Flauger:2020qyi, Hartwig:2021mzw, Nam:2022rqg} for details). While each noise component describes a different physical effect that propagates differently through Time-Delay Interferometry (TDI), leading to a different transfer function, all the signal components are GW signals, thus sharing a common response function.

\subsection{TDI, signal, and noise description }
\label{sec:TDI_noise_foregrounds}

The laser frequency contribution~\cite{LISA_performance} is the dominant source of instrumental noise for LISA. To suppress this large noise component and to allow any GW detection, Time-Delay Interferometry (TDI) is employed~\cite{Tinto:2020fcc}. TDI is a post-processing technique that, combining measurements performed at different times, produces synthesized data streams representing laser-noise-free virtual interferometers. In the following, we denote with $\eta_{ij}(t)$ the phase measurement performed in spacecraft $i$ at time $t$ of a signal emitted from spacecraft $j$ at time $t - L_{ij}$, where $L_{ij}$ is the distance between the two spacecrafts. Moreover, we denote with $D_{ij}$ the delay operator defined as $D_{ij} x(t) \equiv x(t - L_{ij})$. In practice, TDI consists in defining variables as a linear combination of single-link measurements and delay operators. As shown in refs.~\cite{Armstrong_1999, Prince:2002hp, Shaddock:2003bc, Shaddock:2003dj, Tinto:2003vj, Vallisneri:2005ji, Muratore:2020mdf, Muratore:2021uqj}, different TDI combinations lead to laser-noise suppression, each with distinct sensitivities to GW signals and instrumental noise. The most common choice for LISA data analysis is to use three Michelson-like variables\footnote{In this work, we consider ``first-generation'' TDI variables only. These variables achieve laser-noise cancellation in a scenario that respects our working assumptions. More realistic investigations, which involve, \emph{e.g.,} , time-evolving unequal arms, would require ``second-generation'' TDI variables~\cite{Vallisneri:2005ji,Muratore:2020mdf,Muratore:2021uqj,Hartwig:2021mzw}.}, denoted as X, Y, and Z, with X defined as
	\begin{subequations}\label{eq:tdi-definition}
		\begin{equation}
			{\rm X}  = (1 - D_{13}D_{31})(\eta_{12} + D_{12} \eta_{21}) + (D_{12}D_{21} - 1)(\eta_{13} + D_{13} \eta_{31}) \; ,
		\end{equation}
and Y and Z variables are obtained through cyclic permutations of the indexes. The XYZ variables are often recombined into (quasi-)orthogonal channels, typically referred to as A, E, and T defined as~\cite{Prince:2002hp}
	\begin{equation}
		{\rm A} = \frac{{\rm Z} - {\rm X}}{\sqrt{2}}\;, \qquad  {\rm E} = \frac{{\rm X} - 2 {\rm Y} + {\rm Z}}{\sqrt{6}} \;, \qquad  {\rm T}=\frac{{\rm X} + {\rm Y} + {\rm Z}}{\sqrt{3}}  \; .
	\end{equation}
 \end{subequations}
It is known that for configurations with equal arm lengths and the same noise levels for all spacecrafts, the AET combination is perfectly diagonal. Moreover, under these assumptions, the T channel strongly suppresses GW signals compared to instrumental noise, effectively providing a noise monitor. Since in this work, we do not violate these assumptions\footnote{See, \emph{e.g.,} , ref.~\cite{Hartwig:2021mzw} for a recent analysis for a non-equilateral geometry and unequal noise in the three spacecraft using different sets of TDI variables.}, we focus on this particular TDI basis, significantly simplifying the computations required, \emph{e.g.,}  for likelihood evaluation. This feature makes the AET channels particularly appealing for our analysis.  \\

Assuming TDI suppresses laser noise, the main residual noise sources for LISA (also known as secondary noises) are the Test Mass (TM) noise, representing deviations from free-fall in the TM trajectories, and the Optical Metrology System (OMS) noise, representing uncertainties in the determination of the TM positions. The total noise power spectrum for any TDI variable can thus be expressed as
\begin{equation}
	\label{eq:noise_spectrum}
	P_{N,ij}(f) \equiv   \sum_\nu P^\nu_{N,ij}(f)  =  \left[ T^{ \textrm{TM} }_{ij,lk}(f) S^{ \textrm{TM}}_{lk }(f) +T^{ \textrm{OMS} }_{ij,lk}(f) S^{ \textrm{OMS}}_{lk }(f)  \right]\; ,
\end{equation}
where $T^{ \textrm{TM}}_{ij,lk}$ and $T^{ \textrm{OMS}}_{ij,lk}$ are the TM and OMS transfer functions, projecting the individual TM and OMS noise components onto the TDI variable. $S^{ \textrm{TM} }$ and $S^{ \textrm{OMS} }$ are estimated as~\cite{LISA:2017pwj}
\begin{align}
\label{eq:Acc_noise_def}
S^\text{TM}_{lk}(f) & = A_{lk}^2 \;  \left(1 + \left(\frac{0.4 \textrm{mHz}}{f}\right)^2\right)\left(1 + \left(\frac{f}{8  \textrm{mHz}}\right)^4\right) \left(\frac{1}{2 \pi f c}\right)^2 \;  \left( \frac{\mathrm{fm}^2 }{ \mathrm{s}^3 } \right) \;,  \\ 
\label{eq:OMS_noise_def}
S^\text{OMS}_{lk}(f) & = P_{lk}^2 \; \left(1 + \left(\frac{2\times10^{-3} \textrm{Hz}}{f}\right)^4 \right) \times \left(\frac{2 \pi f}{c}\right)^2 \; \times \left(\frac{ \mathrm{pm}^2 }{ \mathrm{Hz} }\right) \;.
\end{align}
Under the assumption of equal arms and noise levels, we can simplify the above coefficients
as products of Kronecker $\delta$s and two constant parameters, $A_{lk}^2 = A^2 \, \delta_{lk}$ and $P_{lk}^2 = P^2 \, \delta_{lk}$. It is worth noting that, for the A and E channels, the TM noise dominates at lower frequencies, while the OMS noise dominates at higher frequencies. On the contrary, the OMS noise prevails across all frequencies for the T channel. In the analyses performed in this paper, we apply Gaussian priors to $A$ and $P$, which are centered on their nominal values, 3 and 15, and have a width of 20\%.\\

Beyond cosmological signals, the \GWB\ signal in the LISA band will have two guaranteed contributions from astrophysical sources. At low frequency, there will be a stochastic contribution from many unresolved Compact Galactic Binaries (CGBs)~\cite{Nissanke:2012eh}. At larger frequencies, there will be a signal due to the incoherent superposition of Stellar Origin Black Hole Binaries (SOBHBs) and Binary Neutron Stars (BNSs)~\cite{Regimbau:2011rp, KAGRA:2021kbb} (see also~\cite{Perigois:2020ymr, Babak:2023lro, Lehoucq:2023zlt}). As a consequence, the signal part of~\cref{eq:data_statistics} can be expressed as
\begin{equation}
	\label{eq:signal_spectrum}
	P_{S,ij}(f) \equiv   \sum_\sigma P^\sigma_{S,ij}(f)  =  \mathcal{R}_{ij}(f)  \left[ S_{ \textrm{Gal} }(f) +S_{ \textrm{Ext} }(f)  + S_{ \textrm{Cosmo} }(f) \right]\; , 
\end{equation}
where $\mathcal{R}_{ij}(f)$ is the LISA response function (see refs.~\cite{Flauger:2020qyi} and Appendix A of~\cite{Caprini:2024hue} for more details), $S_{ \textrm{Gal} }(f)$ and $S_{ \textrm{Ext} }(f)$ represent the galactic and extragalactic foreground contribution respectively, and the last contribution is given by all the signal templates listed in~\cref{sec:models} and related to the underlying inflationary mechanism. Furthermore, in order not to introduce degeneracies between different inflationary templates, we will assume $S_{ \textrm{Cosmo} }(f)$ to be represented only by one template at a time. The GW strain power spectral density (PSD) is related to $\Omega_{\textrm{GW}}(f)$, energy density per logarithmic frequency interval normalized by the critical density, defined in eq.~\eqref{eq:energy density}, through
\begin{equation}
h^2  \Omega_{\textrm{GW}}(f) \equiv \frac{4\pi^2 f^3}{3H_0^2} S_{h}(f )  \;. 
\end{equation}
This provides a rescaling between $S$ and $\Omega$, which we employ also for the noise. 

The two foreground models we adopt in this paper are the state-of-the-art spectral models. In particular, for the CGB we use the model of ref.~\cite{Karnesis:2021tsh}:
\begin{equation}
\label{eq:SGWB_gal}
 S_{\textrm{Gal}}(f)=A_{\textrm{Gal}}\left(\frac{f}{ 1\,\textrm{Hz}}\right)^{-\frac{7}{3}} \times  e^{-(f/f_1)^\alpha}\times 
 \frac{1}{2}\left[1+\tanh{\frac{f_{\textrm{knee}}-f}{f_2}}\right]\;,
 \end{equation}
 where the first factor comes from the superposition of many inspiraling signals~\cite{Phinney:2001di}, the second factor is due to the loss of stochasticity at higher frequencies, and the last factor, which produces a sharp cut-off in the spectrum, models the complete removal of binaries emitting at sufficiently large frequency. For what concerns the values of the parameters used for the injection, we use the time-dependent parameterisation introduced in ref.~\cite{Karnesis:2021tsh}:   
 \begin{eqnarray}
 \log_{10} (f_1) &=& a_1 \log_{10}(T_{\mathrm obs}) + b_1\,,\nonumber\\
 \log_{10} (f_{\textrm{knee}}) &=& a_k \log_{10}(T_{\mathrm obs}) + b_k\,,
 \end{eqnarray}
and set $T_{\mathrm obs}=4$ years with $100\%$ duty cycle, $a_1 = -0.15;~b_1=-2.72;  ~a_k = -0.37;~b_k=-2.49$, together with $A_{\textrm{Gal}} = 1.15\cdot 10^{-44};~\alpha = 1.56; f_2 = 6.7 \times 10^{-4}$Hz. In principle, all these should be measured together with the parameters of the SGWB of cosmological origin. In practice, we restrict our analysis by varying only the amplitude parameter $A_{\textrm{Gal}}$. In reality, since we work in $\Omega_{\rm{GW}}$ units, the parameter we use in the analysis is $\log_{10} (h^2 \Omega_{\rm Gal}) \equiv \log_{10} [ 4 \pi^2 A_{\textrm{Gal}} / ( 3 H_0^2) ]$. Finally, we impose a Gaussian prior on $\log_{10} (h^2 \Omega_{\rm Gal})$ with central value $\simeq -7.8412$ and standard deviation $\simeq 0.21$. \\

For the extragalactic contribution, the background signal can be adequately described by a power-law model with a fixed value for the tilt:
\begin{equation}
    \label{eq:Ext_template}
  h^2 \Omega_{\textrm{Ext}}  = 10^{\log_{10} (h^2 \Omega_{\rm Ext})}\left(\frac{f}{0.001 {\mathrm Hz}}\right)^{2/3} \; . 
  \end{equation}
  As for the galactic template in~\cref{eq:SGWB_gal}, the tilt comes from the superposition of many signals in the inspiral phase~\cite{Phinney:2001di}. While the subtraction of sufficiently loud events might lead to deviations from this behavior, this effect is expected to be small~\cite{Babak:2023lro}. For this reason, in the present work, we assume the parameterisation in~\cref{eq:Ext_template}, controlled by the amplitude parameter only, to suffice. Recent observations by the LVK collaboration~\cite{KAGRA:2021duu} suggest that the \GWB\ due to SOBHB and BNS should have $\Omega_{\textrm{SOBHB+BNS}} (25 \textrm{Hz})= 7.2^{+3.3}_{-2.3} \times 10^{-10}$ which, rescaled at LISA frequency $(f_*=10^{-3}\textrm{Hz})$ implies $\log_{10} (h^2 \Omega_{\rm Ext}) \simeq -12.38 $. In our analyses, we impose a Gaussian prior on this parameter, centered around such a value, and with a standard deviation equal to $\simeq 0.17 $. 

\subsection{Data generation, likelihood, and Fisher matrix}
\label{subsec:data_generation}

LISA is scheduled to function for at least 4.5 years up to a maximum of 10 years. Operations such as antenna repointing introduce  
(periodic) data gaps in the schedule, resulting in a duty cycle of about $82\,\%$~\cite{Colpi:2024xhw}. We assume an intermediate scenario, setting the \texttt{SGWBinner} code to work with $N_d=126$ data segments of duration $\tau=11.4$ days each. This sums up to $T_{\rm obs} =4$ effective years of data. Then, we perform a Fourier transform in each of the segments to get the frequency domain data $\tilde{d}^s_i(f_\textmd{k})$, where $s$ indexes segments, $\textmd{k}$ indexes frequencies within the detector range and $i$ indexes the TDI channels. As mentioned above, we assume a segment duration of approximately 11.5 days, corresponding to $\Delta f = N_d/T_{\rm obs} \sim 10^{-6}$ Hz. Under these assumptions, we generate $N_d$ Gaussian realisations for the signal and all noise components, with zero mean and variances defined by their respective power spectral densities. To lower the numerical complexity of the problem we perform two operations. Firstly, we define a new set $\bar{D}^\textmd{k}_{ij} \equiv \sum_{s = 1}^{N_d} \tilde{d}^s_i(f_\textmd{k}) \tilde{d}^s_j(f_\textmd{k}) / N_d$, by averaging over segments. Then, we down-sample these data using the coarse-graining procedure introduced in~\cite{Caprini:2019pxz, Flauger:2020qyi}. By applying these techniques, we obtain a new data set $D^k_{ij}$, where $k$ now indexes a sparser set of frequencies $f^{k}_{ij}$. These frequencies are weighted according to $w^k_{ij}$, which corresponds to the number of points averaged over during the coarse-graining procedure. The down-sampled data set retains statistical properties similar to those of $\bar{D}^\textmd{k}_{ij}$,  while being computationally more manageable.\\

The full likelihood employed in our analyses reads
\begin{equation}
\label{eq:likelihood}
\ln \mathcal{L} (\vec{\theta} ) = \frac{1}{3} \ln \mathcal{L}_{\mathrm G} (\vec{\theta} | D^k_{ij}) +  \frac{2}{3} \ln \mathcal{L}_{\mathrm{LN}} (\vec{\theta} | D^k_{ij}) \; ,
\end{equation}
given by the sum of a Gaussian
\begin{equation}
\ln \mathcal{L}_{\mathrm G} (\vec{\theta} | D^k_{ij}) = -\frac{N_d}{2} \sum_{k} \sum_{i,j} w^k_{ij} \left[ 1 - D^k_{ij} / D^{\mathrm Th}_{ij} (f^{k}_{ij}, \vec{\theta}) \right]^2  \; ,
\end{equation}
and of a log-normal component
\begin{equation}
\ln \mathcal{L}_{\mathrm{LN}} (\vec{\theta} | D^k_{ij}) = -\frac{N_d}{2} \sum_{k} \sum_{i,j} w^k_{ij} \ln^2 \left[ D^{\mathrm Th}_{ij} (f^{k}_{ij}, \vec{\theta}) / D^k_{ij}   \right] \; .
\end{equation}
The latter  is included to take into account the mild non-Gaussianity introduced by the data generation, and to avoid biased results~\cite{Bond:1998qg, Sievers:2002tq, WMAP:2003pyh, Hamimeche:2008ai}. Here 
$\vec{\theta} = \{\vec{\theta}_s, \vec{\theta}_n\} $ is the vector of signal and noise parameters
  \begin{equation}
      \vec{\theta}_n=\{A,P\}\;,    \qquad   \quad \vec{\theta}_s=\{\vec{\theta}_\textrm{fg},\vec{\theta}_\textrm{cosmo}\} \;,
      \label{eq:vec_param}
  \end{equation}
  with  
  \begin{eqnarray}
      &&\vec{\theta}_\textrm{fg} = \{ h^2 \Omega_{\rm Gal}, h^2 \Omega_{\rm Ext}\} 
      \;, \nonumber
      \\
        &&\vec{\theta}_{\textrm{cosmo}}   
        \qquad\textrm{(template dependent)}\;,
         \label{eq:vec_param_explicit}
    \end{eqnarray}
 while $D^{\mathrm Th}_{ij} (f_k, \vec{\theta}) $ is the theoretical model for the data (containing both signal and noise). As
 explained above,  under the assumptions discussed in the previous section, the AET TDI basis is diagonal, so that $D^k_{ij} \neq 0$ only for $i =j$, which further simplifies the analysis. The priors for the signal, foreground, and noise parameters are added to~\cref{eq:likelihood} to get the posterior distribution. While the noise and foreground priors are discussed in~\cref{sec:TDI_noise_foregrounds}, the priors for the signal parameters of each template are set according to the discussion in~\cref{sec:models}. Finally, we sample the parameter space using the nested  sampling algorithm implemented in \texttt{Polychord}~\cite{Handley:2015vkr, Handley:2015fda}, via its \texttt{Cobaya}~\cite{Torrado:2020dgo} interface, and visualise results using \texttt{GetDist}~\cite{Lewis:2019xzd}.\\

We conclude this section by recalling the main ingredients of the Fisher Information Matrix (FIM) formalism. The log-likelihood for Gaussian and zero mean data $\tilde{d}_i^k$, with $i$ running over TDI channels and $k$ running over frequency bins $f_k$, described only by their variance, say $C_{ij}(f_k,\vec{\theta})$, can be written as
\begin{equation}
- \ln  \mathcal{L} (\tilde{d}^k_i \vert \vec{\theta} ) \propto \sum_k \left\{ \ln \left\{ \det[C_{ij}(f_k,\vec{\theta})] \right\} + \tilde{d}^k_i \, C^{-1}_{ij} (f_k,\vec{\theta}) \, \tilde{d}_j^{k*} \right\}\; , 
\end{equation}
which is also known as Whittle likelihood. The FIM $F_{\alpha \beta}$, representing the information on the model parameters, is defined as
\begin{equation}
\label{eq:FIM_definition}
F_{\alpha \beta} \equiv - \left. \frac{\partial^2 \ln  \mathcal{L} }{ \partial \theta^\alpha \partial \theta^\beta } \right|_{\vec{\theta} =  \vec{\theta}_0} = \sum_k \textrm{Tr} \left[ C^{-1}  \frac{\partial C}{\partial \theta^\alpha} C^{-1} \frac{\partial C}{\partial \theta^\beta} \right]_{\vec{\theta} =  \vec{\theta}_0} \; ,
\end{equation}
where, $\vec{\theta}_0$ represent the best-fit parameter(s), determined by solving
\begin{equation}
   \left. \frac{\partial \ln \mathcal{L} }{ \partial \theta^\alpha  } \right|_{\vec{\theta} =  \vec{\theta}_0 } \propto \sum_k \frac{ \partial  C_{ji}}{\partial \theta^\alpha} \left[ C^{-1}_{ij} -  C^{-1}_{im} \;\tilde{d}_m^k \tilde{d}_n^{k*} \; C^{-1}_{nj}  \right]   = 0  \; ,
\end{equation}
under the assumption $C_{ij}(f_k, \theta_0) = \tilde{d}^k_i \tilde{d}_{j}^{k*}$. In practice, the discrete sum over finite frequencies can be replaced with a continuous integral over the frequency range as
\begin{equation}
    \label{eq:FIM_final}
    F_{\alpha \beta} \equiv T_{\rm obs} \sum_{i \in \{ {\rm AET} \} }\int_{f_{\mathrm{min}}}^{f_{\mathrm{max}}} \frac{\partial \ln C_{ii}}{\partial \theta^\alpha} \frac{\partial \ln C_{ii}}{\partial \theta^\beta} \, \textrm{d}f \; , 
\end{equation}
where $f_{\mathrm{min}}$, $f_{\mathrm{max}}$ are the minimal and maximal frequencies measured by the detector, which we assume to be $f_{\mathrm{min}} = 3 \times 10^{-5}$\,Hz and $f_{\mathrm{max}} = 0.5$\,Hz~\cite{Colpi:2024xhw}, while we remind that $T_{\rm obs}$ is the total observation time and that we have exploited the fact that the AET basis is diagonal. If non-trivial (log-)priors are included in the analysis, their derivatives should be consistently added to~\cref{eq:FIM_final} to get the full FIM. Finally, the covariance matrix $C_{\alpha \beta}$, which provides estimates on the determination and (on the covariance) of the model parameters, is computed by inverting the FIM.  When we provide the errors on a given parameter, we marginalize over all the other ones, which amounts in taking the square root of the diagonal element of the inverse of the correlation matrix for that parameter as its error.

Given its computational efficiency, in the following section we employ the FIM approach to scan the parameter space of the templates introduced in~\cref{sec:models}, and to assess the prospect of reconstructing the template parameters with some level of accuracy.  It is worth stressing that the FIM formalism only works under the assumption that the likelihood is well approximated by a Gaussian distribution in the model parameters around the best fit. 
 Typically, a rule of thumb to assess the quality of the FIM approximation is through the Signal-to-Noise ratio (SNR), defined as~\cite{Romano:2016dpx}
\begin{equation}
\label{eq:SNR_def}
\textrm{SNR} \equiv \sqrt{T_{\rm obs} \;\sum_{i \in \{ {\rm AET} \} } \int_{f_{\rm min}}^{f_{\rm max}} \left( \frac{S_{i, \mathrm{ GW}}}{S_{i, \mathrm N}} \right)^2 \; \textrm{d}f }  \; . 
\end{equation}
The SNR scales linearly with the signal amplitude and with the square root of the observation time. Given the limitations of the FIM, as already mentioned, we will test the validity of the FIM results by directly sampling the likelihood using Nested sampling (see also refs.~\cite{Boileau:2021gbr,Boileau:2021sni,Boileau:2022ter} for a similar approach).

\section{Reconstruction forecasts with the SGWBinner}
\label{sec:results}

In this section we apply the methodology outlined in \cref{sec:SGWBinner} to forecast the  capabilities of LISA to reconstruct Gaussian, isotropic SGWB signals characterised by the templates  discussed in \cref{sec:models}.  We focus our analysis on the frequency structure of the templates and the parameters characterising them. In
\cref{sec:interpr} we then discuss the physical consequences of our findings for inflationary models.

The results we present are based on several assumptions (some of them anticipated in \cref{sec:intro}). We assume that any potential discrepancy between instrumental noise and the noise model, as well as between the SGWB signal and the chosen template, introduces systematic errors below the level of statistical uncertainties. 

The analysis of each template follows the same rationale. 
We initially employ a FIM formalism to forecast the reconstruction errors on parameters of each inflationary template, taking into account the reconstruction uncertainties of the instrumental noise and foregrounds. 
In this way, for every inflationary template parameter $\theta_{\textrm{cosmo},i}$, we compute its Fisher reconstruction error marginalized over all the reconstruction uncertainties on the remaining signal and noise parameters, namely $\vec{\theta}_n$, $\vec{\theta}_\textrm{fg}$ and $\theta_{\textrm{cosmo},j}$ with every $j\ne i$ (see \cref{eq:vec_param}). We then present our Fisher results in two-dimensional color maps displaying how each error changes when varying the injected values of a pair of inflationary template parameters and leaving the others parameters at some given fixed injected values (see, \emph{e.g.,}  \cref{fig:Fisher_PL}).
To facilitate the interpretation we also highlight specific contour lines for the errors $1 \%$ and  $30 \%$, or $0.01$ and $0.3$, 
depending on whether the relative or absolute error is reported in the maps. To investigate the impact of the astrophysical foregrounds, such contour lines are determined both in the case where the foreground parameters are known a priori and in the case they are reconstructed together with the inflationary and noise template parameters.

The Fisher analysis has a drawback since it assumes the likelihood to be Gaussian in the inflationary signal, foregrounds, and noise parameters. Due to numerical precision limitations, it also struggles in dealing with strong multidimensional correlations among parameters differing by many orders of magnitude.
In our analysis
we test the Fisher approximation on signals following our templates. In most of our tests, we find that the Fisher approximation reliably estimates the order of magnitude of the errors when 
the signals have ${\rm SNR}\gtrsim 10$ and the estimated errors are much smaller than the injected values. 
Therefore, the results of our Fisher analysis must  be interpreted with caution when these two conditions are not met.

We then pursue the analysis of each template focusing on benchmarks.
For each benchmark, we perform a complementary Bayesian template-based 
analysis by means of  the 
SGWBinner \cite{Caprini:2019pxz, Flauger:2020qyi}, and we present triangle plots showing 1D and 2D posterior distributions for the parameters of the primordial signal obtained by directly sampling the likelihood using nested sampling (see \emph{e.g.,} ~\cref{fig:PL_MCMC}). 
For these benchmarks, in order to elucidate the Fisher analysis accuracy, we also plot the 1D and 2D contours from the Fisher results. This helps capture some features that the Fisher analysis does not catch. 
For our purposes, we consider one simulation per benchmark, but it is worth noting that finer details of the results can exhibit some realisation dependencies. In addition to the triangle plots, we use the samples from the nested sampling analyses to plot the functional posterior distribution of noise, foregrounds and primordial signal. The latter, often called predictive posterior distribution, provides complementary insights into the ability of LISA to reconstruct the benchmark signals. For these plots, we use the publicly available \texttt{fgivenx} code~\cite{fgivenx,Handley:2019fll}.

The anisotropic nature of the galactic foreground gives rise to a  time-dependent modulation of the signal, as the detector moves through space~\cite{Allen:1996gp,Cornish:2003tz}. The improvement is however limited~\cite{Mentasti:2023uyi}, and our analysis does not leverage this feature, opting instead for considering the average signal integrated over the mission duration. 
 Therefore, our work may be suboptimal in this respect. For this reason, we present comparisons between the reconstruction forecasts with and without astrophysical components.
In the latter case, the results can serve as  reference for the  separation of the astrophysical and cosmological components of the  SGWB.
In general, we expect that
 the presence of foregrounds impacts the reconstruction in two ways: foregrounds can introduce degeneracies among parameters, and the galactic (extragalactic) foreground can cover the primordial signal in a large (small) fraction of the low-frequency (high-frequency) sensitivity region.  We will meet examples of these phenomena in our analysis below. 

\subsection{Forecasts for the power law template}
\label{sec_forPL}
\begin{figure}
	\centering
	\includegraphics[width=0.49\textwidth]{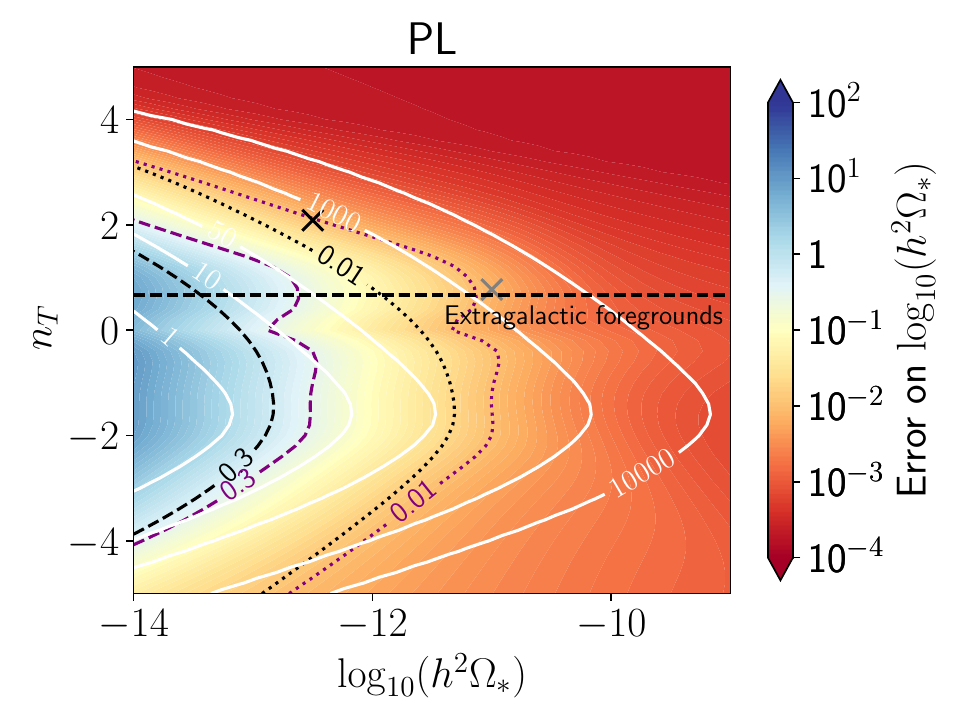} 
	\includegraphics[width=0.49\columnwidth]{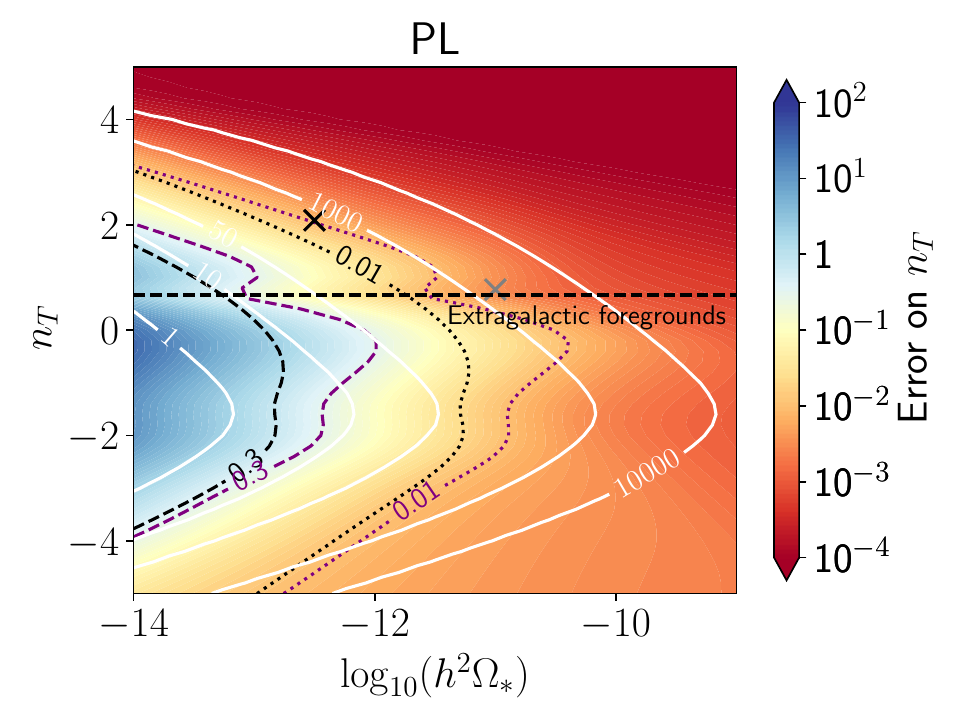} 
	\caption{\small Fisher forecast for the PL template. The color map shows the 68 \% CL errors on the amplitude (left) and spectral index (right) as a function of the value of the injected parameters. SNR contour lines  are plotted in white. The pairs of dashed (dotted) lines mark the $\sigma =0.3$ ($\sigma =0.01$) contours, respectively in the absence (black) and in the presence (purple) of foregrounds. The black and gray crosses display the benchmarks PL-BNK\_1 and PL-BNK\_2, respectively. We also highlight the line $n_T=2/3$, which corresponds to the spectral index of extra-galactic foregrounds, with a black dashed line.
	}
	\label{fig:Fisher_PL}
\end{figure}

\begin{figure}
	\begin{center}
		\includegraphics[width=.495\columnwidth]{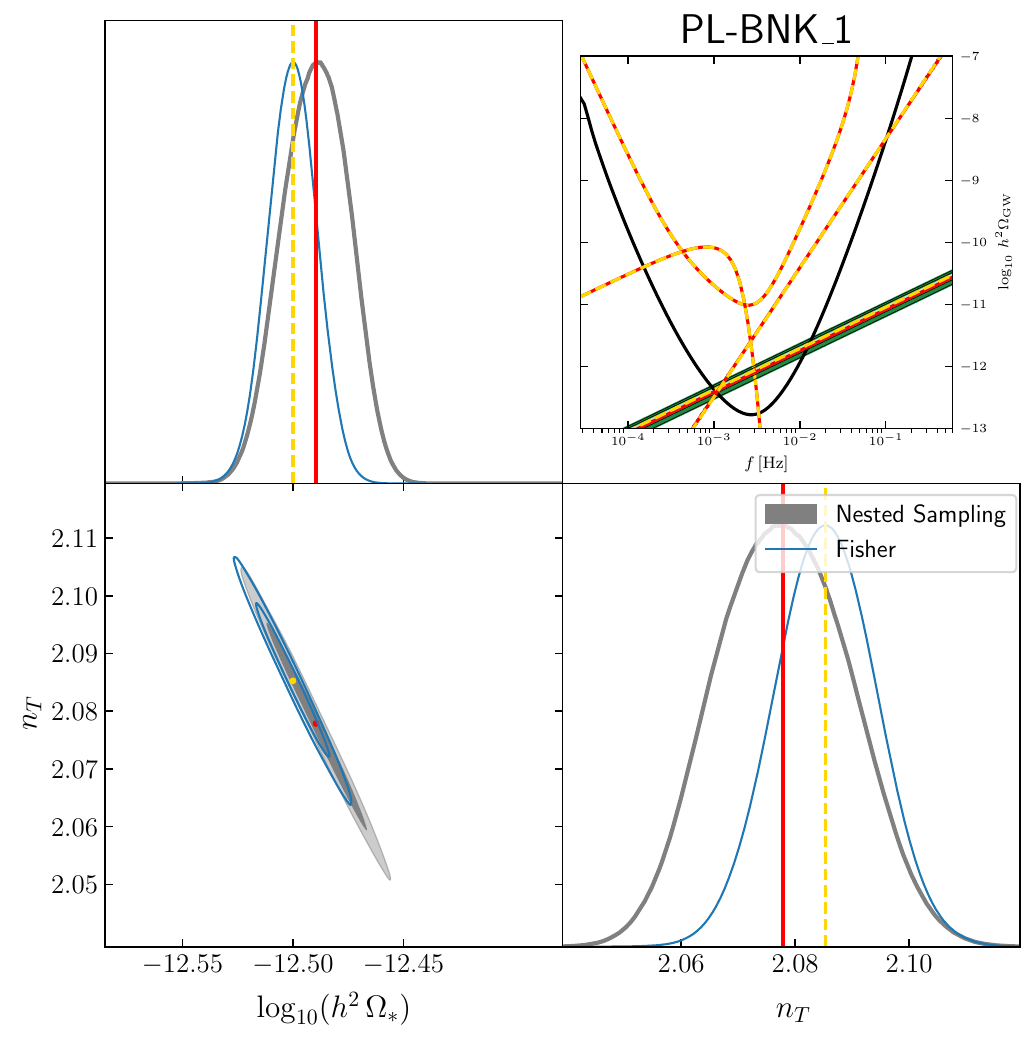}
		\includegraphics[width=.495\columnwidth]{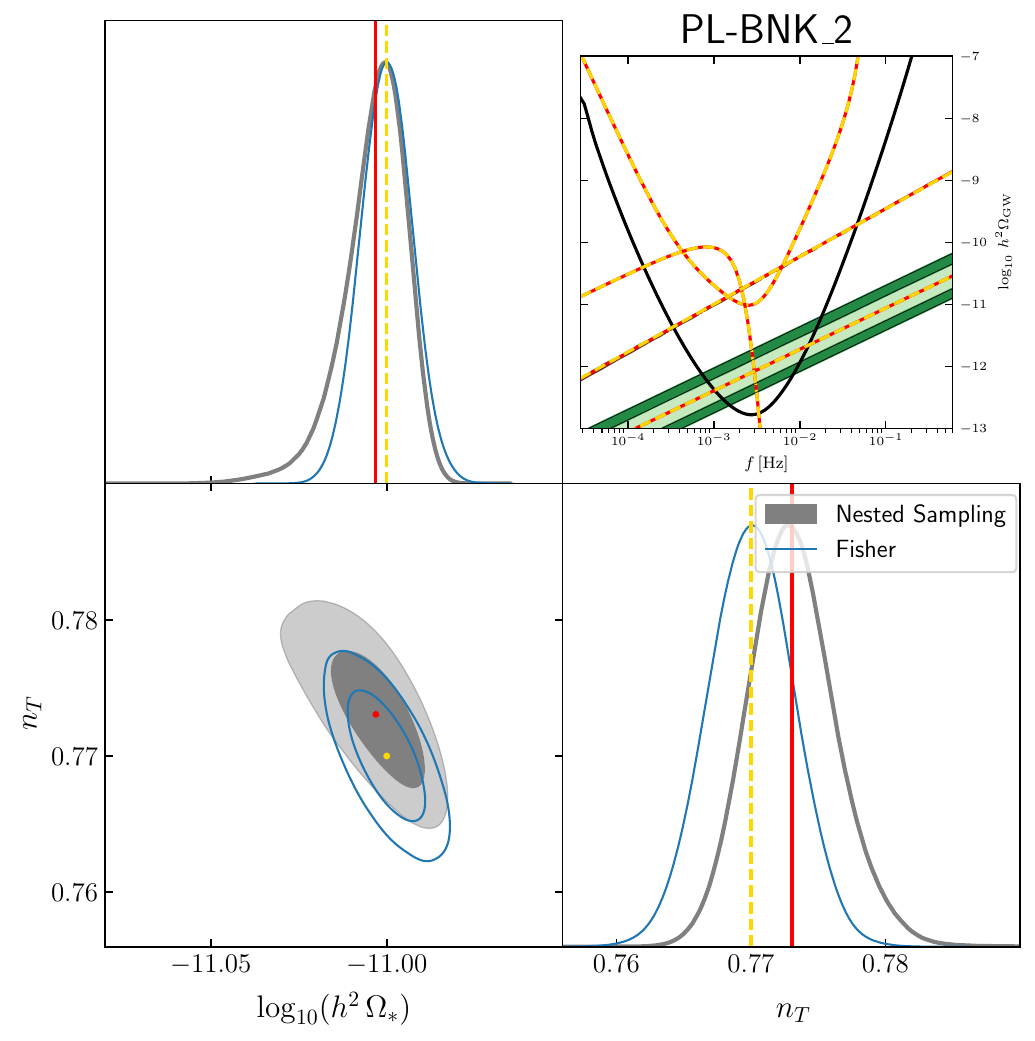}
        \includegraphics[width=\columnwidth]{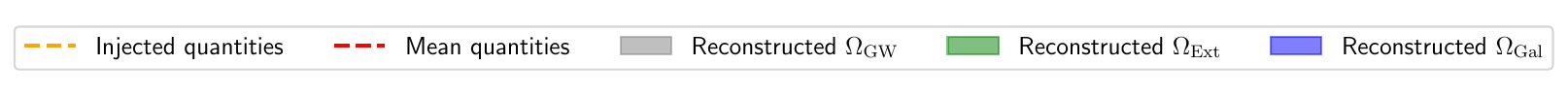}
	\end{center}
	\caption{\label{fig:PL_MCMC} \small \emph{Left panel}: 1D and 2D posterior distributions 
derived from the PL template reconstruction 
of the benchmark PL-BNK\_1 ($\{\log_{10}(h^2\,\Omega_*),  n_T\}=\{-12.5,\,2.085\}$) in the presence of galactic and extragalactic foregrounds. The posteriors of foregrounds and noise parameters are omitted for clarity. In the corner plot,  the gold and red dots, and corresponding vertical lines, show the injected parameters and their reconstructed mean values. The 68\% C.L.~(95\% C.L.) reconstruction region is displayed in dark (light) gray. The equivalent regions and 1D posteriors obtained with the Fisher approximation are in blue. 
    The top-right inset visualises the injected and reconstructed signals, with 68 and 95 \% CL error bands. The error bands on the galactic foreground and instrumental noise are too small to be visible. The LISA PLS is plotted in solid black. \emph{Right panel}: Like the left panel but for PL-BNK\_2 ($\{\log_{10}(h^2\,\Omega_*),  n_T\}=\{-11,\,0.77\}$). In the inset plots of both panels, the error bands on the galactic foreground and instrumental noise are too small to be visible.}
\end{figure}

In \cref{fig:Fisher_PL} and \cref{fig:PL_MCMC}  we present the forecasts for the PL template discussed in \cref{sec:pl}. The color maps show that the errors on both the amplitude and the tilt of the spectrum decrease as the amplitude $\Omega_*$ increases, as the power law becomes steeper, \emph{\emph{i.e.,}} $\lvert n_T\rvert$ increases, and, more in general, as the SNR of the primordial signal becomes larger. \Cref{fig:Fisher_PL} shows that the SNR contours of the primordial signal generally (although not perfectly) follows the reconstruction error contours.\footnote{The SNR is evaluated with respect to the nominal LISA sensitivity, which is the one we inject, and takes into account the existence of three TDI channels and the foregrounds as an additional source of nuisance. Several studies have adopted the criterion SNR\,$\gtrsim 10$ as a proxy for the condition of SGWB detectability and reconstruction \cite{Caprini:2019pxz, Flauger:2020qyi}.} Comparing the black and purple dashed/dotted lines, we learn how the presence of foregrounds degrades the measurements of the two signal parameters. For example, for a flat signal (\emph{\emph{i.e.,}} $n_T=0$), an accuracy $\sigma\simeq \mathcal{O}(0.01)$ on the logarithm of the amplitude requires $h^2 \Omega_* \simeq 10^{-12}$ without foregrounds, and is only slightly  larger when foregrounds are included. 
Achieving the same level of accuracy on the tilt  requires slightly larger values for the signal amplitude. Notice a peculiar behaviour along the line at $n_T=2/3$, where the primordial SGWB is degenerate with the foreground due to the extragalactic compact binaries.  
The possibility to separate a primordial signal with $n_T\simeq 2/3$ from extragalactic foregrounds crucially depends on our prior knowledge about the amplitude of the latter, which we have in practice implemented through a Gaussian prior, as explained in the previous section.

\noindent \textbf{Benchmarks. ---} We consider two benchmarks: PL-BNK\_1 and PL-BNK\_2. In the former, the PL parameters are set as $\{\log_{10}(h^2\,\Omega_*),  n_T\}=\{-12.5,\,\,2.085\}$,\, while in the latter they are set as $\{\log_{10}(h^2\,\Omega_*),  n_T\}=\{-11,\,0.77\}$. Both benchmarks can be produced within the axion inflation scenario, while the first of them is consistent with models of inflation with broken space diffeomorphisms, but does not respect the so called Higuchi bound for massive gravitons~\cite{Higuchi:1986py} (more on this in \cref{sec_int_PL}). 

We run the PL-template-based \texttt{SGWBinner} analysis on our two benchmarks, and display the obtained 1D and 2D posteriors in
\Cref{fig:PL_MCMC}. 
As the corner plots in the figure show, the injected values of both benchmarks are reconstructed well within the 68 \% CL contours (the foreground and noise reconstruction are omitted for clarity). In each panel, the inset plot highlights the injected and reconstructed benchmark signal, noise the foregrounds (injected and modelled as explained in \cref{sec:TDI_noise_foregrounds}) with their \snCL error bands. For the galactic foreground, the reconstruction is very accurate, with the reconstructed amplitude within the \seCL error band (recall that we vary only the amplitude in our analysis, keeping the spectral shape of foregrounds fixed) while the error bands on the extragalactic foreground are larger, but still within the \seCL error band.
For the PL-BNK\_2,  the degeneracy between the PL signal and the extragalactic foreground leads to a slightly less accurate reconstruction. For our two injected signals, and in particular for PL-BNK\_1, we   notice a very good agreement with the Fisher analysis, which captures very well the shape of the posterior distribution of the tilt and amplitude of the power law.
We conclude that  for a PL template with a sufficiently large injected signal with respect to the LISA sensitivity, the signal reconstruction degrades in the presence of foregrounds, but it is still very accurate. As expected, the impact of the foregrounds is less pronounced when the amplitude (or the tilt) of the PL is large.

\subsection{Forecasts for the log-normal template}
\label{subsec:log-normal}
\begin{figure}
	\centering
	\includegraphics[width=0.49\textwidth]{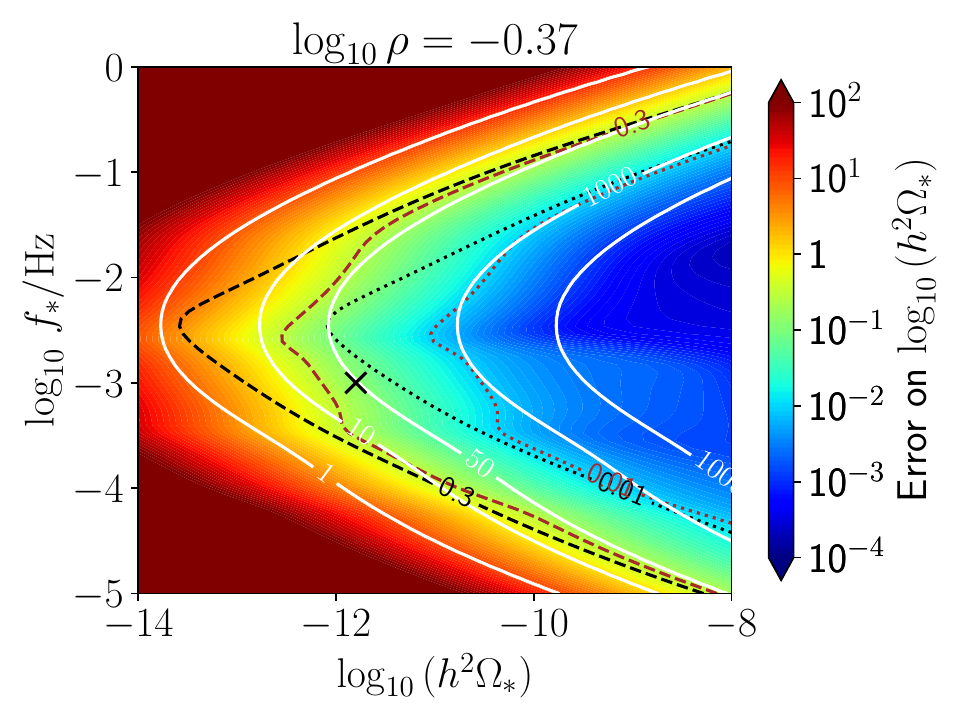} 
 \includegraphics[width=0.49\textwidth]{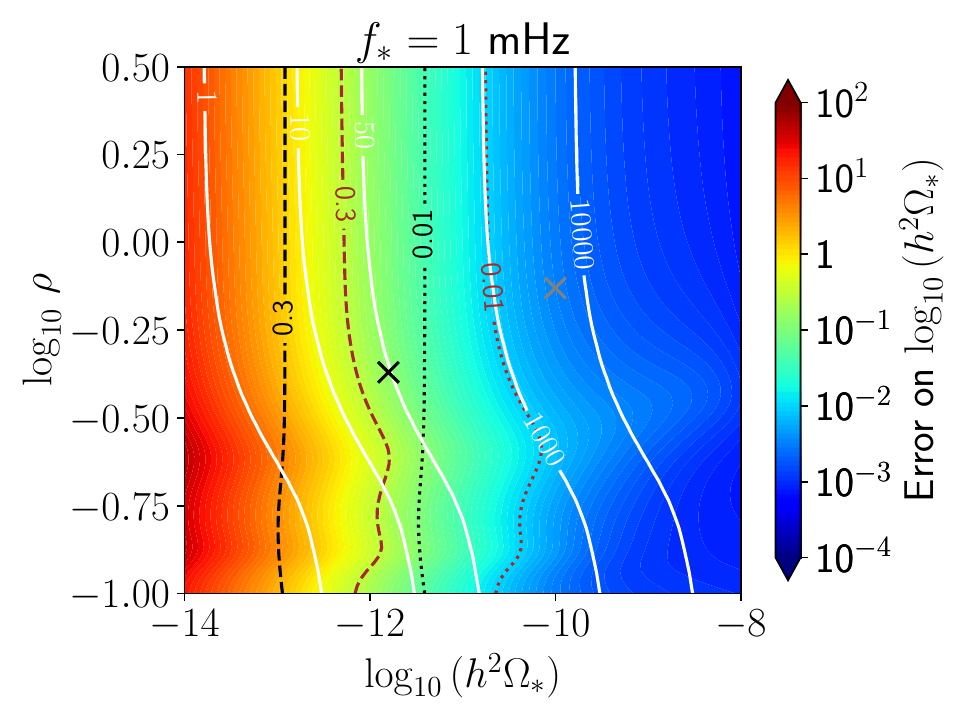} 
	\includegraphics[width=0.49\columnwidth]{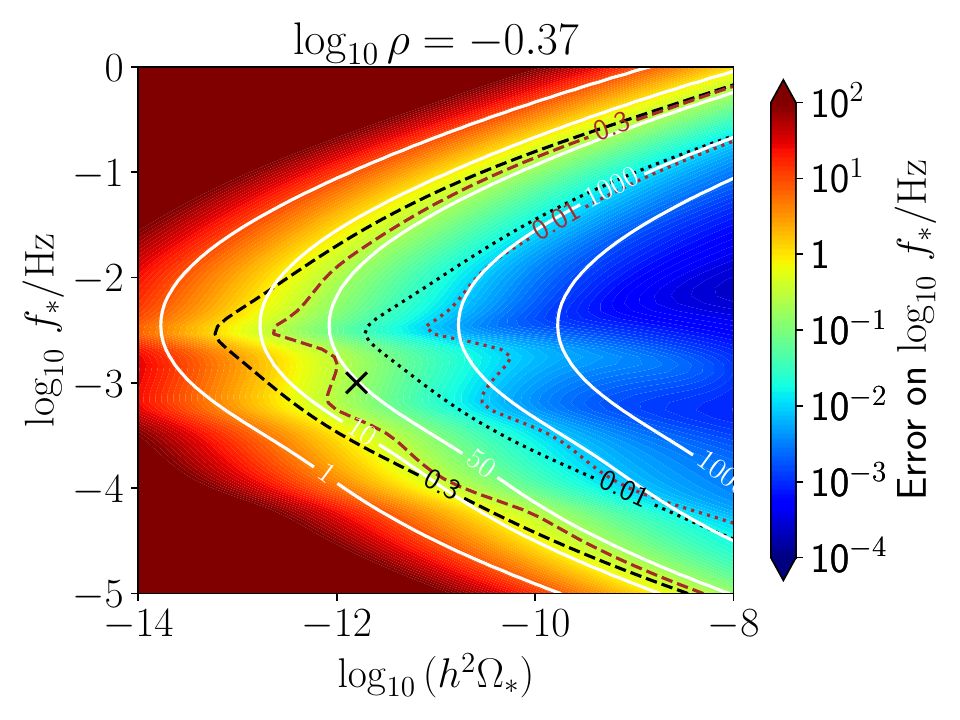} 
 \includegraphics[width=0.49\columnwidth]{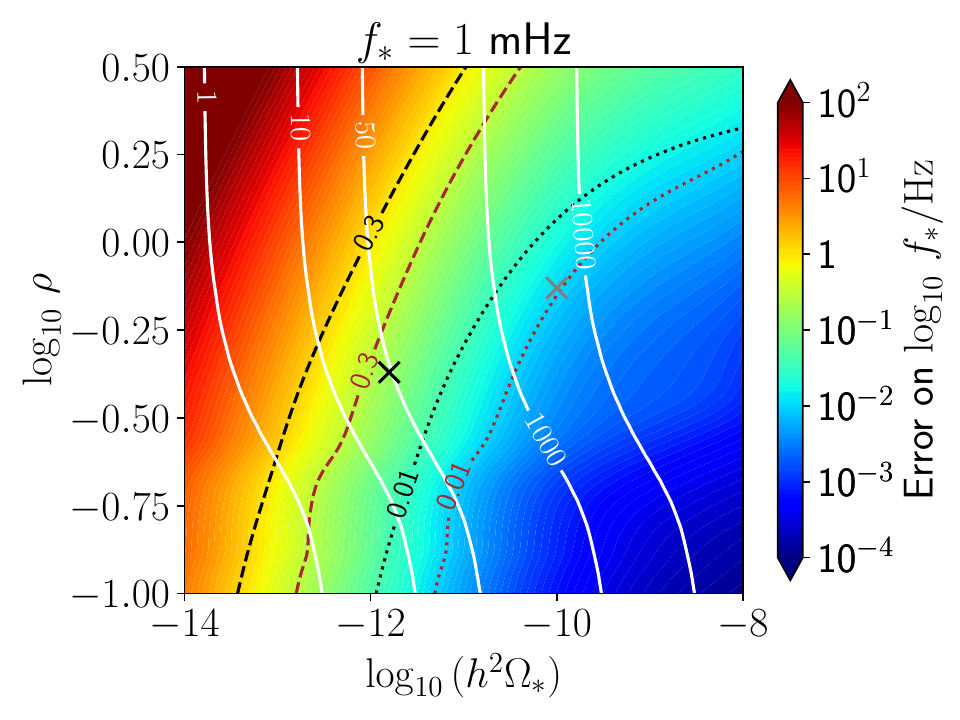} 
	\includegraphics[width=0.49\textwidth]{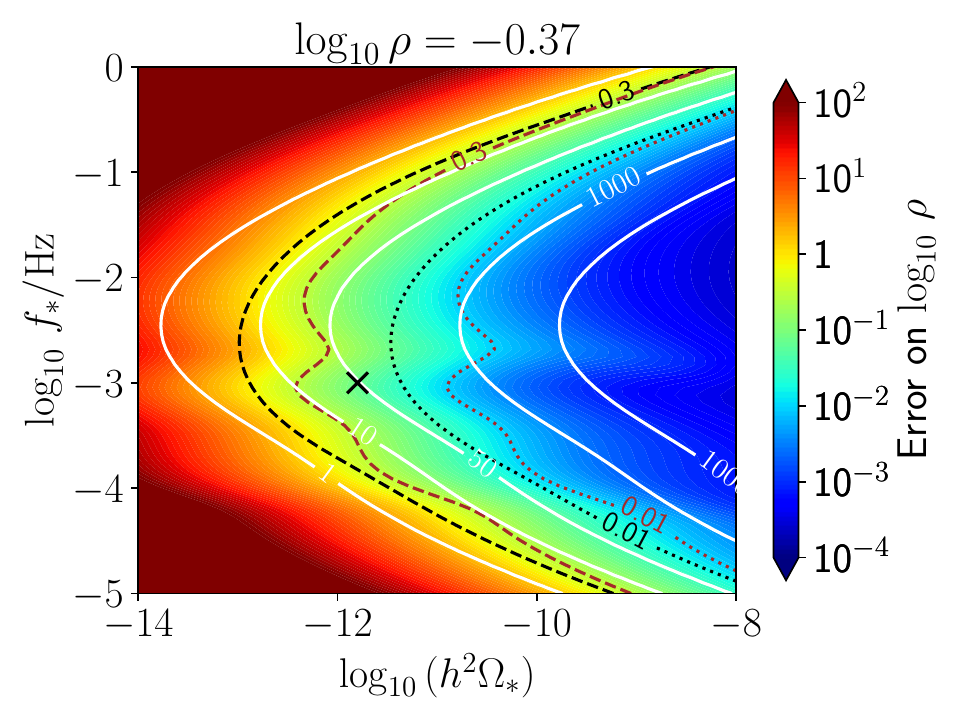} 
\includegraphics[width=0.49\columnwidth]{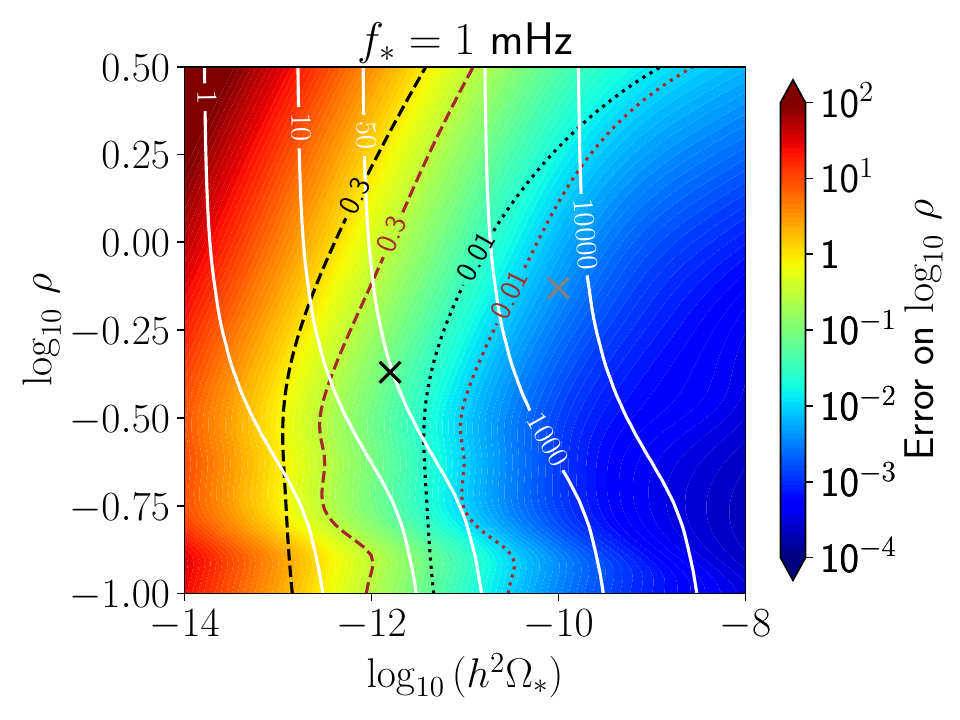} 
	\caption{\small Fisher forecasts for the LN template. The panels show the \seCL reconstruction error on $h^2\Omega_*$ (top panels), $f_*$ (central panels) and $\rho$ (bottom panels) as a function of the injected values of $f_*$, $h^2\Omega_*$ and $\rho$ set as specified in the axes and the title of each panel. 
SNR countour lines are plotted in white. The pairs of dashed (dotted) lines mark the $ \sigma =0.3$ ($\sigma =0.01$) contours, respectively in the absence (black) and in the presence (purple) of foregrounds. The gray and black cross display the benchmarks LN-BNK\_1 and LN-BNK\_2 respectively.
 }
	\label{fig:Fisher_LN}
\end{figure}

We now investigate the LN  template given in \cref{eq:master_modelIII_bum}.
\Cref{fig:Fisher_LN} displays the Fisher-approximation reconstruction errors obtained  in the parameter plane with peak frequency $f_{*}=1\,$mHz (left column) and the parameter plane with bump width $\log_{10} \rho =-0.37$ (right column).
 When  fixing  the width of the bump, the impact of the foregrounds on the reconstruction is less important with respect  to the case where we fix the peak frequency, at least for what concerns the accuracy on the error on the amplitude. Overall, an accuracy of $\sigma\sim 0.1$  can be  obtained  with a signal characterised by SNR\,$\gtrsim \,10$ in the absence of foregrounds. When foregrounds are included, to reach the same accuracy, an order of magnitude larger amplitude is required.  The minimum error on $\Omega_*$, $f_*$, and $\rho$ is reached for $10^{-3} \lesssim f_{*}/{\rm Hz} \lesssim 10^{-2}$, where the peak of the signal is in the best sensitivity region of LISA. When we fix the frequency peak of the bump (say at $1$ {\rm mHz}) and we vary the width and the amplitude, the error on the latter is at percent level for $h^2 \Omega_* \gtrsim 10^{-12}$, while for $f_*$ the accuracy improves for signals with smaller width (at fixed amplitude). Finally, when we fix the frequency peak, a better accuracy on the width
requires larger amplitude signals with  smaller width. Overall, 
the constraints on all parameters degrade when the signal peak approaches the borders of 
the LISA frequency band.

\begin{figure}
	\begin{center}
		\includegraphics[width=.495\columnwidth]{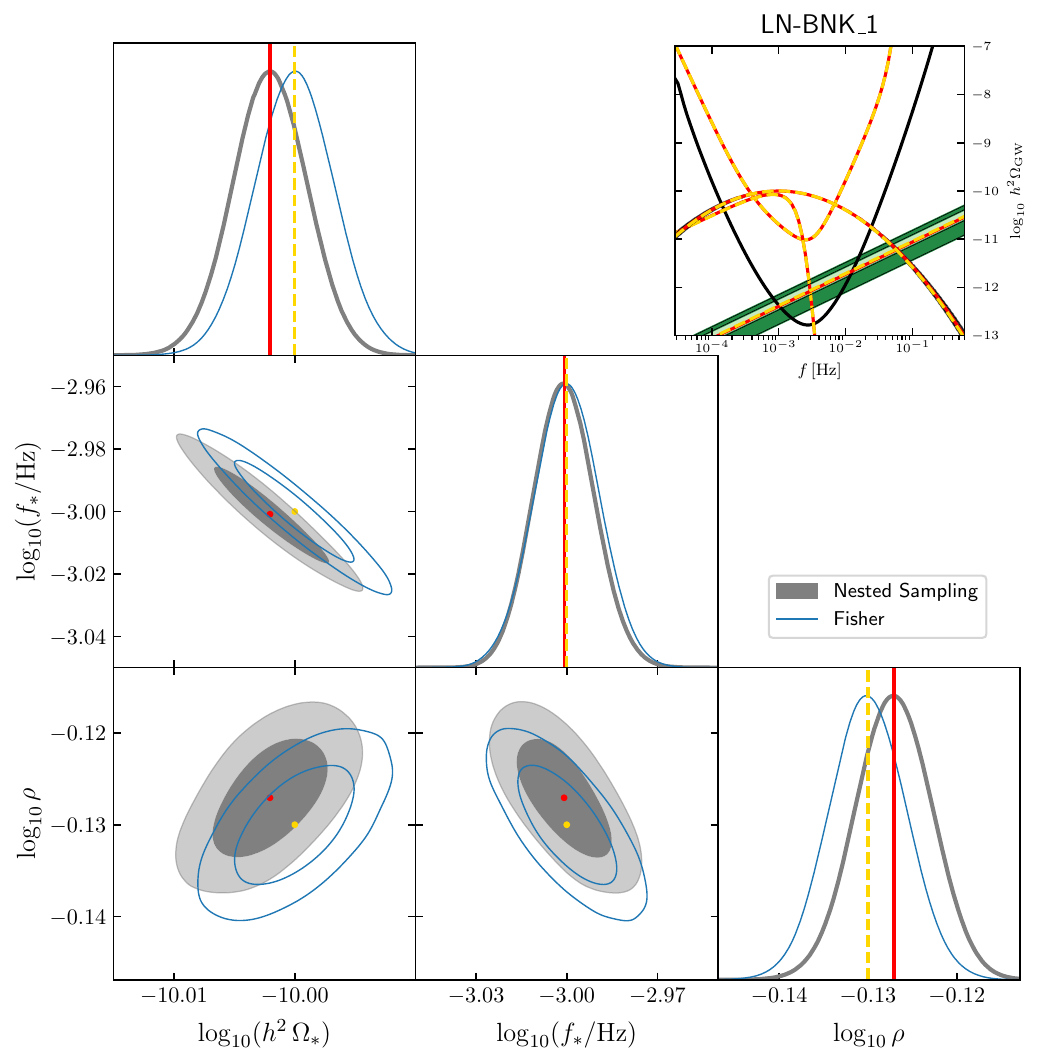}
		\includegraphics[width=.495\columnwidth]{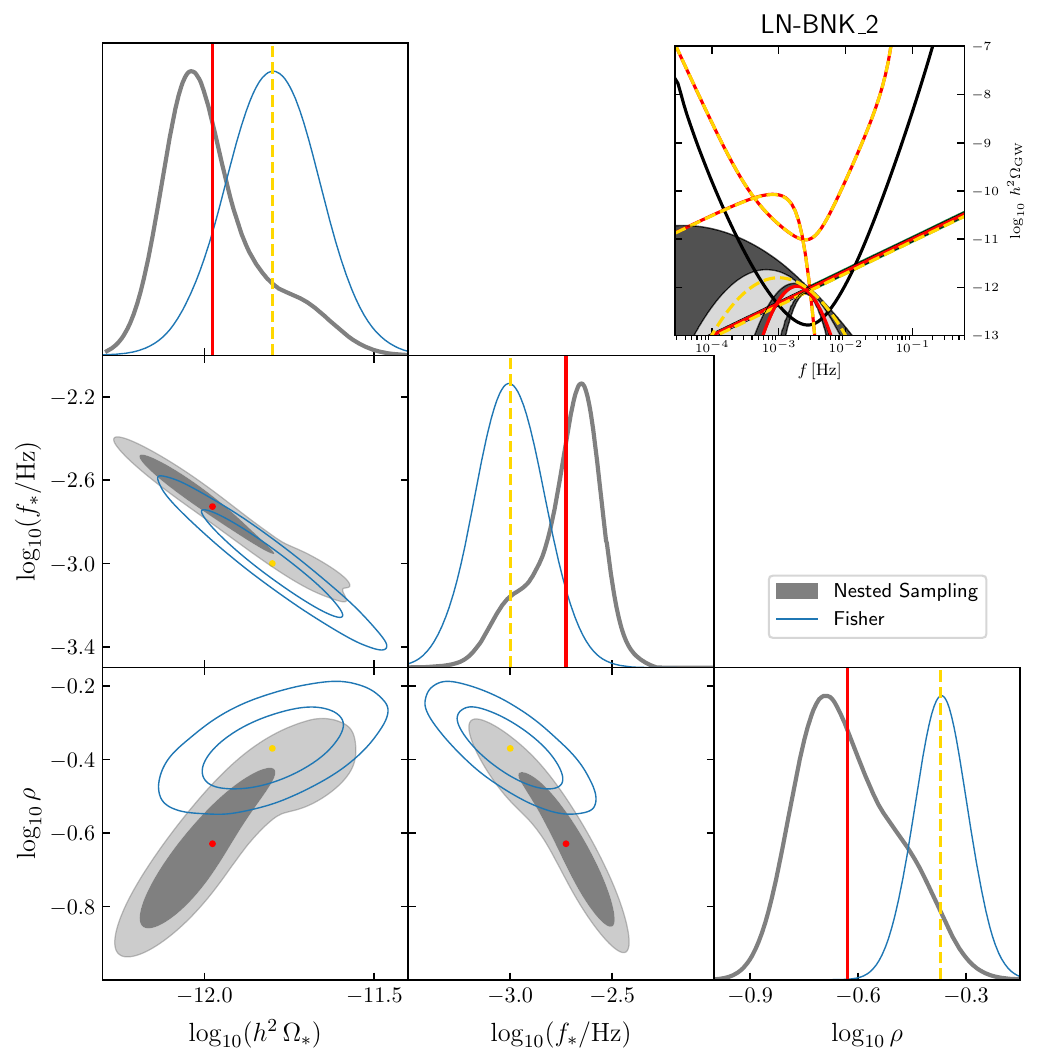}
        \includegraphics[width=\columnwidth]{Legend_Reconstruction.pdf}
	\end{center}
\caption{\small\label{fig:LN_MCMC}  1D and 2D posterior distributions 
derived from the LN template reconstruction 
of the benchmarks LN-BNK\_1 (left panel) and LN-BNK\_2 (right panel). Lines' styles and color codes
are as in \cref{fig:PL_MCMC}.
LN-BNK\_1 and LN-BNK\_2 are defined as in \cref{eq:master_modelIII_bum} with $\{\log_{10}(h^2\,\Omega_*),\,\log_{10}\rho,\,f_*/{\rm mHz}\}$ set to $\{-9.9,\,-0.13,\, 11\}$ and  $\{-11.8,\,-0.37,\,1\}$, respectively.
 }
\end{figure}

\noindent\textbf{Benchmarks. ---} We consider the benchmarks LN-BNK\_1 and LN-BNK\_2, which are defined as in \cref{eq:master_modelIII_bum}  with $\{\log_{10}(h^2\,\Omega_*),\,\log_{10}\rho,\,f_*\,{\rm [Hz]}\}$ set at $\{-9.9,\,-0.13,\,10^{-3}\}$ and  $\{-11.8,\,-0.37,\,10^{-3}\}$, respectively. These parameter choices are based on the setups detailed in \cref{sec_LNB}. 

In  \cref{fig:LN_MCMC} we plot the 1D and 2D posterior distributions of the LN parameters obtained by the template-based \texttt{SGWBinner} analysis run on LN-BNK\_1 (left panel) and LN-BNK\_2 (right panel).
For the LN-BNK\_1, the reconstructed mean values for the amplitude, peak frequency, and width parameters are within \seCL. The marginalized posterior distribution for all the parameters appears Gaussian with the reconstructed parameters all within 1\% accuracy, and in good agreement with the Fisher analysis presented above. Therefore signal, noise, and foregrounds are all well reconstructed for this specific benchmark. Only for the extragalactic foreground the reconstruction is less accurate, as it is partly obscured by the loud signal and might be prior-dominated. The posterior distributions of the signal parameters agree very well with the Fisher analysis; we only notice an almost negligible shift of the mean values compared to the injected signal.

The situation is different for LN-BNK\_2 (right plot), which is characterised by an amplitude smaller than the foregrounds, but still large enough for the bump to be detected. However, the bump being below the foreground level, its frequency shape reconstruction has wide error bars.
This can also be noticed by the fact that the best fits are different from the injected signal, but still fall within the \nfCL contours of the posterior distributions. Constraints are definitely looser than those for  LN-BNK\_1, but the signal parameters can still be constrained. Despite the high SNR ($\sim 50$), we observe a difference between the posterior distributions obtained through the nested sampling and the contours obtained using the Fisher approach. Such a difference may arise both from the non-Gaussian posteriors and from a certain level of realisation dependence. As a result, the \seCL region tends to shift towards amplitudes slightly smaller than the injected values, leading to worsened constraints. Nevertheless, the constraints on parameters  remain nearly of the same order of magnitude as those obtained through the Fisher approach, demonstrating its reliability in providing a 
reasonable good estimate of error order of magnitudes.   Interestingly, the foregrounds being larger than the cosmological signal, they are reconstructed with very high accuracy.

\subsection{Forecasts for the broken power law template}
We now discuss the BPL template defined in eq.~\eqref{eq:master_modelIV_stBBN}. As the template depends on numerous parameters, we only report on the analysis covering some parameter-space slices crossing the benchmarks  BPL-BNK\_1 and BPL-BNK\_2 for brevity. The template parameters 
$\log_{10}(h^2\,\Omega_*), f_* /{\rm mHz},\, n_{t,1},\, n_{t,2},\,\delta$ are defined to be equal to $-10.5,  1,\, 4, 0,\,1$ for BPL-BNK\_1 and  equal to $-9.3,  1,\, 2.65, -2.1,\,5.3$ for BPL-BNK\_2. These benchmarks are motivated by the theory scenarios discussed in \cref{subsec:BPL}.\footnote{The BPL template also describes the SGWB that cosmological first-order phase transitions produce in some regimes~\cite{Caprini:2015zlo, Caprini:2019egz}. Reference~\cite{Caprini:2024hue} analyses the LISA BPL-template-based reconstruction dedicated to such a cosmological SGWB source. That analysis differs from the present one as theoretical knowledge of the first-order phase transition dynamics permits fixing several BPL parameters a priori. This suppresses many degeneracies that arise in the present BPL reconstruction where we keep all template parameters a priori unknown to cover the huge variety of inflationary mechanisms leading to a BPL signal.} Since the two scenarios lead to qualitatively different parameter choices --  despite being described by the same template -- the discussion in this section will differ from the other sections, and we will address each scenario individually.

\begin{figure}
	\centering
	\includegraphics[width=0.49\textwidth]{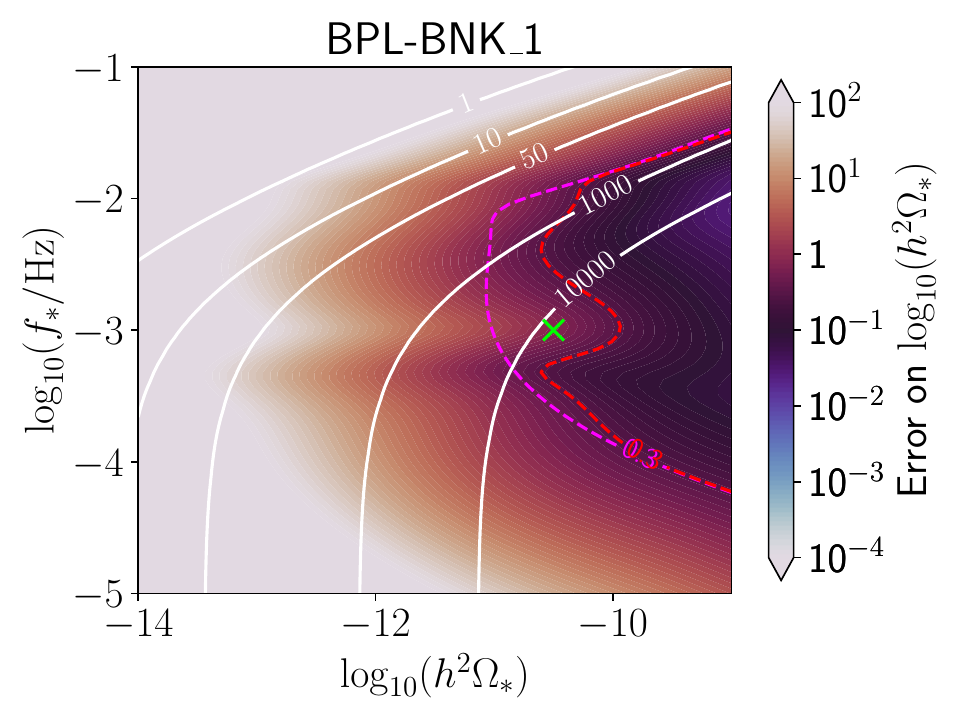} 
	\includegraphics[width=0.49\textwidth]{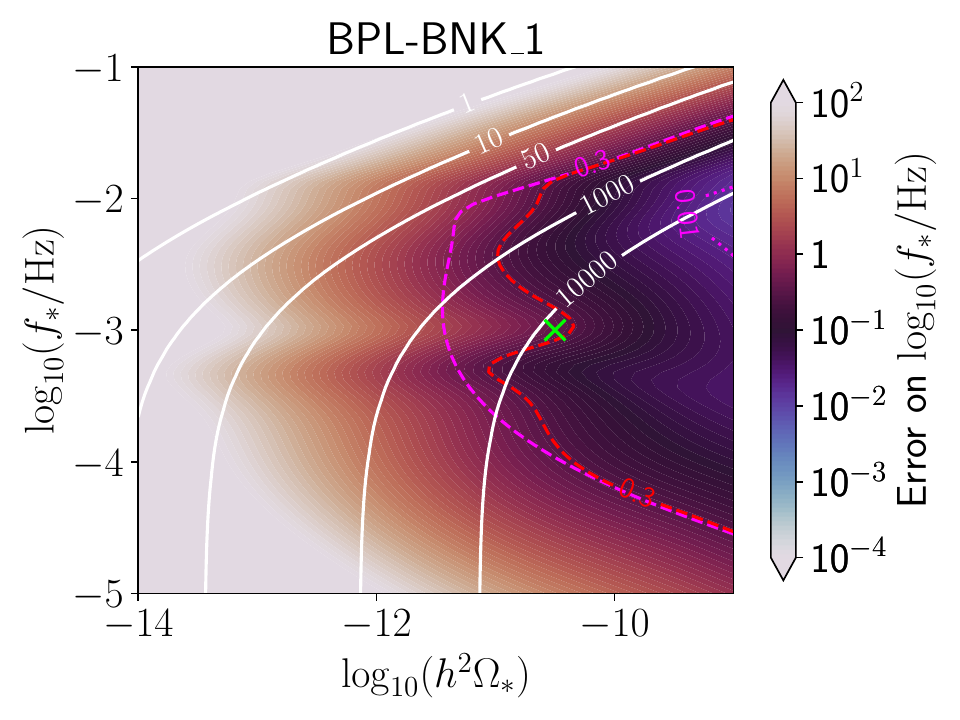} \includegraphics[width=0.49\textwidth]{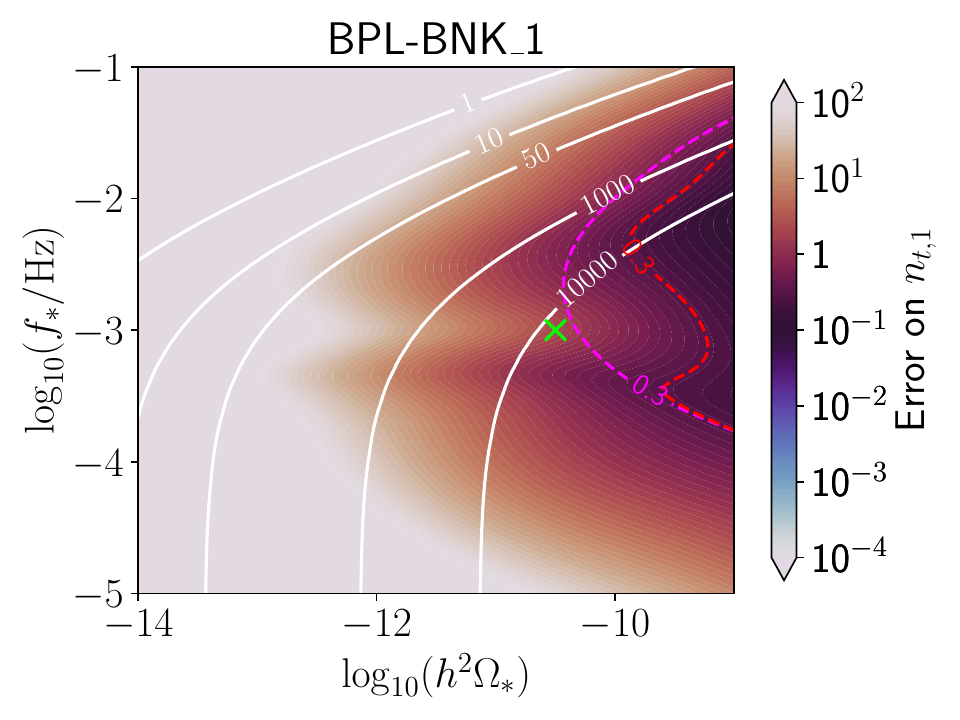} 
	\includegraphics[width=0.49\textwidth]{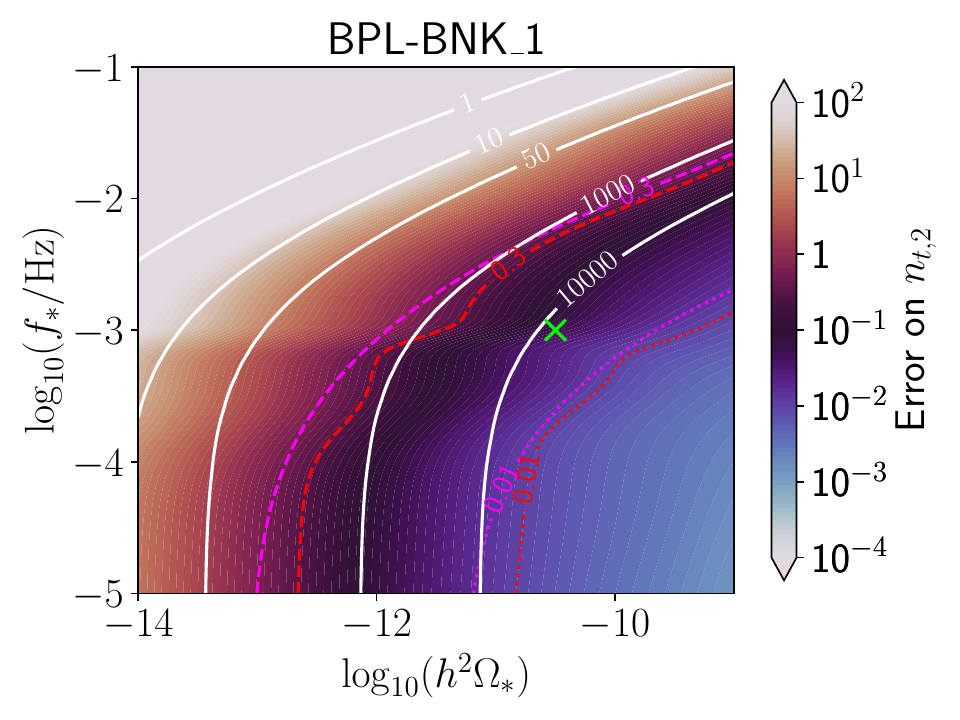} 
	\includegraphics[width=0.49\textwidth]{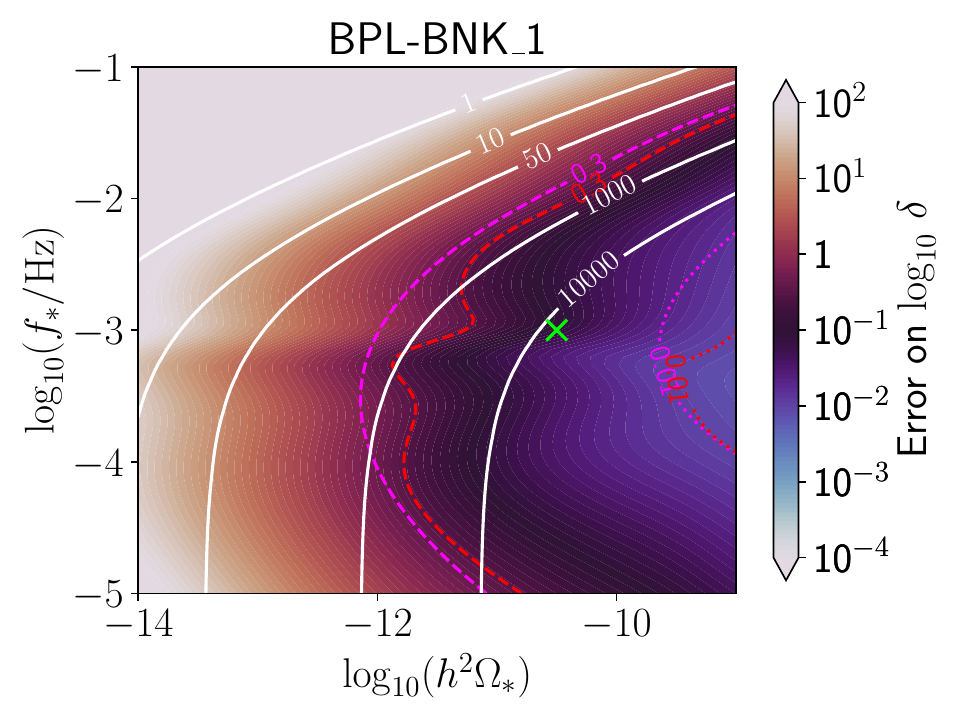} 
	
	\caption{\small Fisher forecasts for the BPL template. The panels show the \seCL reconstruction error on $h^2\Omega_*$ (top left), $f_*$ (top right), $n_{t,1}$ (central left),  $n_{t,2}$ (central right)  and $\delta$ (bottom) as a function of the injected values of $f_*$, $h^2\Omega_*$ specified in the axes and $n_{t,1}$, $n_{t,2}$ and $\delta$ specified in the title of each panel. 
SNR contour lines are plotted in white. The pairs of dashed (dotted) lines mark the $ \sigma =0.3$ ($\sigma =0.01$) contours, respectively in the absence (magenta) and in the presence (red) of foregrounds. The green crosses display the benchmarks BPL-BNK\_1. 
	\label{fig:Fisher_BPL_1}
 }
\end{figure}

\begin{figure}
	\begin{center}
		\includegraphics[width=\columnwidth]{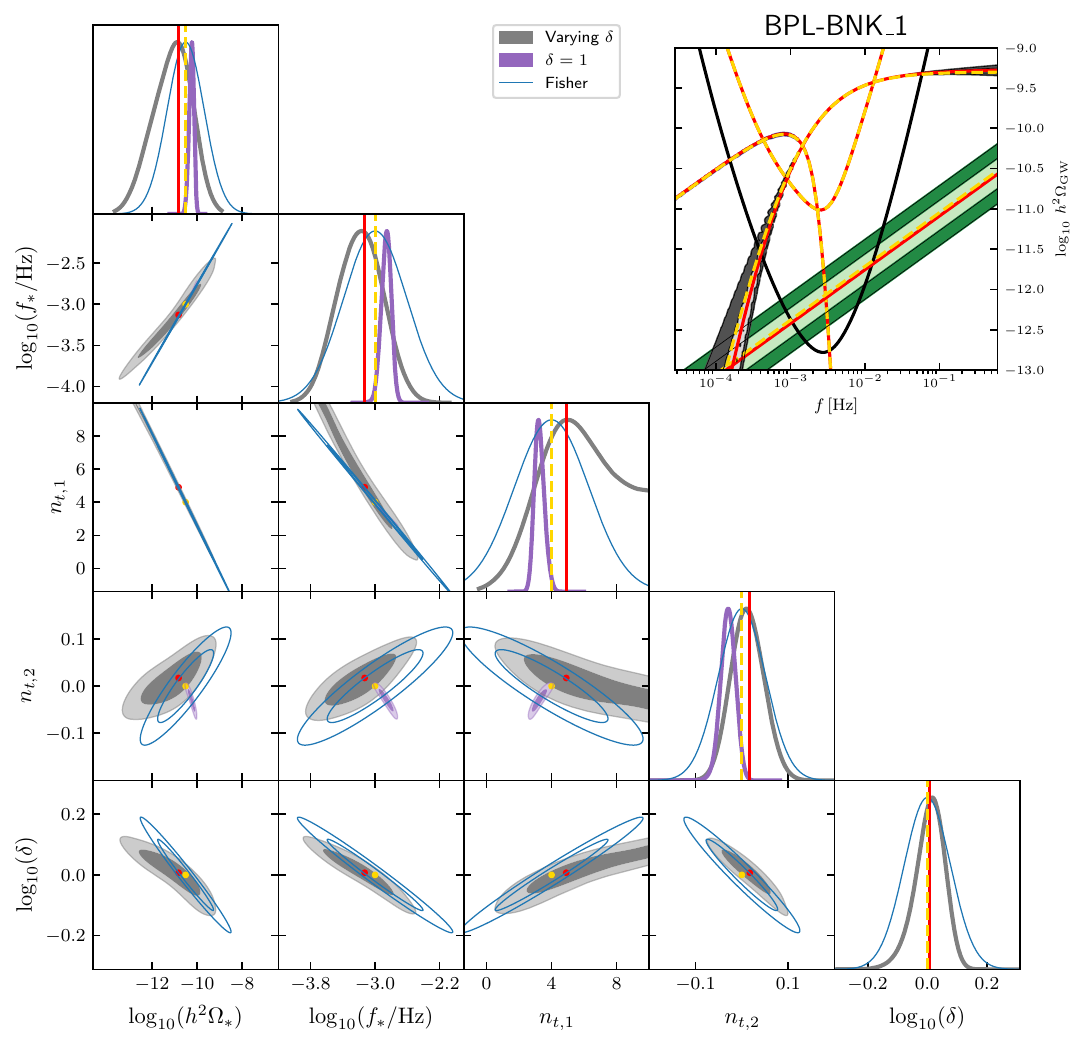}
        \includegraphics[width=\columnwidth]{Legend_Reconstruction.pdf}
	\end{center}
\caption{\label{fig:BPL1_MCMC} \small  
 1D and 2D posterior distributions 
derived from the BPL template reconstruction 
of the benchmark BPL-BNK\_1. Lines' styles and color codes
are as in \cref{fig:PL_MCMC}.
In addition, the purple contours display the 1D and 2D posteriors recovered by assuming the injected parameter $\delta=1$ to be known a priori. 
BPL-BNK\_1 is defined as in eq.~\eqref{eq:master_modelIV_stBBN} with $\log_{10}(h^2\,\Omega_*), f_* /{\rm mHz},\, n_{t,1},\, n_{t,2},\,\delta$ equal to $-10.5,  1,\, 4, 0,\,1$, respectively. 
}
\end{figure}

\begin{figure}
	\begin{center}
		\includegraphics[width=.9\columnwidth]{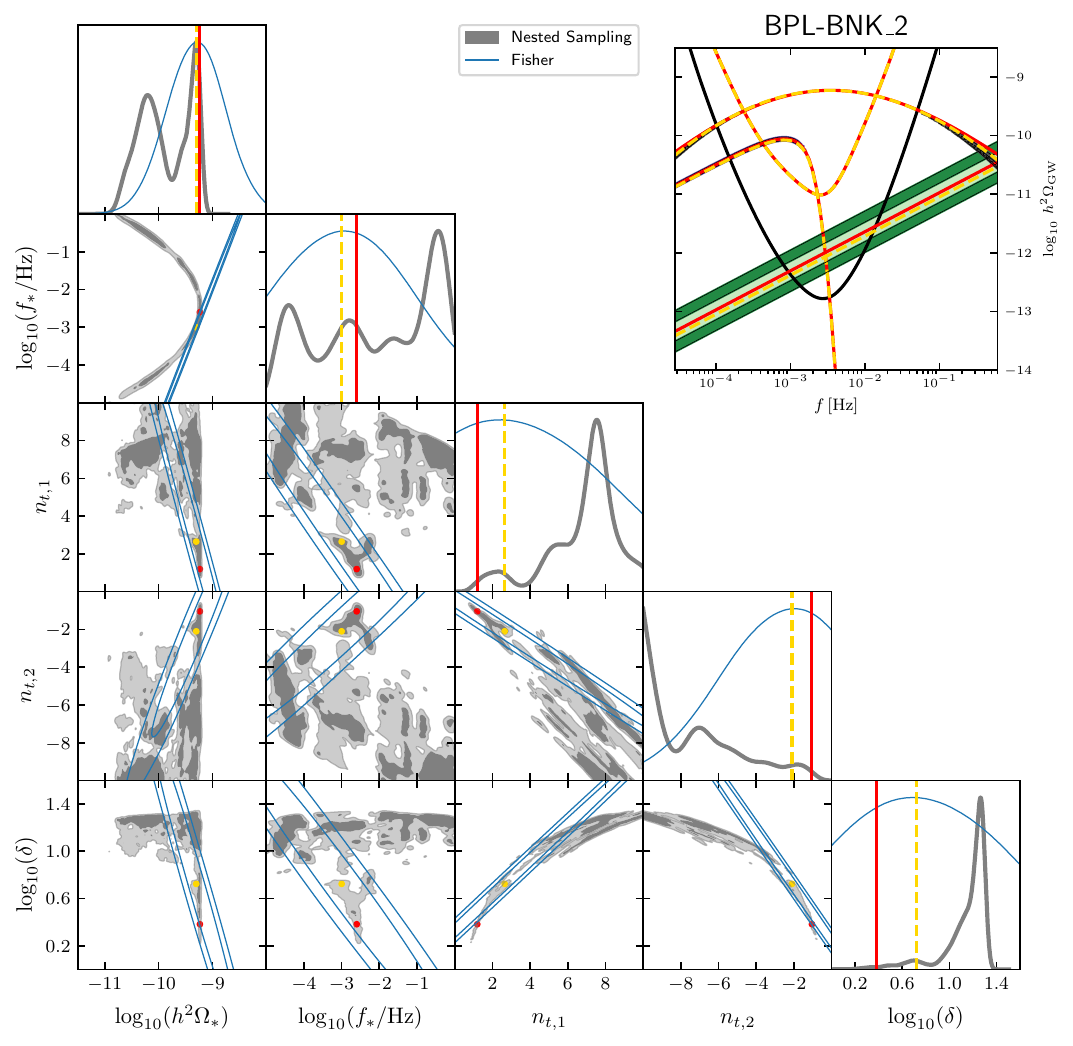}
        \includegraphics[width=\columnwidth]{Legend_Reconstruction.pdf}
	\end{center}
	\caption{\label{fig:BPL2_MCMC} \small  
 1D and 2D posterior distributions 
derived from the BPL template reconstruction 
of the benchmark BPL-BNK\_2. Lines' styles and color codes
are as in \cref{fig:PL_MCMC}.
BPL-BNK\_2 is defined as in eq.~\eqref{eq:master_modelIV_stBBN} with $\log_{10}(h^2\,\Omega_*), f_* /{\rm mHz},\, n_{t,1},\, n_{t,2},\,\delta$ equal to $-9.3,  1,\, 2.65, -2.1,\,5.3$, respectively.
 }
\end{figure}

{\bf  BPL-BNK\_1.}  \Cref{fig:Fisher_BPL_1} presents the Fisher reconstruction forecast maps in the case that the template parameters $f_*$ and $h^2 \Omega_*$ are injected with values varying in the intervals $[10^{-5},10^{-1}]$ and $[10^{-14},10^{-9}]$, respectively, while the other template parameters are set as for BPL-BNK\_1.
It turns out that, in this parameter-space slice,
requiring all parameters to be reconstructed with $\sigma\lesssim 0.3$ typically implies SNR\,$\gtrsim\!\! \,10^3$. 
An error of the order $\sigma\sim 0.01$ on the parameters $\log_{10}(h^2\,\Omega_*)$, $f_{*}$  and $n_{t,1}$ requires $\log_{10}(h^2\,\Omega_*) \gtrsim -9$, while for $n_{t,2}$ an amplitude  $\log_{10}(h^2\,\Omega_*) \gtrsim -11$ is sufficient. We  notice that, for a given signal amplitude, the SNR is independent of $f_*$ for  $f_*\lesssim 10^{-4}{\rm Hz}$. This is due to the fact that for $f_*\lesssim 10^{-4}{\rm Hz}$ the part of the signal on the left of $f_*$ is outside the LISA sensitivity, whereas the one on the right is just a constant. As soon as $f_*$ enters the LISA frequency band, the SNR starts showing a dependence on both the amplitude and the other parameters. The errors on the amplitude and peak frequency have their minima when the amplitude has a peak in correspondence with the LISA maximum sensitivity. A similar dependence is shown for the error on the low frequency tilt $n_{t,1}$, although we note  that the order of magnitude of this error is generically quite large. Constraints on this parameter are bad (only a loose bound $0<n_{t,\,1}<10$ at 95\% CL, where the upper bound is set by the prior)  mainly because of the particular choice $n_{t,1}=4$.  The growth of the spectrum being quite steep, bounds on this parameter are driven by a limited number of bins in frequency, making it harder to constrain large values of $n_{t,1}$ than small values. Instead, the error on the high frequency tilt is smaller when the amplitude has a peak at low frequency ($f_*<10^{-4}$ {\rm Hz}). In fact, in this case, the flat part of the spectrum would entirely lie within the LISA band.

The impact of the foregrounds on the reconstruction seems less pronounced with respect to the PL, except for a choice of $f_*$ close to the LISA characteristic frequency. Such impact is even more reduced for $\delta$ and $n_{t,2}$. 
The corner plot in \cref{fig:BPL1_MCMC} shows the reconstructed mean value parameters for the BPL-BNK\_1. All mean values result within the 95\% C.L., which reflects the accuracy of the reconstruction for the BPL case. In particular, the high-frequency spectral tilt $n_{t,2}$ and the width $\delta$ can be measured with very high accuracy.\\
We notice a banana-shaped degeneracy among some parameters, in particular for 2D posteriors in the planes containing the smoothing parameter $\delta$. For comparison, we also show the constraints obtained by fixing $\delta=1$, which corresponds to the injected value (see purple contours in the figure). The constraints are significantly better, although the example is for illustration purposes only: we would need to develop theoretical arguments to fix such (otherwise free) parameter.  Nevertheless, the degeneracies do not impact the reconstruction of the signal, of the noise, and of foregrounds. Importantly, despite the non-Gaussian shape of the posterior distribution,  the blue contours show that our estimates for the errors, as obtained with the Fisher analysis, are in good agreement with the MC analysis.  Finally,  notice that the \snCL error bands on the galactic foreground and instrumental noise are small, while for the extragalactic foreground  the errors are larger and prior-dominated.  \\

{\bf  BPL-BNK\_2.} 
The benchmark BPL-BNK\_2 represents a broad bump, asymmetric around its peak. We find that the BPL recostrunction of this kind of signal  
 introduces several complexities
that the Fisher approximation does not capture unless the SNR is exceedingly large. For this reason, we opt not to show the Fisher reconstruction maps but directly inspect the corner plot for our nested sampling analysis run on BPL-BNK\_2 (see \cref{fig:BPL2_MCMC}).
We can notice that LISA can accurately reconstruct the frequency shape of the injected signal.  However, the posterior distributions are extremely non-Gaussian in several parameter directions.
As expected, the Fisher formalism does not  capture 
such complex degeneracies. 

In general, BPL-BNK\_2 exhibits posteriors  that  are challenging to sample. Their complexity likely require strategies that go beyond the scope of this paper. For instance, broad bump profiles within the sensitivity window of LISA may be over-parameterised by our BPL template, and might potentially be better characterised using other templates.\footnote{Although model selection considerations go beyond the scope of this work, we checked that the BPL and LN bump templates both fit the BPL-BNK\_2 signal with similar $\chi^2$.}
An alternative may be to expand the BPL template in eq.~\eqref{eq:master_modelIV_stBBN} around $f=f_*$, and build combinations of $\left\{\delta,\,n_{{t},1},\,n_{{t},2}	\right\}$ that control linear and quadratic departures from $f_*$. This procedure would likely reduce the degeneracy between parameters, but the new parameterisation would not necessarily be optimal for the search of benchmarks like BPL-BNK\_1.

\subsection{Forecasts for the double peak template}

\begin{figure}
\centering
\includegraphics[width=0.4\textwidth]{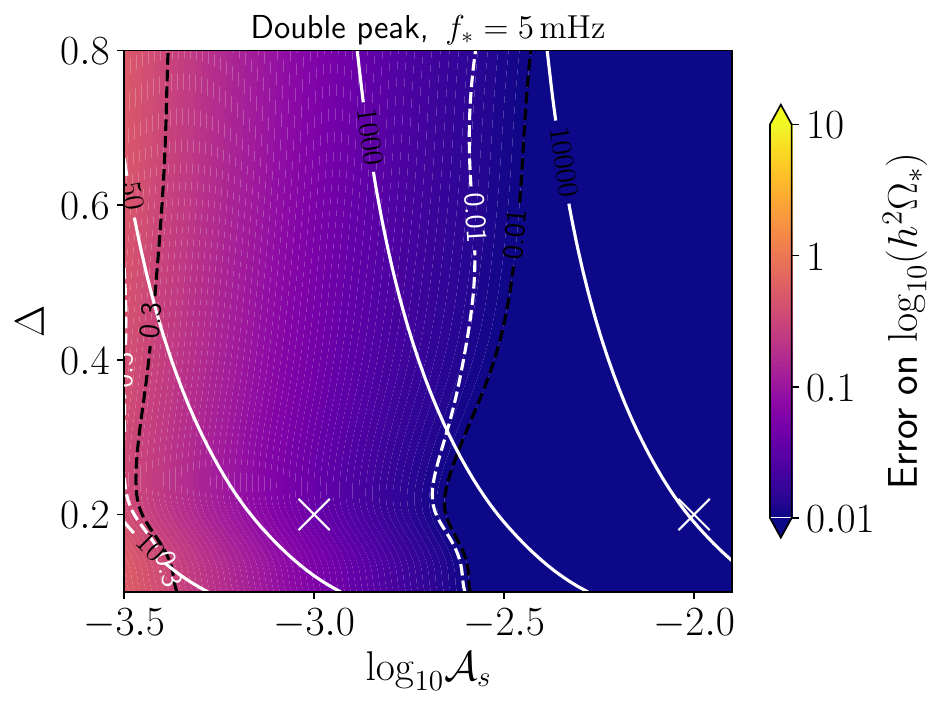}
\includegraphics[width=0.4\textwidth]{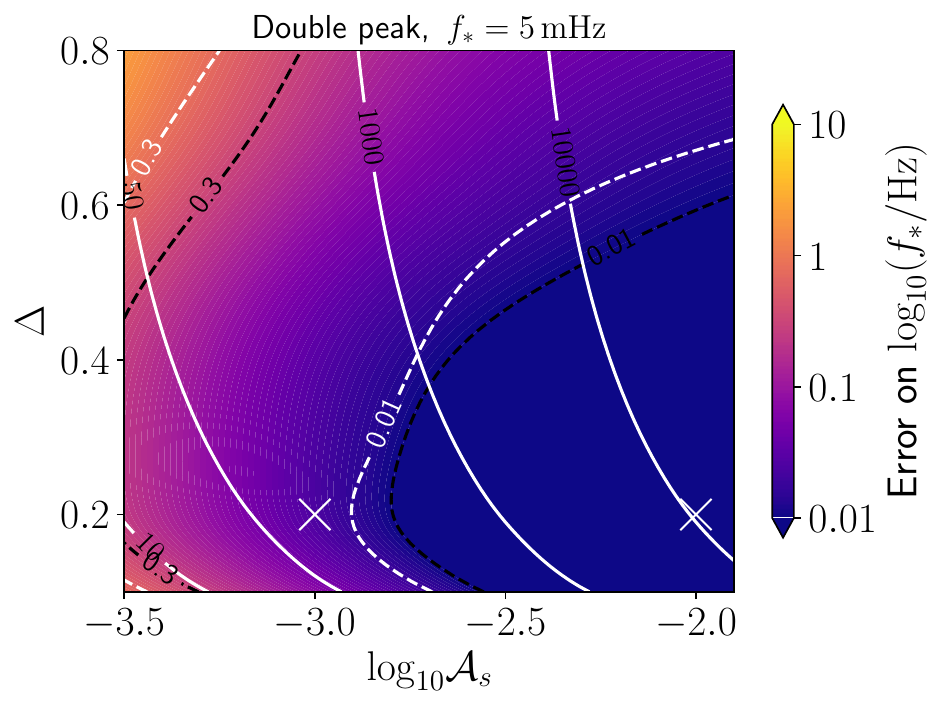}\\
\includegraphics[width=0.4\columnwidth]{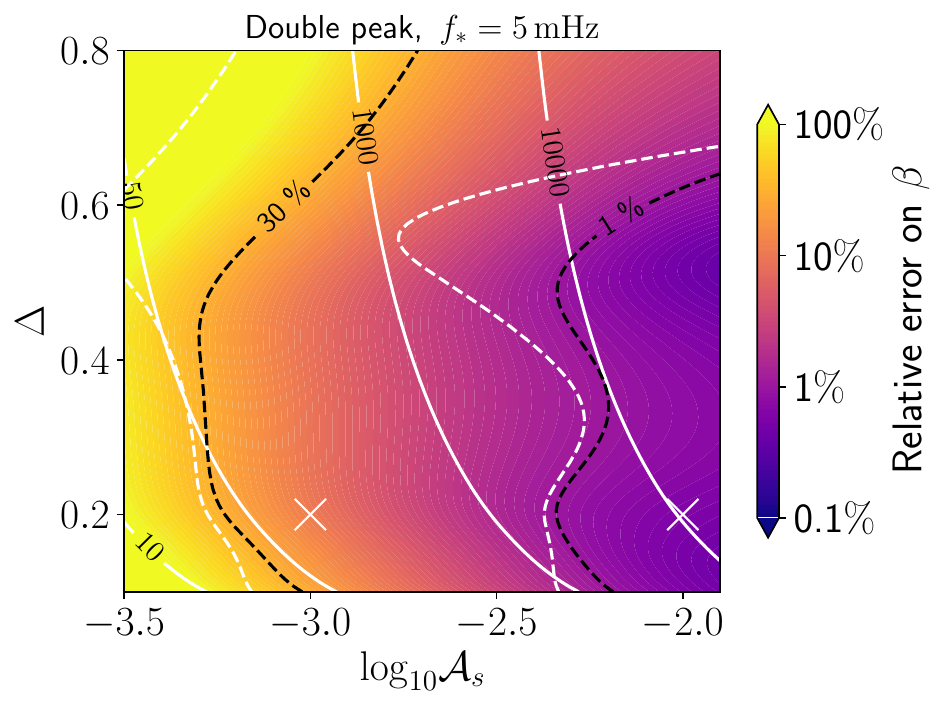}
\includegraphics[width=0.4\columnwidth]{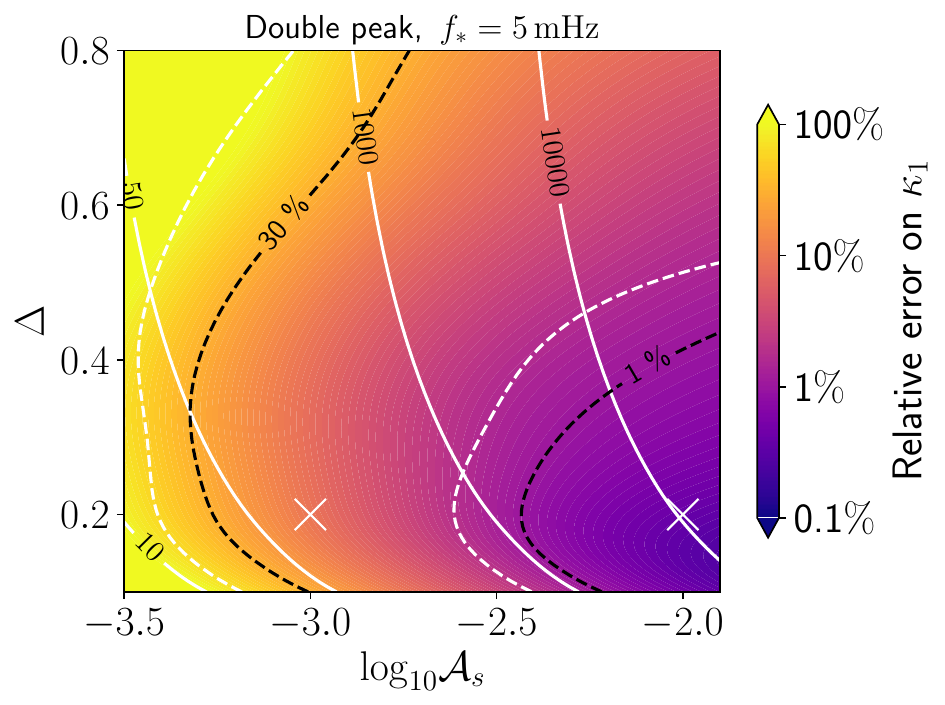}\\
\includegraphics[width=0.4\columnwidth]{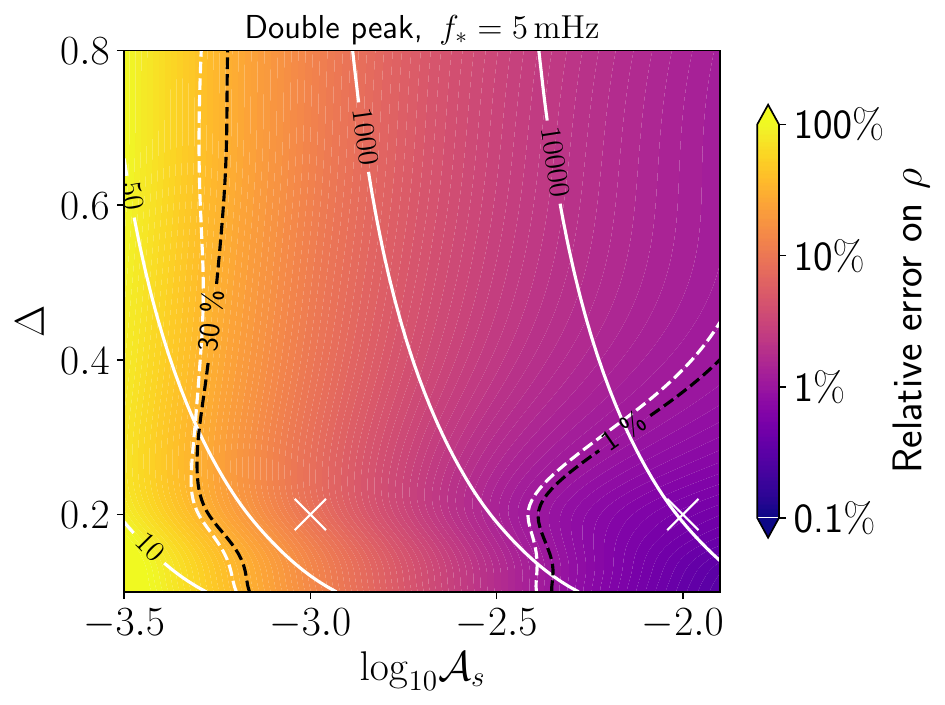}
\includegraphics[width=0.4\columnwidth]{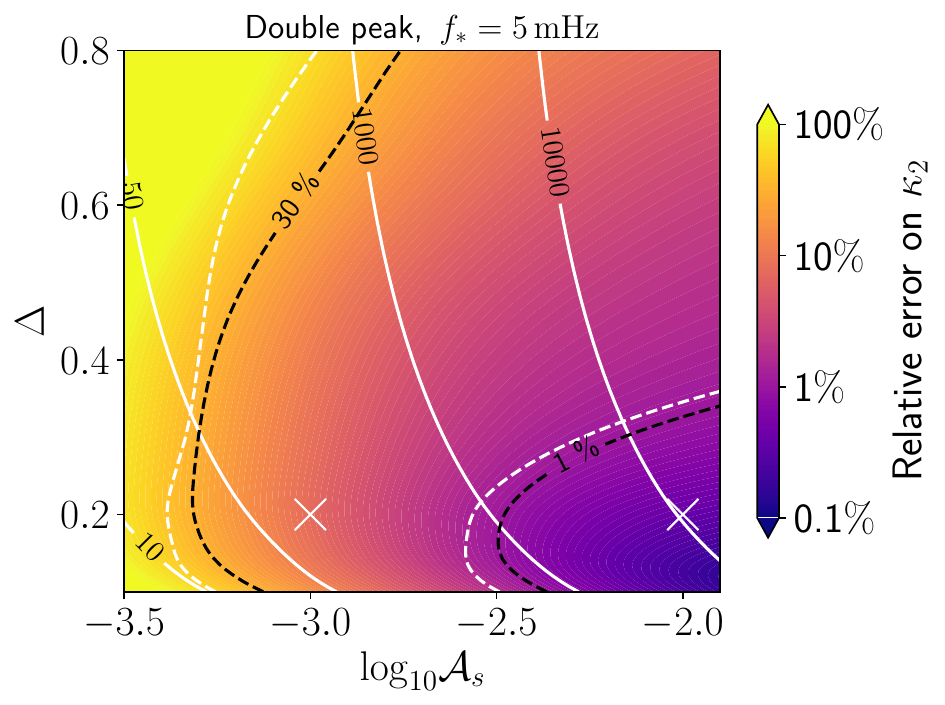}
\includegraphics[width=0.38\columnwidth]{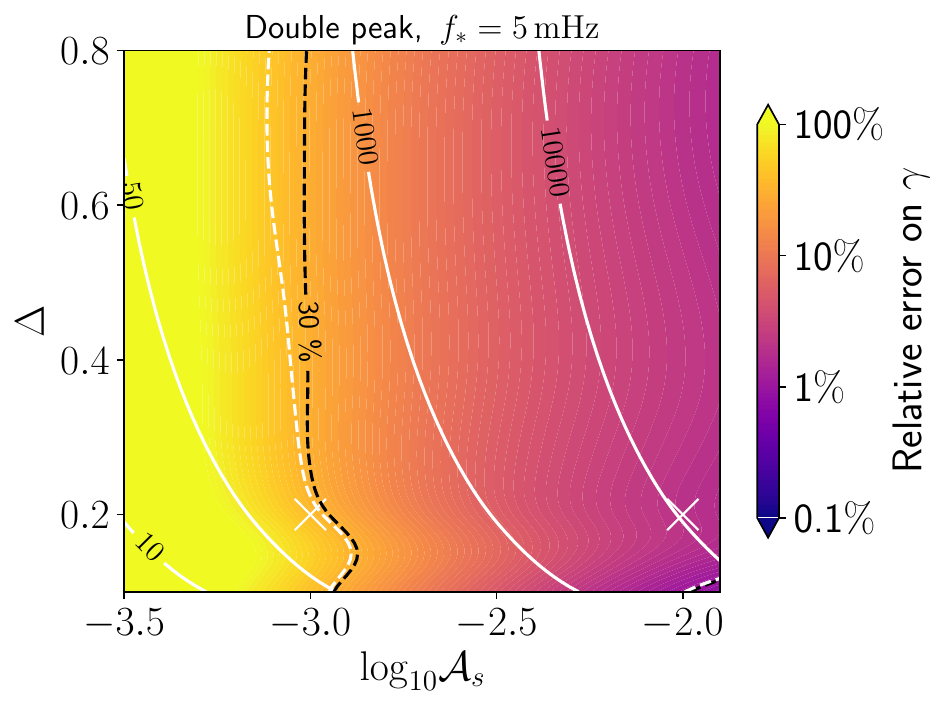}
\caption{\small \label{Fisher_double} 
Fisher forecasts for the DP template. The panels show the \seCL reconstruction error on $h^2\Omega_*$ (top left), $f_*$ (top right), $\beta$ (next-to-top left),  $\kappa_1$ (next-to-top right),  $\rho$ (next-to-bottom left),  $\kappa_2$ (next-to-top right) and $\gamma$ (bottom) as a function of the $\mathcal{P}_\zeta^{\rm ln}$ parameters $\Delta$ and $\mathcal{A}_s$ (and $k_*$ adjusted to yield $f_*=5\,$mHz).
SNR contour lines are plotted in white. Depending on the panel, the pairs of dashed lines mark the absolute error $ \sigma =0.3$ ($\sigma =0.01$) and relative error 30\% ($1\%$) contours, respectively in the absence [white] and in the presence [black] of foregrounds. The white crosses display the benchmarks DP-BNK\_1 and DP-BNK\_2. 
}
\end{figure}

\begin{figure}
\centering
\includegraphics[width=0.4\textwidth]{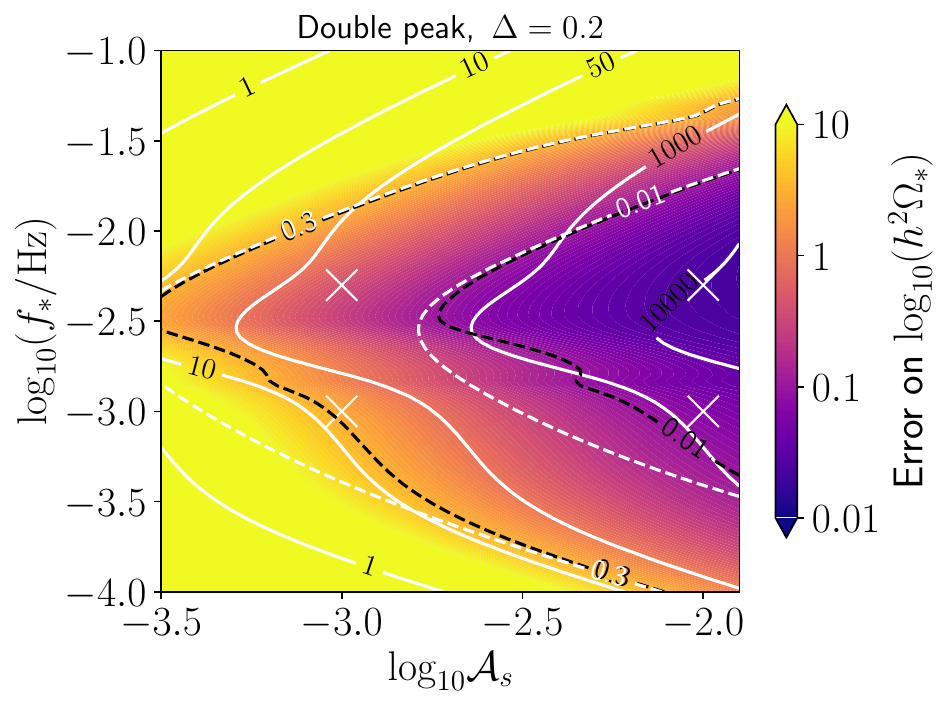}
\includegraphics[width=0.4\textwidth]{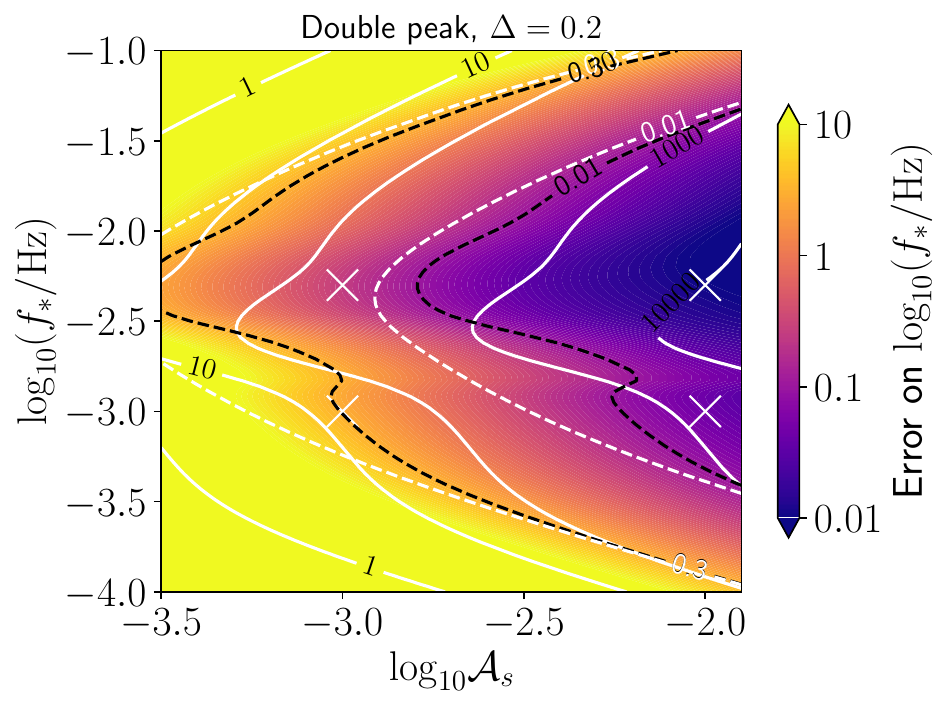}\\
\includegraphics[width=0.4\columnwidth]{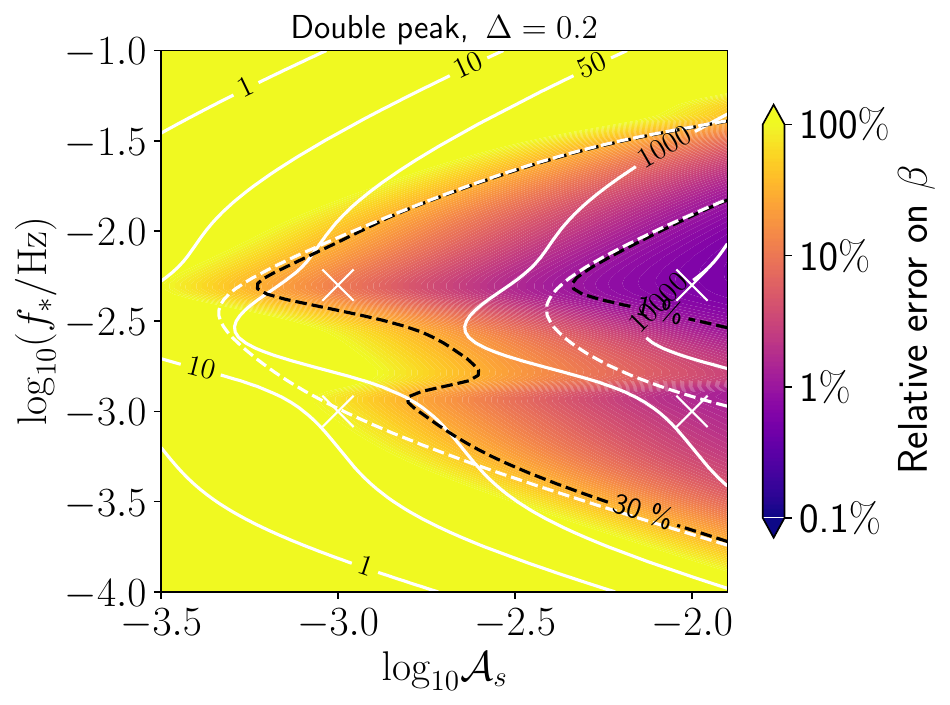}
\includegraphics[width=0.4\columnwidth]{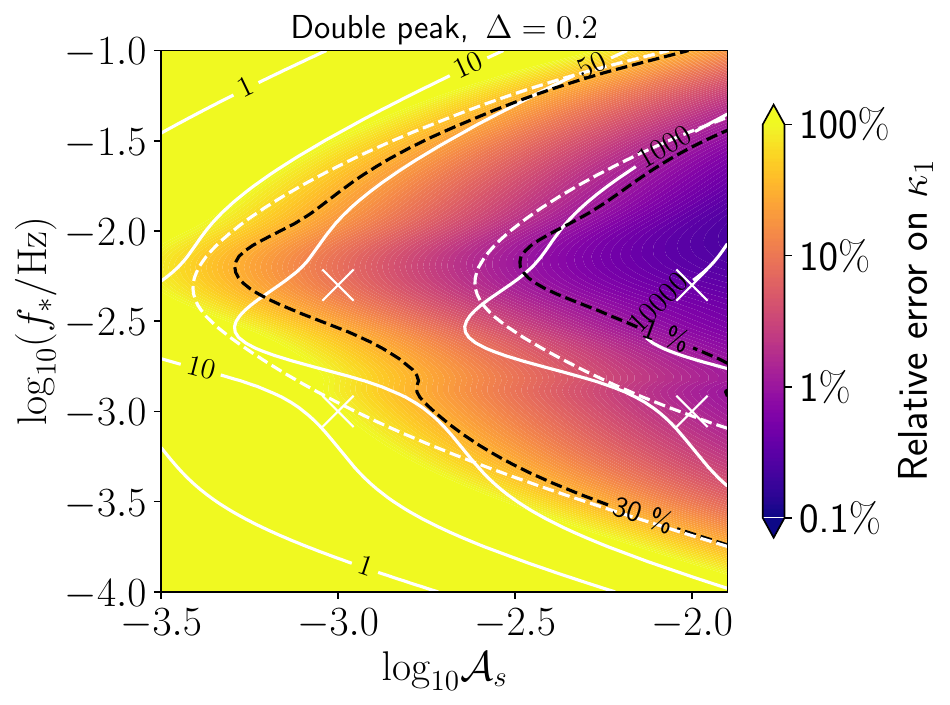}\\
\includegraphics[width=0.4\columnwidth]{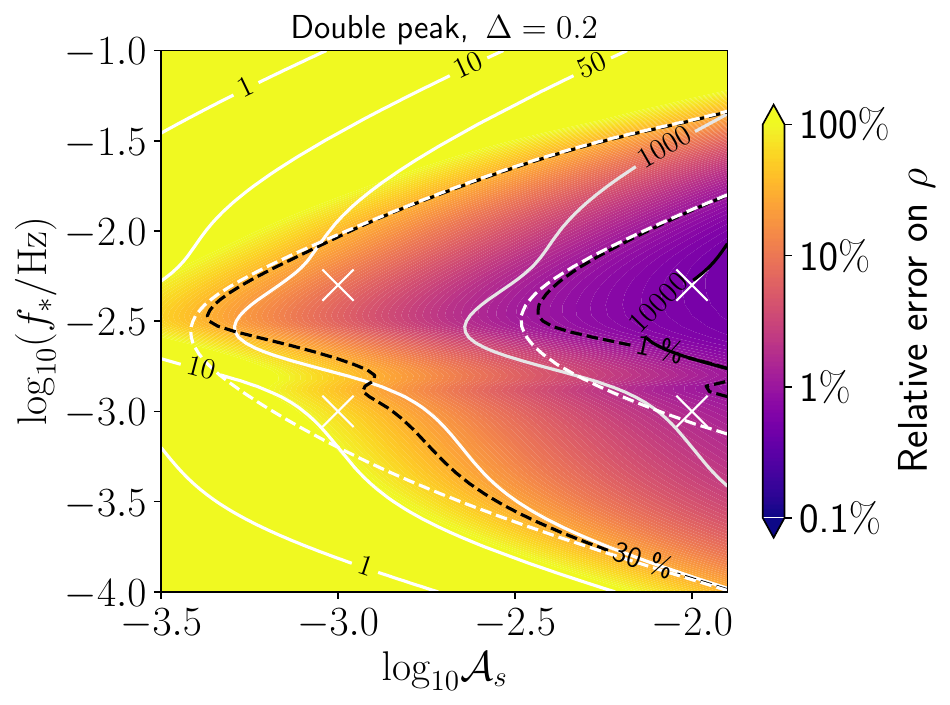}
\includegraphics[width=0.4\columnwidth]{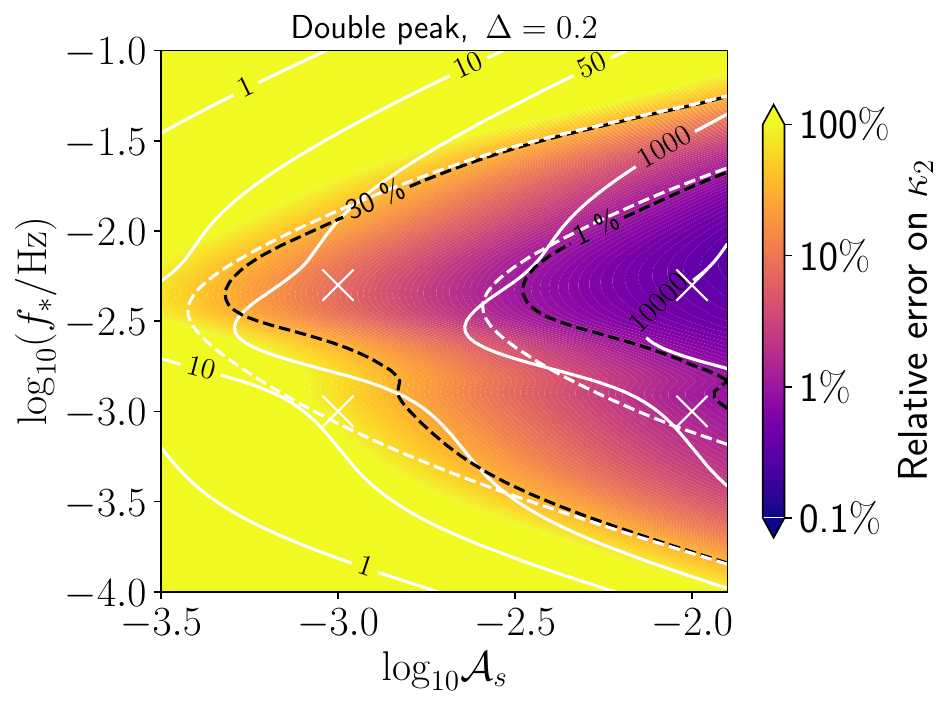}
\includegraphics[width=0.38\columnwidth]{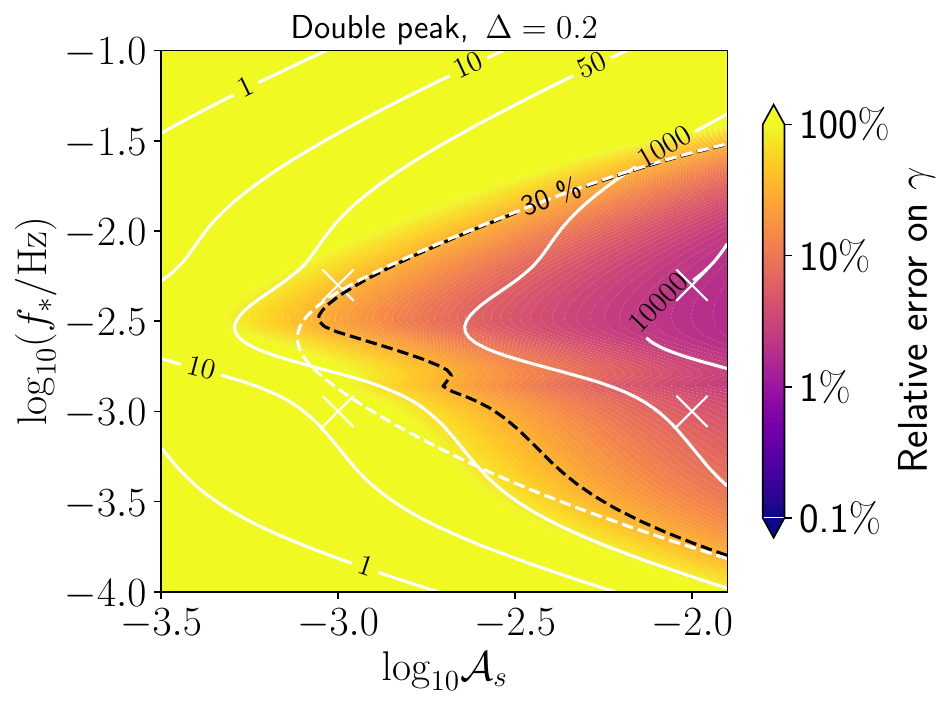}
\caption{\small Like in \cref{Fisher_double} but with errors  
as a function of the $\mathcal{P}_\zeta^{\rm ln}$ parameters $k_*$ (expressed in terms of $f_*$) and $\mathcal{A}_s$ with fixed $\Delta =0.2$.
\label{Fisher_double_pivot}
}
\end{figure}

We now discuss the Fisher forecasts for a SGWB characterised by a DP template,  presented in \cref{sec:double_peak}. The DP template is given by \cref{eq:templateDP} and is characterised by seven parameters, whose geometrical meaning is explained in \cref{sec:double_peak}. Their  impact on the signal shape is visualised in  \cref{fig:doublepeak_parameters} in \cref{app:variation_of_template_parameters}. 
 
As previously stressed, SGWB profiles featuring a double peak structure can be produced by curvature fluctuations at second order in perturbations if their spectrum is sufficiently narrow. This implies that a given narrow curvature power spectrum $\mathcal{P}_\zeta$ (with all its parameters fixed) univocally defines the values of the seven DP parameters and, in turn, varying the parameters of that given $\mathcal{P}_\zeta$ defines a hypersurface in the DP parameter space.
 Thus, unlike the Fisher forecast maps in the previous sections, for which we scan the space of the template parameters by varying them directly, here we find it instructive to slice the DP template parameter space by assuming the log-normal curvature power spectrum $\mathcal{P}_\zeta^{\rm ln}$ of eq.~\eqref{eq:Pk_LN1}, which only depends on three parameters, \emph{i.e.,}~the overall amplitude $\mathcal{A}_s$, the peak width $\Delta$ and the peak position moved via $k_*$.\footnote{Note that the pivot scale $f_*$ in the DP template marks the transition between the two peaks and it is not exactly equal to the peak of the primordial power spectrum. However, the two are trivially related through the rescaling $f_* = 2 /(\kappa_2\sqrt{3})f_{\rm PS}$ (where $f_{\rm PS}$ is the frequency associated with the peak of the power spectrum and we have assumed standard radiation domination after inflation).} 
 \Cref{app:lognormal} provides the numerical mapping relating the three  $\mathcal{P}_\zeta^{\rm ln}$ parameters to the seven parameters of the DP template.

\begin{figure}
	\begin{center}
	\includegraphics[width=.48\columnwidth]{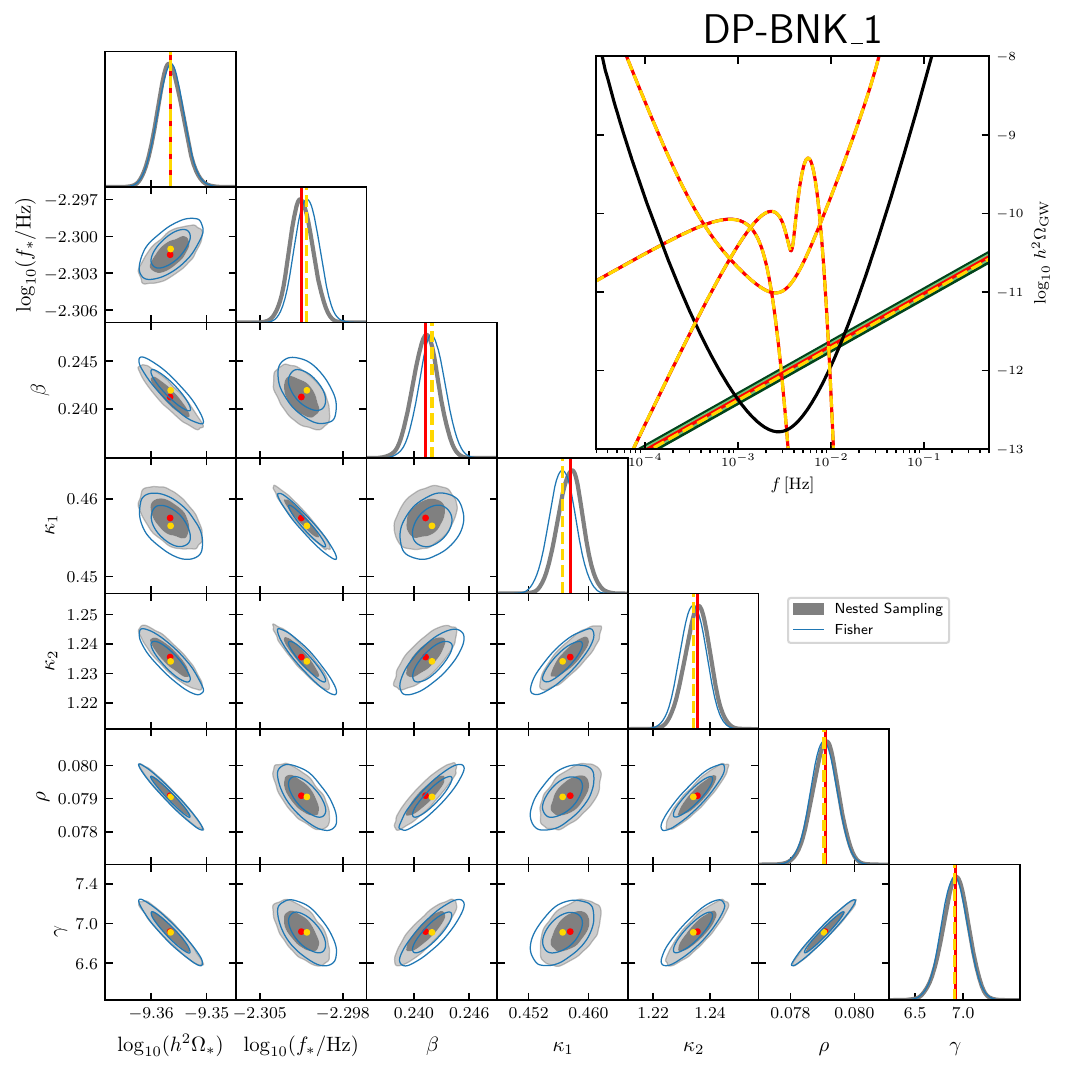}
		\includegraphics[width=.48\columnwidth]{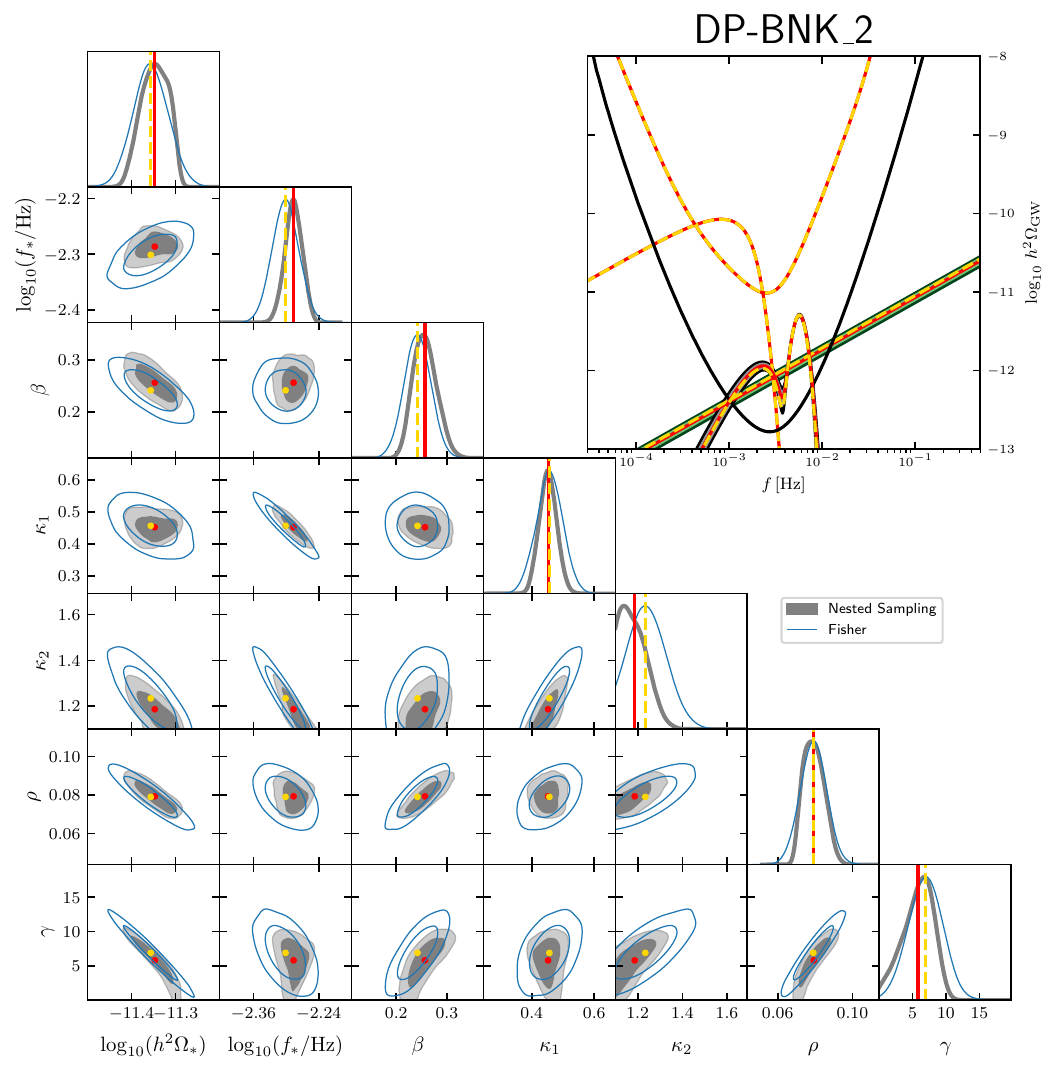}
   \includegraphics[width=.495\columnwidth]{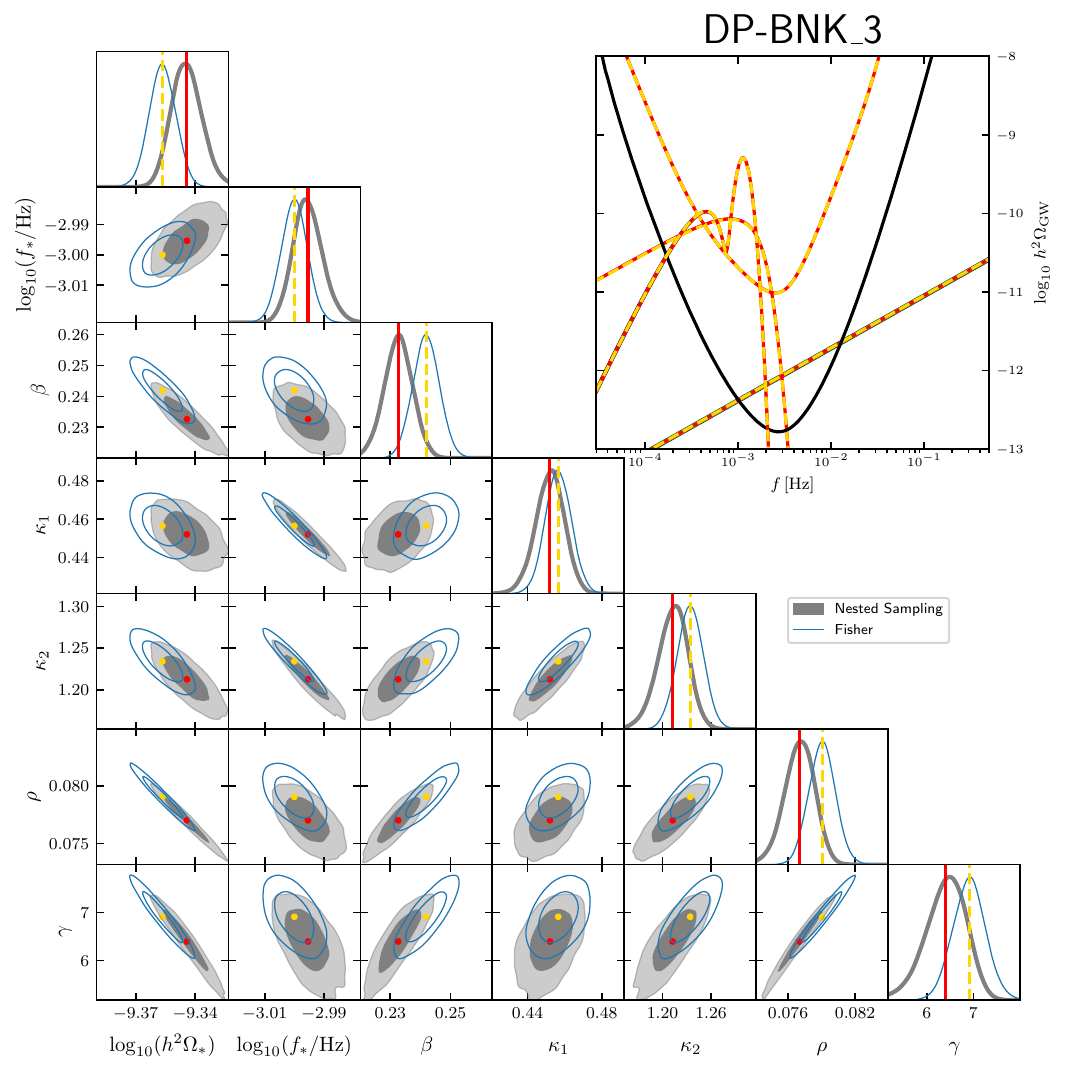}
		\includegraphics[width=.48\columnwidth]{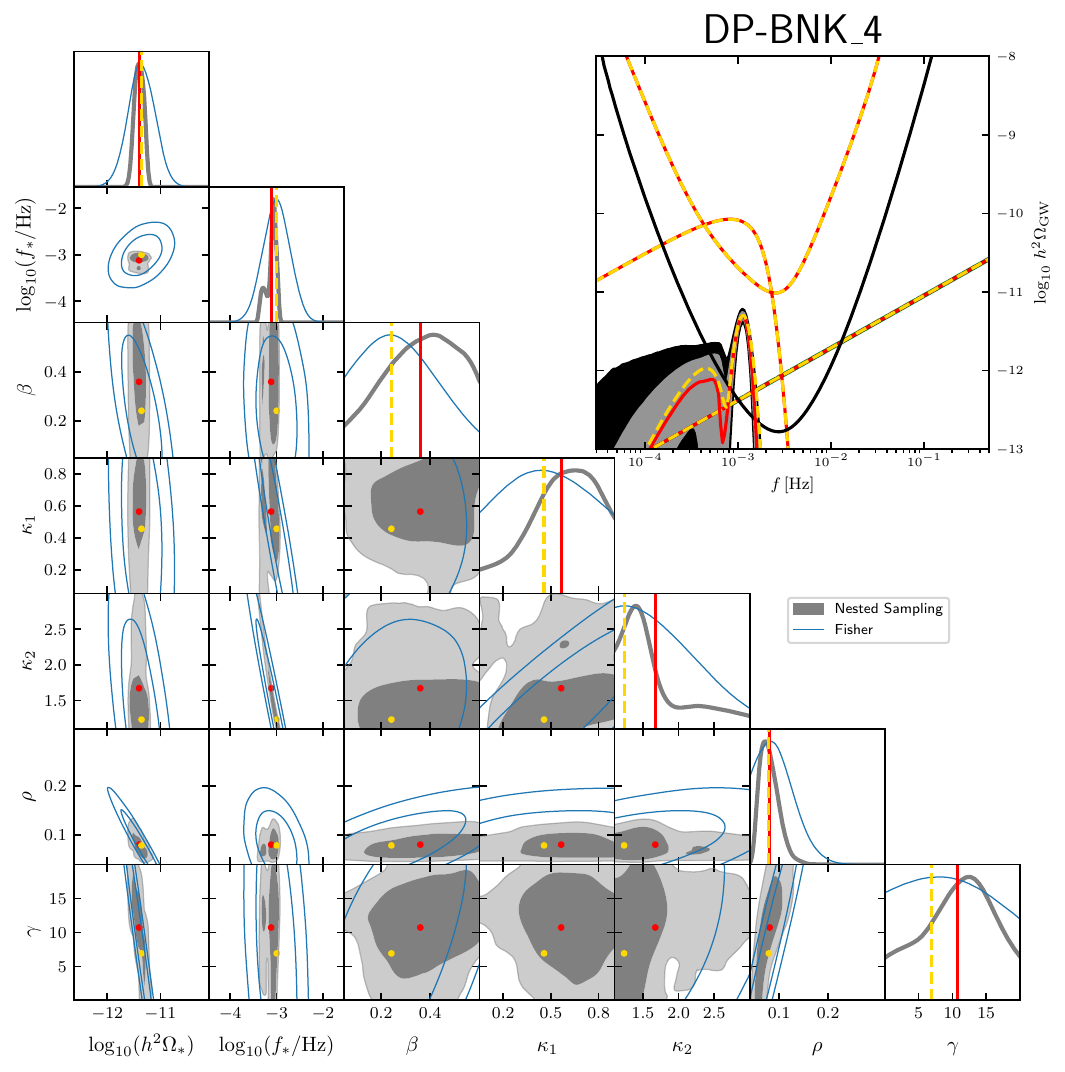}
        \includegraphics[width=\columnwidth]{Legend_Reconstruction.pdf}
	\end{center}
\caption{ \small 
 1D and 2D posterior distributions 
derived from the DP template reconstruction
of the benchmarks DP-BNK\_1 (top left),  DP-BNK\_2 (top right), DP-BNK\_3 (bottom left) and DP-BNK\_4 (bottom right) defined as in \cref{eq:templateDP} with $\{\log_{10}h^2\Omega_* ,\,f_*/\textrm{Hz}\}$ equal to $\{-9.35, 5\}$,  $\{-11.35, 5\}$,  $\{-9.35, 1\}$ and  $\{-11.35, 1\}$, respectively, and $\{\beta , \,\kappa_1,\, \kappa_2,\, \rho, \,\gamma \} = \{0.242, \,0.456,\, 1.234,\, 0.08, \,6.91 \}$ fixed for all benchmarks. Lines' styles and color codes
are as in \cref{fig:PL_MCMC}.
}
\label{doublepeak1_MCMC}
\end{figure}

 In  \cref{Fisher_double} we report the Fisher reconstruction uncertainties on the DP template parameters as a function of the values of $\{\mathcal{A}_s,\Delta\}$ when $f_* = 5 \,\mathrm{mHz}$. To produce the figure, we fix the template parameter $f_*$ and for each pair of values of $\{\mathcal{A}_s,\Delta\}$ we compute the six left DP template parameters. 
 Subsequently, we input the values of these seven parameters into our Fisher code that determines the \seCL reconstruction errors on the DP parameters. 
With this choice of $f_*$, the first peak happens to be close to the one of the galactic foreground (around $1 \,\mathrm{mHz}$) while the second and higher peak is always above the galactic foreground and comparable to the extragalactic one only for $\log_{10} h^2\Omega_* \lesssim -12$, approximately corresponding to $\log_{10} \mathcal{A}_s \lesssim -3$ (cf.~top panel in~\cref{doublepeak1_MCMC}). As a consequence, including (or excluding) the foreground reconstruction in the analysis largely influences the estimate of the parameters associated with the first and second peak, as discussed in detail below.

The top-left panel
of \cref{Fisher_double}
highlights that errors associated with the overall amplitude only weakly depend on $\Delta$, and the 0.01 and 0.3 error contour lines computed in the presence (dashed white) or the absence (dashed black) of foregrounds slightly differ from each other for $\log_{10}\mathcal{A}_s \lesssim -3.3$.  
This can be explained by the fact that, in the vicinity of such a value of $\mathcal{A}_s$, the extragalactic foreground starts covering the second and higher peak of the template.
For the same reason, the parameters associated with the second peak all display a similar behaviour (see below). 

The other panels show the errors on the parameters controlling the double peak shape. 
For a given amplitude $\mathcal{A}_s$, they reach minimal values at  $\Delta\simeq 0.2$ -- $0.4$. Departing from this range of $\Delta$ implies larger errors. Indeed, the SNR decreases towards small $\Delta$, while the two peaks tend to merge towards high $\Delta$, so that the overall structure becomes more problematic to reconstruct (and several degeneracies among parameters are expected). 
Only the error on $\gamma$ has a mild dependence on $\Delta$. The parameter determines the behaviour of the very steep ultraviolet tail which fast falls below the LISA sensitivity. 
Only a very loud signal with $\log_{10}\mathcal{A}_s \gtrsim -2$ (SNR $\gtrsim 10^5$) would allow us to reconstruct $\gamma$ below the percent level (for this case, $1\,\%$ error lines end outside the panel for the plotted range).

\Cref{Fisher_double} moreover shows how foregrounds deteriorate the reconstruction by including error lines from an analysis done in absence of them. 
As a rule of thumb, for all parameters but  $\gamma$,
a reconstruction error below $30\%$ in the range $\Delta\simeq 0.35$ requires SNR\,$\gtrsim \,50$ when foregrounds are absent (\emph{i.e.,}~perfectly known a priori). 
The inclusion of foregrounds 
significantly affects the errors on the parameters associated with the first peak, \emph{i.e.,} $f_*,\beta,\kappa_1$, already for $\log_{10}\mathcal{A}_s \lesssim -2.0$ (roughly corresponding to $h^2\Omega_*$ in the range  $[10^{-10}, 10^{-9}]$) and in particular for values of $\Delta\ge 0.5$. As already mentioned, 
 for $f_* = 5\,\mathrm{mHz}$ and large $\Delta$, the first peak of the template competes with the peak of the galactic foreground, resulting in a more challenging reconstruction for this part of the primordial signal.
In contrast, error lines on the parameters associated with the second peak, \emph{i.e.,} $\rho,\kappa_2,\gamma$, feel the presence of the foregrounds in the analysis only when the overall amplitude is decreased, \emph{i.e.,} $\log_{10}\mathcal{A}_s\lesssim -3.3$, and the second peak amplitude becomes comparable with the one of the extragalactic foreground.
\\

To complete our analysis on the Fisher forecast, in \cref{Fisher_double_pivot} we show the uncertainties on the DP template parameters as a function of the values  $\{\mathcal{A}_s,f_*\}$; namely we fix the template shape by setting the log-normal power spectrum parameter to $\Delta = 0.2$ and we change the overall amplitude of the signal and the scale setting its position in the LISA band. Thus, in all panels, error lines follow the behaviour of the ones of constant SNR.\footnote{Note the similarity between the panels in  \cref{Fisher_double_pivot} and the first three panels in \cref{fig:Fisher_LN}. As in the current case, there we scan overall amplitude and pivot scale of a template with fixed shape.} Furthermore, for all parameters, the error has its minimum, for a fixed overall amplitude, around $\log_{10}f_* \simeq -2.5$ with a small shift towards higher $f_*$ if the parameter is related to the first peak (note that $\log_{10}(h^2\Omega_*)$ is determined by the second peak),
the reason being that thanks to this shift, the first peak is centered where LISA has its maximum sensitivity.
 The parameters which are easily reconstructed are the overall amplitude and the pivot scale, while $\gamma$ remains the most elusive, although large errors on $\gamma$ do not compromise the reconstruction of the overall structure of the template. Errors on the parameters $\beta,\kappa_1,\kappa_2,\rho$ have  similar behaviours; for the sweetest spot in $f_*$ errors are below $30\%$ for $\log_{10}\mathcal{A}_s\gtrsim -3.3$. Finally, the effects of  foregrounds are relevant for $\log_{10}(f_*/\mathrm{Hz})\lesssim -2.4$ and for $-3.4\lesssim\log_{10}\mathcal{A}_s\lesssim -2.5$.
 As before,
  for this range of $f_*$ the extragalactic and (mainly) the galactic foreground can bury the primordial signal at some LISA frequencies,  depending on the amplitude of the primordial signal.\\

\noindent {\bf Benchmarks.---} 
Inflationary scenarios with a narrowly peaked primordial power spectrum are among the main motivations for an SGWB following the DP template. We focus on the narrow log-normal power spectrum $\mathcal{P}_\zeta^{\rm ln}$ with $\Delta = 0.2$ to set our DB benchmarks.
The choice $\Delta = 0.2$ fixes 
five out of seven parameters of the DP template, namely $\beta = 0.242, \,\kappa_1=0.456,\, \kappa_2=1.234,\, \rho=0.08$ and $\gamma=6.91$. 
We then consider the case $\mathcal{A}_s =10^{-2}$  ($\mathcal{A}_s =10^{-3}$) 
as a regime where the curvature power spectrum seeds a PBH abundance around (much below) the observed dark matter relic density (see \cref{sec:PBH}). These two values of $\mathcal{A}_s$, combined with $\Delta = 0.2$, set the DP parameter $h^2\Omega_*$ to $-9.35$ and $-11.35$. In addition, we consider the cases $f_*=1\,\mathrm{mHz}$ and $f_* = 5\,\mathrm{mHz}$ to illustrate how the galactic foreground, peaking around $1\, \mathrm{mHz}$, impacts the parameter reconstruction. We denote as DP-BNK\_\{1,\,2,\,3,\,4\} the DP benchmarks with  $\{\log_{10}(h^2\,\Omega_*),\,f_*/\mathrm{mHz}\}=\{(-9.35,\,5),\,(-11.35,\,5),\,(-9.35,\,1),\,(-11.35,\,1)\}$, and 
$\beta, \kappa_1, \kappa_2, \rho$ and $ \gamma$ fixed as above.

In  \cref{doublepeak1_MCMC} we present the DB-template-based reconstruction for these benchmarks. 
For the DP-BNK\_1 and DP-BNK\_3, which have large amplitude,  all the parameters are well reconstructed and consistent with the injected values. The consistency is at $95\%\ \rm{C.L.}$ when $f_{*}= 5 \,\rm{mHz}$ for DP-BNK\_1 and $68\%\ \rm{C.L.}$ for DP-BNK\_3.
In both cases, some degeneracies are present, but they are expected given the complex template parameter dependencies. In particular, the parameter $\gamma$ exhibits several degeneracies with other parameters (mainly $\rho$ and $h^2 \Omega_*$). For instance, 
opportune variations of the width and skewness of the second peak, $\gamma$ and $\rho$, can mimic a change in the overall amplitude $\mathcal{A}_s$ (see \cref{fig:doublepeak_parameters} in \cref{app:variation_of_template_parameters}), and this degeneracy is reflected in the corner plot. 
Nevertheless, we find that all these degeneracies do not jeopardise the reconstruction performance. Remarkably, all the 1D posteriors are Gaussian: this fact, combined with a large SNR signal, reflects  a good agreement between the nested sampling and the Fisher analysis. Despite the large signal, both foregrounds appear very well reconstructed, as well as the noise parameters (\emph{i.e.,} unobservable error bars). This result is likely due to the DP template's peculiar shape which does not resemble the one of the foregrounds.  \\

The situation changes when we consider the benchmarks DP-BNK\_2 and DP-BNK\_4, which have lower amplitudes than the previous benchmarks. For DP-BNK\_2, the pivot frequency $f_*$ is sufficiently large for the signal not to be covered by the galactic foreground despite its small amplitude. As a consequence, all the DP parameters are well reconstructed, with the mean values consistent with the injected values at \seCL
levels. Also, in this case, the degeneracies do not jeopardise the likelihood sampling. Moreover, the Fisher approximation is still able to recover the shapes of the posterior distribution of the parameters, slightly overestimating the errors. 

The reconstruction drastically deteriorates  for DP-BNK\_4. Due to $f_*=1\,\, \mathrm{mHz}$, the signal is entirely covered by the galactic foreground. Several posterior distributions show a non-Gaussian behaviour and  some reconstructions are loose. In this regime, the Fisher formalism is unable to capture the features of the posterior distribution of the model parameters. \\

\subsection{Forecasts for the excited state template}
\label{sec4_excited}
\begin{figure}
\centering
\includegraphics[width=0.32\columnwidth]{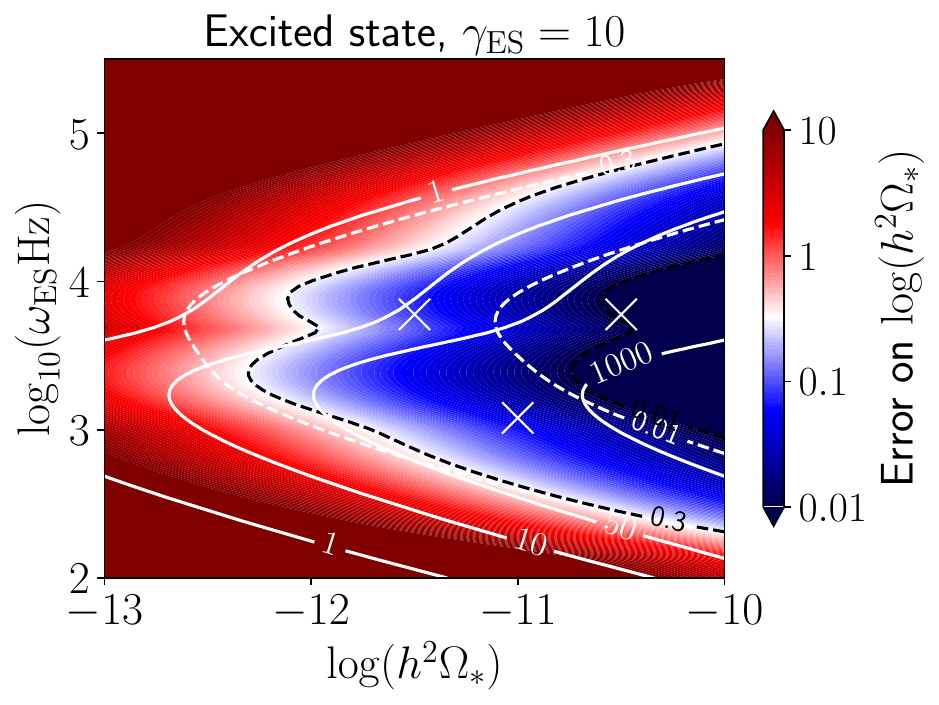}
\includegraphics[width=0.32\columnwidth]{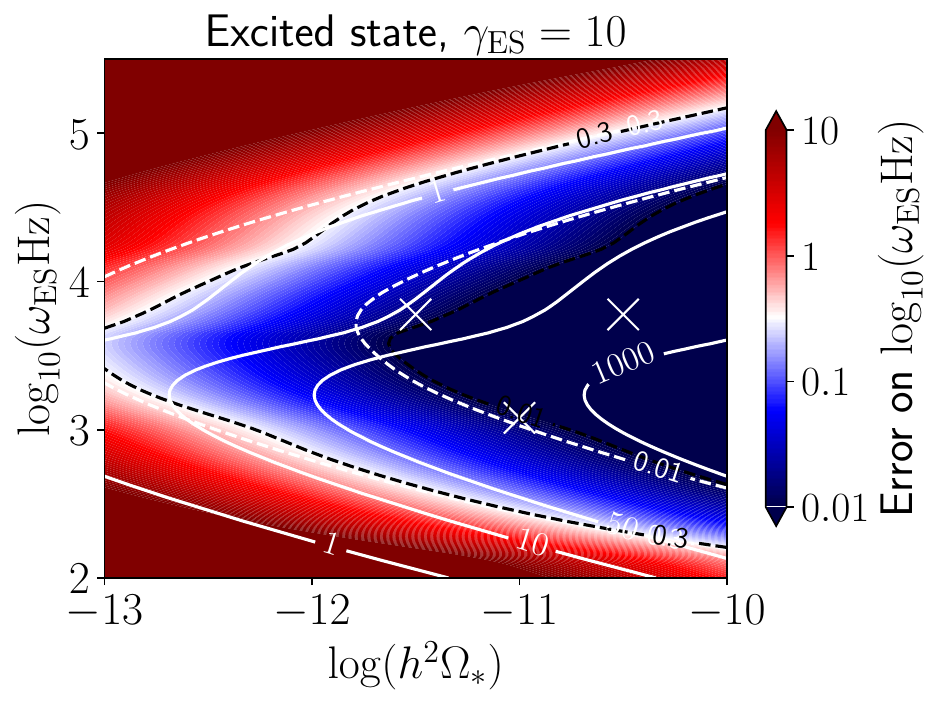}
\includegraphics[width=0.32\columnwidth]{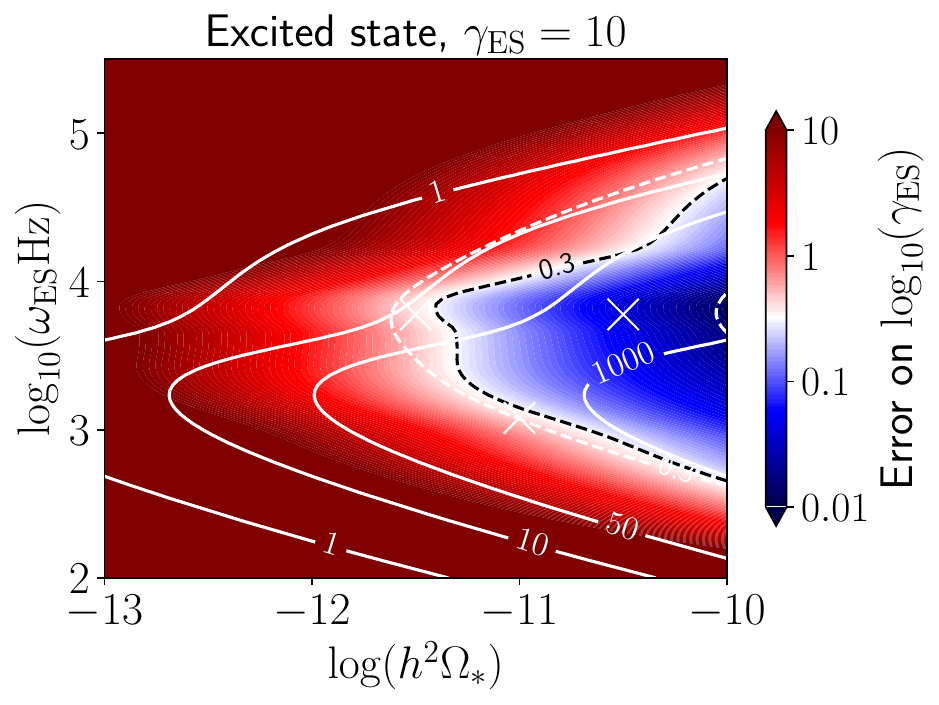}
\caption{\small 
Fisher forecasts for the ES template. The panels show the \seCL reconstruction error on $h^2\Omega_*$ (left), $\omega_{\rm ES}$Hz (middle) and $\gamma_{\rm ES}$ (right) as a function of the injected values
of $h^2\Omega_*$, $\omega_{\rm ES}$Hz and $\gamma_{\rm ES}$  set as specified in the axes and the title of each panel.
SNR contour lines are plotted in white. The pairs of dashed (dotted) lines mark the $ \sigma =0.3$ ($\sigma =0.01$) contours, respectively in the absence (white) and in the presence (black) of foregrounds. The crosses display the benchmarks ES-BNK\_1, ES-BNK\_2 and ES-BNK\_3. 
}
\label{fig:Fisher_excited}
\end{figure}

\begin{figure}
\centering
\includegraphics[width=0.45\linewidth]{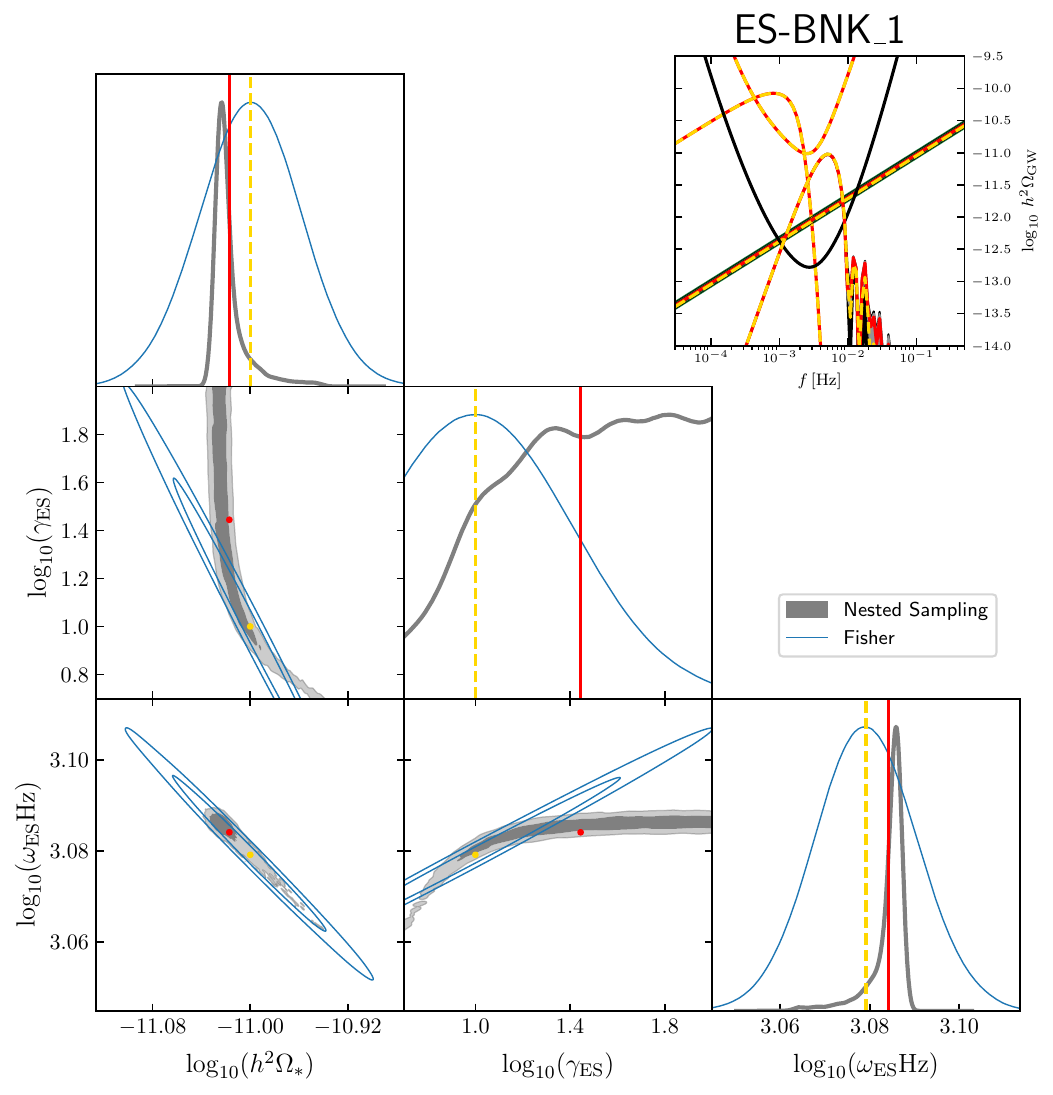}
\includegraphics[width=0.45\linewidth]{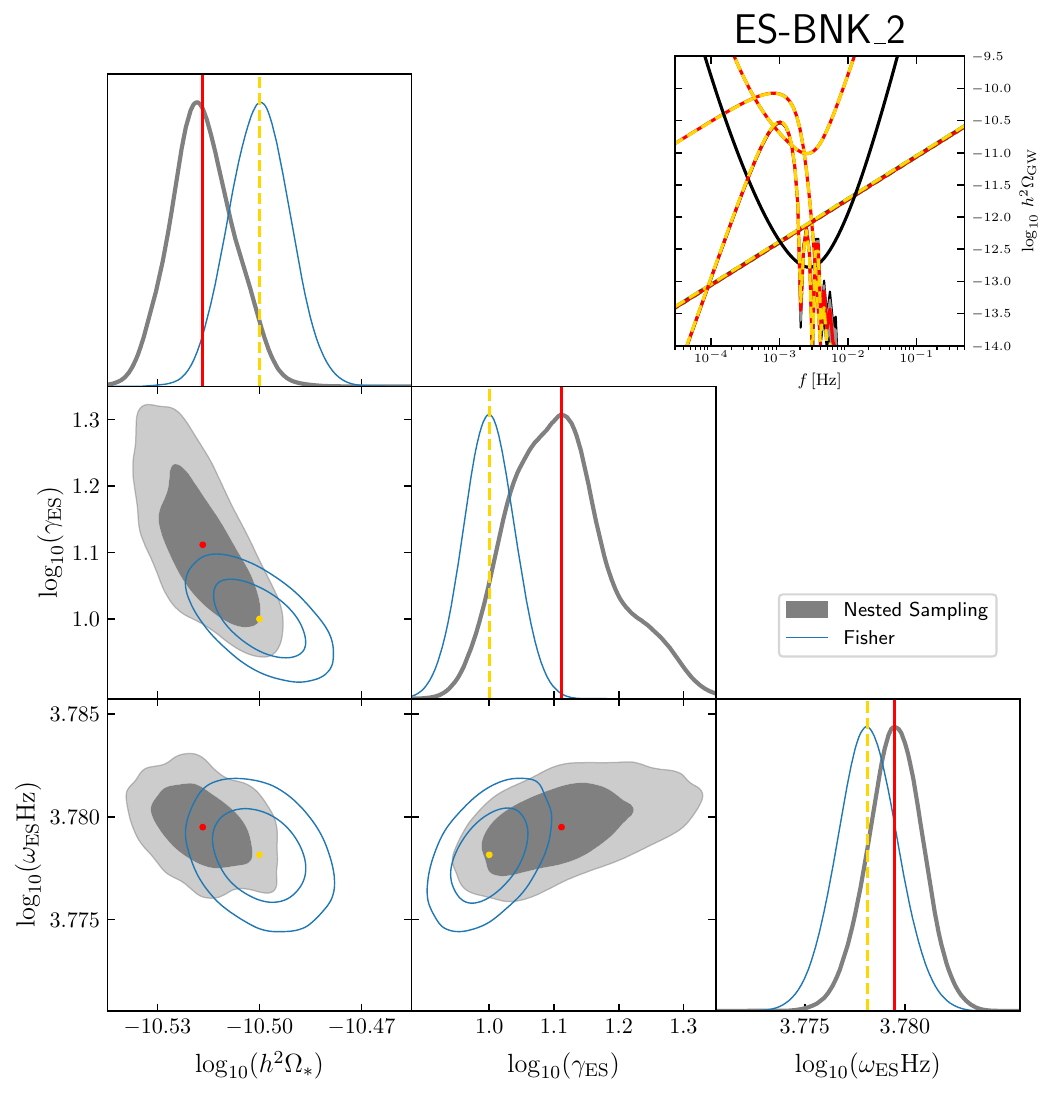}\\
\includegraphics[width=0.45\linewidth]{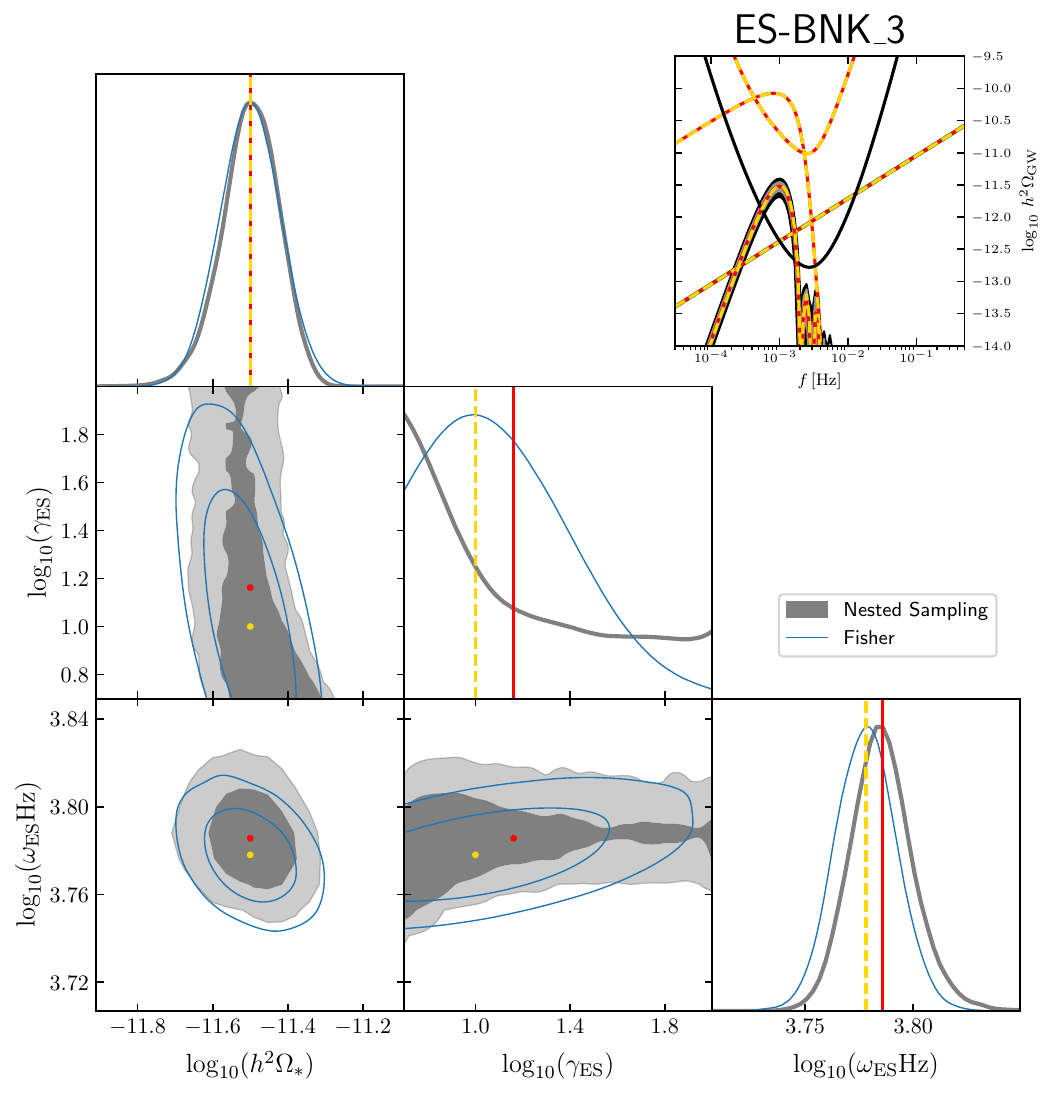}
\includegraphics[width=\columnwidth]{Legend_Reconstruction.pdf}
\caption{\small 
 1D and 2D posterior distributions 
derived from the ES template reconstruction of the benchmarks
ES-BNK\_1 (top left),  ES-BNK\_2 (top right) and ES-BNK\_3 (bottom). Lines' styles and color codes
are as in \cref{fig:PL_MCMC}.
ES-BNK\_1, ES-BNK\_2 and ES-BNK\_3 are defined as in \cref{templateexcited} with 
$\{\log_{10}h^2\Omega_*,\gamma_{\mathrm{ES}},\omega_{\mathrm{ES}}\mathrm{Hz}\}$ set at $\{-11,10,6 \cdot 10^{3} \}$,  $\{-11.5,10,6 \cdot 10^{3} \}$ and $\{-11,10,1.2 \cdot 10^{3}\}$, respectively. 
}
\label{Excited_triangular}
\end{figure}

The ES template described in \cref{templateexcited} is characterised by the three parameters $h^2\Omega_* , \gamma_{\mathrm{ES}}$ and $\omega_{\mathrm{ES}}$. 
Our Fisher reconstruction forecast for this template is presented in \cref{fig:Fisher_excited}.  
Reconstruction errors on $h^2\Omega_*$ and $\omega_{\mathrm{ES}}$ below 1 -- 10\% typically require SNR $\gtrsim$ 50 when $3\le \log_{10}{\omega_{\mathrm{ES}}} \le 4$. We recall that the peak of the template is related to the frequency of the oscillations in the ultraviolet tail by $ f_{\mathrm{max}} \simeq 6/ \omega_{\mathrm{ES}} $ (see \cref{sec:excited}). This relationship explains why the relative errors on the various parameters are minimized at $3\le\log_{10}(\omega_{\rm ES} \,\mathrm{Hz})\le 4$, which range corresponds to signals peaking at $f_{\mathrm{max}} \simeq (0.6- 6)\, \mathrm{mHz}$ where
the sensitivity of LISA is optimal.   Reconstructing the amplitude $h^2\Omega_*$ with the same precision, on the other hand, requires larger amplitude values compared to  $\omega_{\mathrm{ES}}$.
Finally, the parameter $\gamma_{\mathrm{ES}}$, which controls the extension of the oscillations, is harder to measure: at $f_{\rm max}\sim 1\, {\rm mHz}$, the error on $\log_{10}\gamma_{\mathrm{ES}}$ is below 0.3 only if $\log_{10}\, h^2\Omega_*\gtrsim-11.5$. Indeed, if the amplitude is not sufficiently large, the ES oscillatory features are not detectable as they fall outside the LISA sensitivity.
As for the foregrounds, they affect the errors on the parameters primarily in the region $\log_{10}(\omega_{\rm ES}\mathrm{Hz})\gtrsim 3.5 $, for which the signal peaks at a frequency $\lesssim 2 \,\mathrm {mHz}$, overlapping with the  galactic foreground.

\noindent {\bf Benchmarks.---} 
On theoretical grounds the parameter $\gamma_{\mathrm{ES}}$ is expected to be much larger than one, but cannot be arbitrarily large. In view of the microphysical setups motivating the template in \cref{sec:excited}, we fix $\gamma_{\mathrm{ES}}=10$ for all benchmarks. To test ES scenarios peaking at frequencies lower and higher than the maximum of the galactic foreground, we consider the choices $\omega_{\mathrm{ES}}\,= 6\,\mathrm{mHz}^{-1}$ and $\omega_{\mathrm{ES}}\,= 6/5\,\mathrm{mHz}^{-1}$, respectively leading to an ES main peak at $1\,\mathrm{mHz}$ and $5\,\mathrm{mHz}$. For the former choice, in order to investigate the impact of the galactic foreground when the ES signal is just or much below it, we consider two amplitude scenarios, namely $\log_{10}(h^2\Omega_*)=-10.5$ and $\log_{10}(h^2\Omega_*)=-11.5$. This leads us to consider benchmarks ES-BNK\_1, ES-BNK\_2 and 
ES-BNK\_3 whose parameters $\{\log_{10}(h^2\Omega_*),\gamma_{\mathrm{ES}},\,\omega_{\mathrm{ES}}\,\mathrm{mHz}\}$ are respectively set at 
$\{-11,\,10,\,6/5\}$,
$\{-10.5,\,10,\,6\}$ and 
$\{-11.5,\,10,\,6\}$.

We first discuss the reconstruction of ES-BNK\_1 (see top left panel in \cref{Excited_triangular}). The signal peaks at $5\,{\rm mHz}$, where the galactic foreground has small amplitude, and stands above the extragalactic one. The parameters $h^2\Omega_*$ and $\omega_{\rm{ES}}$ are well constrained with the reconstructed mean values within \nfCL error bars. Their 2D posterior is Gaussian, so that the Fisher approximation recovers it. However, a remarkable non-Gaussian correlation arises in the posterior of $\gamma_{\mathrm{ES}}$ with $h^2\Omega_*$ and $\omega_{\mathrm{ES}}$. In particular, the 2D posteriors have 45-degree turns for small $\gamma_{\mathrm{ES}}$. This can be explained as a `second order degeneracy':  simultaneously changing $h^2\Omega_*$ and $\omega_{\mathrm{ES}}$ mimics a variation in $\gamma_{\mathrm{ES}}$ (see \cref{fig:excitedstates} in \cref{app:variation_of_template_parameters}).
In this case, as expected, the Fisher approximation fails, and we can only set a mild lower bound at $68$\% C.L.. Both galactic and extragalactic foregrounds are reconstructed, although constraints on the latter are prior-dominated, as they are masked by the primordial signal within the LISA band.

Concerning ES-BNK\_2 (see top right panel in~\cref{Excited_triangular}), 
the peak lies below the galactic foreground, despite the ES-BNK\_2's having an amplitude larger than ES-BNK\_1. However, constraints on $\omega_{\mathrm{ES}}$ and $\gamma_{\mathrm{ES}}$ are still quite good. In particular,  the parameter  $ \gamma_{\mathrm{ES}}$ can be detected, as the oscillatory part of the signal falls within the LISA sensitivity thanks to its relatively large amplitude. Some degeneracies among the parameters appear, but the non-Gaussianity of the posterior is milder than for ES-BNK\_1, and we can notice a good agreement with the Fisher analysis, which captures quite well the shapes of the posteriors. Noise and foregrounds are also well reconstructed. 

Finally, the benchmark ES-BNK\_3 peaks at $1\, \rm{mHz}$ (see bottom panel in~\cref{Excited_triangular}) like ES-BNK\_2, but has a smaller amplitude, so the peak is well below the galactic foreground. Despite this fact, the overall peak's amplitude and position are well reconstructed, with their mean values within the \seCL error contour. For their 2D posterior, there is good agreement between the Fisher and the nested sampling analyses. As for ES-BNK\_1, we notice a degeneracy between  $\omega_{\mathrm{ES}}$ and $\gamma_{\mathrm{ES}}$. The oscillations fall outside the sensitivity of LISA, and, as a consequence, the parameter $\gamma_{\mathrm{ES}}$ results unconstrained.

\subsection{Forecasts for the linear oscillations template}
\label{sec4_linear}
\begin{figure}
\centering
\includegraphics[width=0.44\columnwidth]{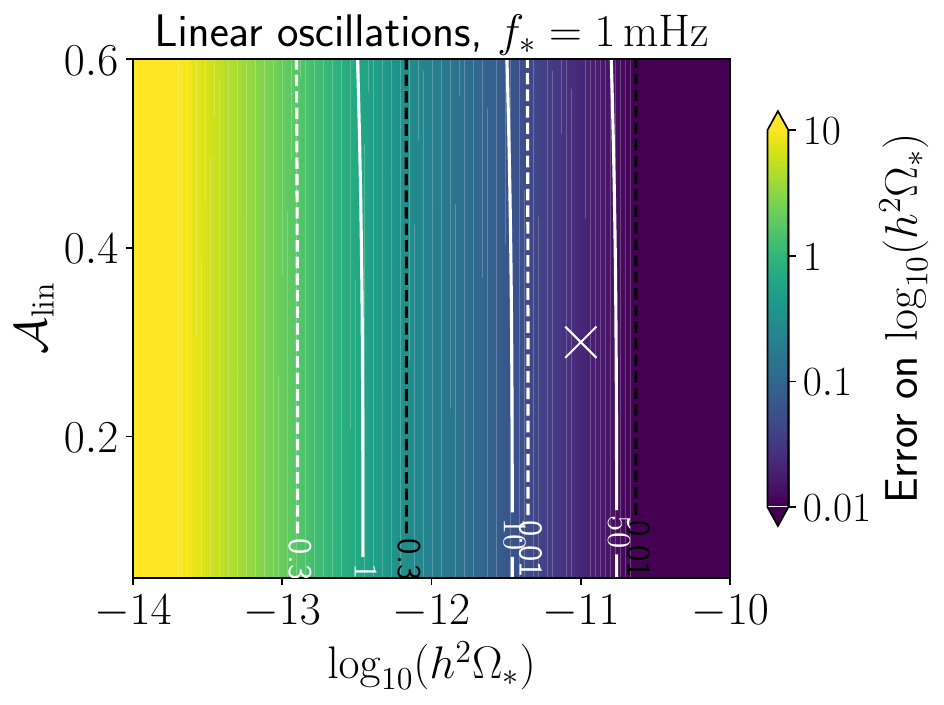}
\includegraphics[width=0.44\columnwidth]{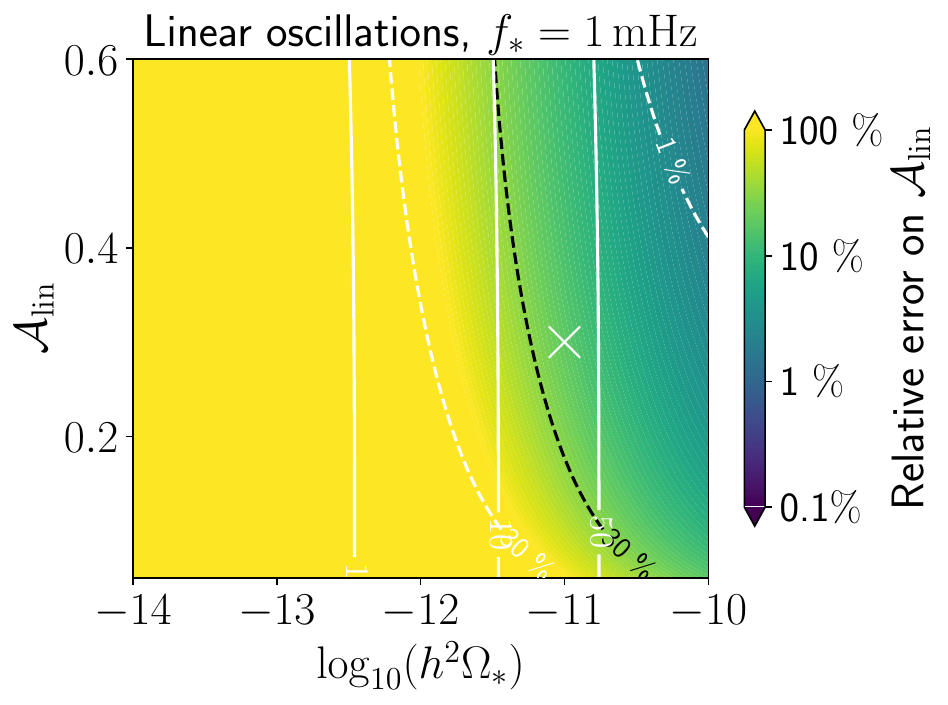}
\includegraphics[width=0.44\columnwidth]{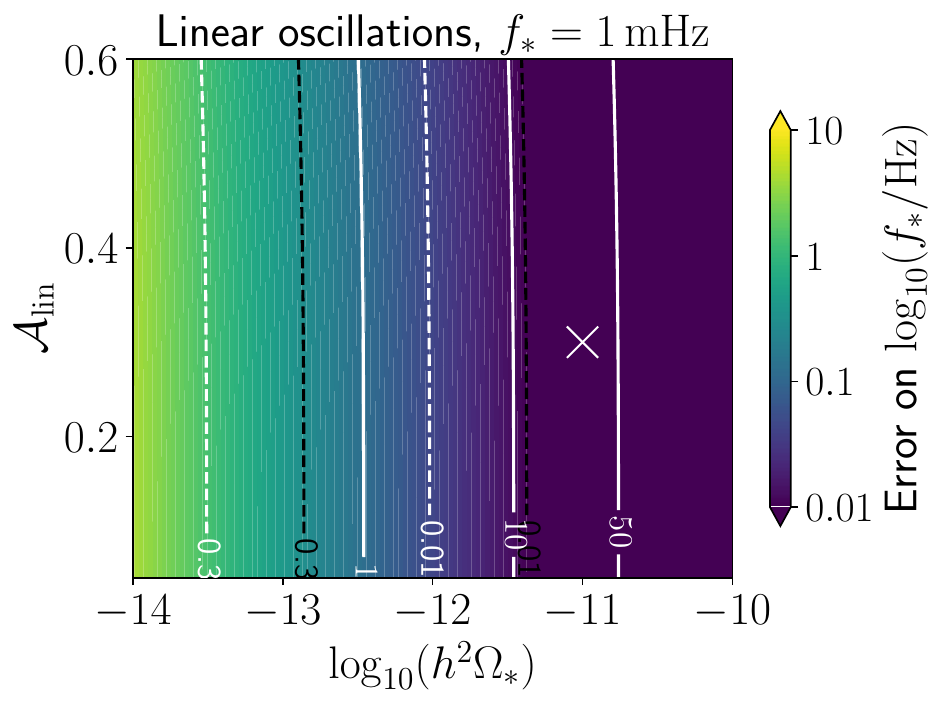}
\includegraphics[width=0.44\columnwidth]{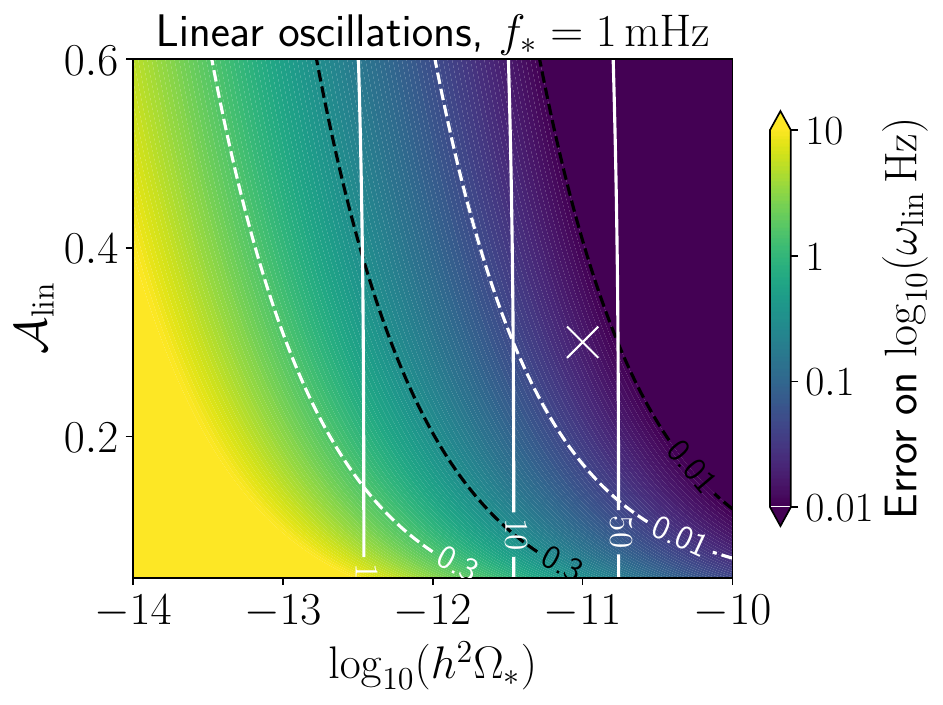}\\
\includegraphics[width=0.44\columnwidth]{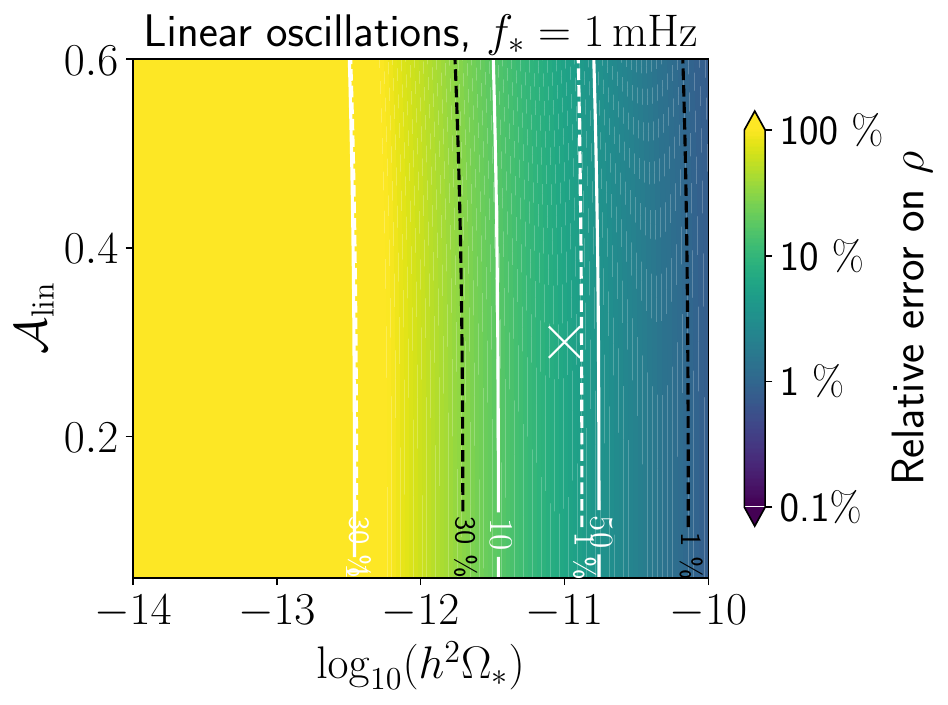}
\includegraphics[width=0.43\columnwidth]{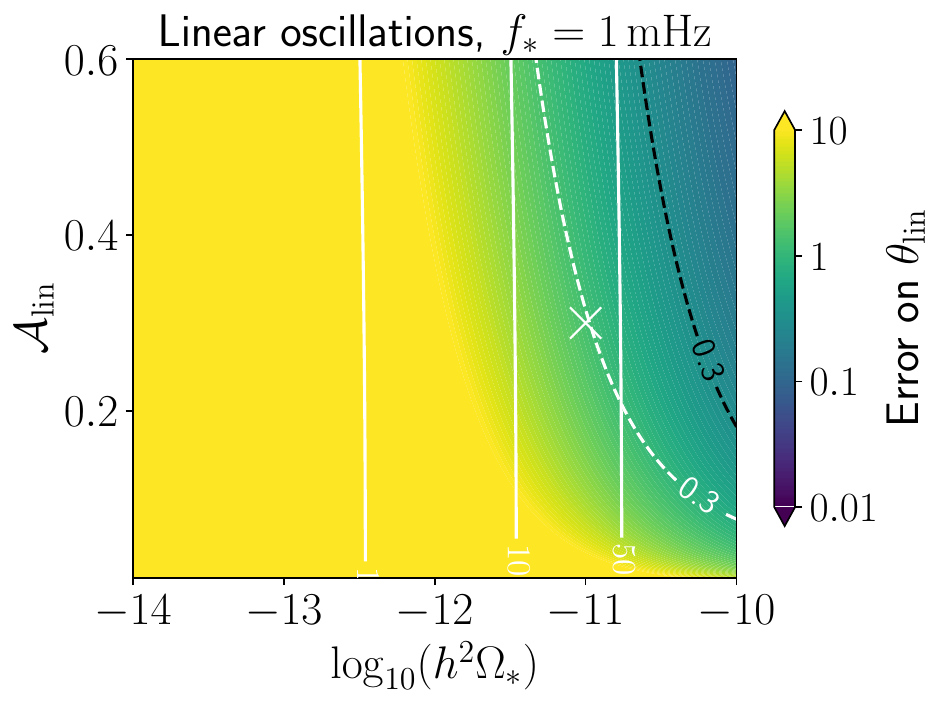}\\
\caption{\small
Fisher forecasts for the LO template with the LN envelope. The panels show the \seCL reconstruction error on $h^2\Omega_*$ (top left), $\mathcal{A}_{\rm lin}$ (top right), $f_*$ (central left), $\omega_{\rm lin}$ (central right), $\rho$ (bottom left) and $\theta_{\rm lin}$ (bottom right) as a function of the injected values $h^2\Omega_*$ and $\mathcal{A}_{\rm lin}$ and by fixing the other parameters to $\{f_*= 1\,\mathrm{mHz} ,\,\rho = 0.08, \,\omega_{\mathrm{lin}}\,\mathrm{Hz}= 5\cdot10^{4},\, \theta_{\mathrm{lin}}= 0 \}$.
SNR contour lines are plotted in white. 
 Depending on the panel, the pairs of dashed lines mark the absolute error $ \sigma =0.3$ ($\sigma =0.01$) and relative error 30\% ($1\%$) contours, respectively in the absence [white] and in the presence [black] of foregrounds.  The cross displays the benchmarks LO-BNK\_1.
 }
\label{fig:Fisher_sharp2}
\end{figure}

We now focus on the LO template in \cref{eq:sharp-template} where, for concreteness, we take the log-normal bump for the envelope $\Omega_{\rm GW}^{\rm env}(f)$. \Cref{fig:Fisher_sharp2} shows the results obtained with the Fisher analysis in a parameter-space slice crossing the benchmarks detailed underneath. The 
left panels show the (either relative or absolute) error on the  
envelope parameters, while the right panels show the errors on the LO parameters, namely the amplitude, frequency and phase of the oscillations. In each panel, we vary the envelope amplitude $h^2 \Omega_{\rm GW}$ as well as the magnitude of the oscillations $\mathcal{A}_{\rm lin}$, while for the other parameters we take $\{f_*/{\rm mHz},\,\rho,\,\omega_{\rm lin}\mathrm{Hz},\theta_{\rm lin}\} = \{ 1, 0.08, 5\cdot 10^4   , 0\}$.

Concerning the envelope parameters, the amplitude and the pivot frequency,  $h^2 \Omega_*$ and $f_*$, have errors below $10\,\%$ for $\log_{10}h^2\Omega_*\gtrsim -13$, whereas a similar error on the width $\rho$ requires $\log_{10}h^2\Omega_*\gtrsim -11.5$, in agreement with the analysis in \cref{subsec:log-normal}. Such errors turn out to be almost independent of the amplitude of the oscillation $\mathcal{A}_{\rm lin}$. Concerning the LO parameters, errors degrade as the amplitude $\mathcal{A}_{\rm lin}$ decreases. Regarding the amplitude of the oscillations, an accuracy below $30\%$ (in the considered parameter space slice) requires $\log_{10}h^2\Omega_* \gtrsim -11$, with a slight dependence on the amplitude of the oscillations, in particular the error deteriorates at $\mathcal{A}_{\rm lin}\lesssim 0.2$.
More surprisingly, pinpointing the frequency $\omega_{\rm lin}$ to good accuracy is possible for signals with smaller amplitude and until very small values of the oscillatory amplitudes. In particular, an absolute error on $\log_{10}\omega_{\rm lin}$ below $0.3$ is granted for $\log_{10}h^2\Omega_* \gtrsim -11.5$ even for very small oscillatory amplitudes, \emph{i.e.,}~$\mathcal{A}_{\rm lin} \lesssim 0.01$. The accuracy on $\theta_{\rm lin}$ is similar to the one on  $\mathcal{A}_{\rm lin}$.

Constraints on all the parameters are clearly affected by astrophysical foregrounds. The impact of the latter varies depending on the specific parameter, but can degrade the accuracy of the error by up to an order of magnitude.\\

\noindent {\bf Benchmarks.---} We consider benchmarks with overall amplitude $h^2 \Omega_* =10^{-11}$. We set the width of the lognormal envelope $\rho = 0.08$ to reproduce the second peak of the benchmarks chosen for the DP template in \cref{doublepeak1_MCMC}. We also fix the amplitude of the oscillations at $\mathcal{A}_{\rm lin} =0.3$ as this value represents a proxy to discriminate between standard radiation domination and a non-standard thermal history (see  \cref{sec:linear}).
Based on theoretical considerations, one expects  $\omega_{\mathrm{lin}} \sim 10 / f_*$, which naturally
provides various oscillations within the range of frequency where the signal peaks. Concretely, we fix it  according to
\cref{strong turn relation} with the microphysical parameters $\eta_\perp =14$ and $\delta = 0.25$. This choice for  $\eta_\perp$ and $\delta$ exemplifies a sharp feature, induced by a strong turn in field space, that gives an amplitude of the curvature power spectrum resulting, within the Gaussian approximation, in an amount of PBHs in the range of the dark matter abundance.
At this point, only $f_*$ is still unfixed, and the two options for its value, $f_*=1\,$mHz and $f_*=5\,$mHz, are the only difference in our benchmarks  LO-BNK\_1 and LO-BNK\_2. Specifically, LO-BNK\_1 and LO-BNK\_2 are defined as the SGWB signals following  \cref{eq:sharp-template} with the LN envelope and the parameters $\{h^2\Omega_*,\,f_*/{\rm mHz },\,\rho,\,{\cal A}_{\rm lin},\,\omega_{\rm lin} {\rm mHz},\,\theta_{\rm lin}\}$ equal to $\{-11,\,1,\,0.08,\,0.3,\,50,\,0\}$ and $\{-11,\,5,\,0.08,\,0.3,\,10,\,0\}$, respectively.

\begin{figure}
	\begin{center}
		\includegraphics[width=.495\columnwidth]{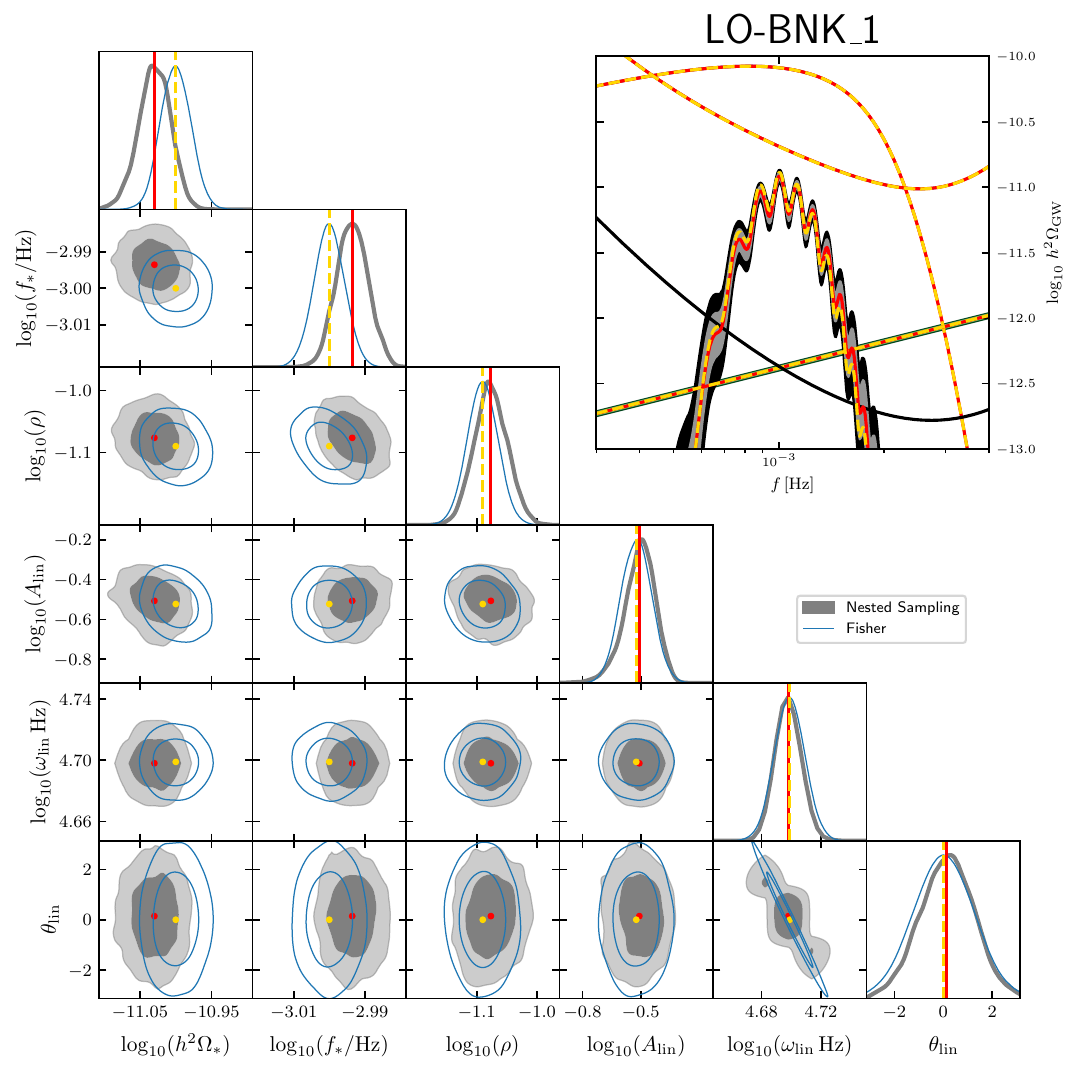}
		\includegraphics[width=.495\columnwidth]{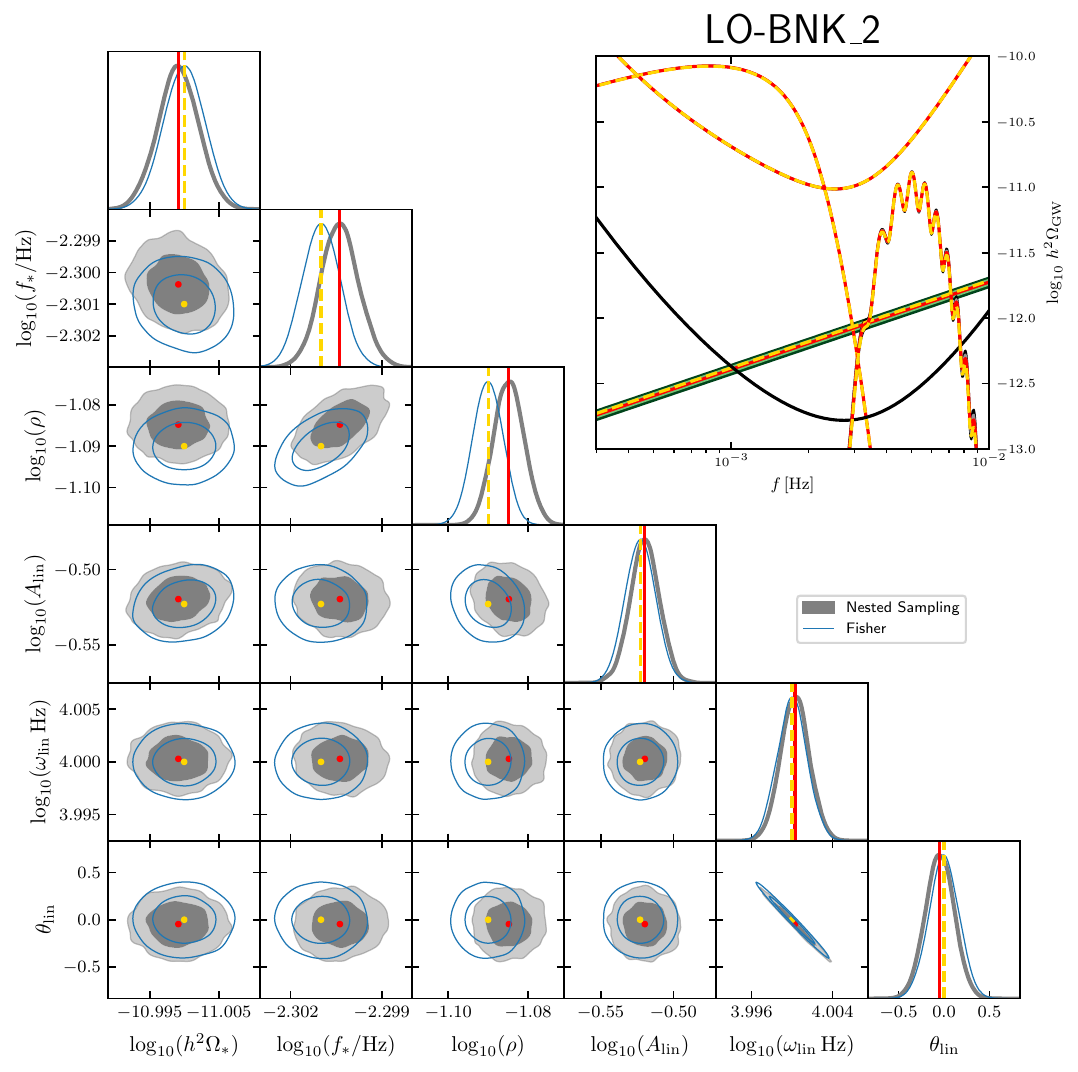}
        \includegraphics[width=\columnwidth]{Legend_Reconstruction.pdf}
	\end{center}

\caption{\small 
1D and 2D posterior distributions 
derived from the LO template reconstruction of the benchmarks LO-BNK\_1 (left) and   LO-BNK\_2 (right).
Lines' styles and color codes
are as in \cref{fig:PL_MCMC}. 
LO-BNK\_1 and   LO-BNK\_2   are defined as in \cref{eq:sharp-template} with $\{f_*/ \mathrm{mHz},\,\log_{10}(\omega_{\mathrm{lin}} \,\mathrm{Hz}) \}= \{5,\,4\} $ and  $\{1,\,4.7\}$ respectively, and fixed
$\{\rho, {\cal A}_{\mathrm {lin}}\} = \{ 0.08, 0.3\}$. 
}  
 \label{sharp_triangular}
\end{figure}

In \cref{sharp_triangular}  (left plot) we show the corner plot of the reconstructed mean value parameters for the LO-BNK\_1.
The oscillatory pattern  turns out to be very well reconstructed for an overall amplitude $\log_{10}(h^2 \Omega_*) =-11$, although the main peak and oscillations are below the galactic foreground. In particular, note the exquisite estimate (order of $0.01\,\%$ error at $1$-sigma) for the frequency of the oscillations. Overall, all the mean values for all the reconstructed parameters are within \seCL from the injected values, while all 2D posteriors are Gaussian and exhibit percent accuracy in very good agreement with the Fisher analysis. A very mild degeneracy appears between the phase $\theta_{\mathrm{lin}}$ and the frequency $\omega_\textrm{lin}$. This degenarecy is more manifest in the reconstruction corner plot of LO-BNK\_2 reported in the right plot of \cref{sharp_triangular}. In this case, the injected signal has the same overall amplitude  as LO-BNK\_1,  but its peak frequency $f_*$ is much larger and consequently,
the signal is shifted to higher frequencies and is not covered by the galactic foreground. Compared to the previous case,  the reconstruction accuracy is even better and the overall oscillatory pattern is  well captured, with relative errors on the various parameters in agreement with the Fisher forecast in \cref{fig:Fisher_sharp2}. We also find that the Fisher analysis captures very well the shape of the posterior distribution. This is true even  for the 2D contour between $\theta_{\mathrm{lin}}$ and the frequency, which, in this case, shows a significant degeneracy.

For both benchmarks, the noise as well as the foregrounds are well reconstructed, with a slightly less accurate reconstruction of the extragalactic foreground when its amplitude is below
the cosmological signal.

\subsection{Forecasts for the resonant oscillations template}
\label{sec:forecastRO}

\begin{figure}
\centering
\includegraphics[width=0.49\textwidth]
{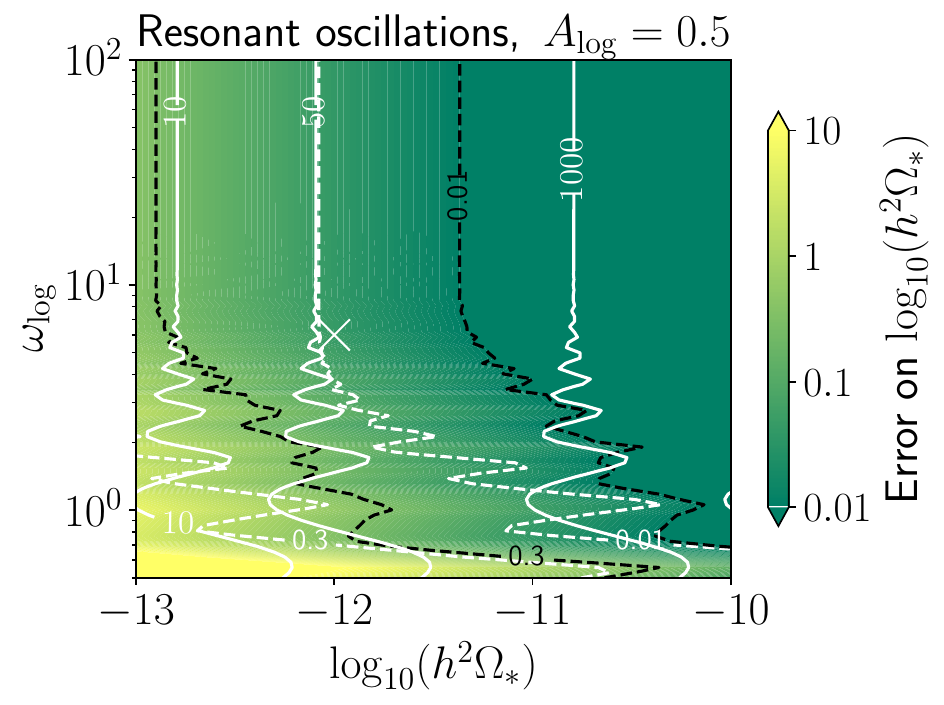}
\includegraphics[width=0.49\columnwidth]{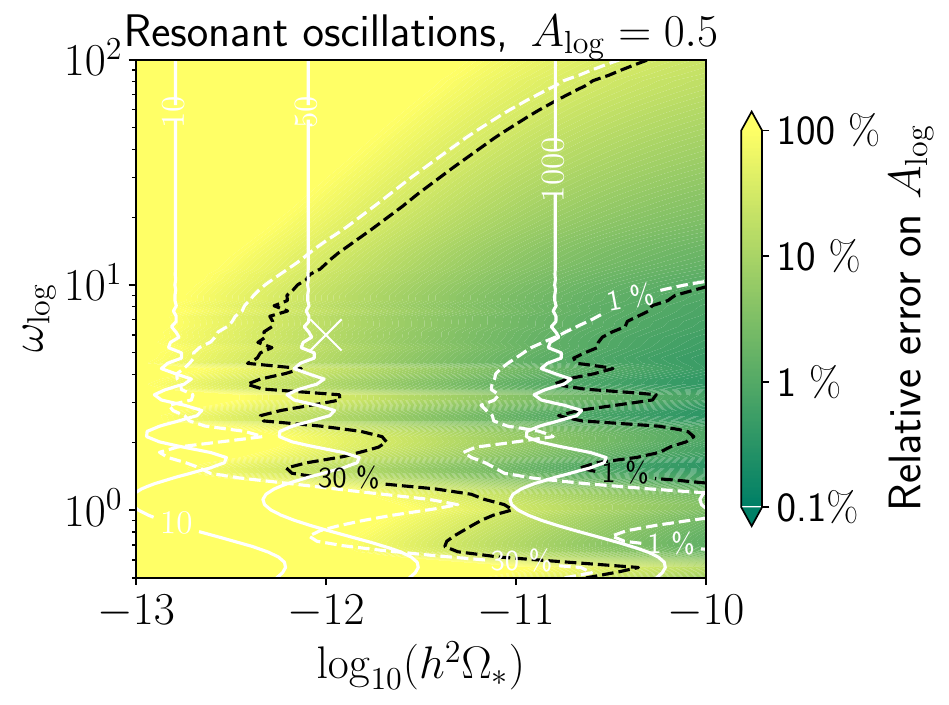}
\includegraphics[width=0.49\columnwidth]{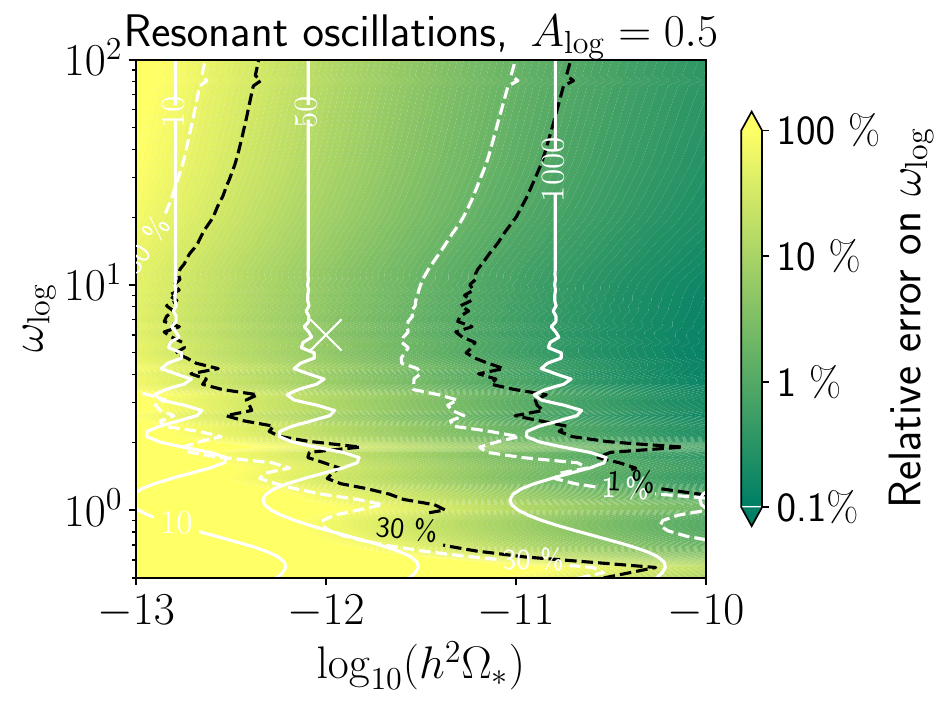}
\includegraphics[width=0.49\columnwidth]{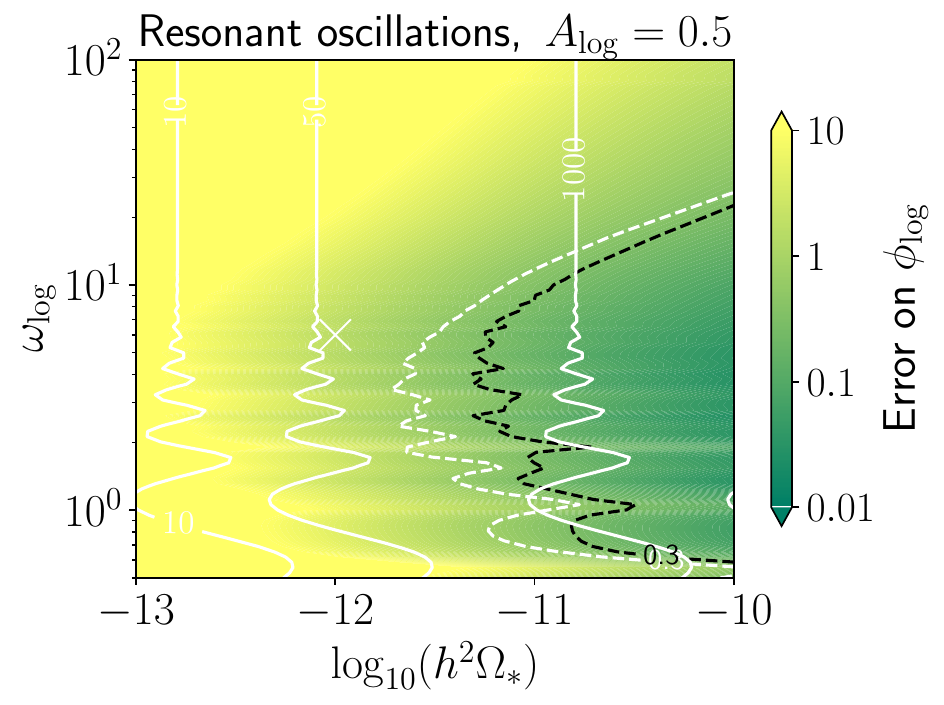}
\caption{
\small
Fisher forecasts for the RO template with the flat envelope $h^2\Omega_{\rm GW}= h^2\Omega_*$. The panels show the \seCL reconstruction error on $\log_{10}(h^2\Omega_*)$ (top left), $A_{\rm log}$ (top right), $\omega_{\rm log}$ (bottom left) and $\phi_{\rm log}$ (bottom right)  as a function of the injected values of $h^2\Omega_*$ and $\omega_{\rm log}$ and by fixing $A_{\rm log}= 0.5$ and $\phi_{\rm log} = 0$.
SNR contour lines are plotted in white. 
 Depending on the panel, the pairs of dashed lines mark the absolute error $ \sigma =0.3$ ($\sigma =0.01$) and relative error 30\% ($1\%$) contours, respectively in the absence [white] and in the presence [black] of foregrounds.  The cross display the benchmark  RO-BNK\_2.
}
\label{Fisher_resonant1}
\end{figure}

The RO template is  defined in eq.~\eqref{eq:resonant-template}. For concreteness, in our RO forecasts we focus on the case where the envelope is a flat spectrum, \emph{i.e.,}~$h^2\Omega_{\rm GW}^{\rm env}=h^2 \Omega_*$. In \cref{Fisher_resonant1} we  forecast the reconstruction errors obtained with the Fisher analysis as a function of $\omega_\textrm{log}$ and the overall amplitude $h^2\Omega_{*}$,  fixing the amplitude of the logarithmic oscillations in the scalar power spectrum  to the representative value $A_\textrm{log}=0.5$.  We recall that   the parameter $\omega_\textrm{log}$  not only controls the frequency  of the two cosines in eq.~\eqref{eq:resonant-template}, but it also affects their amplitudes through the functions $\mathcal{C}_{1,2}$, see  \cref{fig:C012theta12}.
The SNR and reconstruction-error contour lines exhibit an oscillating behavior at $\omega_\textrm{log}\lesssim 10$.
In fact, in such a case a single period of oscillation spans more than the LISA frequency band.
This explains why the error on the reconstruction of $\omega_\textrm{log}$ rapidly degrades  for  $\omega_\textrm{log} \lesssim 4$. 
Furthermore, a small variation in $\omega_\textrm{log}$ can lead to a frequency shift that rapidly modifies the SNR when a maximum is replaced by a minimum at $f\simeq 3\,$mHz, where LISA has its best sensitivity. This of course leads to an important variation of the reconstruction errors.
The same feature is present independently of whether the foregrounds are also reconstructed. On the other hand, for $\omega_\textrm{log}\gtrsim 15$ there are many oscillations within the LISA frequency band and the SNR then becomes independent of $\omega_\textrm{log}$. A relevant role is also played by the function $\mathcal{C}_{1,2}$ (see \cref{sec:resonant} for more details). The minimum relative error is reached for $\omega_\textrm{log}\simeq 8$ for a given amplitude, with an accuracy of $30\%$ (1$\%$) that requires SNR $\gtrsim 8$ (200)  without foregrounds, and larger when foregrounds are taken into account. This behaviour can be ascribed to the dependencies of  $\mathcal{C}_{1,2}$ on $\omega_\textrm{log}$. In particular, for these values of $\omega_\textrm{log}$, $\mathcal{C}_{1}$ and $\mathcal{C}_{2}$ are comparable and the signal becomes the superposition of two harmonics with about the same amplitude ${\cal A}_1\simeq {\cal A}_2$, a case which is easier to identify. In general, the larger $\omega_\textrm{log}$, the smaller the accuracy, simply because the amplitude of the oscillations gets smaller (again see the $\mathcal{C}_{1,2}$ dependencies on $\omega_\textrm{log}$) and eventually one would reach the limit where the signal varies faster than the LISA frequency resolution, as can be argued  from the fact that the error rapidly increases as $\omega_{\rm log}$ grows.\\

\begin{figure}[]
\centering
\includegraphics[width=0.495\columnwidth]{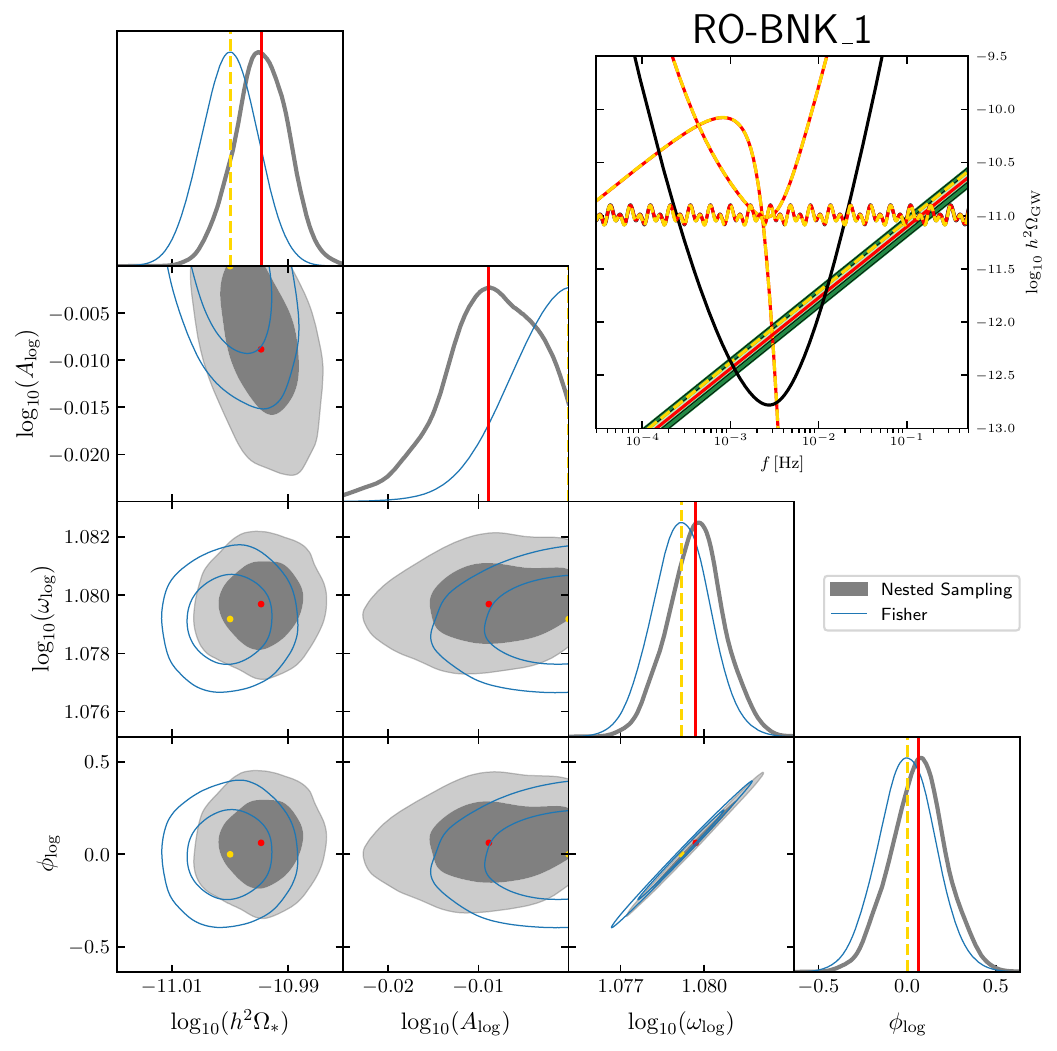}
\includegraphics[width=0.495\columnwidth]{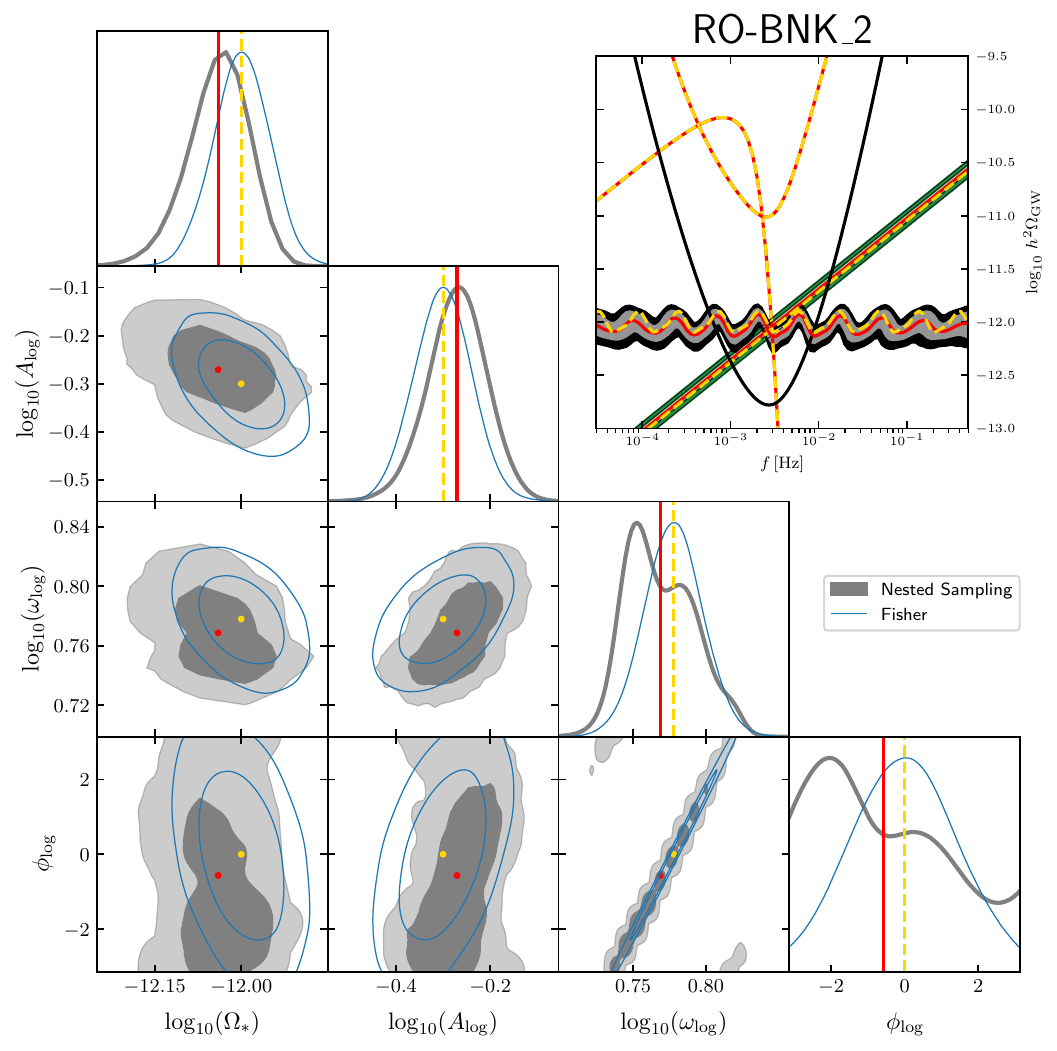}\\\includegraphics[width=\columnwidth]{Legend_Reconstruction.pdf}
\caption{
\small
1D and 2D posterior distributions 
derived from the RO template reconstruction with a flat envelope run on the benchmarks RO-BNK\_1 (left) and   RO-BNK\_2 (right). Lines' styles and color codes
are as in \cref{fig:PL_MCMC}.  
Benchmarks RO-BNK\_1  and  RO-BNK\_2 
are defined as in eq.~(\ref{eq:resonant-template}) with $\{h^2 \Omega_* \, , \ A_\textrm{log} \, , \ \omega_\textrm{log} \, , \  \phi_\textrm{log} \, \}$ set at $\{-11,1,12,0\}$ and $\{-12,0.5,6,0\}$ respectively.
}
\label{Resonant}
\end{figure}

\noindent {\bf Benchmarks.---}  As described in \cref{sec:resonant}, the template is characterised by two harmonics only for $\omega_{\rm log}>\mathcal{O}(1)\omega_c $, with $\omega_c \simeq 4.77$ and the order one parameter depending on $A_{\rm log}$. Thus, we consider two benchmarks, with the rationale that $\omega_{\log}$ provides a signal shape where the double oscillatory pattern does appear, \emph{i.e.,}~RO-BNK\_1 with $\{\log_{10}(h^2 \Omega_*) \, , \ A_\textrm{log} \, , \ \omega_\textrm{log} \, , \  \phi_\textrm{log} \, \}=\{-11,1,12,0\}$, and does not appear, \emph{i.e.,} RO-BNK\_2 with $\{\log_{10}(h^2 \Omega_*) \, , \ A_\textrm{log} \, , \ \omega_\textrm{log} \, , \  \phi_\textrm{log} \, \}=\{-12,0.5,6,0\}$. 
The left panel of \cref{Resonant} shows the \texttt{SGWBinner} RO template-based reconstruction applied to RO-BNK\_1. 
The reconstruction is very accurate, with the injected  parameters well within \seCL contours. The peculiar shape of the template is mainly reconstructed thanks to the part of signal that is above both foregrounds and still in the LISA band. Given the small range of frequency where this happens, 
we interpret it as
 the origin of the strong degeneracy between the parameters $\omega_\textrm{log}$ and $\phi_\textrm{log}$.
  The extragalactic foreground signal is not well reconstructed since, within the LISA frequency band, it is orders of magnitude below RO-BNK\_1.

The right panel of \cref{Resonant} presents the \texttt{SGWBinner} reconstruction for RO-BNK\_2. With its envelope amplitude $h^2\Omega_* = 10^{-12}$, RO-BNK\_2 happens to be right below both foregrounds, and is comparable to them  around $f\sim {\rm mHz}$. 
 We are thus in an envelope amplitude regime where the error accuracy is expected to degrade. However, it turns out that the amplitude is still high enough to permit a clear reconstruction of the overall frequency shape. No sizable biases emerge in the analysis as all the injected values are within the \seCL contours. Notice also that a strong degeneracy appears between the parameters $\omega_\textrm{log}$ and $\phi_\textrm{log}$.  It is interesting to note that LISA seems to constrain all the properties of this benchmark, as can be seen from the reconstructed spectral shape of the SGWB. This, however, is not translated into a constrain on the phase parameter $\phi_{\rm log}$, which is unconstrained within its prior.
This fact is likely due to the implicit dependence of the amplitude of the signal on $\omega_{\rm log}$ and $\phi_{\rm log}$, which enters in the amplitude of the two harmonics in the RO template in  eq.~\eqref{eq:resonant-template}, introducing undesired degeneracies between parameters. 
In both cases,  for the two injected signals, we observe a good agreement between the nested sampling and the Fisher analysis, which captures quite well the shape of the posterior distribution.

\clearpage
\section{Science interpretation highlights}
\label{sec:interpr} 

The analysis in the previous sections  shows the existence of several  well-motivated inflationary models within the LISA detectability reach. Furthermore, for some specific parameter ranges, their signals can be reconstructed with promising small statistical errors.  In this section, we aim at showcasing the scientific breakthroughs the community will achieve in the case of a detection of an SGWB described by our templates.

We consider the inflationary models motivating our template, and the mapping between the template parameters and the inflationary model parameters (see \cref{sec:models}). 
Using this mapping, we take the chains from the analysis of our benchmarks and derive the posterior distribution for the fundamental parameters of the inflationary models.
We finally discuss the scientific implications of the results. For simplicity, and to reduce the computational costs of our analysis, we do not apply this method at such a level of detail for every template. For the less computationally expensive templates and fundamental models, we carry out a thorough study to show how the procedure works, whereas for the others we just discuss how the bounds on template parameters reflect on fundamental parameters of such models.

With the tools and templates provided in this work, constraining fundamental parameters boils down, in practice, to post-processing the outputs of our template-based reconstructions. 
Indeed, in many scenarios, our templates are expected to precisely match (within our assumptions and approximations) the signals induced by a broad class of models, though for each specific model the functional form connecting the template parameters to the fundamental parameters is different. Since only the parameterization is different, but not the shape of the signal, using the fundamental parameters in the iterations of the global fit is not expected to lead to significant changes in the sampling output for neither the primordial SGWB nor the other sources. On the contrary, for many fundamental models, sampling directly in terms of fundamental parameters is expected to worsen the computation-time performances: the parameterizations of our templates are conceived to minimize degeneracies, simplifying the likelihood sampling, and are flexible enough to cover a wide class of fundamental models, so that the global fit (or at least its final module interactions) does not need to be repeated for every fundamental model compatible with the SGWB signal.\footnote{One exception to this is the case where fundamental parameters have stringent theoretical priors. In that case, the results from post-processing our outputs may differ from those used by sampling directly the fundamental parameters; see ref.~\cite{Caprini:2024hue} for a quantitative discussion.}

As discussed in  \cref{sec:deformations}, it is possible that non-standard phases in the early universe  after inflation, or the
presence of relativistic species beyond the standard model, would cause modifications of the frequency shape of all the scenarios.
While this could impact the accuracy of the template search, 
assuming a priori the presence of a given deformation would require to run the analysis for all the templates (and benchmarks) with a set of various assumptions. We choose not to discuss such effects in this section and leave this task to more targeted future work.

\subsection{Inflationary scenarios predicting blue power-law spectra}\label{sec_int_PL}

\begin{figure}[!h]
	\begin{center}
		\includegraphics[width=.5\columnwidth]{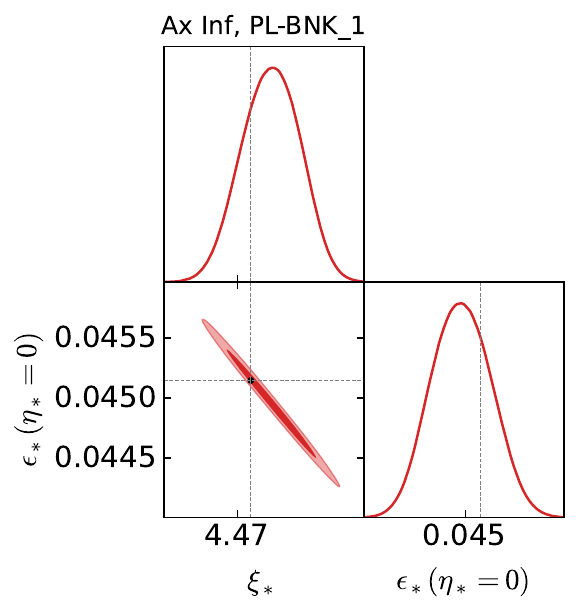}
		\includegraphics[width=.35\columnwidth]{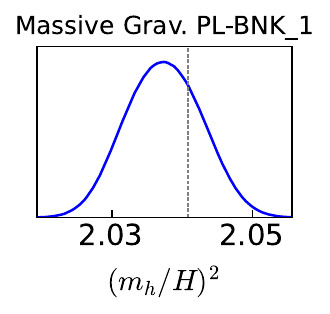}
		\includegraphics[width=.5\columnwidth]{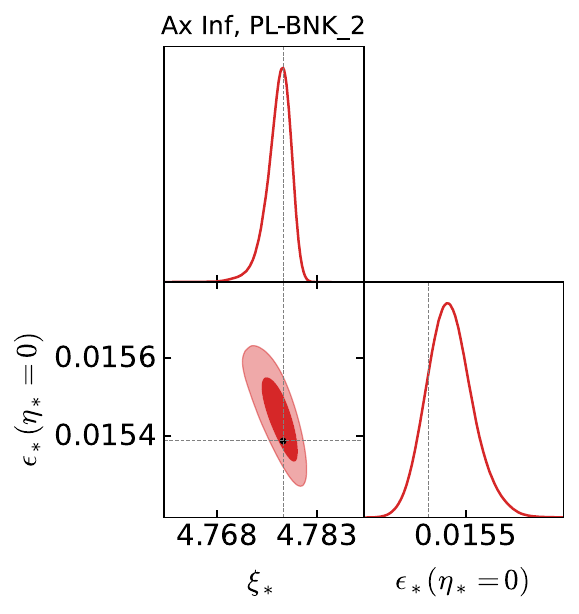}
		\includegraphics[width=.35\columnwidth]{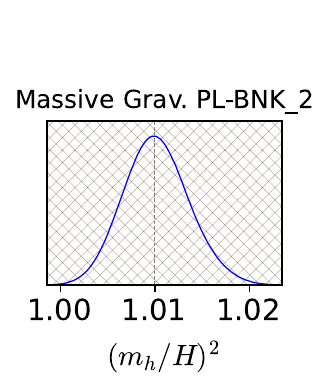}
	\end{center}
	\caption{\label{fig:PL_science_interpretation} \small Constraints on the parameters of axion inflation parameters (left panels) and massive-graviton inflation (right panels) derived from the PL template-based reconstruction of the benchmarks PL-BNK\_1  (top panels) and PL-BNK\_2 (bottom panels). 
 The hatched region  is excluded by the Higuchi bound~\cite{Higuchi:1986py} ($m_h/H<\sqrt{2}$) in models respecting de Sitter isometries.  }
\end{figure}

We start our discussion by examining the implications of a detection of the simplest template among those considered here, \emph{i.e.,}~the  PL template of \cref{sec:pl}.  
 For illustration, we focus on the two benchmarks PL-BNK\_1 and PL-BNK\_2 
 examined in the forecasts of \cref{sec_forPL}, and we analyse the physical implications of our findings for the first two inflationary scenarios discussed in \cref{sec:pl}: 
 axion inflation, and models with a massive graviton during inflation.  For both benchmarks, the constraints on  template parameters summarized in \cref{fig:PL_MCMC} are translated into constraints on  model parameters,  and represented in \cref{fig:PL_science_interpretation}. Due to significant model dependencies in the amplitude of the massive-gravity SGWB spectrum (see \cref{sec:pl}), our scientific interpretation focuses solely on the tilt within that scenario.\\ 

{\bf PL-BNK\_1}. The parameter reconstruction associated with this benchmark would set tight constraints on axion inflation and massive-gravity inflation. In the context of axion inflation, under the assumptions of validity of eq.~\eqref{axion-inf}  discussed in \cref{sec:pl}, the results translate into accurate measurements on the parameters $\xi_* = 4.474\pm0.003$ and $\epsilon_*= 0.0450\pm 0.0003 $ both at 68\% C.L. (we assume $\eta_* = 0$ to derive the second constraint). We note that these constraints are derived assuming a value of $H_*/M_{\rm pl}=10^{-5}$ (consistent with latest constraints on the tensor-to-scalar ratio from large scale measurements) in order to invert the relation~\eqref{axion-inf-inverse} between $\xi_*$ and $\omega_*$. For scenarios with  a massive graviton, we note that our benchmark respects the so called Higuchi bound
$m_h/H\geq\sqrt{2}$~\cite{Higuchi:1986py}. Our results, presented in the top-right panel of \cref{fig:PL_science_interpretation}, indicate very tight marginalized constraints on the graviton mass $(m_h/H)^2 = 2.037\pm 0.005$. \\

{\bf PL-BNK\_2}. Similar to the previous case, the detection of this benchmark would narrow down the viable parameter space of axion inflation, constraining its parameters to $\xi_* = 4.777^{+0.002}_{-0.001}$  and $\epsilon_*(\eta_*=0) = 0.01545^{+0.00005}_{-0.00007}$, still assuming  $H_*/M_{\rm pl}=10^{-5}$. Interestingly, the consequences would be more intriguing for the massive graviton models. As we can learn from the bottom-right panel of \cref{fig:PL_science_interpretation}, the posterior distribution would completely lie within the red hatched region, excluding 
with high statistical significance 
the viability of any massive graviton model that
respects the Higuchi bound. 
Interpreting this detection in terms of massive graviton would thus point to the necessity of more complex theoretical constructions, as for example the one developed in ref.~\cite{Ricciardone:2017kre}.

\subsection{Constraints on small-scale curvature power spectrum and primordial black holes}\label{sec:PBH}

One of the main motivations behind the second scenarios considered for the BPL and the DP template is the enhancement of curvature perturbations at small scales. 
We use our reconstruction forecasts in \cref{doublepeak1_MCMC} to estimate the future bounds on the amplitude of ${\cal P}_\zeta$ at LISA scales.

The fundamental relation to consider is the one connecting the frequency $f$ of the SGWB to the comoving wavenumber $k =  2 \pi f$ characterising primordial perturbations, which is 
\begin{align}\label{GW_peak_frequency}
f
\simeq 15 \, {\rm mHz}
\left( \frac{k}{10^{13} {\rm Mpc}^{-1}} \right).
\end{align}
This shows that frequencies accessible by LISA correspond to small scales that CMB and other large-scale observations cannot probe. 
Currently, the bound on $\Delta N_\text{\tiny eff}$ coming from CMB data \cite{Planck:2018vyg} constrains $\Omega_{\rm GW} \leq 1.6 \cdot 10^{-6}$ at 95 \% C.L., which corresponds to the bound ${\cal A}_s\leq 0.69 $.  Primordial black hole overproduction (see, \emph{e.g.,}  refs.~\cite{Saito:2008jc,Carr:2020gox,Yuan:2021qgz,Ozsoy:2023ryl} for reviews and references therein) sets stronger bounds. 
This forces the amplitude to be below ${\cal A}_s\lesssim 10^{-2}$, depending on the assumptions on spectral shape and curvature perturbation statistics (\emph{e.g.,} ref.~\cite{Gow:2020bzo}). LISA will push this constraint to much smaller amplitudes 
in the range of scales $k \in [10^{10}, 10^{14}]\, {\rm Mpc}^{-1}$.\\

{\bf Null observation of a BPL or DP \GWB.} The null detection of a SGWB with broken power-law or double-peak template can be converted into an upper bound on ${\cal P}_\zeta$. For presentation purposes, we estimate this upper bound for the DP from the information available in \cref{Fisher_double} which considers the ${\cal P}_\zeta^{\rm ln}$ scenario.

In the parameter space where the precision on the signal amplitude $h^2 \Omega_*$ degrades below $30\%$, we are practically insensitive to the presence of DP signal at 99.7 \% CL (i.e. 3$\sigma$). 
In the case at hand, the null detection would then lead to the upper bound ${\cal A}_s \leq 10^{-3.5}$ at 99.7 \% CL in the ${\cal P}_\zeta^{\rm ln}$ parameter region with $f_* \simeq 5$ mHz and $\Delta\lesssim 1$.\footnote{Rigorous bounds in the low-SNR limit would eventually be set adopting a Bayesian analysis.} 
Analogously, the upper bound on ${\cal A}_s$ at different $f_*$ can be obtained from \cref{Fisher_double_pivot}, and becomes more severe by departing from $f_*=5\,{\rm mHz}\simeq 10^{-2.3}\,$Hz.
Similar upper bounds are obtained assuming the null detection of induced \GWB \ with a BPL shape (see fig.~\ref{fig:Fisher_BPL_1}).\\

{\bf Detection of a BPL or DP SGWB background.} 
Assuming the detection of a signal compatible with the scalar-induced SGWB scenario, we can use the  
FIM forecast errors to estimate the tight bounds LISA would set on the enhanced curvature spectrum and the required inflationary scenario responsible for the enhancement, see fig.~\ref{fig:Fisher_BPL_1} for the BPL case and fig.~\ref{Fisher_double}, \ref{Fisher_double_pivot} for the DP case.

Let us focus on BPL\_BNK\_1. The most informative constraints are set on:
{\it i)} the characteristic frequency
$f_*$, which controls the location of the USR phase leading to the enhancement of the spectrum.
{\it ii)}  the tilt $n_{{t},2}$,  which 
controls the plateau observed in $\Omega_{\rm GW}$ at high frequencies. 
Assuming a second slow-roll phase following the USR enhancement (see \cref{subsec:BPL} for more details), this is related to the slow roll parameters 
$  n_{{t},2} \equiv {d \ln {\cal P}_\zeta}/{d \ln k} \simeq -6 \epsilon_* +  2 \eta_*$.
Using the results reported in \cref{fig:BPL1_MCMC} for BPL\_BNK\_1, we find $\log_{10}( k_* {\rm Mpc}) = 11.6\pm 0.4 $, which places the USR phase at $\Delta N = 29.7 \pm 1.0$ e-folds after the slow-roll phase controlling CMB perturbations, adopting $k_{\rm CMB} = 0.05/{\rm Mpc}$ and approximating the Hubble rate to a constant during that part of the inflationary phase. 
Furthermore, if  $\epsilon_*$ is negligible after the USR phase (\emph{e.g.,}~ref.~\cite{Franciolini:2022pav}), we obtain the constraint $\eta_* = 0.01\pm 0.11$ for the parameters of the subsequent slow-roll phase. \\

{\bf Constraints on Primordial Black Hole population.}
Enhanced curvature power spectra are associated with the formation of PBHs. Therefore, we can forecast the constraining power of LISA on this dark matter candidate. We are considering test case scenarios in which ${\cal P}_\zeta $ features a narrow bump in the mHz range, corresponding to a PBH population with asteroid typical mass.
Uncertainties on template parameters translate into two observables: {\it i)} $f_*$ controls the typical mass of the PBHs population;
{\it ii)} the curvature power spectral amplitude ${\cal A}_s$ translates (highly non-linearly) into uncertainties on $f_\text{\tiny PBH} \equiv \Omega_\text{\tiny PBH} / \Omega_\text{\tiny DM}$ which gives the PBH abundance in terms of the dark matter abundance.
For narrow spectra, the typical PBH mass is related to the mass contained in the Hubble scale $M_H$ at the time of mode crossing as $\langle m_\text{\tiny PBH} \rangle  \simeq 0.6 M_H$~\cite{Musco:2023dak}, where the angular brackets indicate average PBH mass computed accounting for  critical collapse, and $M_H$ is related to the curvature wavenumber by $M_H/M_\odot \simeq 7 \times 10^{-11} [k/(10^{12} {\rm Mpc}^{-1})]$.

Let us consider the case of a DP detection in more detail. From the results in the upper-left panel of \cref{doublepeak1_MCMC}, i.e. considering DP-BNK\_1, we find that the mean PBH mass would be measured with exquisite precision to be $\log_{10} \langle m_\text{\tiny PBH}/M_\odot \rangle \approx -11.407 \pm 0.003 $.
Finally, considering the same benchmark scenario, the overall amplitude translates into an uncertainty on the PBH abundance $\log_{10}(f_\text{\tiny PBH}) = 0.01 \pm 0.06$. For the computation of the abundance we assumed threshold statistics and Gaussian curvature perturbations, see e.g. \cite{Young:2019yug,DeLuca:2019qsy,Ferrante:2022mui,Gow:2022jfb}.
Similar conclusions are reached in the case of DP-BNK\_3, also associated with the production of large fraction of the dark matter in the form of asteroidal mass PBHs.
Regardless of the well-known exponential dependence of the PBH abundance on the power spectral amplitude,
assuming emission in a radiation-dominated universe and Gaussian perturbations, LISA observations would be able to constrain PBH dark matter abundance with better than $10\%$ precision, provided $f_\text{\tiny PBH}$ saturates the upper bound from PBH overproduction. 
Therefore, LISA constraints will likely be limited by the large model dependence of the PBH abundance \cite{LISACosmologyWorkingGroup:2023njw} as well as remaining systematic uncertainties on their computation (see, \emph{e.g.,}   ref.~\cite{DeLuca:2023tun}). 
Importantly, in case the LISA reconstruction precision is  degraded compared to the one forecasted in this work, it would drastically decrease the capabilities to pin-down the PBH abundance, 
 due to the strong dependence of $f_{\rm PBH}$ on the amplitude ${\cal A}_s$. In the cases of DP-BNK\_2 and DP-BNK\_4, the assumed amplitude of perturbations is small and associated with a negligible abundance of PBHs. Upper bounds derived from LISA data would be strong enough to rule out a large contribution to the dark matter from this PBH formation scenario.  

A more thorough derivation of constraints on the scalar-induced SGWB scenarios requires more complex post-processing of the forecast presented in the previous sections, based on solving the inverse problem accounting for the full fundamental-parameter degeneracies and model dependence. This is part of an ongoing LISA cosmology working group activities and its results will be reported elsewhere. 

\subsection{Small-scale primordial features}

Gravitational-wave cosmology has the potential to shed light on primordial features during inflation. These are characterised by various types of oscillations in the SGWB frequency profile that are distinctive of specific mechanisms active during inflation. LISA has the ability to reconstruct parameters associated with primordial features with a striking precision and thus opens the possibility to detect them at scales much shorter than the CMB-LSS ones, where current constraints are based.   \\

{\bf Excited states.} 
The most informative parameter to retrieve from the detection of the excited states template (see \cref{sec:excited})  is $\omega_{\rm ES}$. This parameter determines the location of the primary peak of the signal as well as the periodicity of the higher-frequency peaks. Measuring it enables us to deduce the existence of a dynamically generated excited state during inflation, and to identify when this particle production event occurred, or equivalently the momentum  scale of modes crossing the Hubble radius at that time, $k_{\rm ES}= 4 \pi /\omega_{\rm ES}$ \cite{Fumagalli:2021mpc}. 

More concretely, let us call $\Delta N$ the number of e-folds between the time the CMB pivot scale exits the Hubble radius and the time particle are copiously produced. 
Using the previous expression for $k_{\rm ES}$ and  
the relation \eqref{GW_peak_frequency}, we obtain
\begin{equation}
\label{eq:efolds}
\Delta N = \log \left[  \left(\frac{4}{3}\cdot 10^{15}\right)
\left( \frac{1}{\omega_{\rm ES} \mathrm{Hz}}\right) 
\left(\frac{\mathrm{Mpc}^{-1}}{k_\mathrm{CMB}}\right)%
\left(\frac{H_{\mathrm{CMB}}}{H_{\mathrm{f}}}\right) \right]. 
\end{equation}
Thus, by using error propagation in \cref{eq:efolds}, we find that the detection of either ES-BNK\_1, ES-BNK\_2 or ES-BNK\_3 with the uncertainties on $\omega_{\rm ES}$ reported in \cref{sharp_triangular} would respectively lead to $\Delta N =30.7 \pm 0.1 $, $\Delta N =29.12 \pm 0.01 $ and $\Delta N =29.13 \pm 0.07 $ with the (mild) assumption $H_{\mathrm{CMB}} = H_{\mathrm{f}}$ and using $k_{\mathrm{CMB}} =0.05\,\mathrm{Mpc}^{-1}$.

The second parameter of interest is $\gamma_{\rm ES}$. It controls the high-frequency behaviour of the signal, in particular the range of frequencies where the signal dies off. Its knowledge teaches us how much inside the horizon particle production took place or, equivalently, the energy scale   $E_{\rm ES}\equiv \gamma_{\rm ES} H$ (in Hubble units) of the produced particles. Unfortunately, as we have learned from section \ref{sec4_excited}, this parameter is intrinsically related to the low-signal part of the template, and thus is hard to reconstruct, unless the signal is  loud, \emph{i.e.,} $\log_{10}h^2\Omega_* \gtrsim -11$. Thus, among the three benchmarks chosen in \cref{Excited_triangular}, only with the detection of ES-BNK\_2 (with an injected $\gamma_{\rm ES} = 10$)  can LISA retrieve relevant information on $E_{\rm ES}$ thanks to the reconstruction $\gamma_{\rm ES} =12.6 \pm 1.3$.

Note that a sharp feature phenomenon generating an excited state with copious particle production generates both the signal described by the excited state template \eqref{templateexcited} and the one described by the linear oscillations template \eqref{eq:sharp-template} discussed 
below.
As the relative amplitudes of the two signals is model-dependent, and as their peak frequencies differ, we treated the two templates separately. It is nonetheless instructive to compare the two. The time of the sharp feature is the most robust physical information, in the sense that it can be identified  well from each template. By contrast, how much inside the Hubble radius the particle production was efficient is retrieved very easily from the linear oscillations template, as it is then related to the peak frequency (see below). On the other hand, in the case of the excited state template, this information is  retrieved only if the signal is very loud as the information is then contained in the high-frequency, but low amplitude part of the signal.\\

{\bf Sharp features, i.e. linear oscillations.} Let us now discuss what can be learned in a relatively model-independent manner from the detection of linear oscillations in the SGWB, using the LO template in \cref{eq:sharp-template}. The most informative parameter is the oscillation frequency $\omega_{\rm lin}$. Measuring it enables us to prove the existence of a sharp feature during inflation, and to identify when it occurred. Sharp features refer to localized events --- of duration smaller than one $e$-fold --- independently of their precise realisation like, \emph{e.g.,}  a step in the inflationary potential or a sharp turn in the inflaton trajectory. They generate density fluctuations with a power spectrum characterised by linear oscillations (see eq.~\eqref{Power-spectrum-sharp-feature}) with frequency $\omega=2/f_\textrm{f}$ directly related to the scale crossing the Hubble radius at the time of occurence of the feature, $k_\textrm{f}=2 \pi  f_\textrm{f}$  (setting the scale factor today and using natural units). These oscillations are then transferred to the scalar-induced GW density profile if their amplitude is sufficently large, \emph{i.e.,} for
 a significant particle production generated by the sharp feature, with occupation numbers of order one or larger \cite{Fumagalli:2020nvq}. Hence, pinpointing the frequency of the template $\omega_\textrm{lin}=c_s^{-1} \omega$ gives us $k_\textrm{f}$ or, equivalently, the time of the feature. There is a theoretical degeneracy in the mapping between the measured $\omega_\textrm{lin}$ and the inferred $k_\textrm{f}$ due to the uncertainty about the propagation speed of density fluctuations $c_s$, equal to $1/\sqrt{3}$ in the conventional scenario of radiation domination at horizon reentry. This degeneracy may be lifted, however, as we comment below.

As in the previous section, let us call $\Delta N$ the number of e-folds between the time the CMB pivot scale exits the Hubble radius and the time of the feature. 
Using $\omega_{\rm lin}  = 2 c_s^{-1}/ f_{\rm f} $ and the relation \eqref{GW_peak_frequency}, we obtain the analogous of \cref{eq:efolds}. This yields
$\Delta N = \log \left(  \frac{4\cdot 10^{17}}{15} c_s^{-1}/(\omega_{\rm lin} \mathrm{Hz})\right) $
for $H_{\mathrm{CMB}} = H_{\mathrm{f}}$ and $k_{\mathrm{CMB}} =0.05\,\mathrm{Mpc}^{-1}$.
Thus, by using error propagation, we find that the detection of either LO-BNK\_1 or LO-BNK\_2 with the uncertainties on $\omega_{\rm lin}$ reported in \cref{sharp_triangular} would respectively lead to $\Delta N =27.55 \pm 0.02 $ and $\Delta N =29.161 \pm 0.002 $ for the typical radiation domination scenarios with
$c_s = 1/\sqrt{3}$. 

Further information can be derived from the precise measurement of the peak frequency $f_*$. Combined with $\omega_\textrm{lin}$, it establishes the quantity $E_{\rm f} = f_* \omega_\textrm{lin}/2$, which gives the characteristic energy scale of the feature in Hubble-scale units. LISA would measure it as $E_{\rm f} = 25 \pm 1$ for LO-BNK\_1 and $E_{\rm f} = 25.05 \pm 0.05$ for LO-BNK\_2. 
Intriguingly, recasting this bound on two-field inflation setups with strong and sharp turns yields a measurement of the rate $\eta_\perp $ at which the inflaton abruptly changes direction in the exponential expansion of the universe close to the Big Bang.\footnote{In these setups, one finds $\eta_\perp \simeq E_{\rm f} /2$ from \cref{strong turn relation}.}
In turn, the measurement of $\eta_\perp $ sheds light on the number of e-folds $\delta$ during which the feature was active. In fact, from a given value ${\cal P}_0$ of the scalar power spectrum away from the feature, measured or assumed at \emph{e.g.,} ~the CMB scale, one can extrapolate the product $\eta_\perp \delta$ via  eq.~\eqref{sharp-turn-result} and, subsequently, infer $\delta$ from $\eta_\perp$.

Finally, in a wide parameter space of our forecast, LISA can reconstruct ${\cal A}_\textrm{lin}$ with an accuracy that permits imposing a lower bound above  $0.2$.
With such a lower bound on ${\cal A}_\textrm{lin}$, LISA would prove not only the existence of an inflationary sharp transition
but also of a non-standard thermal history at the time of Hubble reentry \cite{Witkowski:2021raz}.
Indeed, for a universe dominated by a perfect fluid, a measurement ${\cal A}_\textrm{lin} > 0.2$ indicates that the fluid's equation of state at the Hubble reentry time was stiffer than the one of radiation, $w > 1/3$. 
From the analysis in section \ref{sec4_linear}, it is rather remarkable  that identifying the oscillation frequency requires signals with an amplitude only an order of magnitude above the detection threshold of the overall peak itself, and that this identification can be performed with oscillations of relative amplitude ${\cal A}_\textrm{lin}$ as tiny as  $0.01$. \\

{\bf Resonant features, i.e. logarithmic oscillations.} Detecting a signal fulfilling the RO template of  \eqref{eq:resonant-template2} would prove the existence of periodic modulations of the inflationary Lagrangian with frequency $\omega_\textrm{log}$ in e-fold unit. 
Based on our forecast, LISA has a resolution on $\omega_\textrm{log}$ that permits ruling out the presence of oscillations in a wide parameter region of the RO template (c.f.~the parameter region with a robust upper value on $\omega_\textrm{log}$ in \cref{Fisher_resonant1}).

As explained  in \cref{sec:resonant}, the literature has started investigating explicit setups leading to resonant oscillations in the SGWB only recently. 
A precise relationship between $\omega_{\rm log}$ and the microphysics of inflation thus requires further investigation. 
On cosmological scales, oscillations resulting from a heavy field of mass $M$ being displaced from its minimum have been studied in detail \cite{Slosar:2019gvt}. 
Considering such a mechanism for definiteness, the physics interpretation of the detection of the benchmarks in 
\cref{Resonant} would bound $\omega_{\mathrm{log}}=M/H$ within
$12\pm 0.2$ for RO-BNK\_1 (injected $\omega_{\rm log} =12$) and $5.89 \pm 0.24$ for RO-BNK\_2 (injected $\omega_{\rm log} =6$) at \seCL, thus pinpointing with high precision the mass of new particles well above what can be detected with terrestrial colliders. Actually, the accuracy of the reconstruction of $\omega_\textrm{log}$  mildly depends on the frequency for $\omega_\textrm{log} \gtrsim 6$: from $\omega_\textrm{log}=6$ to  $\omega_\textrm{log}=100$, the same relative error on $\omega_\textrm{log}$ is attained by having a signal louder by $\log_{10}(h^2\Omega_*) \simeq 0.7$ only, for all relevant values of the overall signal amplitude $h^2\Omega_*$. This is quite remarkable, as the relative amplitude of the oscillations in the template roughly goes from 100\% to 1\% between these two values of $\omega_\textrm{log}$ (for a fixed order-one value of $A_\textrm{log}$) \cite{Fumagalli:2021cel}. This highlights the
striking discovery potential of LISA, which can identify resonant features during inflation with characteristic energy scales at least two orders of magnitude above the Hubble scale. This result  prompts further studies by theorists, as this quantitative analysis reveals that LISA will be sensitive to much smaller effects than is hitherto acknowledged \cite{Calcagni:2023vxg}.
\section{Conclusions}
\label{sec:conclusions}

In this work, we demonstrated the scientific potential of a template-based analysis of cosmological SGWB with the LISA detector. Our focus is on inflationary scenarios producing an \GWB\, in the milli-Hertz regime. 
We showed that LISA will provide very accurate constraints on parameters of templates which characterise several inflationary scenarios.

The simplest vanilla single-field models of inflation generate a slightly red-tilted \GWB\ spectrum. 
Within this  minimal framework,
extrapolating to milli-Hertz LISA frequencies  the  current bounds on the non-observation of primordial B-modes from BICEP-Keck and Planck~\cite{Planck:2018vyg,BICEP:2021xfz, Galloni:2022mok,Paoletti:2022anb}, obtained at femto-Hertz scales,  leads to a spectrum which is well below the sensitivity of LISA and other existing and planned interferometers.
 However,
 there are several  theoretical realisations
 of inflation which go beyond the simplest setup described above,
 and which change its predictions towards large frequencies.
 GW observations thus provide a unique avenue to obtain direct experimental information on frequency scales much larger than the CMB ones, which map to the inflationary dynamics closer to the final stages of inflation. With the SGWB measurement, LISA then holds the potential to 
 unveil the physics of inflation and, at worst, in the event of a non-observation of a primordial SGWB,  provide constraints ruling out many inflationary setups and helping prioritize the theoretical efforts towards the surviving models.

As a first step towards testing inflation with LISA observations, we identified seven representative templates, which encompass the spectral shapes of several spectra of SGWBs motivated by inflationary models. The collection of such a template set, 
 together with corresponding ones developed in companion papers for early universe phase-transitions~\cite{Caprini:2024hue} and cosmic strings~\cite{Blanco-Pillado:2024aca} scenarios, is designed to start a centralized LISA repository of cosmological signals, enabling extensions, updates and applications for future LISA studies. Furthermore, to enhance the capabilities of our template repository, we have prototyped a template-based data analysis pipeline integrated with the \texttt{SGWBinner} code \cite{Caprini:2019pxz,Flauger:2020qyi}.

As a second step, we performed a Fisher matrix analysis on each of our templates. We identified the parameter regions that can be probed by LISA, and forecasted the accuracy at which each template  parameter can be constrained.  Furthermore, we developed a full template-based reconstruction, adapting the \texttt{SGWBinner} code to simulate the data which include instrumental noise, foreground signals and the inflationary signals, and performed parameter inference based on the full likelihood.\footnote{The implementation of the template-based approach within \texttt{SGWBinner} goes  beyond the mere validation of the Fisher results. Rather, it represents a significant advance toward the development of a prototype \GWB\ module for the comprehensive LISA global fit.} For each template, we discussed appropriate theoretical priors for its parameters, and
 we identified a few representative injection points, or {\em benchmarks}, for which we simulated LISA data.
   Then, we ran the full template-based reconstruction of the \texttt{SGWBinner}, which uses  a nested sampling algorithm  to explore the likelihood, and  we computed the posterior distribution of the template parameters. Such analysis helped to clarify the degree of realisation dependence of our data and eventual degeneracies as well as non-Gaussianities of the posterior distributions that cannot be fully captured by the Fisher analysis, especially for low SNR injections. As such, we found the two approaches, which we compared against each other for each benchmark, highly complementary. 

Our results show that LISA 
 has the potential  to test primordial SGWBs to a high level of precision.  Reducing the experimental and theoretical uncertainties below the statistical errors quantified in our work is a challenge that the community will have to face in order to avoid
  undermining the exquisite reconstruction precision that LISA   offers us.
  For example, if primordial \GWB\ described by a power-law with amplitude $\Omega_{\rm GW}\sim 10^{-11}$ does exist at mHz frequencies, corresponding to the minimum of its sensitivity, LISA will constrain the template parameters with an accuracy better than percent level. Remarkably, a similar accuracy is achieved also for richer 
frequency shapes, such as those with broken power-law, log-normal and double-peak profiles. If the \GWB\ spectrum have distinctive oscillatory features, the accuracy can even improve below few percent. 
Such impressive precision will enable the distinction between inflationary mechanisms that produce the signal, paving the way for groundbreaking discoveries.

Finally, we discussed the results of our template-based analysis and translated them into constraints on the fundamental parameters of inflationary models that produce the template. We presented the analysis using fundamental parameters for several illustrative inflationary scenarios where a (semi-)analytic relationship between the template and the fundamental model parameters was available. We found that the qualitative interpretation of the reconstructed signals is valuable and informative. We demonstrated that with the projected reconstruction accuracy, LISA would be capable of testing various aspects of inflationary physics, such as the couplings of inflationary axions to gauge fields, the graviton mass during inflation, the primordial fluctuations that could have given rise to primordial black holes, small-scale primordial features like linear or logarithmic oscillations, and the effects of excited states during inflation.

The potential of LISA to detect primordial SGWBs has been evident for some time. Our results demonstrate that such a detection can be achieved with remarkable accuracy. This hinges on accurately characterising the noise sources for the LISA detector,  precisely modelling astrophysical contributions to the foreground, and ensuring that binary waveforms have residuals that do not mimic the SGWB signal.
Our paper may serve as a call to action for the global scientific community, urging collaborative efforts to address these challenges.

\section*{Acknowledgments}
We acknowledge the LISA Cosmology Working Group members for seminal discussions. We especially thank the authors of refs.~\cite{Caprini:2024hue,Blanco-Pillado:2024aca} for the collaborative developments of the \texttt{SGWBinner} code shared among this and those works. We are also grateful to Ema Dimastrogiovanni, Matteo Fasiello,  Sachiko Kuroyanagi and Lukas Witkowski for initial contributions and hints on some inflationary models motivating our templates.
GC is supported by grant PID2020-118159GB-C41 funded by the Spanish Ministry of Science, Innovation and Universities MCIN/AEI/10.13039/501100011033. JF is supported by the grant PID2022-136224NB-C22, funded by MCIN/AEI/10.13039/5011000110 33/FEDER, UE, and by the grant 2021-SGR00872. GN is partly supported by the grant Project No.~302640 of the Research Council of Norway.
MPe acknowledges support from Istituto Nazionale di Fisica
Nucleare (INFN) through the Theoretical Astroparticle Physics (TAsP) project, and from the MIUR Progetti di Ricerca di Rilevante Interesse Nazionale
(PRIN) Bando 2022 - grant 20228RMX4A. MPi acknowledges the hospitality of Imperial College London, which provided office space during some parts of this project. SRP is supported by the European Research Council under the European Union’s Horizon 2020 research and innovation programme (grant agreement No.~758792, Starting Grant project GEODESI).
AR acknowledges financial support from the Supporting TAlent in ReSearch@University of Padova (STARS@UNIPD) for the project ``Constraining Cosmology and Astrophysics with Gravitational Waves, Cosmic Microwave Background and Large-Scale Structure cross-correlations''.  GT is partially funded by the STFC grants ST/T000813/1 and ST/X000648/1. 
VV is supported by the European Union's Horizon Europe research and innovation program under the Marie Sk\l{}odowska-Curie grant agreement No.~101065736, and by the Estonian Research Council grants PRG803, RVTT3 and RVTT7 and the Center of Excellence program TK202.

\appendix
\section{Visualising the effects of the template parameters}
\label{app:variation_of_template_parameters}

In this appendix, we show how variations of the parameters of each template affect the SGWB spectral shape. This is intended to provide a visual interpretation of the dependence of the template to the various parameters. In a few cases, we exaggerate the range of variation of some parameters, to better display their effect. 
For presentation purposes, in all plots contained in this appendix, we arbitrarily fixed $f_* = 1\,$mHz (remembering that a change in $f_*$ simply shifts the spectrum $\Omega_{*}(f/f_*)$ horizontally in frequency).

\begin{itemize}[leftmargin = 0.3cm]

\item {\bf Power law template.} In \cref{fig:PL_parameters}, we show the effect of the amplitude $h^2\Omega_*$ (left panel) and of the tilt $n_t$ (right panel) on the power-law template of eq.~\eqref{eq:PL_template}.
Positive (negative) values of $n_t$ correspond to blue (red) tilted spectra. Notice that the pivot frequency is completely degenerate with the amplitude in this template, and therefore it is removed from the set of parameters. 
\begin{figure}[t!]
\centering
\includegraphics[width=0.7\columnwidth]{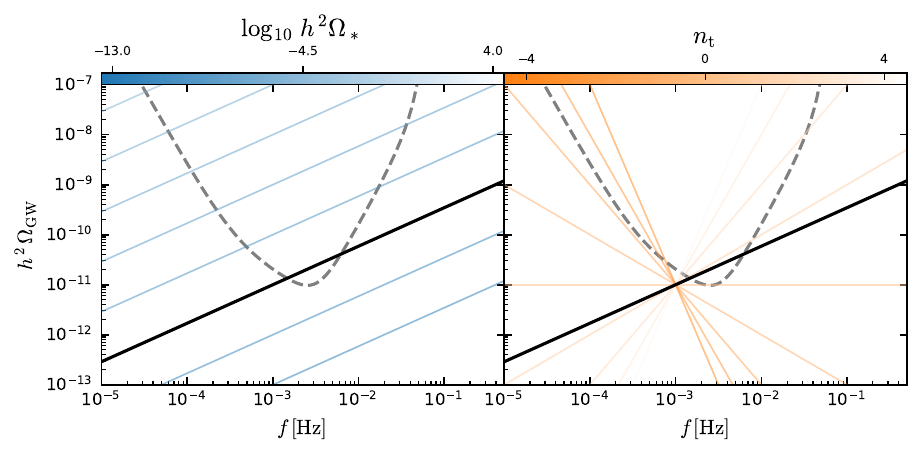} 
\caption{ \small 
Effect on the SGWB spectrum of the power law parameters 
$\Omega_*$ (left) and  $n_t$ (right). For reference, we indicate with a gray dashed line the LISA power-law integrated sensitivity curve \cite{Babak:2021mhe,Colpi:2024xhw}. The black solid line corresponds to  PL-BNK\_2, whose parameters are 
$\{ \log_{10}\Omega_*, n_t\} = \{ -11, \,0.77\}$.
} 
\label{fig:PL_parameters}
\end{figure}

\item {\bf Broken power law.} 
In \cref{fig:BPL_parameters}, we show the effect of the two tilts $n_{t,1}$ and $n_{t,2}$ as well as the transition parameter $\delta $, as defined in eq.~(\ref{eq:master_modelIV_stBBN}).
The two tilts control the slope of the spectrum at small and large frequencies, respectively. If the two tilts are equal to each other, the peak of the spectrum is located at $f_*$, while the peak shifts to smaller (larger) frequencies, when $n_{t,1}<n_{t,2}$ ($n_{t,1}>n_{t,2})$. 
The transition parameter $\delta$ controls the sharpness of the peak and how quickly the power law behavior is attained in the tails.

\begin{figure}[t!]
\centering
\includegraphics[width=1\columnwidth]{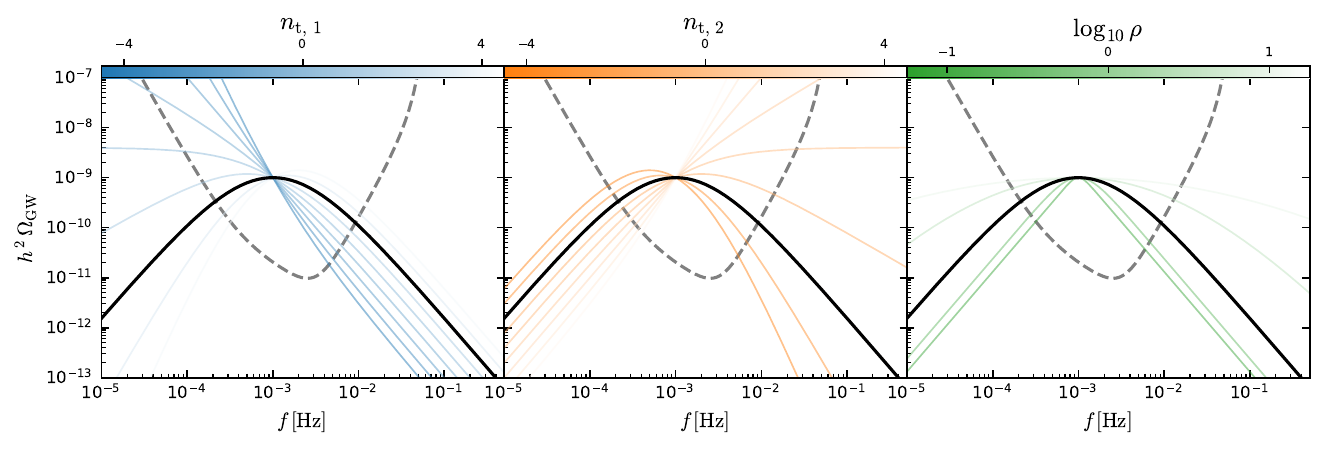} 
\caption{\small 
Effect on the SGWB spectrum of the broken power law template parameters.
 The black line indicates the spectrum assuming $\{\log_{10}\Omega_*, n_{t,1}, n_{t,2}, \delta \} = \{-10,\,2,\,-2,\,1\}$.} 
\label{fig:BPL_parameters}
\end{figure}

\item {\bf Log-normal bump.} 
In \cref{fig:LN_parameters} we show the spectral variations obtained by modifying the log-normal template parameters as defined in eq.~\eqref{eq:master_modelIII_bum}.
The shape of this template is controlled by $\rho$ that determines the width of the spectrum. In the limit $\rho \gg1$, this approaches a flat spectrum.

\begin{figure}[t!]
\centering
\includegraphics[width=.7\columnwidth]{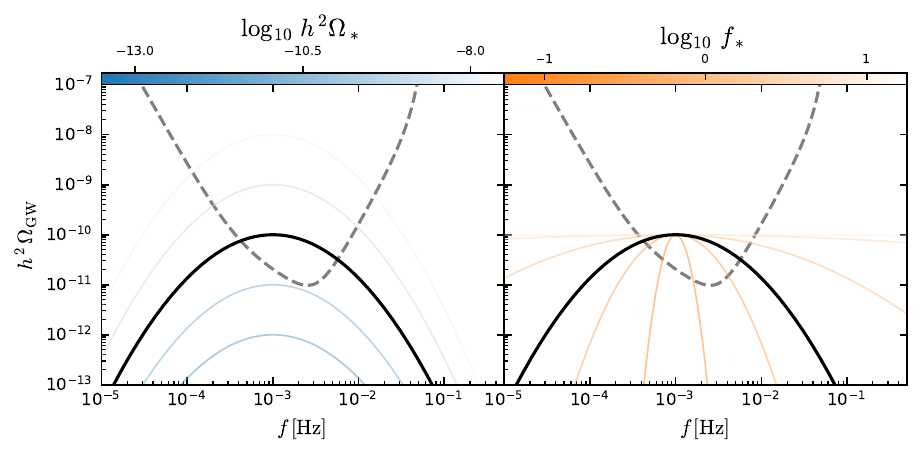} 
\caption{\small 
Effect on the SGWB spectrum of the log-normal template parameters.
The black line indicates the spectrum assuming  $\{\log_{10}\Omega_*,\,f_*\,[{\rm Hz}],\, \rho \} = \{-10,\, 10^{-3},\,0.5 \}$.} 
\label{fig:LN_parameters}
\end{figure}

\begin{figure}[t!]
\centering
\includegraphics[width=1\columnwidth]{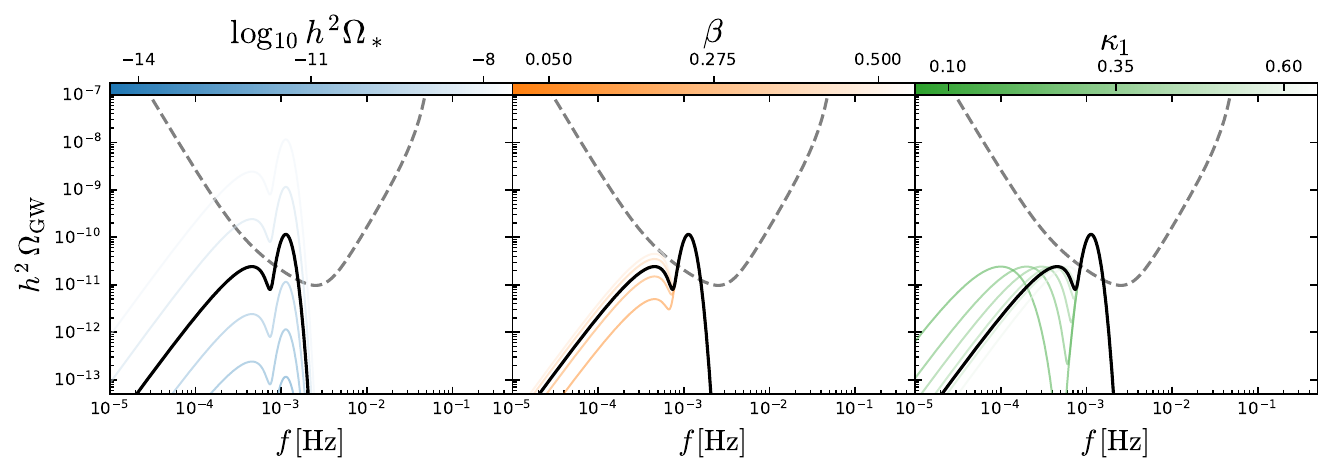} 
\includegraphics[width=1\columnwidth]{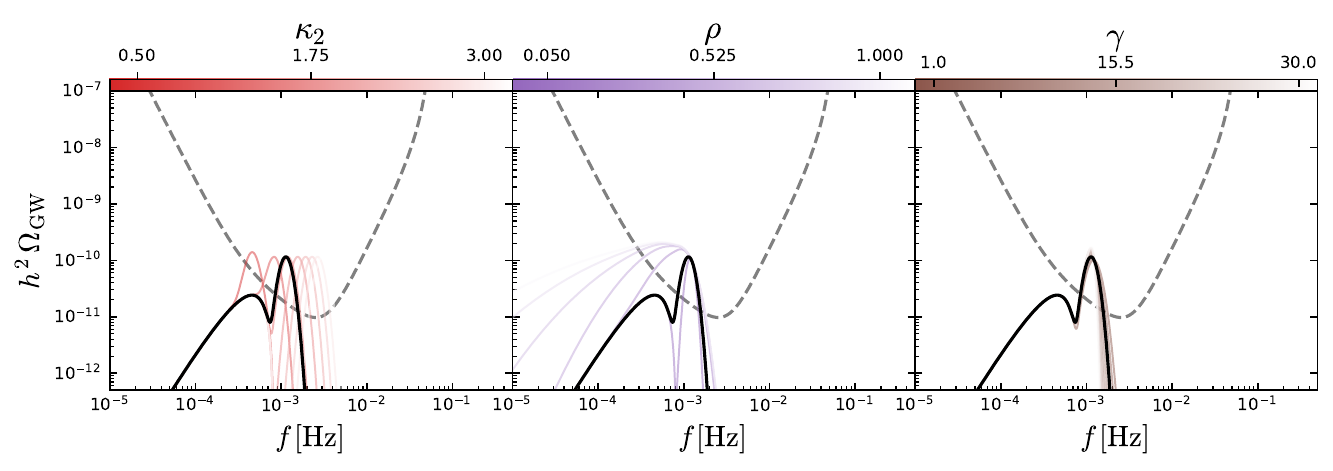} 
\caption{\small 
Effect on the SGWB spectrum of the double peak template parameters.
The black line indicates the spectrum assuming $\{\log_{10}\Omega_*,  \beta, \kappa_1, \kappa_2, \rho, \gamma \} = \{ -10,0.242,0.456,1.234,0.08,6.96\}$ (corresponding to a log-normal primordial power spectrum with $\Delta=0.2$ as the benchmarks in the main text).} 
\label{fig:doublepeak_parameters}
\end{figure}

\item {\bf Double peak template.} 
In \cref{fig:doublepeak_parameters}, we show the effect of varying the parameters of the double peak template of eq.~\eqref{eq:templateDP}. A first set of parameters $\{ \beta, \kappa_1\}$ controls the properties of the first (low-frequency) peak: $\beta$ parameterises its the amplitude compared to the dominant peak, while $\kappa_1$ shifts it compared to $f_*$ (smaller values of $\kappa_1$ would place the peak at smaller frequencies). 
A second set of parameters $\{ \kappa_2, \rho, \gamma \}$ controls the shape of the dominant peak: $\kappa_2$ shifts its frequency, $\rho$ controls its width and $\gamma $ parameterises its skewness by controlling the large frequency cut-off. 

\item {\bf Excited states.} 
In \cref{fig:excitedstates}, 
we show the spectral variations obtained by modifying
the excited states template parameters as defined in eq.~\eqref{templateexcited}. The parameter $\omega_{\rm ES}$ controls the shift in frequencies of the entire template (analogously to $f_*$). The shape is affected only by varying $\gamma_{\rm ES}$. In particular, larger values of $\gamma_{\rm ES}$ move the cut-off of the template to larger frequencies, allowing more oscillation cycles to appear close to the dominant peak.

\begin{figure}[t!]
\centering
\includegraphics[width=1\columnwidth]{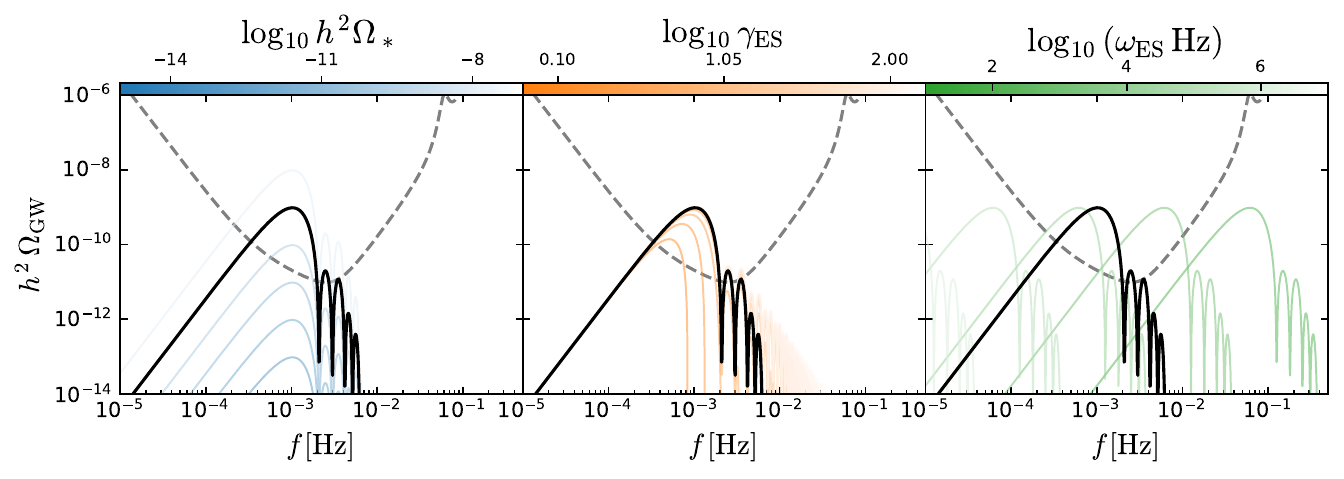} 
\caption{\small 
Effect on the SGWB spectrum of the excited states template parameters.
The black line indicates the spectrum assuming $\{\log_{10}h^2\Omega_*, \log_{10}\gamma_{\rm ES}, \log_{10}\omega_{\rm ES}\} = \{ -9, 1, 3.778\}$. (The latter corresponding to $\omega_{\mathrm{ES}} = 6\cdot 10^3 \,\mathrm{Hz}^{-1}$).
} 
\label{fig:excitedstates}
\end{figure}

\item {\bf Linear oscillations.} 
In \cref{fig:linearoscillations} we show the effect of linear oscillations applied to a log-normal envelope as a function of the parameters of the LO template in eq.~\eqref{eq:sharp-template}. The parameter $\rho$ determines the with of the log-normal envelope while $\mathcal{A}_{\mathrm{lin}} $ controls the amplitude of the oscillations, $\omega_{\mathrm{lin}}$ their frequency in linear scale, and $\theta_{\mathrm{lin}} $ their phase. 

\begin{figure}[t!]
\centering
\includegraphics[width=1\columnwidth]{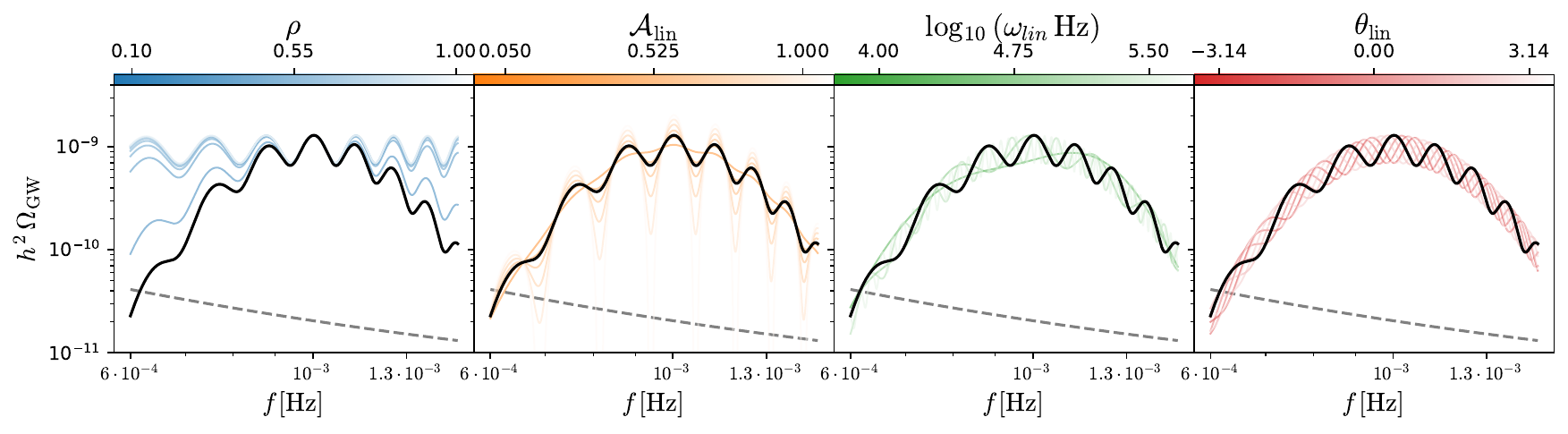} 
\caption{\small 
Effect on the SGWB spectrum of the linear oscillations template parameters.
The black line indicates the spectrum assuming $\{\mathcal{A}_{\mathrm{lin}},\log_{10}(\omega_{\mathrm{lin}}\,\mathrm{Hz}),\theta_{\mathrm{lin}} \} = \{0.3, 4.7 ,0\}$, where the envelope is assumed to be the log-normal with $\vec{\theta}_{\rm env} = \{\log_{10}\Omega_*, \rho \} = \{-9, 0.08\}$. 
In these plots, the range of frequencies was zoomed in close to the peak of the signal, to highlight spectral oscillations. For this reason only a small portion of the LISA power-law integrated sensitivity is visible (gray dashed line). }
\label{fig:linearoscillations}
\end{figure}

\begin{figure}[t!]
\centering
\includegraphics[width=1\columnwidth]{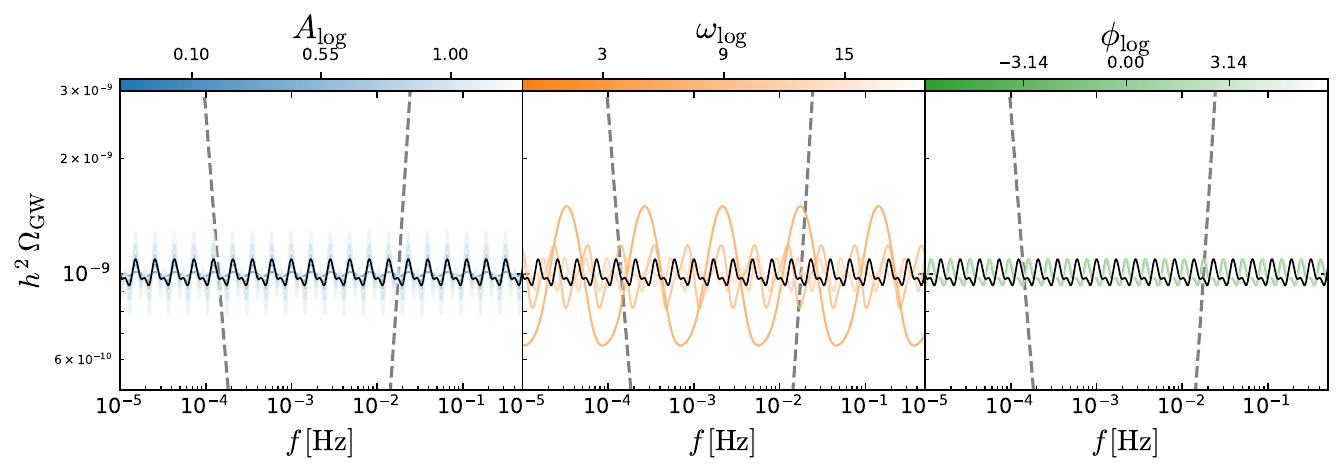} 
\caption{\small 
Effect on the SGWB spectrum of the logarithmic resonant oscillation template parameters.
The black lines corresponds to $\{\log_{10}\Omega_*, A_\textrm{log}, \omega_\textrm{log}, \phi_\textrm{log}\} = \{-9, 0.5, 12 , 0\}$ and a flat envelope. 
} 
\label{fig:flatresonant}
\end{figure}

\item {\bf Logarithmic resonant oscillations.} 
In \cref{fig:flatresonant}, we show the effect of changing the parameters of the logarithmic resonant oscillations template of eq.~(\ref{eq:resonant-template}), assuming a flat underlying envelope. The parameters $A_\textrm{log}$ and $\phi_\textrm{log}$ affect, respectively, the amplitude and phase of the oscillations, and $\omega_\textrm{log}$ modifies nontrivially the behaviour of the spectrum; while for sufficiently large values of $\omega_\textrm{log}$ a regular double feature is observed, only a monochromatic oscillation is observed in the opposite limit, with much larger overall amplitude. This is induced by the complex dependencies 
of the coefficients $\mathcal{A}_1$ and $\mathcal{A}_2$ defined in \cref{eq:resonant-template-const}.

\end{itemize}

\section{Induced GWs at second order: from $\mathcal{P_\zeta}(k)$ to $\Omega_{\rm GW}(f)$}
\label{app:from_P_to_Omega}

Within single- and multi-field models of inflation  featuring an enhancement of the scalar fluctuations by a short period of nonattractor evolution, one expects the generation of a potentially sizeable \GWB\ at frequencies $f$ when perturbations of comoving wavenumber $k \simeq 6.47\times 10^{14}\, {\rm Mpc}^{-1} \left( {f}/{{\rm Hz}} \right)$ re-enter the Hubble sphere (see ref.~\cite{Domenech:2021ztg} for a recent review). These scales are related to the formation of primordial black holes with asteroidal masses, which attracted ample attention due to the possibility that they could provide candidates for (part of) dark matter~\cite{Sasaki:2018dmp,Carr:2020gox}. LISA will be able to test and constrain this scenario~\cite{Saito:2008jc,Bartolo:2018evs}. 

One can compute the
GW signal sourced by scalar perturbations \cite{Tomita:1975kj,Matarrese:1993zf,Matarrese:1997ay,Acquaviva:2002ud,Mollerach:2003nq,Nakamura:2004rm,Ananda:2006af,Baumann:2007zm,Espinosa:2018eve, Kohri:2018awv,Inomata:2019yww,Domenech:2019quo} by solving the Einstein equations at second order in perturbation theory.
Assuming the emission takes place in a radiation-dominated universe, as is the case for those scales in standard cosmologies, one can derive the transfer functions
\begin{eqnarray}\label{eq:IcIs}
    {\cal I}_c(d,s) &=& -36\pi\frac{(s^2+d^2-2)^2}{(s^2-d^2)^3}\Theta(s-1)\,, \\
    {\cal I}_s(d,s) &=& -36\frac{(s^2+d^2-2)}{(s^2-d^2)^2}\left[\frac{(s^2+d^2-2)}{(s^2-d^2)}\ln\frac{(1-d^2)}{|s^2-1|+\epsilon}+2\right] \,,
\end{eqnarray}
by integrating the equation of motion for the emission of GWs at second order with the Green's function method. In order to find concise analytical solutions, such as the one presented in eq.~\eqref{eq:IcIs}, in the literature the observer is typically located at the asymptotic future, $\eta_{\rm obs}\rightarrow \infty$. This, however, generates a fictitious divergence in the limit of $s \rightarrow 1$, which is also retained in the spectrum produced by monochromatic scalar perturbations (see also discussions in ref.~\cite{Ananda:2006af}). The regulator $\epsilon$ is introduced in order to avoid this divergence. A value of order $\epsilon = 10^{-13}$ is sufficient to recover the physically regularized solution assuming the emission does not continue after the  matter-radiation equality for modes with frequency around the LISA peak sensitivity ($f\simeq 10^{-3} $Hz). Given the logarithmic nature of the divergent term, the observables related to $\Omega_{\rm GW}$ are, in practice, insensitive to slightly different choices of the regulator. For concreteness, we will adopt $\epsilon = 10^{-13}$. The main peak in the scalar-induced GW spectrum roughly coincides with the peak in the curvature power spectrum, and its amplitude is $\Omega_{\rm GW} = \mathcal{O}(10^{-5}) {\cal A}_s^2$, where ${\cal A}_s$ is the characteristic scalar spectral amplitude. 

Different models of inflation can give rise to different kinds of peaks in the curvature power spectrum, depending on the detailed shape of the curvature spectrum.\footnote{Also, post-inflationary physics, as an unconventional phase of expansion prior to radiation domination epoch, can change the slope of the induced spectrum, see, \emph{e.g.},~refs.~\cite{Domenech:2019quo,Domenech:2020kqm}. In this sense, induced SGWB can probe the early thermal history of our universe.} We do not attempt to explore the full model dependence. Instead, in the following subsections we motivate general shapes of the curvature power spectrum that are expected in a general class of inflationary dynamics giving rise to enhanced spectra at small scales, and derive what values of the parameters characterise the template \eqref{eq:templateDP}.
\subsection{The spectral index in the infrared}

Although it is known that, for the sources studied in this section, we have a universal  $\Omega_{\rm GW} \propto f^3$ scaling in the deep infrared \cite{Caprini:2009fx,Cai:2019cdl,Hook:2020phx}, the slope of the spectrum changes when approaching the first peak. Interestingly, it still has a universal behavior, with a spectral index
$
n_p
\equiv 
{{\rm d}\ln\Omega_{\rm GW}}/{{\rm d}\ln f }\simeq2.5. 
$
We confirm this statement by following ref.~\cite{Pi:2020otn} in the case of narrow curvature peaks (see also refs.~\cite{Yuan:2019wwo,Adshead:2021hnm,Atal:2021jyo,Franciolini:2023pbf,Yuan:2023ofl,Xie:2024cwp} for an alternative derivation). 

\begin{figure}[t]
\centering
\includegraphics[width = 0.49\textwidth]{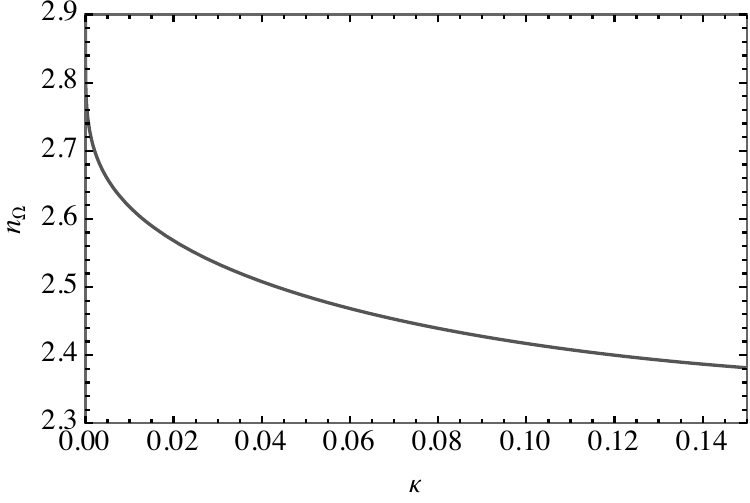}
\caption{\small 
Plot of eq.~\eqref{calcnom}, choosing $\Delta=0.2$. The spectral index of the \GWB\ approaches $n_\Omega=3$ for $\kappa\to0$, but at larger $\kappa$ it is more like $2.5$, in agreement with the numerical results.}
 \label{plotexnom}
\end{figure}

We assume that $\mathcal{P}_\zeta(k)$ has a pronounced peak with maximum at $\kappa \equiv k/k_*=1$. We assume that the peak has a narrow width controlled by a dimensionless quantity dubbed $\Delta$, much smaller than one. We Taylor expand $\ln \mathcal{P}_\zeta$ around the maximum as
\begin{eqnarray} \label{TayExp1}
    \ln \mathcal{P}_\zeta(\kappa)\,=\,\ln \left(\frac{{\cal A}_\zeta}{\sqrt{2 \pi}\,\Delta} \right)-\frac{\ln^2 \kappa}{2 \,\Delta^2}+\frac{\alpha_3}{3!\,\Delta^3}\,\ln^3 \kappa +\frac{\alpha_4}{4!\,\Delta^4}\,\ln^4 \kappa+\dots,
\end{eqnarray}
where, for convenience, in this section we define ${\cal A}_\zeta=\sqrt{2 \pi}\,\Delta {\cal A}_s$, following the notation introduced in ref.~\cite{Pi:2020otn}. The coefficients $\alpha_i$, $i\ge3$ in the Taylor expansion are real numbers, generally expected to be of ${\cal O}(1)$. The pure log-normal case corresponds to $\alpha_i=0$. 
We expand $\Omega_\text{GW}(\kappa)$ in the limit of small $\kappa$, at leading order in $\Delta$ and $\kappa$, which corresponds to the infrared limit, going  beyond the limit of ref.~\cite{Pi:2020otn}. 
With some manipulations, we can write the \GWB\ spectrum as
\begin{align}\label{Omega3}
\Omega_\text{GW}&=\frac{3}{\pi}\frac{\mathcal{A}_\zeta^2}{\Delta^2}\kappa^2e^{\Delta^2}
\int^\infty_{-\infty}\,d s\,
\int^{\xi(s)}_{\chi(s)}\,
dt\,
\mathcal{T}\left(s,t\right)
\exp\left[-\frac{F_0(s,t)}{2\Delta^2}\right],
\end{align}
with
\begin{align}
F_0(s,t)&=\left( s+\sqrt{2} (\ln \kappa+\Delta^2)\right)^2+t^2
\nonumber
\\&
-\frac{\alpha_3}{6 \Delta} \left[\sqrt{2} s (s^2+3 t^2) +6 (s^2+t^2) \ln \kappa
+ 6 
\sqrt{2} s \ln^2\kappa+4 \ln^4\kappa \right]
+\alpha_4(\dots)
\end{align}
and
\begin{eqnarray}
\mathcal{T}(s,t)
&=&\frac14\left(\cosh(\sqrt2t)-\frac14e^{-\sqrt2s}-e^{\sqrt2s}\sinh^2(\sqrt2t)\right)^2
\left(\cosh(\sqrt2t)-\frac32e^{-\sqrt2s}\right)^4
\nonumber\\
&&\times\left\{\left[\ln\left|\frac{3-4e^{\sqrt2s}\cosh^2\left(t/\sqrt2\right)}{3-4e^{\sqrt2s}\sinh^2\left(t/\sqrt2\right)}\right|
-\frac2{\cosh(\sqrt2t)-\frac32e^{-\sqrt2s}}
\right]^2\right.\nonumber\\
&&\qquad\left.+\pi^2\Theta\left(2e^{\frac{s}{\sqrt2}}\cosh\frac{t}{\sqrt2}-\sqrt3\right)\vphantom{\ln\left|\frac{3-4e^{\sqrt2s}\cosh^2\left(t/\sqrt2\right)}{3-4e^{\sqrt2s}\sinh^2\left(t/\sqrt2\right)}\right|^2} \right\}.\label{def:Tst} 
\end{eqnarray}  
The extrema of integration along the coordinate $t$ are defined as 
\begin{align}
    \chi(s)&=\mathbf{Re}\left[\sqrt{2} \text{arccosh}\left(\frac{e^{-s/\sqrt2}}{2}\right)\right],
    \qquad 
    \xi(s)=\sqrt{2} \text{arcsinh}\left(\frac{e^{-s/\sqrt2}}{2}\right),
\end{align}
where the real part is taken to ensure that $\chi(s)=0$ for $s>-\sqrt2\ln2$. 
Within the working assumption of small $\Delta$, we make use of the
 saddle-point approximation to evaluate the integral. 
We substitute $t=\tau\,\Delta$ and 
 $s=-\sqrt{2}\,\left(\ln \kappa+\Delta^2\right)+\sigma\,\Delta$. 
 We keep only the leading contributions in $\Delta$, to obtain
\begin{equation}
 \frac{-F_0(\sigma,\tau)}{2 \Delta^2} =-\frac12 \sigma^2-\frac12 \tau^2
 +\frac{\alpha_3}{6\,\sqrt{2}} \sigma \left(\sigma^2+3 \tau^2\right)+\frac{\alpha_4}{48}
 \left(
 \sigma^4+ 6 \sigma^2 \tau^2+\tau^4 
  \right)+\Delta(\dots).
\end{equation}
Finally, in the IR limit $\kappa/\Delta \ll 1$, we can expand the integral over $\tau$ in \cref{Omega3}, to find
 \begin{align}
 \label{Omega5a}
\frac{\Omega_\text{GW}(k)}{\mathcal A_\zeta^2}=&
\frac{3 \kappa^3}{{2 \pi} \Delta} \,\left[ \frac{\pi^2}{4}+ \ln^2 \left( \frac{\sqrt 3\,\kappa\,e^{\Delta^2}}{2}   \right) 
\right]
\int^\infty_{-\infty}\,d \sigma\,
\exp^{\left[
-\frac12 \sigma^2
+\frac{\alpha_3}{6\,\sqrt{2}} \sigma^3+\frac{\alpha_4 }{48} \sigma^4+\dots
  \right]}.
\end{align}
Importantly, the integral along $\sigma$ is independent of $\kappa$,
hence
it only affects the overall normalization and not the scale dependence in the IR. 
At small $\kappa$, the spectrum scales as $\kappa^3$ (as expected) but there
are log corrections that modify the slope.
We can compute the spectral index associated
with eq.~\eqref{Omega5a}, to obtain 
 \begin{eqnarray} \label{calcnom}
n_\Omega
\,=\,3+\frac{8\, \ln \left( {\sqrt 3\,\kappa\,e^{\Delta^2}}/{2} \right)}{\pi^2+4
 \ln^2 \left( {\sqrt 3\,\kappa\,e^{\Delta^2}}/{2} \right)
}\,.
\end{eqnarray}
For very small $\kappa$, the second term is negligible, but as $\kappa$ increases
 the log corrections change $n_\Omega$ and bring it nearby $n_\Omega=2.5$. See \cref{plotexnom} for a plot
 of formula \eqref{calcnom}.
These results justify fixing spectral index appearing in eq.~\eqref{eq:templateDP} to $n_p = 2.5$ at the intermediate frequencies before reaching the far-infrared tail.

\subsection{DP template from log-normal curvature spectrum}
\label{app:lognormal}

\begin{figure}[h!]
\centering
\includegraphics[width=0.9\columnwidth]{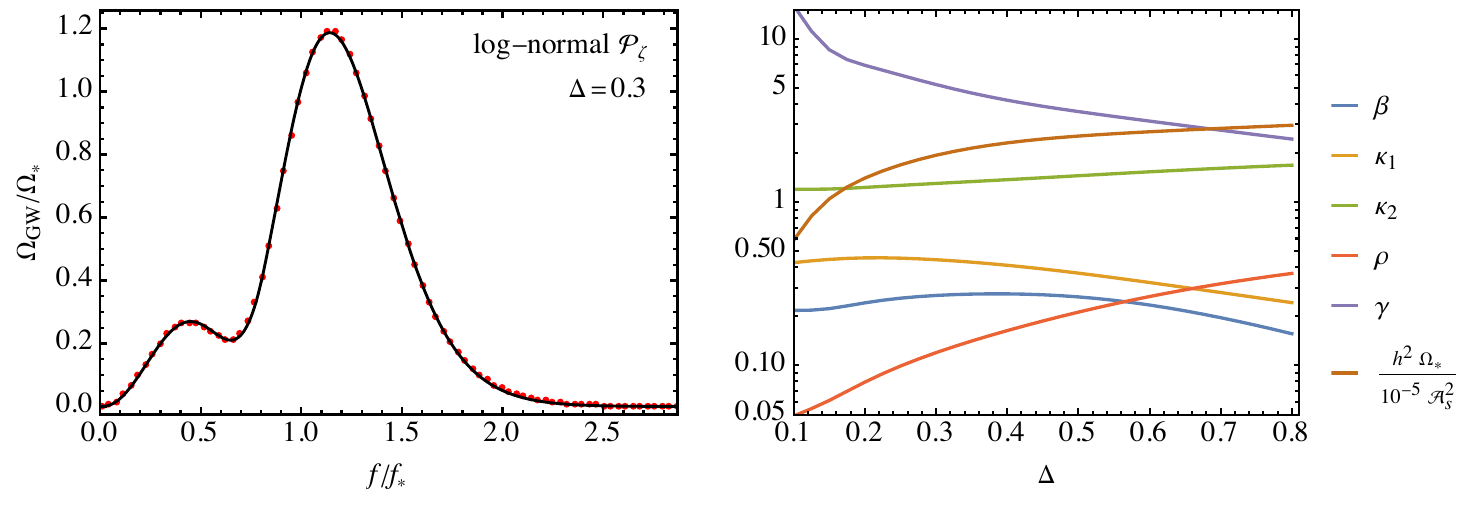} 
\caption{\small \textit{Left panel:} The red points indicate the scalar-induced GW spectrum from a log-normal curvature power spectrum and the solid curves show the fit by the DP template to these points. \textit{Right panel:} Best-fit parameters characterising the DP template when matched with the GW spectrum induced from a log-normal curvature power spectrum of the form \eqref{eq:Pk_LN} as a function of $\Delta$.} 
\label{fig:OmegaSF2a}
\end{figure}

We discuss here details of the connection between the DP template \eqref{eq:templateDP} and the log-normal scalar power spectrum scenario.
For definiteness, we take the log-normal primordial curvature power spectrum as
\begin{equation} \label{eq:Pk_LN}
    \mathcal{P}_\zeta(k)\!=\!{\cal A}_s \exp\!\left[-\frac{1}{2\Delta^2}\ln^2\!\left(\frac{k}{k_*}\right)\right],
\end{equation}
where ${\cal A}_s$ parameterises the peak amplitude while $\Delta>0$ describes the width of the peak. In the left panel of \cref{fig:OmegaSF2a} we show how the template presented in eq.~\eqref{eq:templateDP} fits an example case with $\Delta=0.3$. We report the scaling of the parameters of the template, as a function of $\Delta$, in the right panel of \cref{fig:OmegaSF2a}. These fits, alongside the ones derived in the following section, motivate the choice of prior ranges for the parameters that characterise the DP template.

\subsection{DP template from a broken power-law curvature spectrum}
\label{app:UVcutoff}

\begin{figure}[h!]
\centering
\includegraphics[width=0.9\columnwidth]{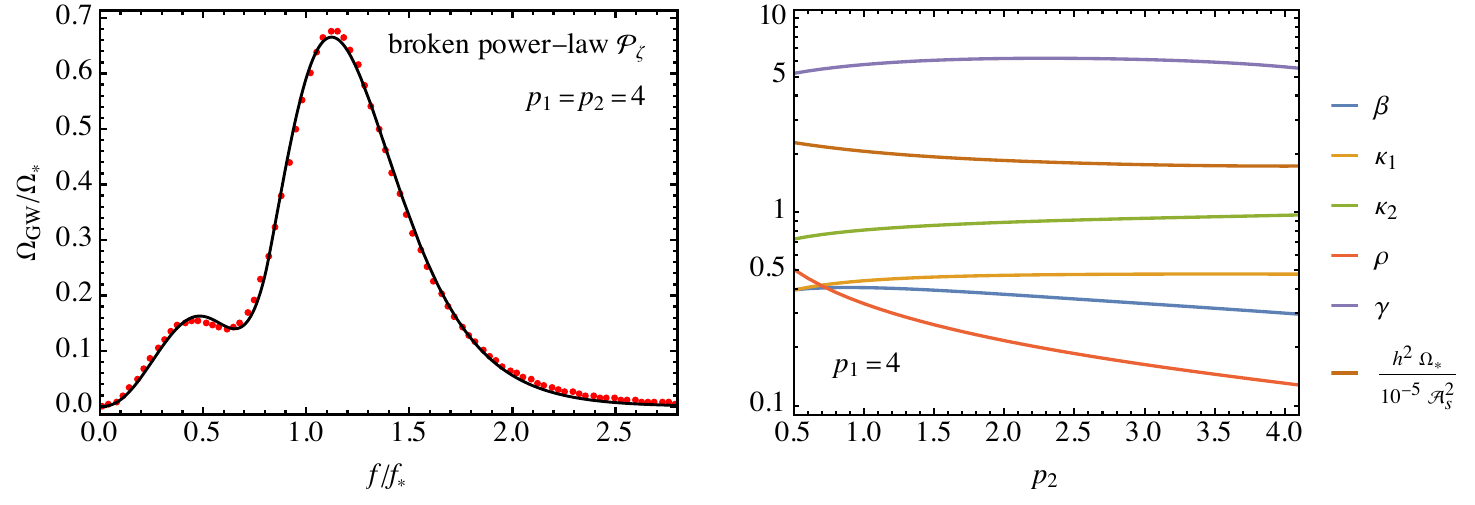} 
\caption{\small \textit{Left panel:} The red points indicate the scalar-induced GW spectrum from a broken power-law curvature power spectrum and the solid curves show the fit by the DP template to these points. \textit{Right panel:} Fit of the DP template to the scalar-induced GW spectrum from a broken power-law curvature power spectrum as a function of $p_2$.}
\label{fig:Omega-broken-power-law}
\end{figure}

Following the same structure of the previous section, we report here details of the connection between the DP template \eqref{eq:templateDP} and the broken power-law scalar power spectrum defined in eq.~\eqref{eq:Pk_bpl1}. In the left panel of \cref{fig:Omega-broken-power-law} we show how the template presented in eq.~\eqref{eq:templateDP} fits an example case with $p_1 = p_2 = 4$. We report the scaling of the parameters of the template, as a function of $p_2$, in the right panel of \cref{fig:Omega-broken-power-law}, where we have fixed $p_1 = 4$ as motivated by USR models of inflation (see also related discussion in the main text).

A transient nonattractor phase in single-field inflationary scenarios leads to a spectral growth which is bounded from above.  In particular, ref.~\cite{Byrnes:2018txb} (see also ref.~\cite{Ozsoy:2019lyy}) showed that for {\it all} single-field models which include a short nonattractor phase the curvature spectrum cannot grow faster than ${\cal P}_\zeta(k) \,\propto \,k^4$. However, including a further prolonged phase of non-slow-roll evolution, the spectrum can grow as ${\cal P}_\zeta(k) \,\propto \,k^5\,(\ln k)^2$ \cite{Carrilho:2019oqg}. If we consider multiple phases of non-attractor evolution, then ${\cal P}_\zeta(k) \,\propto \,k^8$ can be reached~\cite{Tasinato:2020vdk,Davies:2021loj}. Instead, after the peak, the decay of the spectrum can occur with an arbitrarily steep (negative) slope. 

\begin{figure}[h!]
\centering
\includegraphics[width=0.9\textwidth]{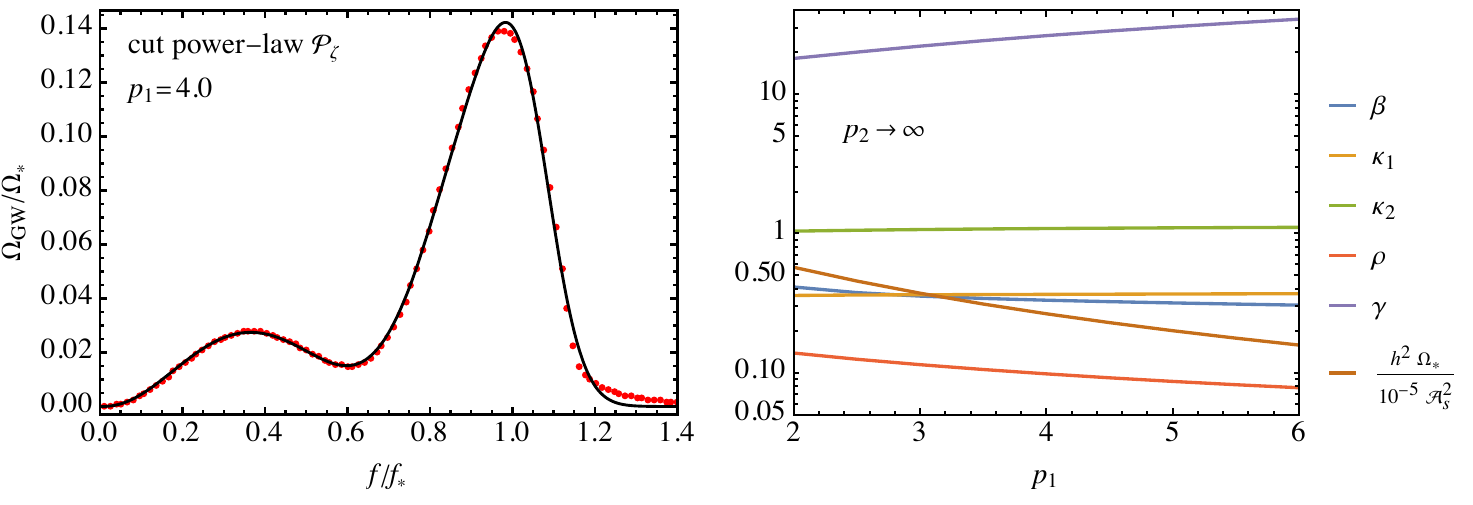}
\caption{\small Fit of the DP template to the scalar-induced GW spectrum from a broken power-law curvature power spectrum as a function of $p_1$ in the limit $p_2\to \infty$.}
\label{fig:OmegaSF2b}
\end{figure}

In light of these results, we verified  that the DP template is able to capture also the \GWB\ spectral features produced by models with a very steep spectral growth. 
In \cref{fig:OmegaSF2b}, we show the corresponding best-fit template parameters  as a function of $p_1$, where, for simplicity, we have fixed the spectral drop  after the peak $p_2$ to very large values.

\bibliographystyle{JHEP}
\bibliography{biblio_arXiv}

\providecommand{\href}[2]{#2}\begingroup\raggedright\begin{thebibliography}{100}

\bibitem{LISA:2017pwj}
{\scshape LISA} collaboration, \emph{{Laser Interferometer Space Antenna}},
  \href{https://arxiv.org/abs/1702.00786}{{\ttfamily 1702.00786}}.

\bibitem{Colpi:2024xhw}
M.~Colpi et~al., \emph{{LISA Definition Study Report}},
  \href{https://arxiv.org/abs/2402.07571}{{\ttfamily 2402.07571}}.

\bibitem{Braglia:2024siw}
M.~Braglia, M.~Pieroni and S.~Marsat, \emph{{The impact of a primordial
  gravitational wave background on LISA resolvable sources}},
  \href{https://arxiv.org/abs/2406.10048}{{\ttfamily 2406.10048}}.

\bibitem{Cornish:2005qw}
N.~J. Cornish and J.~Crowder, \emph{{LISA data analysis using MCMC methods}},
  \href{https://doi.org/10.1103/PhysRevD.72.043005}{\emph{Phys. Rev. D}
  {\bfseries 72} (2005) 043005}
  [\href{https://arxiv.org/abs/gr-qc/0506059}{{\ttfamily gr-qc/0506059}}].

\bibitem{Vallisneri:2008ye}
M.~Vallisneri, \emph{{A LISA Data-Analysis Primer}},
  \href{https://doi.org/10.1088/0264-9381/26/9/094024}{\emph{Class. Quant.
  Grav.} {\bfseries 26} (2009) 094024}
  [\href{https://arxiv.org/abs/0812.0751}{{\ttfamily 0812.0751}}].

\bibitem{MockLISADataChallengeTaskForce:2009wir}
{\scshape Mock LISA Data Challenge Task Force} collaboration, \emph{{The Mock
  LISA Data Challenges: From Challenge 3 to Challenge 4}},
  \href{https://doi.org/10.1088/0264-9381/27/8/084009}{\emph{Class. Quant.
  Grav.} {\bfseries 27} (2010) 084009}
  [\href{https://arxiv.org/abs/0912.0548}{{\ttfamily 0912.0548}}].

\bibitem{Littenberg:2023xpl}
T.~B. Littenberg and N.~J. Cornish, \emph{{Prototype global analysis of LISA
  data with multiple source types}},
  \href{https://doi.org/10.1103/PhysRevD.107.063004}{\emph{Phys. Rev. D}
  {\bfseries 107} (2023) 063004}
  [\href{https://arxiv.org/abs/2301.03673}{{\ttfamily 2301.03673}}].

\bibitem{Caprini:2019pxz}
C.~Caprini, D.~G. Figueroa, R.~Flauger, G.~Nardini, M.~Peloso, M.~Pieroni
  et~al., \emph{{Reconstructing the spectral shape of a stochastic
  gravitational wave background with LISA}},
  \href{https://doi.org/10.1088/1475-7516/2019/11/017}{\emph{JCAP} {\bfseries
  11} (2019) 017} [\href{https://arxiv.org/abs/1906.09244}{{\ttfamily
  1906.09244}}].

\bibitem{Flauger:2020qyi}
R.~Flauger, N.~Karnesis, G.~Nardini, M.~Pieroni, A.~Ricciardone and J.~Torrado,
  \emph{{Improved reconstruction of a stochastic gravitational wave background
  with LISA}}, \href{https://doi.org/10.1088/1475-7516/2021/01/059}{\emph{JCAP}
  {\bfseries 01} (2021) 059}
  [\href{https://arxiv.org/abs/2009.11845}{{\ttfamily 2009.11845}}].

\bibitem{Karnesis:2019mph}
N.~Karnesis, M.~Lilley and A.~Petiteau, \emph{{Assessing the detectability of a
  Stochastic Gravitational Wave Background with LISA, using an excess of power
  approach}}, \href{https://doi.org/10.1088/1361-6382/abb637}{\emph{Class.
  Quant. Grav.} {\bfseries 37} (2020) 215017}
  [\href{https://arxiv.org/abs/1906.09027}{{\ttfamily 1906.09027}}].

\bibitem{Caprini:2024hue}
{\scshape LISA Cosmology Working Group} collaboration, \emph{{Gravitational
  waves from first-order phase transitions in LISA: reconstruction pipeline and
  physics interpretation}},  \href{https://arxiv.org/abs/2403.03723}{{\ttfamily
  2403.03723}}.

\bibitem{Blanco-Pillado:2024aca}
{\scshape LISA Cosmology Working Group} collaboration, \emph{{Gravitational
  waves from cosmic strings in LISA: reconstruction pipeline and physics
  interpretation}},  \href{https://arxiv.org/abs/2405.03740}{{\ttfamily
  2405.03740}}.

\bibitem{Guth:1980zm}
A.~H. Guth, \emph{{The Inflationary Universe: A Possible Solution to the
  Horizon and Flatness Problems}},
  \href{https://doi.org/10.1103/PhysRevD.23.347}{\emph{Phys. Rev. D} {\bfseries
  23} (1981) 347}.

\bibitem{Starobinsky:1980te}
A.~A. Starobinsky, \emph{{A New Type of Isotropic Cosmological Models Without
  Singularity}},
  \href{https://doi.org/10.1016/0370-2693(80)90670-X}{\emph{Phys. Lett. B}
  {\bfseries 91} (1980) 99}.

\bibitem{Linde:1981mu}
A.~D. Linde, \emph{{A New Inflationary Universe Scenario: A Possible Solution
  of the Horizon, Flatness, Homogeneity, Isotropy and Primordial Monopole
  Problems}}, \href{https://doi.org/10.1016/0370-2693(82)91219-9}{\emph{Phys.
  Lett. B} {\bfseries 108} (1982) 389}.

\bibitem{Albrecht:1982wi}
A.~Albrecht and P.~J. Steinhardt, \emph{{Cosmology for Grand Unified Theories
  with Radiatively Induced Symmetry Breaking}},
  \href{https://doi.org/10.1103/PhysRevLett.48.1220}{\emph{Phys. Rev. Lett.}
  {\bfseries 48} (1982) 1220}.

\bibitem{Mukhanov:1981xt}
V.~F. Mukhanov and G.~V. Chibisov, \emph{{Quantum Fluctuations and a
  Nonsingular Universe}}, {\emph{JETP Lett.} {\bfseries 33} (1981) 532}.

\bibitem{Hawking:1982cz}
S.~W. Hawking, \emph{{The Development of Irregularities in a Single Bubble
  Inflationary Universe}},
  \href{https://doi.org/10.1016/0370-2693(82)90373-2}{\emph{Phys. Lett. B}
  {\bfseries 115} (1982) 295}.

\bibitem{Starobinsky:1982ee}
A.~A. Starobinsky, \emph{{Dynamics of Phase Transition in the New Inflationary
  Universe Scenario and Generation of Perturbations}},
  \href{https://doi.org/10.1016/0370-2693(82)90541-X}{\emph{Phys. Lett. B}
  {\bfseries 117} (1982) 175}.

\bibitem{Guth:1982ec}
A.~H. Guth and S.~Y. Pi, \emph{{Fluctuations in the New Inflationary
  Universe}}, \href{https://doi.org/10.1103/PhysRevLett.49.1110}{\emph{Phys.
  Rev. Lett.} {\bfseries 49} (1982) 1110}.

\bibitem{Bardeen:1983qw}
J.~M. Bardeen, P.~J. Steinhardt and M.~S. Turner, \emph{{Spontaneous Creation
  of Almost Scale - Free Density Perturbations in an Inflationary Universe}},
  \href{https://doi.org/10.1103/PhysRevD.28.679}{\emph{Phys. Rev. D} {\bfseries
  28} (1983) 679}.

\bibitem{Bartolo:2016ami}
N.~Bartolo et~al., \emph{{Science with the space-based interferometer LISA. IV:
  Probing inflation with gravitational waves}},
  \href{https://doi.org/10.1088/1475-7516/2016/12/026}{\emph{JCAP} {\bfseries
  12} (2016) 026} [\href{https://arxiv.org/abs/1610.06481}{{\ttfamily
  1610.06481}}].

\bibitem{LISACosmologyWorkingGroup:2022jok}
{\scshape LISA Cosmology Working Group} collaboration, \emph{{Cosmology with
  the Laser Interferometer Space Antenna}},
  \href{https://doi.org/10.1007/s41114-023-00045-2}{\emph{Living Rev. Rel.}
  {\bfseries 26} (2023) 5} [\href{https://arxiv.org/abs/2204.05434}{{\ttfamily
  2204.05434}}].

\bibitem{LISACosmologyWorkingGroup:2022kbp}
{\scshape LISA Cosmology Working Group} collaboration, \emph{{Probing
  anisotropies of the Stochastic Gravitational Wave Background with LISA}},
  \href{https://doi.org/10.1088/1475-7516/2022/11/009}{\emph{JCAP} {\bfseries
  11} (2022) 009} [\href{https://arxiv.org/abs/2201.08782}{{\ttfamily
  2201.08782}}].

\bibitem{Domcke:2019zls}
V.~Domcke, J.~Garcia-Bellido, M.~Peloso, M.~Pieroni, A.~Ricciardone, L.~Sorbo
  et~al., \emph{{Measuring the net circular polarization of the stochastic
  gravitational wave background with interferometers}},
  \href{https://doi.org/10.1088/1475-7516/2020/05/028}{\emph{JCAP} {\bfseries
  05} (2020) 028} [\href{https://arxiv.org/abs/1910.08052}{{\ttfamily
  1910.08052}}].

\bibitem{LISA:2022yao}
{\scshape LISA} collaboration, \emph{{Astrophysics with the Laser
  Interferometer Space Antenna}},
  \href{https://doi.org/10.1007/s41114-022-00041-y}{\emph{Living Rev. Rel.}
  {\bfseries 26} (2023) 2} [\href{https://arxiv.org/abs/2203.06016}{{\ttfamily
  2203.06016}}].

\bibitem{LISAConsortiumWaveformWorkingGroup:2023arg}
{\scshape LISA Consortium Waveform Working Group} collaboration,
  \emph{{Waveform Modelling for the Laser Interferometer Space Antenna}},
  \href{https://arxiv.org/abs/2311.01300}{{\ttfamily 2311.01300}}.

\bibitem{Pagano:2015hma}
L.~Pagano, L.~Salvati and A.~Melchiorri, \emph{{New constraints on primordial
  gravitational waves from Planck 2015}},
  \href{https://doi.org/10.1016/j.physletb.2016.07.078}{\emph{Phys. Lett. B}
  {\bfseries 760} (2016) 823}
  [\href{https://arxiv.org/abs/1508.02393}{{\ttfamily 1508.02393}}].

\bibitem{Barnaby:2010vf}
N.~Barnaby and M.~Peloso, \emph{{Large Nongaussianity in Axion Inflation}},
  \href{https://doi.org/10.1103/PhysRevLett.106.181301}{\emph{Phys. Rev. Lett.}
  {\bfseries 106} (2011) 181301}
  [\href{https://arxiv.org/abs/1011.1500}{{\ttfamily 1011.1500}}].

\bibitem{Sorbo:2011rz}
L.~Sorbo, \emph{{Parity violation in the Cosmic Microwave Background from a
  pseudoscalar inflaton}},
  \href{https://doi.org/10.1088/1475-7516/2011/06/003}{\emph{JCAP} {\bfseries
  06} (2011) 003} [\href{https://arxiv.org/abs/1101.1525}{{\ttfamily
  1101.1525}}].

\bibitem{Garcia-Bellido:2023ser}
J.~Garcia-Bellido, A.~Papageorgiou, M.~Peloso and L.~Sorbo, \emph{{A flashing
  beacon in axion inflation: recurring bursts of gravitational waves in the
  strong backreaction regime}},
  \href{https://doi.org/10.1088/1475-7516/2024/01/034}{\emph{JCAP} {\bfseries
  01} (2024) 034} [\href{https://arxiv.org/abs/2303.13425}{{\ttfamily
  2303.13425}}].

\bibitem{Caravano:2022epk}
A.~Caravano, E.~Komatsu, K.~D. Lozanov and J.~Weller, \emph{{Lattice
  simulations of axion-U(1) inflation}},
  \href{https://doi.org/10.1103/PhysRevD.108.043504}{\emph{Phys. Rev. D}
  {\bfseries 108} (2023) 043504}
  [\href{https://arxiv.org/abs/2204.12874}{{\ttfamily 2204.12874}}].

\bibitem{Figueroa:2023oxc}
D.~G. Figueroa, J.~Lizarraga, A.~Urio and J.~Urrestilla, \emph{{Strong
  Backreaction Regime in Axion Inflation}},
  \href{https://doi.org/10.1103/PhysRevLett.131.151003}{\emph{Phys. Rev. Lett.}
  {\bfseries 131} (2023) 151003}
  [\href{https://arxiv.org/abs/2303.17436}{{\ttfamily 2303.17436}}].

\bibitem{Peloso:2022ovc}
M.~Peloso and L.~Sorbo, \emph{{Instability in axion inflation with strong
  backreaction from gauge modes}},
  \href{https://doi.org/10.1088/1475-7516/2023/01/038}{\emph{JCAP} {\bfseries
  01} (2023) 038} [\href{https://arxiv.org/abs/2209.08131}{{\ttfamily
  2209.08131}}].

\bibitem{Endlich:2012pz}
S.~Endlich, A.~Nicolis and J.~Wang, \emph{{Solid Inflation}},
  \href{https://doi.org/10.1088/1475-7516/2013/10/011}{\emph{JCAP} {\bfseries
  10} (2013) 011} [\href{https://arxiv.org/abs/1210.0569}{{\ttfamily
  1210.0569}}].

\bibitem{Ricciardone:2016lym}
A.~Ricciardone and G.~Tasinato, \emph{{Primordial gravitational waves in
  supersolid inflation}},
  \href{https://doi.org/10.1103/PhysRevD.96.023508}{\emph{Phys. Rev. D}
  {\bfseries 96} (2017) 023508}
  [\href{https://arxiv.org/abs/1611.04516}{{\ttfamily 1611.04516}}].

\bibitem{Bartolo:2015qvr}
N.~Bartolo, D.~Cannone, A.~Ricciardone and G.~Tasinato, \emph{{Distinctive
  signatures of space-time diffeomorphism breaking in EFT of inflation}},
  \href{https://doi.org/10.1088/1475-7516/2016/03/044}{\emph{JCAP} {\bfseries
  03} (2016) 044} [\href{https://arxiv.org/abs/1511.07414}{{\ttfamily
  1511.07414}}].

\bibitem{Cannone:2014uqa}
D.~Cannone, G.~Tasinato and D.~Wands, \emph{{Generalised tensor fluctuations
  and inflation}},
  \href{https://doi.org/10.1088/1475-7516/2015/01/029}{\emph{JCAP} {\bfseries
  01} (2015) 029} [\href{https://arxiv.org/abs/1409.6568}{{\ttfamily
  1409.6568}}].

\bibitem{Cai:2014uka}
Y.-F. Cai, J.-O. Gong, S.~Pi, E.~N. Saridakis and S.-Y. Wu, \emph{{On the
  possibility of blue tensor spectrum within single field inflation}},
  \href{https://doi.org/10.1016/j.nuclphysb.2015.09.025}{\emph{Nucl. Phys. B}
  {\bfseries 900} (2015) 517}
  [\href{https://arxiv.org/abs/1412.7241}{{\ttfamily 1412.7241}}].

\bibitem{Lin:2015cqa}
C.~Lin and L.~Z. Labun, \emph{{Effective Field Theory of Broken Spatial
  Diffeomorphisms}}, \href{https://doi.org/10.1007/JHEP03(2016)128}{\emph{JHEP}
  {\bfseries 03} (2016) 128}
  [\href{https://arxiv.org/abs/1501.07160}{{\ttfamily 1501.07160}}].

\bibitem{Cannone:2015rra}
D.~Cannone, J.-O. Gong and G.~Tasinato, \emph{{Breaking discrete symmetries in
  the effective field theory of inflation}},
  \href{https://doi.org/10.1088/1475-7516/2015/08/003}{\emph{JCAP} {\bfseries
  08} (2015) 003} [\href{https://arxiv.org/abs/1505.05773}{{\ttfamily
  1505.05773}}].

\bibitem{Graef:2015ova}
L.~Graef and R.~Brandenberger, \emph{{Breaking of Spatial Diffeomorphism
  Invariance, Inflation and the Spectrum of Cosmological Perturbations}},
  \href{https://doi.org/10.1088/1475-7516/2015/10/009}{\emph{JCAP} {\bfseries
  10} (2015) 009} [\href{https://arxiv.org/abs/1506.00896}{{\ttfamily
  1506.00896}}].

\bibitem{Domenech:2017kno}
G.~Dom\`enech, T.~Hiramatsu, C.~Lin, M.~Sasaki, M.~Shiraishi and Y.~Wang,
  \emph{{CMB Scale Dependent Non-Gaussianity from Massive Gravity during
  Inflation}}, \href{https://doi.org/10.1088/1475-7516/2017/05/034}{\emph{JCAP}
  {\bfseries 05} (2017) 034}
  [\href{https://arxiv.org/abs/1701.05554}{{\ttfamily 1701.05554}}].

\bibitem{Ricciardone:2017kre}
A.~Ricciardone and G.~Tasinato, \emph{{Anisotropic tensor power spectrum at
  interferometer scales induced by tensor squeezed non-Gaussianity}},
  \href{https://doi.org/10.1088/1475-7516/2018/02/011}{\emph{JCAP} {\bfseries
  02} (2018) 011} [\href{https://arxiv.org/abs/1711.02635}{{\ttfamily
  1711.02635}}].

\bibitem{Celoria:2017idi}
M.~Celoria, D.~Comelli and L.~Pilo, \emph{{Self-gravitating $\Lambda$-media}},
  \href{https://doi.org/10.1088/1475-7516/2019/01/057}{\emph{JCAP} {\bfseries
  01} (2019) 057} [\href{https://arxiv.org/abs/1712.04827}{{\ttfamily
  1712.04827}}].

\bibitem{Dimastrogiovanni:2018uqy}
E.~Dimastrogiovanni, M.~Fasiello and G.~Tasinato, \emph{{Probing the
  inflationary particle content: extra spin-2 field}},
  \href{https://doi.org/10.1088/1475-7516/2018/08/016}{\emph{JCAP} {\bfseries
  08} (2018) 016} [\href{https://arxiv.org/abs/1806.00850}{{\ttfamily
  1806.00850}}].

\bibitem{Fujita:2018ehq}
T.~Fujita, S.~Kuroyanagi, S.~Mizuno and S.~Mukohyama, \emph{{Blue-tilted
  Primordial Gravitational Waves from Massive Gravity}},
  \href{https://doi.org/10.1016/j.physletb.2018.12.025}{\emph{Phys. Lett. B}
  {\bfseries 789} (2019) 215}
  [\href{https://arxiv.org/abs/1808.02381}{{\ttfamily 1808.02381}}].

\bibitem{Higuchi:1986py}
A.~Higuchi, \emph{{Forbidden Mass Range for Spin-2 Field Theory in De Sitter
  Space-time}}, \href{https://doi.org/10.1016/0550-3213(87)90691-2}{\emph{Nucl.
  Phys. B} {\bfseries 282} (1987) 397}.

\bibitem{Kobayashi:2011nu}
T.~Kobayashi, M.~Yamaguchi and J.~Yokoyama, \emph{{Generalized G-inflation:
  Inflation with the most general second-order field equations}},
  \href{https://doi.org/10.1143/PTP.126.511}{\emph{Prog. Theor. Phys.}
  {\bfseries 126} (2011) 511}
  [\href{https://arxiv.org/abs/1105.5723}{{\ttfamily 1105.5723}}].

\bibitem{Creminelli:2014wna}
P.~Creminelli, J.~Gleyzes, J.~Nore\~na and F.~Vernizzi, \emph{{Resilience of
  the standard predictions for primordial tensor modes}},
  \href{https://doi.org/10.1103/PhysRevLett.113.231301}{\emph{Phys. Rev. Lett.}
  {\bfseries 113} (2014) 231301}
  [\href{https://arxiv.org/abs/1407.8439}{{\ttfamily 1407.8439}}].

\bibitem{Iacconi:2020yxn}
L.~Iacconi, M.~Fasiello, H.~Assadullahi and D.~Wands, \emph{{Small-scale Tests
  of Inflation}},
  \href{https://doi.org/10.1088/1475-7516/2020/12/005}{\emph{JCAP} {\bfseries
  12} (2020) 005} [\href{https://arxiv.org/abs/2008.00452}{{\ttfamily
  2008.00452}}].

\bibitem{Dimastrogiovanni:2021mfs}
E.~Dimastrogiovanni, M.~Fasiello, A.~Malhotra, P.~D. Meerburg and G.~Orlando,
  \emph{{Testing the early universe with anisotropies of the gravitational wave
  background}},
  \href{https://doi.org/10.1088/1475-7516/2022/02/040}{\emph{JCAP} {\bfseries
  02} (2022) 040} [\href{https://arxiv.org/abs/2109.03077}{{\ttfamily
  2109.03077}}].

\bibitem{Raveri:2014eea}
M.~Raveri, C.~Baccigalupi, A.~Silvestri and S.-Y. Zhou, \emph{{Measuring the
  speed of cosmological gravitational waves}},
  \href{https://doi.org/10.1103/PhysRevD.91.061501}{\emph{Phys. Rev. D}
  {\bfseries 91} (2015) 061501}
  [\href{https://arxiv.org/abs/1405.7974}{{\ttfamily 1405.7974}}].

\bibitem{Namba:2015gja}
R.~Namba, M.~Peloso, M.~Shiraishi, L.~Sorbo and C.~Unal, \emph{{Scale-dependent
  gravitational waves from a rolling axion}},
  \href{https://doi.org/10.1088/1475-7516/2016/01/041}{\emph{JCAP} {\bfseries
  01} (2016) 041} [\href{https://arxiv.org/abs/1509.07521}{{\ttfamily
  1509.07521}}].

\bibitem{Campeti:2022acx}
P.~Campeti, O.~\"Ozsoy, I.~Obata and M.~Shiraishi, \emph{{New constraints on
  axion-gauge field dynamics during inflation from Planck and BICEP/Keck data
  sets}}, \href{https://doi.org/10.1088/1475-7516/2022/07/039}{\emph{JCAP}
  {\bfseries 07} (2022) 039}
  [\href{https://arxiv.org/abs/2203.03401}{{\ttfamily 2203.03401}}].

\bibitem{Dimastrogiovanni:2016fuu}
E.~Dimastrogiovanni, M.~Fasiello and T.~Fujita, \emph{{Primordial Gravitational
  Waves from Axion-Gauge Fields Dynamics}},
  \href{https://doi.org/10.1088/1475-7516/2017/01/019}{\emph{JCAP} {\bfseries
  01} (2017) 019} [\href{https://arxiv.org/abs/1608.04216}{{\ttfamily
  1608.04216}}].

\bibitem{Thorne:2017jft}
B.~Thorne, T.~Fujita, M.~Hazumi, N.~Katayama, E.~Komatsu and M.~Shiraishi,
  \emph{{Finding the chiral gravitational wave background of an axion-SU(2)
  inflationary model using CMB observations and laser interferometers}},
  \href{https://doi.org/10.1103/PhysRevD.97.043506}{\emph{Phys. Rev. D}
  {\bfseries 97} (2018) 043506}
  [\href{https://arxiv.org/abs/1707.03240}{{\ttfamily 1707.03240}}].

\bibitem{Putti:2024uyr}
M.~Putti, N.~Bartolo, S.~Bhattacharya and M.~Peloso, \emph{{CMB spectral
  distortions from enhanced primordial perturbations: the role of spectator
  axions}},  \href{https://arxiv.org/abs/2403.08594}{{\ttfamily 2403.08594}}.

\bibitem{Wands:1998yp}
D.~Wands, \emph{{Duality invariance of cosmological perturbation spectra}},
  \href{https://doi.org/10.1103/PhysRevD.60.023507}{\emph{Phys. Rev. D}
  {\bfseries 60} (1999) 023507}
  [\href{https://arxiv.org/abs/gr-qc/9809062}{{\ttfamily gr-qc/9809062}}].

\bibitem{Leach:2000yw}
S.~M. Leach and A.~R. Liddle, \emph{{Inflationary perturbations near horizon
  crossing}}, \href{https://doi.org/10.1103/PhysRevD.63.043508}{\emph{Phys.
  Rev. D} {\bfseries 63} (2001) 043508}
  [\href{https://arxiv.org/abs/astro-ph/0010082}{{\ttfamily
  astro-ph/0010082}}].

\bibitem{Leach:2001zf}
S.~M. Leach, M.~Sasaki, D.~Wands and A.~R. Liddle, \emph{{Enhancement of
  superhorizon scale inflationary curvature perturbations}},
  \href{https://doi.org/10.1103/PhysRevD.64.023512}{\emph{Phys. Rev. D}
  {\bfseries 64} (2001) 023512}
  [\href{https://arxiv.org/abs/astro-ph/0101406}{{\ttfamily
  astro-ph/0101406}}].

\bibitem{Biagetti:2018pjj}
M.~Biagetti, G.~Franciolini, A.~Kehagias and A.~Riotto, \emph{{Primordial Black
  Holes from Inflation and Quantum Diffusion}},
  \href{https://doi.org/10.1088/1475-7516/2018/07/032}{\emph{JCAP} {\bfseries
  07} (2018) 032} [\href{https://arxiv.org/abs/1804.07124}{{\ttfamily
  1804.07124}}].

\bibitem{Franciolini:2022pav}
G.~Franciolini and A.~Urbano, \emph{{Primordial black hole dark matter from
  inflation: The reverse engineering approach}},
  \href{https://doi.org/10.1103/PhysRevD.106.123519}{\emph{Phys. Rev. D}
  {\bfseries 106} (2022) 123519}
  [\href{https://arxiv.org/abs/2207.10056}{{\ttfamily 2207.10056}}].

\bibitem{DeLuca:2020agl}
V.~De~Luca, G.~Franciolini and A.~Riotto, \emph{{NANOGrav Data Hints at
  Primordial Black Holes as Dark Matter}},
  \href{https://doi.org/10.1103/PhysRevLett.126.041303}{\emph{Phys. Rev. Lett.}
  {\bfseries 126} (2021) 041303}
  [\href{https://arxiv.org/abs/2009.08268}{{\ttfamily 2009.08268}}].

\bibitem{NANOGrav:2023gor}
{\scshape NANOGrav} collaboration, \emph{{The NANOGrav 15 yr Data Set: Evidence
  for a Gravitational-wave Background}},
  \href{https://doi.org/10.3847/2041-8213/acdac6}{\emph{Astrophys. J. Lett.}
  {\bfseries 951} (2023) L8}
  [\href{https://arxiv.org/abs/2306.16213}{{\ttfamily 2306.16213}}].

\bibitem{EPTA:2023fyk}
{\scshape EPTA, InPTA:} collaboration, \emph{{The second data release from the
  European Pulsar Timing Array - III. Search for gravitational wave signals}},
  \href{https://doi.org/10.1051/0004-6361/202346844}{\emph{Astron. Astrophys.}
  {\bfseries 678} (2023) A50}
  [\href{https://arxiv.org/abs/2306.16214}{{\ttfamily 2306.16214}}].

\bibitem{Reardon:2023gzh}
D.~J. Reardon et~al., \emph{{Search for an Isotropic Gravitational-wave
  Background with the Parkes Pulsar Timing Array}},
  \href{https://doi.org/10.3847/2041-8213/acdd02}{\emph{Astrophys. J. Lett.}
  {\bfseries 951} (2023) L6}
  [\href{https://arxiv.org/abs/2306.16215}{{\ttfamily 2306.16215}}].

\bibitem{Xu:2023wog}
H.~Xu et~al., \emph{{Searching for the Nano-Hertz Stochastic Gravitational Wave
  Background with the Chinese Pulsar Timing Array Data Release I}},
  \href{https://doi.org/10.1088/1674-4527/acdfa5}{\emph{Res. Astron.
  Astrophys.} {\bfseries 23} (2023) 075024}
  [\href{https://arxiv.org/abs/2306.16216}{{\ttfamily 2306.16216}}].

\bibitem{Sasaki:2018dmp}
M.~Sasaki, T.~Suyama, T.~Tanaka and S.~Yokoyama, \emph{{Primordial black
  holes\textemdash{}perspectives in gravitational wave astronomy}},
  \href{https://doi.org/10.1088/1361-6382/aaa7b4}{\emph{Class. Quant. Grav.}
  {\bfseries 35} (2018) 063001}
  [\href{https://arxiv.org/abs/1801.05235}{{\ttfamily 1801.05235}}].

\bibitem{Vaskonen:2020lbd}
V.~Vaskonen and H.~Veerm\"ae, \emph{{Did NANOGrav see a signal from primordial
  black hole formation?}},
  \href{https://doi.org/10.1103/PhysRevLett.126.051303}{\emph{Phys. Rev. Lett.}
  {\bfseries 126} (2021) 051303}
  [\href{https://arxiv.org/abs/2009.07832}{{\ttfamily 2009.07832}}].

\bibitem{Linde:1993cn}
A.~D. Linde, \emph{{Hybrid inflation}},
  \href{https://doi.org/10.1103/PhysRevD.49.748}{\emph{Phys. Rev. D} {\bfseries
  49} (1994) 748} [\href{https://arxiv.org/abs/astro-ph/9307002}{{\ttfamily
  astro-ph/9307002}}].

\bibitem{Garcia-Bellido:1996mdl}
J.~Garcia-Bellido, A.~D. Linde and D.~Wands, \emph{{Density perturbations and
  black hole formation in hybrid inflation}},
  \href{https://doi.org/10.1103/PhysRevD.54.6040}{\emph{Phys. Rev. D}
  {\bfseries 54} (1996) 6040}
  [\href{https://arxiv.org/abs/astro-ph/9605094}{{\ttfamily
  astro-ph/9605094}}].

\bibitem{Clesse:2015wea}
S.~Clesse and J.~Garc\'\i{}a-Bellido, \emph{{Massive Primordial Black Holes
  from Hybrid Inflation as Dark Matter and the seeds of Galaxies}},
  \href{https://doi.org/10.1103/PhysRevD.92.023524}{\emph{Phys. Rev. D}
  {\bfseries 92} (2015) 023524}
  [\href{https://arxiv.org/abs/1501.07565}{{\ttfamily 1501.07565}}].

\bibitem{Fumagalli:2020adf}
J.~Fumagalli, S.~Renaux-Petel, J.~W. Ronayne and L.~T. Witkowski,
  \emph{{Turning in the landscape: A new mechanism for generating primordial
  black holes}},
  \href{https://doi.org/10.1016/j.physletb.2023.137921}{\emph{Phys. Lett. B}
  {\bfseries 841} (2023) 137921}
  [\href{https://arxiv.org/abs/2004.08369}{{\ttfamily 2004.08369}}].

\bibitem{Braglia:2020eai}
M.~Braglia, D.~K. Hazra, F.~Finelli, G.~F. Smoot, L.~Sriramkumar and A.~A.
  Starobinsky, \emph{{Generating PBHs and small-scale GWs in two-field models
  of inflation}},
  \href{https://doi.org/10.1088/1475-7516/2020/08/001}{\emph{JCAP} {\bfseries
  08} (2020) 001} [\href{https://arxiv.org/abs/2005.02895}{{\ttfamily
  2005.02895}}].

\bibitem{Braglia:2020taf}
M.~Braglia, X.~Chen and D.~K. Hazra, \emph{{Probing Primordial Features with
  the Stochastic Gravitational Wave Background}},
  \href{https://doi.org/10.1088/1475-7516/2021/03/005}{\emph{JCAP} {\bfseries
  03} (2021) 005} [\href{https://arxiv.org/abs/2012.05821}{{\ttfamily
  2012.05821}}].

\bibitem{Clesse:2018ogk}
S.~Clesse, J.~Garc\'\i{}a-Bellido and S.~Orani, \emph{{Detecting the Stochastic
  Gravitational Wave Background from Primordial Black Hole Formation}},
  \href{https://arxiv.org/abs/1812.11011}{{\ttfamily 1812.11011}}.

\bibitem{Spanos:2021hpk}
V.~C. Spanos and I.~D. Stamou, \emph{{Gravitational waves and primordial black
  holes from supersymmetric hybrid inflation}},
  \href{https://doi.org/10.1103/PhysRevD.104.123537}{\emph{Phys. Rev. D}
  {\bfseries 104} (2021) 123537}
  [\href{https://arxiv.org/abs/2108.05671}{{\ttfamily 2108.05671}}].

\bibitem{Braglia:2022phb}
M.~Braglia, A.~Linde, R.~Kallosh and F.~Finelli, \emph{{Hybrid
  \ensuremath{\alpha}-attractors, primordial black holes and gravitational wave
  backgrounds}},
  \href{https://doi.org/10.1088/1475-7516/2023/04/033}{\emph{JCAP} {\bfseries
  04} (2023) 033} [\href{https://arxiv.org/abs/2211.14262}{{\ttfamily
  2211.14262}}].

\bibitem{Tomita:1975kj}
K.~Tomita, \emph{{Evolution of Irregularities in a Chaotic Early Universe}},
  \href{https://doi.org/10.1143/PTP.54.730}{\emph{Prog. Theor. Phys.}
  {\bfseries 54} (1975) 730}.

\bibitem{Matarrese:1993zf}
S.~Matarrese, O.~Pantano and D.~Saez, \emph{{General relativistic dynamics of
  irrotational dust: Cosmological implications}},
  \href{https://doi.org/10.1103/PhysRevLett.72.320}{\emph{Phys. Rev. Lett.}
  {\bfseries 72} (1994) 320}
  [\href{https://arxiv.org/abs/astro-ph/9310036}{{\ttfamily
  astro-ph/9310036}}].

\bibitem{Matarrese:1997ay}
S.~Matarrese, S.~Mollerach and M.~Bruni, \emph{{Second order perturbations of
  the Einstein-de Sitter universe}},
  \href{https://doi.org/10.1103/PhysRevD.58.043504}{\emph{Phys. Rev. D}
  {\bfseries 58} (1998) 043504}
  [\href{https://arxiv.org/abs/astro-ph/9707278}{{\ttfamily
  astro-ph/9707278}}].

\bibitem{Acquaviva:2002ud}
V.~Acquaviva, N.~Bartolo, S.~Matarrese and A.~Riotto, \emph{{Second order
  cosmological perturbations from inflation}},
  \href{https://doi.org/10.1016/S0550-3213(03)00550-9}{\emph{Nucl. Phys. B}
  {\bfseries 667} (2003) 119}
  [\href{https://arxiv.org/abs/astro-ph/0209156}{{\ttfamily
  astro-ph/0209156}}].

\bibitem{Mollerach:2003nq}
S.~Mollerach, D.~Harari and S.~Matarrese, \emph{{CMB polarization from
  secondary vector and tensor modes}},
  \href{https://doi.org/10.1103/PhysRevD.69.063002}{\emph{Phys. Rev. D}
  {\bfseries 69} (2004) 063002}
  [\href{https://arxiv.org/abs/astro-ph/0310711}{{\ttfamily
  astro-ph/0310711}}].

\bibitem{Nakamura:2004rm}
K.~Nakamura, \emph{{Second-order gauge invariant cosmological perturbation
  theory: Einstein equations in terms of gauge invariant variables}},
  \href{https://doi.org/10.1143/PTP.117.17}{\emph{Prog. Theor. Phys.}
  {\bfseries 117} (2007) 17}
  [\href{https://arxiv.org/abs/gr-qc/0605108}{{\ttfamily gr-qc/0605108}}].

\bibitem{Ananda:2006af}
K.~N. Ananda, C.~Clarkson and D.~Wands, \emph{{The Cosmological gravitational
  wave background from primordial density perturbations}},
  \href{https://doi.org/10.1103/PhysRevD.75.123518}{\emph{Phys. Rev. D}
  {\bfseries 75} (2007) 123518}
  [\href{https://arxiv.org/abs/gr-qc/0612013}{{\ttfamily gr-qc/0612013}}].

\bibitem{Baumann:2007zm}
D.~Baumann, P.~J. Steinhardt, K.~Takahashi and K.~Ichiki, \emph{{Gravitational
  Wave Spectrum Induced by Primordial Scalar Perturbations}},
  \href{https://doi.org/10.1103/PhysRevD.76.084019}{\emph{Phys. Rev. D}
  {\bfseries 76} (2007) 084019}
  [\href{https://arxiv.org/abs/hep-th/0703290}{{\ttfamily hep-th/0703290}}].

\bibitem{Espinosa:2018eve}
J.~R. Espinosa, D.~Racco and A.~Riotto, \emph{{A Cosmological Signature of the
  SM Higgs Instability: Gravitational Waves}},
  \href{https://doi.org/10.1088/1475-7516/2018/09/012}{\emph{JCAP} {\bfseries
  09} (2018) 012} [\href{https://arxiv.org/abs/1804.07732}{{\ttfamily
  1804.07732}}].

\bibitem{Kohri:2018awv}
K.~Kohri and T.~Terada, \emph{{Semianalytic calculation of gravitational wave
  spectrum nonlinearly induced from primordial curvature perturbations}},
  \href{https://doi.org/10.1103/PhysRevD.97.123532}{\emph{Phys. Rev. D}
  {\bfseries 97} (2018) 123532}
  [\href{https://arxiv.org/abs/1804.08577}{{\ttfamily 1804.08577}}].

\bibitem{Inomata:2019yww}
K.~Inomata and T.~Terada, \emph{{Gauge Independence of Induced Gravitational
  Waves}}, \href{https://doi.org/10.1103/PhysRevD.101.023523}{\emph{Phys. Rev.
  D} {\bfseries 101} (2020) 023523}
  [\href{https://arxiv.org/abs/1912.00785}{{\ttfamily 1912.00785}}].

\bibitem{Domenech:2019quo}
G.~Dom\`enech, \emph{{Induced gravitational waves in a general cosmological
  background}}, \href{https://doi.org/10.1142/S0218271820500285}{\emph{Int. J.
  Mod. Phys. D} {\bfseries 29} (2020) 2050028}
  [\href{https://arxiv.org/abs/1912.05583}{{\ttfamily 1912.05583}}].

\bibitem{Saito:2008jc}
R.~Saito and J.~Yokoyama, \emph{{Gravitational wave background as a probe of
  the primordial black hole abundance}},
  \href{https://doi.org/10.1103/PhysRevLett.102.161101}{\emph{Phys. Rev. Lett.}
  {\bfseries 102} (2009) 161101}
  [\href{https://arxiv.org/abs/0812.4339}{{\ttfamily 0812.4339}}].

\bibitem{Cai:2018tuh}
Y.-F. Cai, X.~Tong, D.-G. Wang and S.-F. Yan, \emph{{Primordial Black Holes
  from Sound Speed Resonance during Inflation}},
  \href{https://doi.org/10.1103/PhysRevLett.121.081306}{\emph{Phys. Rev. Lett.}
  {\bfseries 121} (2018) 081306}
  [\href{https://arxiv.org/abs/1805.03639}{{\ttfamily 1805.03639}}].

\bibitem{Cai:2019jah}
Y.-F. Cai, C.~Chen, X.~Tong, D.-G. Wang and S.-F. Yan, \emph{{When Primordial
  Black Holes from Sound Speed Resonance Meet a Stochastic Background of
  Gravitational Waves}},
  \href{https://doi.org/10.1103/PhysRevD.100.043518}{\emph{Phys. Rev. D}
  {\bfseries 100} (2019) 043518}
  [\href{https://arxiv.org/abs/1902.08187}{{\ttfamily 1902.08187}}].

\bibitem{Chen:2019zza}
C.~Chen and Y.-F. Cai, \emph{{Primordial black holes from sound speed resonance
  in the inflaton-curvaton mixed scenario}},
  \href{https://doi.org/10.1088/1475-7516/2019/10/068}{\emph{JCAP} {\bfseries
  10} (2019) 068} [\href{https://arxiv.org/abs/1908.03942}{{\ttfamily
  1908.03942}}].

\bibitem{Chen:2020uhe}
C.~Chen, X.-H. Ma and Y.-F. Cai, \emph{{Dirac-Born-Infeld realization of sound
  speed resonance mechanism for primordial black holes}},
  \href{https://doi.org/10.1103/PhysRevD.102.063526}{\emph{Phys. Rev. D}
  {\bfseries 102} (2020) 063526}
  [\href{https://arxiv.org/abs/2003.03821}{{\ttfamily 2003.03821}}].

\bibitem{Addazi:2022ukh}
A.~Addazi, S.~Capozziello and Q.~Gan, \emph{{Induced gravitational waves from
  multi-sound speed resonances during cosmological inflation}},
  \href{https://doi.org/10.1088/1475-7516/2022/08/051}{\emph{JCAP} {\bfseries
  08} (2022) 051} [\href{https://arxiv.org/abs/2204.07668}{{\ttfamily
  2204.07668}}].

\bibitem{Karam:2022nym}
A.~Karam, N.~Koivunen, E.~Tomberg, V.~Vaskonen and H.~Veerm\"ae, \emph{{Anatomy
  of single-field inflationary models for primordial black holes}},
  \href{https://doi.org/10.1088/1475-7516/2023/03/013}{\emph{JCAP} {\bfseries
  03} (2023) 013} [\href{https://arxiv.org/abs/2205.13540}{{\ttfamily
  2205.13540}}].

\bibitem{Byrnes:2018txb}
C.~T. Byrnes, P.~S. Cole and S.~P. Patil, \emph{{Steepest growth of the power
  spectrum and primordial black holes}},
  \href{https://doi.org/10.1088/1475-7516/2019/06/028}{\emph{JCAP} {\bfseries
  06} (2019) 028} [\href{https://arxiv.org/abs/1811.11158}{{\ttfamily
  1811.11158}}].

\bibitem{Carrilho:2019oqg}
P.~Carrilho, K.~A. Malik and D.~J. Mulryne, \emph{{Dissecting the growth of the
  power spectrum for primordial black holes}},
  \href{https://doi.org/10.1103/PhysRevD.100.103529}{\emph{Phys. Rev. D}
  {\bfseries 100} (2019) 103529}
  [\href{https://arxiv.org/abs/1907.05237}{{\ttfamily 1907.05237}}].

\bibitem{Tasinato:2020vdk}
G.~Tasinato, \emph{{An analytic approach to non-slow-roll inflation}},
  \href{https://doi.org/10.1103/PhysRevD.103.023535}{\emph{Phys. Rev. D}
  {\bfseries 103} (2021) 023535}
  [\href{https://arxiv.org/abs/2012.02518}{{\ttfamily 2012.02518}}].

\bibitem{Davies:2021loj}
M.~W. Davies, P.~Carrilho and D.~J. Mulryne, \emph{{Non-Gaussianity in
  inflationary scenarios for primordial black holes}},
  \href{https://doi.org/10.1088/1475-7516/2022/06/019}{\emph{JCAP} {\bfseries
  06} (2022) 019} [\href{https://arxiv.org/abs/2110.08189}{{\ttfamily
  2110.08189}}].

\bibitem{Dimopoulos:2019wew}
K.~Dimopoulos, T.~Markkanen, A.~Racioppi and V.~Vaskonen, \emph{{Primordial
  Black Holes from Thermal Inflation}},
  \href{https://doi.org/10.1088/1475-7516/2019/07/046}{\emph{JCAP} {\bfseries
  07} (2019) 046} [\href{https://arxiv.org/abs/1903.09598}{{\ttfamily
  1903.09598}}].

\bibitem{Fumagalli:2021mpc}
J.~Fumagalli, G.~A. Palma, S.~Renaux-Petel, S.~Sypsas, L.~T. Witkowski and
  C.~Zenteno, \emph{{Primordial gravitational waves from excited states}},
  \href{https://doi.org/10.1007/JHEP03(2022)196}{\emph{JHEP} {\bfseries 03}
  (2022) 196} [\href{https://arxiv.org/abs/2111.14664}{{\ttfamily
  2111.14664}}].

\bibitem{Palma:2020ejf}
G.~A. Palma, S.~Sypsas and C.~Zenteno, \emph{{Seeding primordial black holes in
  multifield inflation}},
  \href{https://doi.org/10.1103/PhysRevLett.125.121301}{\emph{Phys. Rev. Lett.}
  {\bfseries 125} (2020) 121301}
  [\href{https://arxiv.org/abs/2004.06106}{{\ttfamily 2004.06106}}].

\bibitem{Inomata:2021uqj}
K.~Inomata, E.~McDonough and W.~Hu, \emph{{Primordial black holes arise when
  the inflaton falls}},
  \href{https://doi.org/10.1103/PhysRevD.104.123553}{\emph{Phys. Rev. D}
  {\bfseries 104} (2021) 123553}
  [\href{https://arxiv.org/abs/2104.03972}{{\ttfamily 2104.03972}}].

\bibitem{Inomata:2021tpx}
K.~Inomata, E.~McDonough and W.~Hu, \emph{{Amplification of primordial
  perturbations from the rise or fall of the inflaton}},
  \href{https://doi.org/10.1088/1475-7516/2022/02/031}{\emph{JCAP} {\bfseries
  02} (2022) 031} [\href{https://arxiv.org/abs/2110.14641}{{\ttfamily
  2110.14641}}].

\bibitem{Peng:2021zon}
Z.-Z. Peng, C.~Fu, J.~Liu, Z.-K. Guo and R.-G. Cai, \emph{{Gravitational waves
  from resonant amplification of curvature perturbations during inflation}},
  \href{https://doi.org/10.1088/1475-7516/2021/10/050}{\emph{JCAP} {\bfseries
  10} (2021) 050} [\href{https://arxiv.org/abs/2106.11816}{{\ttfamily
  2106.11816}}].

\bibitem{Inomata:2022yte}
K.~Inomata, M.~Braglia, X.~Chen and S.~Renaux-Petel, \emph{{Questions on
  calculation of primordial power spectrum with large spikes: the resonance
  model case}},
  \href{https://doi.org/10.1088/1475-7516/2023/04/011}{\emph{JCAP} {\bfseries
  04} (2023) 011} [\href{https://arxiv.org/abs/2211.02586}{{\ttfamily
  2211.02586}}].

\bibitem{Cai:2021wzd}
R.-G. Cai, C.~Chen and C.~Fu, \emph{{Primordial black holes and stochastic
  gravitational wave background from inflation with a noncanonical spectator
  field}}, \href{https://doi.org/10.1103/PhysRevD.104.083537}{\emph{Phys. Rev.
  D} {\bfseries 104} (2021) 083537}
  [\href{https://arxiv.org/abs/2108.03422}{{\ttfamily 2108.03422}}].

\bibitem{Inomata:2021zel}
K.~Inomata, \emph{{Bound on induced gravitational waves during inflation era}},
  \href{https://doi.org/10.1103/PhysRevD.104.123525}{\emph{Phys. Rev. D}
  {\bfseries 104} (2021) 123525}
  [\href{https://arxiv.org/abs/2109.06192}{{\ttfamily 2109.06192}}].

\bibitem{Fumagalli:2020nvq}
J.~Fumagalli, S.~Renaux-Petel and L.~T. Witkowski, \emph{{Oscillations in the
  stochastic gravitational wave background from sharp features and particle
  production during inflation}},
  \href{https://doi.org/10.1088/1475-7516/2021/08/030}{\emph{JCAP} {\bfseries
  08} (2021) 030} [\href{https://arxiv.org/abs/2012.02761}{{\ttfamily
  2012.02761}}].

\bibitem{Witkowski:2021raz}
L.~T. Witkowski, G.~Dom\`enech, J.~Fumagalli and S.~Renaux-Petel,
  \emph{{Expansion history-dependent oscillations in the scalar-induced
  gravitational wave background}},
  \href{https://doi.org/10.1088/1475-7516/2022/05/028}{\emph{JCAP} {\bfseries
  05} (2022) 028} [\href{https://arxiv.org/abs/2110.09480}{{\ttfamily
  2110.09480}}].

\bibitem{Slosar:2019gvt}
A.~Slosar et~al., \emph{{Scratches from the Past: Inflationary Archaeology
  through Features in the Power Spectrum of Primordial Fluctuations}},
  {\emph{Bull. Am. Astron. Soc.} {\bfseries 51} (2019) 98}
  [\href{https://arxiv.org/abs/1903.09883}{{\ttfamily 1903.09883}}].

\bibitem{Dalianis:2021iig}
I.~Dalianis, G.~P. Kodaxis, I.~D. Stamou, N.~Tetradis and A.~Tsigkas-Kouvelis,
  \emph{{Spectrum oscillations from features in the potential of single-field
  inflation}}, \href{https://doi.org/10.1103/PhysRevD.104.103510}{\emph{Phys.
  Rev. D} {\bfseries 104} (2021) 103510}
  [\href{https://arxiv.org/abs/2106.02467}{{\ttfamily 2106.02467}}].

\bibitem{Aragam:2023adu}
V.~Aragam, S.~Paban and R.~Rosati, \emph{{Primordial stochastic gravitational
  wave backgrounds from a sharp feature in three-field inflation. Part~I. The
  radiation era}},
  \href{https://doi.org/10.1088/1475-7516/2023/11/014}{\emph{JCAP} {\bfseries
  11} (2023) 014} [\href{https://arxiv.org/abs/2304.00065}{{\ttfamily
  2304.00065}}].

\bibitem{Fumagalli:2021cel}
J.~Fumagalli, S.~e. Renaux-Petel and L.~T. Witkowski, \emph{{Resonant features
  in the stochastic gravitational wave background}},
  \href{https://doi.org/10.1088/1475-7516/2021/08/059}{\emph{JCAP} {\bfseries
  08} (2021) 059} [\href{https://arxiv.org/abs/2105.06481}{{\ttfamily
  2105.06481}}].

\bibitem{Calcagni:2023vxg}
G.~Calcagni and S.~Kuroyanagi, \emph{{Log-periodic gravitational-wave
  background beyond Einstein gravity}},
  \href{https://doi.org/10.1088/1361-6382/ad1123}{\emph{Class. Quant. Grav.}
  {\bfseries 41} (2023) 015031}
  [\href{https://arxiv.org/abs/2308.05904}{{\ttfamily 2308.05904}}].

\bibitem{Fumagalli:2021dtd}
J.~Fumagalli, M.~Pieroni, S.~Renaux-Petel and L.~T. Witkowski, \emph{{Detecting
  primordial features with LISA}},
  \href{https://doi.org/10.1088/1475-7516/2022/07/020}{\emph{JCAP} {\bfseries
  07} (2022) 020} [\href{https://arxiv.org/abs/2112.06903}{{\ttfamily
  2112.06903}}].

\bibitem{Chen:2008wn}
X.~Chen, R.~Easther and E.~A. Lim, \emph{{Generation and Characterization of
  Large Non-Gaussianities in Single Field Inflation}},
  \href{https://doi.org/10.1088/1475-7516/2008/04/010}{\emph{JCAP} {\bfseries
  04} (2008) 010} [\href{https://arxiv.org/abs/0801.3295}{{\ttfamily
  0801.3295}}].

\bibitem{Silverstein:2008sg}
E.~Silverstein and A.~Westphal, \emph{{Monodromy in the CMB: Gravity Waves and
  String Inflation}},
  \href{https://doi.org/10.1103/PhysRevD.78.106003}{\emph{Phys. Rev. D}
  {\bfseries 78} (2008) 106003}
  [\href{https://arxiv.org/abs/0803.3085}{{\ttfamily 0803.3085}}].

\bibitem{Flauger:2009ab}
R.~Flauger, L.~McAllister, E.~Pajer, A.~Westphal and G.~Xu, \emph{{Oscillations
  in the CMB from Axion Monodromy Inflation}},
  \href{https://doi.org/10.1088/1475-7516/2010/06/009}{\emph{JCAP} {\bfseries
  06} (2010) 009} [\href{https://arxiv.org/abs/0907.2916}{{\ttfamily
  0907.2916}}].

\bibitem{Parameswaran:2016qqq}
S.~Parameswaran, G.~Tasinato and I.~Zavala, \emph{{Subleading Effects and the
  Field Range in Axion Inflation}},
  \href{https://doi.org/10.1088/1475-7516/2016/04/008}{\emph{JCAP} {\bfseries
  04} (2016) 008} [\href{https://arxiv.org/abs/1602.02812}{{\ttfamily
  1602.02812}}].

\bibitem{Ozsoy:2018flq}
O.~\"Ozsoy, S.~Parameswaran, G.~Tasinato and I.~Zavala, \emph{{Mechanisms for
  Primordial Black Hole Production in String Theory}},
  \href{https://doi.org/10.1088/1475-7516/2018/07/005}{\emph{JCAP} {\bfseries
  07} (2018) 005} [\href{https://arxiv.org/abs/1803.07626}{{\ttfamily
  1803.07626}}].

\bibitem{Bhattacharya:2022fze}
S.~Bhattacharya and I.~Zavala, \emph{{Sharp turns in axion monodromy:
  primordial black holes and gravitational waves}},
  \href{https://doi.org/10.1088/1475-7516/2023/04/065}{\emph{JCAP} {\bfseries
  04} (2023) 065} [\href{https://arxiv.org/abs/2205.06065}{{\ttfamily
  2205.06065}}].

\bibitem{Mavromatos:2022yql}
N.~E. Mavromatos, V.~C. Spanos and I.~D. Stamou, \emph{{Primordial black holes
  and gravitational waves in multiaxion-Chern-Simons inflation}},
  \href{https://doi.org/10.1103/PhysRevD.106.063532}{\emph{Phys. Rev. D}
  {\bfseries 106} (2022) 063532}
  [\href{https://arxiv.org/abs/2206.07963}{{\ttfamily 2206.07963}}].

\bibitem{Boutivas:2022qtl}
K.~Boutivas, I.~Dalianis, G.~P. Kodaxis and N.~Tetradis, \emph{{The effect of
  multiple features on the power spectrum in two-field inflation}},
  \href{https://doi.org/10.1088/1475-7516/2022/08/021}{\emph{JCAP} {\bfseries
  08} (2022) 021} [\href{https://arxiv.org/abs/2203.15605}{{\ttfamily
  2203.15605}}].

\bibitem{Seto:2003kc}
N.~Seto and J.~Yokoyama, \emph{{Probing the equation of state of the early
  universe with a space laser interferometer}},
  \href{https://doi.org/10.1143/JPSJ.72.3082}{\emph{J. Phys. Soc. Jap.}
  {\bfseries 72} (2003) 3082}
  [\href{https://arxiv.org/abs/gr-qc/0305096}{{\ttfamily gr-qc/0305096}}].

\bibitem{Boyle:2005se}
L.~A. Boyle and P.~J. Steinhardt, \emph{{Probing the early universe with
  inflationary gravitational waves}},
  \href{https://doi.org/10.1103/PhysRevD.77.063504}{\emph{Phys. Rev. D}
  {\bfseries 77} (2008) 063504}
  [\href{https://arxiv.org/abs/astro-ph/0512014}{{\ttfamily
  astro-ph/0512014}}].

\bibitem{Boyle:2007zx}
L.~A. Boyle and A.~Buonanno, \emph{{Relating gravitational wave constraints
  from primordial nucleosynthesis, pulsar timing, laser interferometers, and
  the CMB: Implications for the early Universe}},
  \href{https://doi.org/10.1103/PhysRevD.78.043531}{\emph{Phys. Rev. D}
  {\bfseries 78} (2008) 043531}
  [\href{https://arxiv.org/abs/0708.2279}{{\ttfamily 0708.2279}}].

\bibitem{Nakayama:2008ip}
K.~Nakayama, S.~Saito, Y.~Suwa and J.~Yokoyama, \emph{{Space laser
  interferometers can determine the thermal history of the early Universe}},
  \href{https://doi.org/10.1103/PhysRevD.77.124001}{\emph{Phys. Rev. D}
  {\bfseries 77} (2008) 124001}
  [\href{https://arxiv.org/abs/0802.2452}{{\ttfamily 0802.2452}}].

\bibitem{Kuroyanagi:2011fy}
S.~Kuroyanagi, K.~Nakayama and S.~Saito, \emph{{Prospects for determination of
  thermal history after inflation with future gravitational wave detectors}},
  \href{https://doi.org/10.1103/PhysRevD.84.123513}{\emph{Phys. Rev. D}
  {\bfseries 84} (2011) 123513}
  [\href{https://arxiv.org/abs/1110.4169}{{\ttfamily 1110.4169}}].

\bibitem{Figueroa:2019paj}
D.~G. Figueroa and E.~H. Tanin, \emph{{Ability of LIGO and LISA to probe the
  equation of state of the early Universe}},
  \href{https://doi.org/10.1088/1475-7516/2019/08/011}{\emph{JCAP} {\bfseries
  08} (2019) 011} [\href{https://arxiv.org/abs/1905.11960}{{\ttfamily
  1905.11960}}].

\bibitem{ValbusaDallArmi:2020ifo}
L.~Valbusa~Dall'Armi, A.~Ricciardone, N.~Bartolo, D.~Bertacca and S.~Matarrese,
  \emph{{Imprint of relativistic particles on the anisotropies of the
  stochastic gravitational-wave background}},
  \href{https://doi.org/10.1103/PhysRevD.103.023522}{\emph{Phys. Rev. D}
  {\bfseries 103} (2021) 023522}
  [\href{https://arxiv.org/abs/2007.01215}{{\ttfamily 2007.01215}}].

\bibitem{Domenech:2020kqm}
G.~Dom\`enech, S.~Pi and M.~Sasaki, \emph{{Induced gravitational waves as a
  probe of thermal history of the universe}},
  \href{https://doi.org/10.1088/1475-7516/2020/08/017}{\emph{JCAP} {\bfseries
  08} (2020) 017} [\href{https://arxiv.org/abs/2005.12314}{{\ttfamily
  2005.12314}}].

\bibitem{Kuroyanagi:2013ns}
S.~Kuroyanagi, C.~Ringeval and T.~Takahashi, \emph{{Early universe tomography
  with CMB and gravitational waves}},
  \href{https://doi.org/10.1103/PhysRevD.87.083502}{\emph{Phys. Rev. D}
  {\bfseries 87} (2013) 083502}
  [\href{https://arxiv.org/abs/1301.1778}{{\ttfamily 1301.1778}}].

\bibitem{DEramo:2019tit}
F.~D'Eramo and K.~Schmitz, \emph{{Imprint of a scalar era on the primordial
  spectrum of gravitational waves}},
  \href{https://doi.org/10.1103/PhysRevResearch.1.013010}{\emph{Phys. Rev.
  Research.} {\bfseries 1} (2019) 013010}
  [\href{https://arxiv.org/abs/1904.07870}{{\ttfamily 1904.07870}}].

\bibitem{Kuroyanagi:2020sfw}
S.~Kuroyanagi, T.~Takahashi and S.~Yokoyama, \emph{{Blue-tilted inflationary
  tensor spectrum and reheating in the light of NANOGrav results}},
  \href{https://doi.org/10.1088/1475-7516/2021/01/071}{\emph{JCAP} {\bfseries
  01} (2021) 071} [\href{https://arxiv.org/abs/2011.03323}{{\ttfamily
  2011.03323}}].

\bibitem{Nakayama:2009ce}
K.~Nakayama and J.~Yokoyama, \emph{{Gravitational Wave Background and
  Non-Gaussianity as a Probe of the Curvaton Scenario}},
  \href{https://doi.org/10.1088/1475-7516/2010/01/010}{\emph{JCAP} {\bfseries
  01} (2010) 010} [\href{https://arxiv.org/abs/0910.0715}{{\ttfamily
  0910.0715}}].

\bibitem{Durrer:2011bi}
R.~Durrer and J.~Hasenkamp, \emph{{Testing Superstring Theories with
  Gravitational Waves}},
  \href{https://doi.org/10.1103/PhysRevD.84.064027}{\emph{Phys. Rev. D}
  {\bfseries 84} (2011) 064027}
  [\href{https://arxiv.org/abs/1105.5283}{{\ttfamily 1105.5283}}].

\bibitem{Kuroyanagi:2014nba}
S.~Kuroyanagi, T.~Takahashi and S.~Yokoyama, \emph{{Blue-tilted Tensor Spectrum
  and Thermal History of the Universe}},
  \href{https://doi.org/10.1088/1475-7516/2015/02/003}{\emph{JCAP} {\bfseries
  02} (2015) 003} [\href{https://arxiv.org/abs/1407.4785}{{\ttfamily
  1407.4785}}].

\bibitem{Giovannini:1998bp}
M.~Giovannini, \emph{{Gravitational waves constraints on postinflationary
  phases stiffer than radiation}},
  \href{https://doi.org/10.1103/PhysRevD.58.083504}{\emph{Phys. Rev. D}
  {\bfseries 58} (1998) 083504}
  [\href{https://arxiv.org/abs/hep-ph/9806329}{{\ttfamily hep-ph/9806329}}].

\bibitem{Peebles:1998qn}
P.~J.~E. Peebles and A.~Vilenkin, \emph{{Quintessential inflation}},
  \href{https://doi.org/10.1103/PhysRevD.59.063505}{\emph{Phys. Rev. D}
  {\bfseries 59} (1999) 063505}
  [\href{https://arxiv.org/abs/astro-ph/9810509}{{\ttfamily
  astro-ph/9810509}}].

\bibitem{Giovannini:1999bh}
M.~Giovannini, \emph{{Production and detection of relic gravitons in
  quintessential inflationary models}},
  \href{https://doi.org/10.1103/PhysRevD.60.123511}{\emph{Phys. Rev. D}
  {\bfseries 60} (1999) 123511}
  [\href{https://arxiv.org/abs/astro-ph/9903004}{{\ttfamily
  astro-ph/9903004}}].

\bibitem{Giovannini:1999qj}
M.~Giovannini, \emph{{Spikes in the relic graviton background from
  quintessential inflation}},
  \href{https://doi.org/10.1088/0264-9381/16/9/308}{\emph{Class. Quant. Grav.}
  {\bfseries 16} (1999) 2905}
  [\href{https://arxiv.org/abs/hep-ph/9903263}{{\ttfamily hep-ph/9903263}}].

\bibitem{Tashiro:2003qp}
H.~Tashiro, T.~Chiba and M.~Sasaki, \emph{{Reheating after quintessential
  inflation and gravitational waves}},
  \href{https://doi.org/10.1088/0264-9381/21/7/004}{\emph{Class. Quant. Grav.}
  {\bfseries 21} (2004) 1761}
  [\href{https://arxiv.org/abs/gr-qc/0307068}{{\ttfamily gr-qc/0307068}}].

\bibitem{Giovannini:2008tm}
M.~Giovannini, \emph{{Thermal history of the plasma and high-frequency
  gravitons}},
  \href{https://doi.org/10.1088/0264-9381/26/4/045004}{\emph{Class. Quant.
  Grav.} {\bfseries 26} (2009) 045004}
  [\href{https://arxiv.org/abs/0807.4317}{{\ttfamily 0807.4317}}].

\bibitem{Ahmad:2019jbm}
S.~Ahmad, A.~De~Felice, N.~Jaman, S.~Kuroyanagi and M.~Sami,
  \emph{{Baryogenesis in the paradigm of quintessential inflation}},
  \href{https://doi.org/10.1103/PhysRevD.100.103525}{\emph{Phys. Rev. D}
  {\bfseries 100} (2019) 103525}
  [\href{https://arxiv.org/abs/1908.03742}{{\ttfamily 1908.03742}}].

\bibitem{Gouttenoire:2021jhk}
Y.~Gouttenoire, G.~Servant and P.~Simakachorn, \emph{{Kination cosmology from
  scalar fields and gravitational-wave signatures}},
  \href{https://arxiv.org/abs/2111.01150}{{\ttfamily 2111.01150}}.

\bibitem{Duval:2024jsg}
H.~Duval, S.~Kuroyanagi, A.~Mariotti, A.~Romero-Rodr\'\i{}guez and
  M.~Sakellariadou, \emph{{Investigating cosmic histories with a stiff era
  through Gravitational Waves}},
  \href{https://arxiv.org/abs/2405.10201}{{\ttfamily 2405.10201}}.

\bibitem{Caldwell:2018giq}
R.~R. Caldwell, T.~L. Smith and D.~G.~E. Walker, \emph{{Using a Primordial
  Gravitational Wave Background to Illuminate New Physics}},
  \href{https://doi.org/10.1103/PhysRevD.100.043513}{\emph{Phys. Rev. D}
  {\bfseries 100} (2019) 043513}
  [\href{https://arxiv.org/abs/1812.07577}{{\ttfamily 1812.07577}}].

\bibitem{Escriva:2024ivo}
A.~Escriv\`a, R.~Inui, Y.~Tada and C.-M. Yoo, \emph{{The LISA forecast on a
  smooth crossover beyond the Standard Model through the scalar-induced
  gravitational waves}},  \href{https://arxiv.org/abs/2404.12591}{{\ttfamily
  2404.12591}}.

\bibitem{Bartolo:2019oiq}
N.~Bartolo, D.~Bertacca, S.~Matarrese, M.~Peloso, A.~Ricciardone, A.~Riotto
  et~al., \emph{{Anisotropies and non-Gaussianity of the Cosmological
  Gravitational Wave Background}},
  \href{https://doi.org/10.1103/PhysRevD.100.121501}{\emph{Phys. Rev. D}
  {\bfseries 100} (2019) 121501}
  [\href{https://arxiv.org/abs/1908.00527}{{\ttfamily 1908.00527}}].

\bibitem{Bartolo:2019yeu}
N.~Bartolo, D.~Bertacca, S.~Matarrese, M.~Peloso, A.~Ricciardone, A.~Riotto
  et~al., \emph{{Characterizing the cosmological gravitational wave background:
  Anisotropies and non-Gaussianity}},
  \href{https://doi.org/10.1103/PhysRevD.102.023527}{\emph{Phys. Rev. D}
  {\bfseries 102} (2020) 023527}
  [\href{https://arxiv.org/abs/1912.09433}{{\ttfamily 1912.09433}}].

\bibitem{Contaldi:2020rht}
C.~R. Contaldi, M.~Pieroni, A.~I. Renzini, G.~Cusin, N.~Karnesis, M.~Peloso
  et~al., \emph{{Maximum likelihood map-making with the Laser Interferometer
  Space Antenna}},
  \href{https://doi.org/10.1103/PhysRevD.102.043502}{\emph{Phys. Rev. D}
  {\bfseries 102} (2020) 043502}
  [\href{https://arxiv.org/abs/2006.03313}{{\ttfamily 2006.03313}}].

\bibitem{Dimastrogiovanni:2022eir}
E.~Dimastrogiovanni, M.~Fasiello, A.~Malhotra and G.~Tasinato, \emph{{Enhancing
  gravitational wave anisotropies with peaked scalar sources}},
  \href{https://doi.org/10.1088/1475-7516/2023/01/018}{\emph{JCAP} {\bfseries
  01} (2023) 018} [\href{https://arxiv.org/abs/2205.05644}{{\ttfamily
  2205.05644}}].

\bibitem{Malhotra:2022ply}
A.~Malhotra, E.~Dimastrogiovanni, G.~Dom\`enech, M.~Fasiello and G.~Tasinato,
  \emph{{New universal property of cosmological gravitational wave
  anisotropies}},
  \href{https://doi.org/10.1103/PhysRevD.107.103502}{\emph{Phys. Rev. D}
  {\bfseries 107} (2023) 103502}
  [\href{https://arxiv.org/abs/2212.10316}{{\ttfamily 2212.10316}}].

\bibitem{Schulze:2023ich}
F.~Schulze, L.~Valbusa~Dall'Armi, J.~Lesgourgues, A.~Ricciardone, N.~Bartolo,
  D.~Bertacca et~al., \emph{{GW\_CLASS: Cosmological Gravitational Wave
  Background in the cosmic linear anisotropy solving system}},
  \href{https://doi.org/10.1088/1475-7516/2023/10/025}{\emph{JCAP} {\bfseries
  10} (2023) 025} [\href{https://arxiv.org/abs/2305.01602}{{\ttfamily
  2305.01602}}].

\bibitem{Mentasti:2023icu}
G.~Mentasti, C.~R. Contaldi and M.~Peloso, \emph{{Intrinsic Limits on the
  Detection of the Anisotropies of the Stochastic Gravitational Wave
  Background}},
  \href{https://doi.org/10.1103/PhysRevLett.131.221403}{\emph{Phys. Rev. Lett.}
  {\bfseries 131} (2023) 221403}
  [\href{https://arxiv.org/abs/2301.08074}{{\ttfamily 2301.08074}}].

\bibitem{Mentasti:2023uyi}
G.~Mentasti, C.~R. Contaldi and M.~Peloso, \emph{{Probing the galactic and
  extragalactic gravitational wave backgrounds with space-based
  interferometers}},
  \href{https://doi.org/10.1088/1475-7516/2024/06/055}{\emph{JCAP} {\bfseries
  06} (2024) 055} [\href{https://arxiv.org/abs/2312.10792}{{\ttfamily
  2312.10792}}].

\bibitem{Perna:2024ehx}
G.~Perna, C.~Testini, A.~Ricciardone and S.~Matarrese, \emph{{Fully
  non-Gaussian Scalar-Induced Gravitational Waves}},
  \href{https://doi.org/10.1088/1475-7516/2024/05/086}{\emph{JCAP} {\bfseries
  05} (2024) 086} [\href{https://arxiv.org/abs/2403.06962}{{\ttfamily
  2403.06962}}].

\bibitem{Adams:2013qma}
M.~R. Adams and N.~J. Cornish, \emph{{Detecting a Stochastic Gravitational Wave
  Background in the presence of a Galactic Foreground and Instrument Noise}},
  \href{https://doi.org/10.1103/PhysRevD.89.022001}{\emph{Phys. Rev. D}
  {\bfseries 89} (2014) 022001}
  [\href{https://arxiv.org/abs/1307.4116}{{\ttfamily 1307.4116}}].

\bibitem{Seto:2007tn}
N.~Seto and A.~Taruya, \emph{{Measuring a Parity Violation Signature in the
  Early Universe via Ground-based Laser Interferometers}},
  \href{https://doi.org/10.1103/PhysRevLett.99.121101}{\emph{Phys. Rev. Lett.}
  {\bfseries 99} (2007) 121101}
  [\href{https://arxiv.org/abs/0707.0535}{{\ttfamily 0707.0535}}].

\bibitem{Seto:2008sr}
N.~Seto and A.~Taruya, \emph{{Polarization analysis of gravitational-wave
  backgrounds from the correlation signals of ground-based interferometers:
  Measuring a circular-polarization mode}},
  \href{https://doi.org/10.1103/PhysRevD.77.103001}{\emph{Phys. Rev. D}
  {\bfseries 77} (2008) 103001}
  [\href{https://arxiv.org/abs/0801.4185}{{\ttfamily 0801.4185}}].

\bibitem{Smith:2016jqs}
T.~L. Smith and R.~Caldwell, \emph{{Sensitivity to a Frequency-Dependent
  Circular Polarization in an Isotropic Stochastic Gravitational Wave
  Background}}, \href{https://doi.org/10.1103/PhysRevD.95.044036}{\emph{Phys.
  Rev. D} {\bfseries 95} (2017) 044036}
  [\href{https://arxiv.org/abs/1609.05901}{{\ttfamily 1609.05901}}].

\bibitem{Crowder:2012ik}
S.~G. Crowder, R.~Namba, V.~Mandic, S.~Mukohyama and M.~Peloso,
  \emph{{Measurement of Parity Violation in the Early Universe using
  Gravitational-wave Detectors}},
  \href{https://doi.org/10.1016/j.physletb.2013.08.077}{\emph{Phys. Lett. B}
  {\bfseries 726} (2013) 66} [\href{https://arxiv.org/abs/1212.4165}{{\ttfamily
  1212.4165}}].

\bibitem{Orlando:2020oko}
G.~Orlando, M.~Pieroni and A.~Ricciardone, \emph{{Measuring Parity Violation in
  the Stochastic Gravitational Wave Background with the LISA-Taiji network}},
  \href{https://doi.org/10.1088/1475-7516/2021/03/069}{\emph{JCAP} {\bfseries
  03} (2021) 069} [\href{https://arxiv.org/abs/2011.07059}{{\ttfamily
  2011.07059}}].

\bibitem{ValbusaDallArmi:2023ydl}
L.~Valbusa~Dall'Armi, A.~Nishizawa, A.~Ricciardone and S.~Matarrese,
  \emph{{Circular Polarization of the Astrophysical Gravitational Wave
  Background}},
  \href{https://doi.org/10.1103/PhysRevLett.131.041401}{\emph{Phys. Rev. Lett.}
  {\bfseries 131} (2023) 041401}
  [\href{https://arxiv.org/abs/2301.08205}{{\ttfamily 2301.08205}}].

\bibitem{Seto:2006hf}
N.~Seto, \emph{{Prospects for direct detection of circular polarization of
  gravitational-wave background}},
  \href{https://doi.org/10.1103/PhysRevLett.97.151101}{\emph{Phys. Rev. Lett.}
  {\bfseries 97} (2006) 151101}
  [\href{https://arxiv.org/abs/astro-ph/0609504}{{\ttfamily
  astro-ph/0609504}}].

\bibitem{Seto:2006dz}
N.~Seto, \emph{{Quest for circular polarization of gravitational wave
  background and orbits of laser interferometers in space}},
  \href{https://doi.org/10.1103/PhysRevD.75.061302}{\emph{Phys. Rev. D}
  {\bfseries 75} (2007) 061302}
  [\href{https://arxiv.org/abs/astro-ph/0609633}{{\ttfamily
  astro-ph/0609633}}].

\bibitem{Robson:2018jly}
T.~Robson and N.~J. Cornish, \emph{{Detecting Gravitational Wave Bursts with
  LISA in the presence of Instrumental Glitches}},
  \href{https://doi.org/10.1103/PhysRevD.99.024019}{\emph{Phys. Rev. D}
  {\bfseries 99} (2019) 024019}
  [\href{https://arxiv.org/abs/1811.04490}{{\ttfamily 1811.04490}}].

\bibitem{Baghi:2021tfd}
Q.~Baghi, N.~Korsakova, J.~Slutsky, E.~Castelli, N.~Karnesis and J.-B. Bayle,
  \emph{{Detection and characterization of instrumental transients in LISA
  Pathfinder and their projection to LISA}},
  \href{https://doi.org/10.1103/PhysRevD.105.042002}{\emph{Phys. Rev. D}
  {\bfseries 105} (2022) 042002}
  [\href{https://arxiv.org/abs/2112.07490}{{\ttfamily 2112.07490}}].

\bibitem{Hartwig:2021mzw}
O.~Hartwig and M.~Muratore, \emph{{Characterization of time delay
  interferometry combinations for the LISA instrument noise}},
  \href{https://doi.org/10.1103/PhysRevD.105.062006}{\emph{Phys. Rev. D}
  {\bfseries 105} (2022) 062006}
  [\href{https://arxiv.org/abs/2111.00975}{{\ttfamily 2111.00975}}].

\bibitem{Nam:2022rqg}
D.~Q. Nam, Y.~Lemiere, A.~Petiteau, J.-B. Bayle, O.~Hartwig, J.~Martino et~al.,
  \emph{{TDI noises transfer functions for LISA}},
  \href{https://arxiv.org/abs/2211.02539}{{\ttfamily 2211.02539}}.

\bibitem{LISA_performance}
{LISA Consortium}, \emph{Lisa performance model and error budget,
  lisa-lcst-inst-tn-003},  tech. rep., 2020.

\bibitem{Tinto:2020fcc}
M.~Tinto and S.~V. Dhurandhar, \emph{{Time-delay interferometry}},
  \href{https://doi.org/10.1007/s41114-020-00029-6}{\emph{Living Rev. Rel.}
  {\bfseries 24} (2021) 1}.

\bibitem{Armstrong_1999}
J.~W. Armstrong, F.~B. Estabrook and M.~Tinto, \emph{Time-delay interferometry
  for space-based gravitational wave searches},
  \href{https://doi.org/10.1086/308110}{\emph{The Astrophysical Journal}
  {\bfseries 527} (1999) 814}.

\bibitem{Prince:2002hp}
T.~A. Prince, M.~Tinto, S.~L. Larson and J.~W. Armstrong, \emph{{The LISA
  optimal sensitivity}},
  \href{https://doi.org/10.1103/PhysRevD.66.122002}{\emph{Phys. Rev. D}
  {\bfseries 66} (2002) 122002}
  [\href{https://arxiv.org/abs/gr-qc/0209039}{{\ttfamily gr-qc/0209039}}].

\bibitem{Shaddock:2003bc}
D.~A. Shaddock, \emph{{Operating LISA as a Sagnac interferometer}},
  \href{https://doi.org/10.1103/PhysRevD.69.022001}{\emph{Phys. Rev. D}
  {\bfseries 69} (2004) 022001}
  [\href{https://arxiv.org/abs/gr-qc/0306125}{{\ttfamily gr-qc/0306125}}].

\bibitem{Shaddock:2003dj}
D.~A. Shaddock, M.~Tinto, F.~B. Estabrook and J.~W. Armstrong, \emph{{Data
  combinations accounting for LISA spacecraft motion}},
  \href{https://doi.org/10.1103/PhysRevD.68.061303}{\emph{Phys. Rev. D}
  {\bfseries 68} (2003) 061303}
  [\href{https://arxiv.org/abs/gr-qc/0307080}{{\ttfamily gr-qc/0307080}}].

\bibitem{Tinto:2003vj}
M.~Tinto, F.~B. Estabrook and J.~W. Armstrong, \emph{{Time delay interferometry
  with moving spacecraft arrays}},
  \href{https://doi.org/10.1103/PhysRevD.69.082001}{\emph{Phys. Rev. D}
  {\bfseries 69} (2004) 082001}
  [\href{https://arxiv.org/abs/gr-qc/0310017}{{\ttfamily gr-qc/0310017}}].

\bibitem{Vallisneri:2005ji}
M.~Vallisneri, \emph{{Geometric time delay interferometry}},
  \href{https://doi.org/10.1103/PhysRevD.76.109903}{\emph{Phys. Rev. D}
  {\bfseries 72} (2005) 042003}
  [\href{https://arxiv.org/abs/gr-qc/0504145}{{\ttfamily gr-qc/0504145}}].

\bibitem{Muratore:2020mdf}
M.~Muratore, D.~Vetrugno and S.~Vitale, \emph{{Revisitation of time delay
  interferometry combinations that suppress laser noise in LISA}},
  \href{https://doi.org/10.1088/1361-6382/ab9d5b}{\emph{Class. Quant. Grav.}
  {\bfseries 37} (2020) 185019}
  [\href{https://arxiv.org/abs/2001.11221}{{\ttfamily 2001.11221}}].

\bibitem{Muratore:2021uqj}
M.~Muratore, D.~Vetrugno, S.~Vitale and O.~Hartwig, \emph{{Time delay
  interferometry combinations as instrument noise monitors for LISA}},
  \href{https://doi.org/10.1103/PhysRevD.105.023009}{\emph{Phys. Rev. D}
  {\bfseries 105} (2022) 023009}
  [\href{https://arxiv.org/abs/2108.02738}{{\ttfamily 2108.02738}}].

\bibitem{Nissanke:2012eh}
S.~Nissanke, M.~Vallisneri, G.~Nelemans and T.~A. Prince,
  \emph{{Gravitational-wave emission from compact Galactic binaries}},
  \href{https://doi.org/10.1088/0004-637X/758/2/131}{\emph{Astrophys. J.}
  {\bfseries 758} (2012) 131}
  [\href{https://arxiv.org/abs/1201.4613}{{\ttfamily 1201.4613}}].

\bibitem{Regimbau:2011rp}
T.~Regimbau, \emph{{The astrophysical gravitational wave stochastic
  background}}, \href{https://doi.org/10.1088/1674-4527/11/4/001}{\emph{Res.
  Astron. Astrophys.} {\bfseries 11} (2011) 369}
  [\href{https://arxiv.org/abs/1101.2762}{{\ttfamily 1101.2762}}].

\bibitem{KAGRA:2021kbb}
{\scshape KAGRA, Virgo, LIGO Scientific} collaboration, \emph{{Upper limits on
  the isotropic gravitational-wave background from Advanced LIGO and Advanced
  Virgo\textquoteright{}s third observing run}},
  \href{https://doi.org/10.1103/PhysRevD.104.022004}{\emph{Phys. Rev. D}
  {\bfseries 104} (2021) 022004}
  [\href{https://arxiv.org/abs/2101.12130}{{\ttfamily 2101.12130}}].

\bibitem{Perigois:2020ymr}
C.~P\'erigois, C.~Belczynski, T.~Bulik and T.~Regimbau, \emph{{StarTrack
  predictions of the stochastic gravitational-wave background from compact
  binary mergers}},
  \href{https://doi.org/10.1103/PhysRevD.103.043002}{\emph{Phys. Rev. D}
  {\bfseries 103} (2021) 043002}
  [\href{https://arxiv.org/abs/2008.04890}{{\ttfamily 2008.04890}}].

\bibitem{Babak:2023lro}
S.~Babak, C.~Caprini, D.~G. Figueroa, N.~Karnesis, P.~Marcoccia, G.~Nardini
  et~al., \emph{{Stochastic gravitational wave background from stellar origin
  binary black holes in LISA}},
  \href{https://doi.org/10.1088/1475-7516/2023/08/034}{\emph{JCAP} {\bfseries
  08} (2023) 034} [\href{https://arxiv.org/abs/2304.06368}{{\ttfamily
  2304.06368}}].

\bibitem{Lehoucq:2023zlt}
L.~Lehoucq, I.~Dvorkin, R.~Srinivasan, C.~Pellouin and A.~Lamberts,
  \emph{{Astrophysical uncertainties in the gravitational-wave background from
  stellar-mass compact binary mergers}},
  \href{https://doi.org/10.1093/mnras/stad2917}{\emph{Mon. Not. Roy. Astron.
  Soc.} {\bfseries 526} (2023) 4378}
  [\href{https://arxiv.org/abs/2306.09861}{{\ttfamily 2306.09861}}].

\bibitem{Karnesis:2021tsh}
N.~Karnesis, S.~Babak, M.~Pieroni, N.~Cornish and T.~Littenberg,
  \emph{{Characterization of the stochastic signal originating from compact
  binary populations as measured by LISA}},
  \href{https://doi.org/10.1103/PhysRevD.104.043019}{\emph{Phys. Rev. D}
  {\bfseries 104} (2021) 043019}
  [\href{https://arxiv.org/abs/2103.14598}{{\ttfamily 2103.14598}}].

\bibitem{Phinney:2001di}
E.~S. Phinney, \emph{{A Practical theorem on gravitational wave backgrounds}},
  \href{https://arxiv.org/abs/astro-ph/0108028}{{\ttfamily astro-ph/0108028}}.

\bibitem{KAGRA:2021duu}
{\scshape KAGRA, VIRGO, LIGO Scientific} collaboration, \emph{{Population of
  Merging Compact Binaries Inferred Using Gravitational Waves through GWTC-3}},
  \href{https://doi.org/10.1103/PhysRevX.13.011048}{\emph{Phys. Rev. X}
  {\bfseries 13} (2023) 011048}
  [\href{https://arxiv.org/abs/2111.03634}{{\ttfamily 2111.03634}}].

\bibitem{Bond:1998qg}
J.~R. Bond, A.~H. Jaffe and L.~E. Knox, \emph{{Radical compression of cosmic
  microwave background data}},
  \href{https://doi.org/10.1086/308625}{\emph{Astrophys. J.} {\bfseries 533}
  (2000) 19} [\href{https://arxiv.org/abs/astro-ph/9808264}{{\ttfamily
  astro-ph/9808264}}].

\bibitem{Sievers:2002tq}
J.~L. Sievers et~al., \emph{{Cosmological parameters from Cosmic Background
  Imager observations and comparisons with BOOMERANG, DASI, and MAXIMA}},
  \href{https://doi.org/10.1086/375510}{\emph{Astrophys. J.} {\bfseries 591}
  (2003) 599} [\href{https://arxiv.org/abs/astro-ph/0205387}{{\ttfamily
  astro-ph/0205387}}].

\bibitem{WMAP:2003pyh}
{\scshape WMAP} collaboration, \emph{{First year Wilkinson Microwave Anisotropy
  Probe (WMAP) observations: Parameter estimation methodology}},
  \href{https://doi.org/10.1086/377335}{\emph{Astrophys. J. Suppl.} {\bfseries
  148} (2003) 195} [\href{https://arxiv.org/abs/astro-ph/0302218}{{\ttfamily
  astro-ph/0302218}}].

\bibitem{Hamimeche:2008ai}
S.~Hamimeche and A.~Lewis, \emph{{Likelihood Analysis of CMB Temperature and
  Polarization Power Spectra}},
  \href{https://doi.org/10.1103/PhysRevD.77.103013}{\emph{Phys. Rev. D}
  {\bfseries 77} (2008) 103013}
  [\href{https://arxiv.org/abs/0801.0554}{{\ttfamily 0801.0554}}].

\bibitem{Handley:2015vkr}
W.~J. Handley, M.~P. Hobson and A.~N. Lasenby, \emph{{polychord:
  next-generation nested sampling}},
  \href{https://doi.org/10.1093/mnras/stv1911}{\emph{Mon. Not. Roy. Astron.
  Soc.} {\bfseries 453} (2015) 4385}
  [\href{https://arxiv.org/abs/1506.00171}{{\ttfamily 1506.00171}}].

\bibitem{Handley:2015fda}
W.~J. Handley, M.~P. Hobson and A.~N. Lasenby, \emph{{PolyChord: nested
  sampling for cosmology}},
  \href{https://doi.org/10.1093/mnrasl/slv047}{\emph{Mon. Not. Roy. Astron.
  Soc.} {\bfseries 450} (2015) L61}
  [\href{https://arxiv.org/abs/1502.01856}{{\ttfamily 1502.01856}}].

\bibitem{Torrado:2020dgo}
J.~Torrado and A.~Lewis, \emph{{Cobaya: Code for Bayesian Analysis of
  hierarchical physical models}},
  \href{https://doi.org/10.1088/1475-7516/2021/05/057}{\emph{JCAP} {\bfseries
  05} (2021) 057} [\href{https://arxiv.org/abs/2005.05290}{{\ttfamily
  2005.05290}}].

\bibitem{Lewis:2019xzd}
A.~Lewis, \emph{{GetDist: a Python package for analysing Monte Carlo samples}},
   \href{https://arxiv.org/abs/1910.13970}{{\ttfamily 1910.13970}}.

\bibitem{Romano:2016dpx}
J.~D. Romano and N.~J. Cornish, \emph{{Detection methods for stochastic
  gravitational-wave backgrounds: a unified treatment}},
  \href{https://doi.org/10.1007/s41114-017-0004-1}{\emph{Living Rev. Rel.}
  {\bfseries 20} (2017) 2} [\href{https://arxiv.org/abs/1608.06889}{{\ttfamily
  1608.06889}}].

\bibitem{Boileau:2021gbr}
G.~Boileau, A.~C. Jenkins, M.~Sakellariadou, R.~Meyer and N.~Christensen,
  \emph{{Ability of LISA to detect a gravitational-wave background of
  cosmological origin: The cosmic string case}},
  \href{https://doi.org/10.1103/PhysRevD.105.023510}{\emph{Phys. Rev. D}
  {\bfseries 105} (2022) 023510}
  [\href{https://arxiv.org/abs/2109.06552}{{\ttfamily 2109.06552}}].

\bibitem{Boileau:2021sni}
G.~Boileau, A.~Lamberts, N.~J. Cornish and R.~Meyer, \emph{{Spectral separation
  of the stochastic gravitational-wave background for LISA in the context of a
  modulated Galactic foreground}},
  \href{https://doi.org/10.1093/mnras/stab2575}{\emph{Mon. Not. Roy. Astron.
  Soc.} {\bfseries 508} (2021) 803}
  [\href{https://arxiv.org/abs/2105.04283}{{\ttfamily 2105.04283}}].

\bibitem{Boileau:2022ter}
G.~Boileau, N.~Christensen, C.~Gowling, M.~Hindmarsh and R.~Meyer,
  \emph{{Prospects for LISA to detect a gravitational-wave background from
  first order phase transitions}},
  \href{https://doi.org/10.1088/1475-7516/2023/02/056}{\emph{JCAP} {\bfseries
  02} (2023) 056} [\href{https://arxiv.org/abs/2209.13277}{{\ttfamily
  2209.13277}}].

\bibitem{fgivenx}
W.~Handley, \emph{fgivenx: Functional posterior plotter},
  \href{https://doi.org/10.21105/joss.00849}{\emph{The Journal of Open Source
  Software} {\bfseries 3} (2018) }.

\bibitem{Handley:2019fll}
W.~J. Handley, A.~N. Lasenby, H.~V. Peiris and M.~P. Hobson, \emph{{Bayesian
  inflationary reconstructions from Planck 2018 data}},
  \href{https://doi.org/10.1103/PhysRevD.100.103511}{\emph{Phys. Rev. D}
  {\bfseries 100} (2019) 103511}
  [\href{https://arxiv.org/abs/1908.00906}{{\ttfamily 1908.00906}}].

\bibitem{Allen:1996gp}
B.~Allen and A.~C. Ottewill, \emph{{Detection of anisotropies in the
  gravitational wave stochastic background}},
  \href{https://doi.org/10.1103/PhysRevD.56.545}{\emph{Phys. Rev. D} {\bfseries
  56} (1997) 545} [\href{https://arxiv.org/abs/gr-qc/9607068}{{\ttfamily
  gr-qc/9607068}}].

\bibitem{Cornish:2003tz}
N.~J. Cornish and R.~W. Hellings, \emph{{The Effects of orbital motion on LISA
  time delay interferometry}},
  \href{https://doi.org/10.1088/0264-9381/20/22/009}{\emph{Class. Quant. Grav.}
  {\bfseries 20} (2003) 4851}
  [\href{https://arxiv.org/abs/gr-qc/0306096}{{\ttfamily gr-qc/0306096}}].

\bibitem{Caprini:2015zlo}
C.~Caprini et~al., \emph{{Science with the space-based interferometer eLISA.
  II: Gravitational waves from cosmological phase transitions}},
  \href{https://doi.org/10.1088/1475-7516/2016/04/001}{\emph{JCAP} {\bfseries
  04} (2016) 001} [\href{https://arxiv.org/abs/1512.06239}{{\ttfamily
  1512.06239}}].

\bibitem{Caprini:2019egz}
C.~Caprini et~al., \emph{{Detecting gravitational waves from cosmological phase
  transitions with LISA: an update}},
  \href{https://doi.org/10.1088/1475-7516/2020/03/024}{\emph{JCAP} {\bfseries
  03} (2020) 024} [\href{https://arxiv.org/abs/1910.13125}{{\ttfamily
  1910.13125}}].

\bibitem{Planck:2018vyg}
{\scshape Planck} collaboration, \emph{{Planck 2018 results. VI. Cosmological
  parameters}},
  \href{https://doi.org/10.1051/0004-6361/201833910}{\emph{Astron. Astrophys.}
  {\bfseries 641} (2020) A6}
  [\href{https://arxiv.org/abs/1807.06209}{{\ttfamily 1807.06209}}].

\bibitem{Carr:2020gox}
B.~Carr, K.~Kohri, Y.~Sendouda and J.~Yokoyama, \emph{{Constraints on
  primordial black holes}},
  \href{https://doi.org/10.1088/1361-6633/ac1e31}{\emph{Rept. Prog. Phys.}
  {\bfseries 84} (2021) 116902}
  [\href{https://arxiv.org/abs/2002.12778}{{\ttfamily 2002.12778}}].

\bibitem{Yuan:2021qgz}
C.~Yuan and Q.-G. Huang, \emph{{A topic review on probing primordial black hole
  dark matter with scalar induced gravitational waves}},
  \href{https://doi.org/10.1016/j.isci.2021.102860}{\emph{iScience} {\bfseries
  24} (2021) 102860} [\href{https://arxiv.org/abs/2103.04739}{{\ttfamily
  2103.04739}}].

\bibitem{Ozsoy:2023ryl}
O.~\"Ozsoy and G.~Tasinato, \emph{{Inflation and Primordial Black Holes}},
  \href{https://doi.org/10.3390/universe9050203}{\emph{Universe} {\bfseries 9}
  (2023) 203} [\href{https://arxiv.org/abs/2301.03600}{{\ttfamily
  2301.03600}}].

\bibitem{Gow:2020bzo}
A.~D. Gow, C.~T. Byrnes, P.~S. Cole and S.~Young, \emph{{The power spectrum on
  small scales: Robust constraints and comparing PBH methodologies}},
  \href{https://doi.org/10.1088/1475-7516/2021/02/002}{\emph{JCAP} {\bfseries
  02} (2021) 002} [\href{https://arxiv.org/abs/2008.03289}{{\ttfamily
  2008.03289}}].

\bibitem{Musco:2023dak}
I.~Musco, K.~Jedamzik and S.~Young, \emph{{Primordial black hole formation
  during the QCD phase transition: Threshold, mass distribution, and
  abundance}}, \href{https://doi.org/10.1103/PhysRevD.109.083506}{\emph{Phys.
  Rev. D} {\bfseries 109} (2024) 083506}
  [\href{https://arxiv.org/abs/2303.07980}{{\ttfamily 2303.07980}}].

\bibitem{Young:2019yug}
S.~Young, I.~Musco and C.~T. Byrnes, \emph{{Primordial black hole formation and
  abundance: contribution from the non-linear relation between the density and
  curvature perturbation}},
  \href{https://doi.org/10.1088/1475-7516/2019/11/012}{\emph{JCAP} {\bfseries
  11} (2019) 012} [\href{https://arxiv.org/abs/1904.00984}{{\ttfamily
  1904.00984}}].

\bibitem{DeLuca:2019qsy}
V.~De~Luca, G.~Franciolini, A.~Kehagias, M.~Peloso, A.~Riotto and C.~\"Unal,
  \emph{{The Ineludible non-Gaussianity of the Primordial Black Hole
  Abundance}}, \href{https://doi.org/10.1088/1475-7516/2019/07/048}{\emph{JCAP}
  {\bfseries 07} (2019) 048}
  [\href{https://arxiv.org/abs/1904.00970}{{\ttfamily 1904.00970}}].

\bibitem{Ferrante:2022mui}
G.~Ferrante, G.~Franciolini, A.~Iovino, Junior. and A.~Urbano,
  \emph{{Primordial non-Gaussianity up to all orders: Theoretical aspects and
  implications for primordial black hole models}},
  \href{https://doi.org/10.1103/PhysRevD.107.043520}{\emph{Phys. Rev. D}
  {\bfseries 107} (2023) 043520}
  [\href{https://arxiv.org/abs/2211.01728}{{\ttfamily 2211.01728}}].

\bibitem{Gow:2022jfb}
A.~D. Gow, H.~Assadullahi, J.~H.~P. Jackson, K.~Koyama, V.~Vennin and D.~Wands,
  \emph{{Non-perturbative non-Gaussianity and primordial black holes}},
  \href{https://doi.org/10.1209/0295-5075/acd417}{\emph{EPL} {\bfseries 142}
  (2023) 49001} [\href{https://arxiv.org/abs/2211.08348}{{\ttfamily
  2211.08348}}].

\bibitem{LISACosmologyWorkingGroup:2023njw}
{\scshape LISA Cosmology Working Group} collaboration, \emph{{Primordial black
  holes and their gravitational-wave signatures}},
  \href{https://arxiv.org/abs/2310.19857}{{\ttfamily 2310.19857}}.

\bibitem{DeLuca:2023tun}
V.~De~Luca, A.~Kehagias and A.~Riotto, \emph{{How well do we know the
  primordial black hole abundance: The crucial role of nonlinearities when
  approaching the horizon}},
  \href{https://doi.org/10.1103/PhysRevD.108.063531}{\emph{Phys. Rev. D}
  {\bfseries 108} (2023) 063531}
  [\href{https://arxiv.org/abs/2307.13633}{{\ttfamily 2307.13633}}].

\bibitem{BICEP:2021xfz}
{\scshape BICEP, Keck} collaboration, \emph{{Improved Constraints on Primordial
  Gravitational Waves using Planck, WMAP, and BICEP/Keck Observations through
  the 2018 Observing Season}},
  \href{https://doi.org/10.1103/PhysRevLett.127.151301}{\emph{Phys. Rev. Lett.}
  {\bfseries 127} (2021) 151301}
  [\href{https://arxiv.org/abs/2110.00483}{{\ttfamily 2110.00483}}].

\bibitem{Galloni:2022mok}
G.~Galloni, N.~Bartolo, S.~Matarrese, M.~Migliaccio, A.~Ricciardone and
  N.~Vittorio, \emph{{Updated constraints on amplitude and tilt of the tensor
  primordial spectrum}},
  \href{https://doi.org/10.1088/1475-7516/2023/04/062}{\emph{JCAP} {\bfseries
  04} (2023) 062} [\href{https://arxiv.org/abs/2208.00188}{{\ttfamily
  2208.00188}}].

\bibitem{Paoletti:2022anb}
D.~Paoletti, F.~Finelli, J.~Valiviita and M.~Hazumi, \emph{{Planck and
  BICEP/Keck Array 2018 constraints on primordial gravitational waves and
  perspectives for future B-mode polarization measurements}},
  \href{https://doi.org/10.1103/PhysRevD.106.083528}{\emph{Phys. Rev. D}
  {\bfseries 106} (2022) 083528}
  [\href{https://arxiv.org/abs/2208.10482}{{\ttfamily 2208.10482}}].

\bibitem{Babak:2021mhe}
S.~Babak, A.~Petiteau and M.~Hewitson, \emph{{LISA Sensitivity and SNR
  Calculations}},  \href{https://arxiv.org/abs/2108.01167}{{\ttfamily
  2108.01167}}.

\bibitem{Domenech:2021ztg}
G.~Dom\`enech, \emph{{Scalar Induced Gravitational Waves Review}},
  \href{https://doi.org/10.3390/universe7110398}{\emph{Universe} {\bfseries 7}
  (2021) 398} [\href{https://arxiv.org/abs/2109.01398}{{\ttfamily
  2109.01398}}].

\bibitem{Bartolo:2018evs}
N.~Bartolo, V.~De~Luca, G.~Franciolini, A.~Lewis, M.~Peloso and A.~Riotto,
  \emph{{Primordial Black Hole Dark Matter: LISA Serendipity}},
  \href{https://doi.org/10.1103/PhysRevLett.122.211301}{\emph{Phys. Rev. Lett.}
  {\bfseries 122} (2019) 211301}
  [\href{https://arxiv.org/abs/1810.12218}{{\ttfamily 1810.12218}}].

\bibitem{Caprini:2009fx}
C.~Caprini, R.~Durrer, T.~Konstandin and G.~Servant, \emph{{General Properties
  of the Gravitational Wave Spectrum from Phase Transitions}},
  \href{https://doi.org/10.1103/PhysRevD.79.083519}{\emph{Phys. Rev. D}
  {\bfseries 79} (2009) 083519}
  [\href{https://arxiv.org/abs/0901.1661}{{\ttfamily 0901.1661}}].

\bibitem{Cai:2019cdl}
R.-G. Cai, S.~Pi and M.~Sasaki, \emph{{Universal infrared scaling of
  gravitational wave background spectra}},
  \href{https://doi.org/10.1103/PhysRevD.102.083528}{\emph{Phys. Rev. D}
  {\bfseries 102} (2020) 083528}
  [\href{https://arxiv.org/abs/1909.13728}{{\ttfamily 1909.13728}}].

\bibitem{Hook:2020phx}
A.~Hook, G.~Marques-Tavares and D.~Racco, \emph{{Causal gravitational waves as
  a probe of free streaming particles and the expansion of the Universe}},
  \href{https://doi.org/10.1007/JHEP02(2021)117}{\emph{JHEP} {\bfseries 02}
  (2021) 117} [\href{https://arxiv.org/abs/2010.03568}{{\ttfamily
  2010.03568}}].

\bibitem{Pi:2020otn}
S.~Pi and M.~Sasaki, \emph{{Gravitational Waves Induced by Scalar Perturbations
  with a Lognormal Peak}},
  \href{https://doi.org/10.1088/1475-7516/2020/09/037}{\emph{JCAP} {\bfseries
  09} (2020) 037} [\href{https://arxiv.org/abs/2005.12306}{{\ttfamily
  2005.12306}}].

\bibitem{Yuan:2019wwo}
C.~Yuan, Z.-C. Chen and Q.-G. Huang, \emph{{Log-dependent slope of scalar
  induced gravitational waves in the infrared regions}},
  \href{https://doi.org/10.1103/PhysRevD.101.043019}{\emph{Phys. Rev. D}
  {\bfseries 101} (2020) 043019}
  [\href{https://arxiv.org/abs/1910.09099}{{\ttfamily 1910.09099}}].

\bibitem{Adshead:2021hnm}
P.~Adshead, K.~D. Lozanov and Z.~J. Weiner, \emph{{Non-Gaussianity and the
  induced gravitational wave background}},
  \href{https://doi.org/10.1088/1475-7516/2021/10/080}{\emph{JCAP} {\bfseries
  10} (2021) 080} [\href{https://arxiv.org/abs/2105.01659}{{\ttfamily
  2105.01659}}].

\bibitem{Atal:2021jyo}
V.~Atal and G.~Dom\`enech, \emph{{Probing non-Gaussianities with the high
  frequency tail of induced gravitational waves}},
  \href{https://doi.org/10.1088/1475-7516/2021/06/001}{\emph{JCAP} {\bfseries
  06} (2021) 001} [\href{https://arxiv.org/abs/2103.01056}{{\ttfamily
  2103.01056}}].

\bibitem{Franciolini:2023pbf}
G.~Franciolini, A.~Iovino, Junior., V.~Vaskonen and H.~Veermae, \emph{{Recent
  Gravitational Wave Observation by Pulsar Timing Arrays and Primordial Black
  Holes: The Importance of Non-Gaussianities}},
  \href{https://doi.org/10.1103/PhysRevLett.131.201401}{\emph{Phys. Rev. Lett.}
  {\bfseries 131} (2023) 201401}
  [\href{https://arxiv.org/abs/2306.17149}{{\ttfamily 2306.17149}}].

\bibitem{Yuan:2023ofl}
C.~Yuan, D.-S. Meng and Q.-G. Huang, \emph{{Full analysis of the scalar-induced
  gravitational waves for the curvature perturbation with local-type
  non-Gaussianities}},
  \href{https://doi.org/10.1088/1475-7516/2023/12/036}{\emph{JCAP} {\bfseries
  12} (2023) 036} [\href{https://arxiv.org/abs/2308.07155}{{\ttfamily
  2308.07155}}].

\bibitem{Xie:2024cwp}
T.~Xie, D.~Zhang, J.~Jiang, J.-R. Li, B.~Wang and Y.-F. Cai, \emph{{Narrow-band
  parametrization for the stochastic gravitational wave background}},
  \href{https://doi.org/10.1103/PhysRevD.109.083529}{\emph{Phys. Rev. D}
  {\bfseries 109} (2024) 083529}
  [\href{https://arxiv.org/abs/2402.02415}{{\ttfamily 2402.02415}}].

\bibitem{Ozsoy:2019lyy}
O.~\"Ozsoy and G.~Tasinato, \emph{{On the slope of the curvature power spectrum
  in non-attractor inflation}},
  \href{https://doi.org/10.1088/1475-7516/2020/04/048}{\emph{JCAP} {\bfseries
  04} (2020) 048} [\href{https://arxiv.org/abs/1912.01061}{{\ttfamily
  1912.01061}}].

\end{thebibliography}\endgroup

\end{document}